\documentclass[twocolumn,showkeys,floatfix,prd,10pt,aps]{revtex4-2}
\usepackage{graphicx,epstopdf}
\usepackage[caption=false]{subfig}
\usepackage{appendix}
\usepackage[T1]{fontenc}
\usepackage{lmodern}
\usepackage[dvipsnames,x11names]{xcolor}
\usepackage[colorlinks=true,linkcolor=Magenta!80!black,citecolor=Magenta!70!black,urlcolor=Magenta!80!black]{hyperref}
\usepackage[sort&compress]{natbib}
\usepackage{lipsum}
\usepackage{morefloats}
\usepackage[pdf]{pstricks}
\usepackage{amsmath,amsthm}
\usepackage{amssymb}
\usepackage{amsfonts}
\usepackage{rotating}
\usepackage{cancel}
\usepackage{mathtools}
\usepackage{bbm}
\usepackage{dsfont}
\usepackage{bbold}
\usepackage{multirow}
\usepackage{ulem}
\usepackage{physics}
\usepackage{orcidlink}
\usepackage{colortbl}

\def\pslash{p\!\!\!\slash }
\def\qslash{q\!\!\!\slash }
\def\xslash{x\!\!\!\slash }
\def\eslash{\varepsilon\!\!\!\slash }

\begin{document}

\title{Magnetic dipole moments as probes of doubly-bottom molecular pentaquarks}

\author{Ula\c{s} \"{O}zdem\orcidlink{0000-0002-1907-2894}}%
\email[]{ulasozdem@aydin.edu.tr}
\affiliation{Health Services Vocational School of Higher Education, Istanbul Aydin University, Sefakoy-Kucukcekmece, 34295 Istanbul, T\"{u}rkiye}

%\date{\today}
\begin{abstract}
We investigate the magnetic dipole moments of doubly-bottom pentaquark
states with spin-parities $J^P=\tfrac{1}{2}^-$ and $\tfrac{3}{2}^-$,
interpreted as hadronic molecules in the $B\Sigma_b$,
$B\Sigma_b^{*}$, and $B^{*}\Sigma_b$ configurations. The analysis is
performed within the framework of QCD light-cone sum rules employing
photon distribution amplitudes. Our results demonstrate a strong
sensitivity of the magnetic dipole moments to the internal spin
structure and quark composition of the states. In particular, the
light-quark sector provides the dominant contribution in the
$B\Sigma_b$ configuration, while the magnetic moment of
$B\Sigma_b^{*}$ is largely governed by the heavy bottom quark. For the
$B^{*}\Sigma_b$ molecular state, both light and heavy sectors
contribute constructively, leading to a significantly enhanced
magnetic dipole moment. These findings indicate that magnetic dipole
moments constitute a sensitive probe of the internal structure of
molecular-type doubly-bottom pentaquarks and provide testable
predictions for future experimental studies.
\end{abstract}

%\keywords{Electromagnetic form factors, doubly-bottom pentaquarks, molecular states, QCD light-cone sum rules}

\maketitle

\section{motivation}\label{motivation}

The conventional quark model, which categorizes hadrons as mesons (quark-antiquark pairs) or baryons (three-quark states), is insufficient to explain the existence of exotic hadrons, such as compact multiquark configurations and hadronic molecules. These unconventional states, lying beyond the standard framework, have become a major focus of modern hadron physics. Over the past three decades, both experimental discoveries and theoretical advances have significantly intensified research efforts in this area.  In particular, several hidden-charm pentaquark states have been discovered, including the $\mathrm{P_{c}(4380)}$, $\mathrm{P_{c}(4440)}$, $\mathrm{P_{c}(4457)}$, $\mathrm{P_{c}(4312)}$, $\mathrm{P_{cs}(4459)}$, and $\mathrm{P_{cs}(4338)}$~\cite{Aaij:2015tga, Aaij:2019vzc, LHCb:2021chn, Collaboration:2022boa}. Since these states contain at least five quarks (such as $c\bar c udd$ or $c\bar c uds$), they are strong candidates for being classified as hidden-charm pentaquarks. In 2021, the LHCb Collaboration further reported the observation of the first doubly-charm tetraquark, $T_{cc}^+(3875)$, in the $D^0 D^0 \pi^+$ invariant mass spectrum~\cite{LHCb:2021vvq, LHCb:2021auc}. The simplest quark model interpretation of this state is a $cc\bar u\bar d$ configuration with quantum numbers $\rm{J^P} = 1^+$. Following these experimental breakthroughs, a variety of theoretical models have been proposed to explain the structure and properties of these exotic states. These include interpretations as compact multiquark states, hadronic molecules, or even manifestations of kinematical effects~\cite{Esposito:2014rxa, Esposito:2016noz, Olsen:2017bmm, Lebed:2016hpi, Nielsen:2009uh, Brambilla:2019esw, Agaev:2020zad, Chen:2016qju, Ali:2017jda, Guo:2017jvc, Liu:2019zoy, Yang:2020atz, Dong:2021juy, Meng:2022ozq, Chen:2022asf, Dong:2021bvy}. Nevertheless, despite significant progress, the internal substructures and the precise quantum numbers of these multiquark states continue to pose fundamental challenges in high-energy hadron physics.

The discovery of the $T_{cc}^+(3875)$ tetraquark state opens a novel avenue for the search for states beyond conventional hadrons. If $T_{cc}^+(3875)$ is indeed a genuine doubly-charm tetraquark, then by analogy, one may expect the existence of doubly-bottom pentaquark states as well. This reasoning mirrors the approach taken to predict hidden-charm pentaquarks based on the discovery of hidden-charm tetraquarks. In the absence of experimental evidence for doubly-bottom pentaquarks, it is imperative to understand the underlying reasons behind their apparent non-observation. To this end, several theoretical investigations have been carried out to explore the possible existence and properties of doubly-charm/bottom pentaquark states~\cite{Chen:2017vai, Liu:2020nil, Chen:2021kad, Chen:2021htr, Dong:2021bvy, Shen:2022zvd, Liu:2023clr, Yang:2020twg, Shimizu:2017xrg, Guo:2017vcf, Zhou:2018bkn, Wang:2018lhz, Xing:2021yid, Duan:2024uuf, Wang:2024brl, Ozdem:2022vip, Zhou:2022gra}. These studies generally suggest that the doubly-bottom pentaquarks lie below the relevant meson–baryon thresholds, making the formation of bound states feasible. Compared to doubly-charm systems, doubly-bottom pentaquarks are anticipated to be heavier. However, due to the complex dynamics of multiquark systems, their actual mass may be reduced, making it difficult to distinguish them from conventional baryons on the basis of mass alone. Therefore, studying other characteristics—particularly their electromagnetic properties—can offer essential insights into their underlying structures.

The electromagnetic multipole moments are measurable quantities that can provide crucial insights into the internal quark-gluon structure and dynamics of hadronic states. Specifically, in photo- and electro-production processes, the magnetic dipole moments (MDMs) of hadrons play an essential role in shaping both the total and differential cross sections. As such, determining the MDMs of hadronic states is vital for understanding their internal configuration.  Although the electromagnetic properties of hadrons have received less attention compared to their mass spectra, production mechanisms, quantum numbers, or decay channels, recent years have seen growing interest in these observables. This is especially true for exotic states with ambiguous internal compositions, where electromagnetic characteristics can serve as critical probes~\cite{Wang:2016dzu, Ozdem:2018qeh, Ortiz-Pacheco:2018ccl, Xu:2020flp, Ozdem:2021btf, Ozdem:2021ugy, Li:2021ryu, Ozdem:2023htj, Wang:2023iox, Ozdem:2022kei, Gao:2021hmv, Ozdem:2024rch, Guo:2023fih, Ozdem:2022iqk, Wang:2022nqs, Wang:2022tib, Ozdem:2024jty, Li:2024wxr, Li:2024jlq, Ozdem:2024yel, Ozdem:2024rqx, Mutuk:2024ltc, Mutuk:2024jxf, Ozdem:2024suc, Ozdem:2025fks, Zhu:2025abk, Ozdem:2025ncd, Ozdem:2025jda, Ozdem:2025ion, Ozdem:2026gmn}.  In this work, we use the QCD light-cone sum rule (LCSR) method to compute the MDMs of the $B^{(*)} \Sigma_b^{(*)}$ states, considering them as molecular configurations: $B \Sigma_b$, $B^* \Sigma_b$, and $B \Sigma_b^*$, with quantum numbers $\mathrm{I(J^P)= \frac{1}{2}(\frac{1}{2}^-)}$, $\mathrm{I(J^P)= \frac{1}{2}(\frac{3}{2}^-)}$, and $\mathrm{I(J^P)= \frac{1}{2}(\frac{3}{2}^-)}$, respectively. 
Among the available tools for calculating non-perturbative QCD quantities, LCSR stands out for its efficiency and reliability~\cite{Chernyak:1990ag, Braun:1988qv, Balitsky:1989ry}. Within this framework, the correlation function is calculated from two complementary perspectives: the hadronic representation and the QCD (operator product expansion) side. By matching these two sides using the quark-hadron duality principle, one derives sum rules for the MDMs at the hadronic level.
 
The structure of this paper is as follows. Following the introduction, Section \ref{formalism} develops the theoretical formalism for MDMs of the \( B \Sigma_b \), \( B^{*} \Sigma_b \), and \( B \Sigma_b^{*} \) pentaquark states. Section \ref{numerical} then provides the numerical analysis and results. Finally, Section \ref{summary} concludes with a summary of the main findings.

\begin{widetext}
 
\section{formalism}\label{formalism}

The QCD light-cone sum rules for the MDMs of the spin-\(\frac{1}{2}\) and spin-\(\frac{3}{2}\) doubly-bottom pentaquarks, \(\mathrm{P_{bb}}\) (\(J^P = \frac{1}{2}^-\)) and \(\mathrm{P_{bb}^*}\) (\(J^P = \frac{3}{2}^-\)) respectively, can be derived from the study of correlation functions evaluated in the presence of a weak external electromagnetic field, denoted by \(\gamma\),
 \begin{align} \label{edmn01}
\Pi(p,q)&=i\int d^4x e^{ip \cdot x} \langle0|T\left\{\rm{J}^{\mathrm{P_{bb}}}(x)\bar{\rm{J}}^{\mathrm{P_{bb}}}(0)\right\}|0\rangle _\gamma \, , \\
%\end{align}
%\begin{align} 
\Pi_{\mu\nu}(p,q)&=i\int d^4x e^{ip \cdot x} \langle0|T\left\{\rm{J}_\mu^{\mathrm{P_{bb}^*}}(x)\bar{\rm{J}}_\nu^{\mathrm{P_{bb}^*}}(0)\right\}|0\rangle _\gamma \,, \label{Pc101}
\end{align} 
where \(\mathrm{J^{P_{bb}}}(x)\) and \(\mathrm{J_\mu^{P^*_{bb}}}(x)\) denote the interpolating currents associated with the respective states. The explicit forms of these currents, which are essential for our analysis within the QCD light-cone sum rule framework, are given by the following expressions~\cite{Wang:2024brl}:
\begin{align}
\label{curpcs2}
\rm{J}^{ B \Sigma_b}(x)& =\frac{1}{\sqrt{3}}\mid  \bar B^0(x) \Sigma_b^0(x) \rangle \, - \sqrt{\frac{2}{3}}\mid B^-(x) \Sigma_b^{+}(x) \rangle ,  \\
%\end{align}
%  \begin{align}
%%%%%%%%%%%%%%%%%%%%%%%%%%%%%%%%%%%%%%%%%%%%%%%%%%%%%%%%%%%%%%%%%%%%%%%%%%%%%%%%%%%%
\rm{J}_{\mu}^{ B \Sigma_b^*}(x)&=\frac{1}{\sqrt{3}}\mid  \bar B^0(x) \Sigma_b^{*0}(x) \rangle \, - \sqrt{\frac{2}{3}}\mid  B^-(x) \Sigma_b^{*+}(x) \rangle , \\
%  \end{align}
%  \begin{align}
%%%%%%%%%%%%%%%%%%%%%%%%%%%%%%%%%%%%%%%%%%%%%%%%%%%%%%%%%%%%%%%%%%%%%%%%%%%%%%%%%%%%%%%%%%%%%%%%%%%%%%%%%%%%%%%%%%%%%%%%%%%%%%%%%%%%%%%%%%%%%%%%%%%%%%%%%%%%%%%%%%%%%%%%
\rm{J}_{\mu}^{ B^* \Sigma_b}(x)& =\frac{1}{\sqrt{3}}\mid  \bar B^{*0}(x) \Sigma_b^0(x) \rangle \, - \sqrt{\frac{2}{3}}\mid  B^{*-}(x) \Sigma_b^{+}(x) \rangle, \label{curpcs3}
  \end{align}
where
  \begin{align}
  & B^-(x)= \bar u^d(x) i \gamma_5 b^d(x), \\
  & \bar B^0(x)= - \bar d^d(x) i \gamma_5 b^d(x), \\
  & B^{*-}(x)= \bar u^d(x)  \gamma_\mu b^d(x), \\
  & \bar B^{*0}(x)= - \bar d^d(x)  \gamma_\mu b^d(x), \\
  & \Sigma_b^0(x)= \varepsilon^{abc} u^{a^T}(x)C \gamma_\mu d^b(x) \gamma^\mu \gamma_5 b^c(x)\\
  & \Sigma_b^+(x)= \varepsilon^{abc} u^{a^T}(x)C \gamma_\mu u^b(x) \gamma^\mu \gamma_5 b^c(x)\\
  & \Sigma_b^{*0}(x)= \varepsilon^{abc} u^{a^T}(x)C \gamma_\mu d^b(x)  b^c(x)\\
  & \Sigma_b^{*+}(x)= \varepsilon^{abc} u^{a^T}(x)C \gamma_\mu u^b(x)  b^c(x).
  \end{align}

Combining the spin of the open-bottom meson with that of the 
singly-bottom baryon, the $S$-wave $B^{(*)}\Sigma_b^{(*)}$ systems 
accommodate, with negative parity throughout, the spin assignments 
$B\Sigma_b \to J^P=\tfrac{1}{2}^-$, 
$B\Sigma_b^{*}\to \tfrac{3}{2}^-$, 
$B^{*}\Sigma_b \to \tfrac{1}{2}^-\oplus\tfrac{3}{2}^-$, and 
$B^{*}\Sigma_b^{*}\to 
\tfrac{1}{2}^-\oplus\tfrac{3}{2}^-\oplus\tfrac{5}{2}^-$. 
In the present work we analyze the $B\Sigma_b\,(\tfrac{1}{2}^-)$, 
$B\Sigma_b^{*}\,(\tfrac{3}{2}^-)$, and 
$B^{*}\Sigma_b\,(\tfrac{3}{2}^-)$ states, while the 
spin-$\tfrac{1}{2}$ partner of $B^{*}\Sigma_b$ and the entire 
$B^{*}\Sigma_b^{*}$ multiplet are left aside. This restriction is 
imposed by the technical considerations detailed below and reflects 
no physical preference among the states. 
Within the light-cone sum rule framework, the mass and residue of 
the state under study enter the magnetic dipole moment as external, 
two-point inputs [cf. Eq.~(\ref{edmn15})]. For the spin-$\tfrac{1}{2}$ 
member of $B^{*}\Sigma_b$ and for the members of the 
$B^{*}\Sigma_b^{*}$ family, these spectroscopic parameters have not 
been determined in the study whose mass and residue predictions we 
adopt~\cite{Wang:2024brl}. In the absence of these inputs the 
corresponding sum rules cannot be evaluated. Once such masses and 
residues become available, the electromagnetic multipole moments of 
these states can be obtained following exactly the procedure 
presented here. 
For the highest-spin member, $B^{*}\Sigma_b^{*}$ with 
$J^P=\tfrac{5}{2}^-$, a further difficulty arises that is specific to 
the electromagnetic analysis and has no counterpart in the mass 
calculation. To the best of our knowledge, a consistent 
electromagnetic vertex and multipole decomposition for a 
spin-$\tfrac{5}{2}$ interpolating current has not been established. Moreover, a spin-$\tfrac{5}{2}$ current couples 
not only to the spin-$\tfrac{5}{2}$ ground state but also to 
lower-lying spin-$\tfrac{1}{2}$ and spin-$\tfrac{3}{2}$ states; 
identifying the appropriate independent Lorentz structures at the 
photon vertex and systematically removing these lower-spin 
admixtures --- a step far more demanding than in the mass sum rule 
--- constitutes a separate and nontrivial challenge that would 
compromise the reliability of the extracted moments. This obstacle 
is independent of the availability of spectroscopic parameters, so 
that the spin-$\tfrac{5}{2}$ case remains beyond the scope of the 
present study in any event. 
Finally, we emphasize that these omissions do not affect the central 
observation that the magnetic dipole moments are strongly sensitive 
to the internal spin structure of the states. The three 
configurations analyzed here already couple the constituent spins in 
distinct ways and yield markedly different magnetic moments --- 
including a change of sign between the $B\Sigma_b^{*}$ and 
$B^{*}\Sigma_b$ states --- which in itself establishes the pronounced 
dependence of the moment on the spin-flavor arrangement. The 
inclusion of the omitted states, once feasible, would supply 
additional data points and can only enrich this pattern rather than 
modify the qualitative conclusion.

Following the QCD light-cone sum rule approach, the correlation function is initially represented in terms of the hadron’s physical parameters. To achieve this, a complete set of hadronic states sharing the same quantum numbers as the interpolating currents \(\mathrm{J^{P_{bb}}}(x)\) and \(\mathrm{J_\mu^{P^*_{bb}}}(x)\) is inserted into the correlation function. Using the dispersion relation, the correlation functions can then be expressed as:
 \begin{align}\label{edmn02}
\Pi^{Had}(p,q)&=\frac{\langle0\mid \rm{J}^{\mathrm{P_{bb}}}(x) \mid
{\mathrm{P_{bb}}}(p, s) \rangle}{[p^{2}-m_{\mathrm{P_{bb}}}^{2}]}
\langle {\mathrm{P_{bb}}}(p, s)\mid
{\mathrm{P_{bb}}}(p+q, s)\rangle_\gamma 
\frac{\langle {\mathrm{P_{bb}}}(p+q, s)\mid
\bar{ \rm{J}}^{\mathrm{P_{bb}}}(0) \mid 0\rangle}{[(p+q)^{2}-m_{\mathrm{P_{bb}}}^{2}]}+ \cdots , \\
%\end{align}
%\nonumber\\ 
%\begin{align}
\Pi^{Had}_{\mu\nu}(p,q)&=\frac{\langle0\mid  \rm{J}_{\mu}^{\mathrm{P_{bb}^*}}(x)\mid
{\mathrm{P_{bb}^*}}(p,s)\rangle}{[p^{2}-m_{{\mathrm{P_{bb}^*}}}^{2}]}
\langle {\mathrm{P_{bb}^*}}(p,s)\mid
{\mathrm{P_{bb}^*}}(p+q,s)\rangle_\gamma 
\frac{\langle {\mathrm{P_{bb}^*}}(p+q,s)\mid
\bar{\rm{J}}_{\nu}^{\mathrm{P_{bb}^*}}(0)\mid 0\rangle}{[(p+q)^{2}-m_{{\mathrm{P_{bb}^*}}}^{2}]}+ \cdots .\label{Pc103}
\end{align}

For the continuation of the calculations, the matrix elements introduced in Eqs.~(\ref{edmn02}) and (\ref{Pc103}) are required. These can be represented in terms of hadronic parameters as follows \cite{Leinweber:1990dv, Weber:1978dh, Nozawa:1990gt}:
%
%\begin{widetext}
%\begin{align}
\begin{align}
\label{edmn005}
\langle0\mid \rm{J}^{\mathrm{P_{bb}}}(x)\mid {\mathrm{P_{bb}}}(p, s)\rangle=&\lambda_{\mathrm{P_{bb}}} \gamma_5 \, \nu(p,s),\\%\label{edmn04}
%\\\nonumber\\
\langle {\mathrm{P_{bb}}}(p+q, s)\mid \rm{\bar J}^{\mathrm{P_{bb}}}(0)\mid 0\rangle=&\lambda_{\mathrm{P_{bb}}}  \, \bar \nu(p+q,s)\, \gamma_5 \label{edmn004}\\%\label{edmn004},\\
%\nonumber\\
%%%%%%%%%%%%%%%%%%%%%%%%%%%%%%%%%%
%\end{align}
%\begin{align}
\langle0\mid \rm{J}_{\mu}^{\mathrm{P_{bb}^*}}(x)\mid {\mathrm{P_{bb}^*}}(p,s)\rangle&=\lambda_{{\mathrm{P_{bb}^*}}}u_{\mu}(p,s),\\
\langle {\mathrm{P_{bb}^*}}(p+q,s)\mid   \rm{\bar{J}}_{\nu}^{\mathrm{P_{bb}^*}}(0)\mid 0\rangle &= \lambda_{{\mathrm{P_{bb}^*}}}\bar u_{\nu}(p+q,s), \\
%%%%%%%%%%%%%%%%%%%%%%%%%%%%%%%%%%%%%%%%%%%%%%%%%%%%%%%%%%%%%%%%%%%%%%%%%%%%%%%%
\langle {\mathrm{P_{bb}}}(p, s)\mid {\mathrm{P_{bb}}}(p+q, s)\rangle_\gamma &=\varepsilon^\mu\,\bar \nu(p, s)\bigg[\big[f_1(q^2)+f_2(q^2)\big] \gamma_\mu +f_2(q^2)\frac{(2p+q)_\mu}{2 m_{\mathrm{P_{bb}}}}\bigg]\,\nu(p+q, s), \\
%%%%%%%%%%%%%%%%%%%%%%%%%%%%%%%%%%%%%%%%%%%%%%%%%%%%%%
%%%%%%%%%%%%%%%%%%%%%%%%%%%%%%%%%%%%%%%%%%%%%%%%%%%%%%
%%%%%%%%%%%%%%%%%%%%%%%%%%%%%%%%%%%%%%%%%%%%%%%%%%%%%%
\langle {\mathrm{P_{bb}^*}}(p,s)\mid {\mathrm{P_{bb}^*}}(p+q,s)\rangle_\gamma &=-\bar u_{\alpha}(p,s)\bigg[F_{1}(q^2)g^{\alpha \beta }\eslash 
-
\frac{1}{2m_{{\mathrm{P_{bb}^*}}}} 
\Big[F_{2}(q^2)g^{\alpha \beta} \eslash\qslash
+F_{4}(q^2)\frac{q^{\alpha}q^{\beta}\eslash\qslash}{(2m_{{\mathrm{P_{bb}^*}}})^2}\Big]
\nonumber\\
&+
F_{3}(q^2)\frac{1}{(2m_{{\mathrm{P_{bb}^*}}})^2}q^{\alpha}q^{\beta}\eslash \bigg] 
u_{\beta}(p+q,s),
\label{matelpar}
\end{align}
where \(\nu(p+q,s)\) and \(u_{\alpha}(p,s)\) denote the spinors associated with the \(\mathrm{P_{bb}}\) and \(\mathrm{P_{bb}^*}\) states, respectively. The quantities \(\lambda_{\mathrm{P_{bb}}}\) and \(\lambda_{\mathrm{P_{bb}^*}}\) represent the corresponding pole residues. Furthermore, \(f_i (q^2)\)  and \(F_i(q^2)\) stand for the form factors governing the respective transition amplitudes.

Using the relations above, the hadronic representation of the 
correlation functions related to the MDMs can be expressed as 
follows:
\begin{align}
\label{edmn05}
\Pi^{Had}(p,q)=&  \lambda^2_{\mathrm{P_{bb}}} \frac{ \varepsilon^\mu \gamma_5 \big(\pslash+m_{\mathrm{P_{bb}}} \big)}{[p^{2}-m_{{\mathrm{P_{bb}}}}^{2}] [(p+q)^{2}-m_{{\mathrm{P_{bb}}}}^{2}]}  \bigg[\big[f_1(q^2)+f_2(q^2)\big] \gamma_\mu
+f_2(q^2)\, \frac{(2p+q)_\mu}{2 m_{\mathrm{P_{bb}}}}\bigg]   
\big(\pslash+\qslash+m_{\mathrm{P_{bb}}}\big) \gamma_5, \\
\Pi^{Had}_{\mu\nu}(p,q)&=- \lambda_{P_{bb}^*}^{2}\frac{ \big(\pslash+m_{P_{bb}^*}\big)}{[(p+q)^{2}-m_{P_{bb}^*}^{2}][p^{2}-m_{P_{bb}^*}^{2}]}
\bigg[g_{\mu\alpha}
-\frac{1}{3}\gamma_{\mu}\gamma_{\alpha}-\frac{2\,p_{\mu}p_{\alpha}}
{3\,m^{2}_{P_{bb}^*}}+\frac{p_{\mu}\gamma_{\alpha}-p_{\alpha}\gamma_{\mu}}{3\,m_{P_{bb}^*}}\bigg] \nonumber\\
& \times \bigg\{F_{1}(q^2)g^{\alpha\beta}\eslash - 
\frac{1}{2m_{P_{bb}^*}}
\Big[F_{2}(q^2)g^{\alpha\beta}\eslash\qslash + \frac{F_{4}(q^2)}{(2m_{P_{bb}^*})^2}q^{\alpha}q^{\beta}\eslash\qslash\Big]+\frac{F_{3}(q^2)}{(2m_{P_{bb}^*})^2}
q^{\alpha}q^{\beta}\eslash\bigg\}
\nonumber\\
&\times
(\pslash+\qslash+m_{P_{bb}^*})
\bigg[g_{\beta\nu}-\frac{1}{3}\gamma_{\beta}\gamma_{\nu}-\frac{2\,(p+q)_{\beta}(p+q)_{\nu}}
{3\,m^{2}_{P_{bb}^*}}+\frac{(p+q)_{\beta}\gamma_{\nu}-(p+q)_{\nu}\gamma_{\beta}}{3\,m_{P_{bb}^*}}\bigg].
\label{final phenpart}
\end{align}

The expressions in Eqs.~\eqref{edmn05} and~\eqref{final phenpart} 
are obtained after inserting complete sets of hadronic states and 
performing the summation over spins. For the spin-$1/2$ case, the 
Dirac spinor summation formula is used, while for the spin-$3/2$ 
case, the Rarita--Schwinger spinor summation is employed. The 
relevant summation formulas read:
\begin{align}
\label{eq:diracsum}
\sum_{s} v(p,s) \bar{v}(p,s) &= \pslash + m_{\mathrm{P_{bb}}}, \\
\sum_{s} v(p+q,s) \bar{v}(p+q,s) &= \pslash + \qslash + m_{\mathrm{P_{bb}}},
\end{align}
\begin{align}
\sum_s u_{\mu}(p,s)\bar{u}_{\alpha}(p,s) &= -(\pslash+m_{P_{bb}^*})\bigg[g_{\mu\alpha}-\frac{1}{3}\gamma_{\mu}\gamma_{\alpha}-\frac{2\,p_{\mu}p_{\alpha}}{3\,m^{2}_{P_{bb}^*}}+\frac{p_{\mu}\gamma_{\alpha}-p_{\alpha}\gamma_{\mu}}{3\,m_{P_{bb}^*}}\bigg],\\
\sum_s u_{\beta}(p+q,s)\bar{u}_{\nu}(p+q,s) &= -(\pslash+\qslash+m_{P_{bb}^*})\bigg[g_{\beta\nu}-\frac{1}{3}\gamma_{\beta}\gamma_{\nu}-\frac{2\,(p+q)_{\beta}(p+q)_{\nu}}{3\,m^{2}_{P_{bb}^*}}+\frac{(p+q)_{\beta}\gamma_{\nu}-(p+q)_{\nu}\gamma_{\beta}}{3\,m_{P_{bb}^*}}\bigg].
\label{eq:raritasum}
\end{align}

Substituting the polarization sums~\eqref{eq:diracsum}--\eqref{eq:raritasum} 
into Eqs.~\eqref{edmn05} and~\eqref{final phenpart} and applying 
the on-shell condition $p^{2}=m^{2}_{\mathrm{P_{bb}^{(*)}}}$ in the 
spinor numerators generates the expanded forms of the correlation 
functions. The spin-$1/2$ case yields a compact expression with a 
small number of Lorentz structures:
\begin{align}
\label{full_pi}
\Pi^{Had}(p,q)&= \frac{\lambda^2_{\mathrm{P_{bb}}}}{[p^{2}-m_{{\mathrm{P_{bb}}}}^{2}][(p+q)^{2}-m_{{\mathrm{P_{bb}}}}^{2}]} \bigg[ (f_1 (q^2) + f_2(q^2))\Big(-2\,m_{\mathrm{P_{bb}}}^2\,\eslash + 2\,(\varepsilon\cdot p)\,\pslash  - m_{\mathrm{P_{bb}}}\,\eslash\,\pslash- m_{\mathrm{P_{bb}}}\,\eslash\,\qslash \nonumber\\
&+ m_{\mathrm{P_{bb}}}\,\pslash\,\eslash + \pslash\,\eslash\,\qslash\Big)
+f_2(q^2)\,\frac{(\varepsilon\cdot p)\,\big(m_{\mathrm{P_{bb}}}\,\qslash - \pslash\,\qslash\big)}{m_{\mathrm{P_{bb}}}}
  \bigg].
\end{align}
The spin-$3/2$ case, by contrast, generates several hundred 
Lorentz structures organized by the four form factors $F_i(q^2)$, 
which are systematically produced by the contraction of the two 
Rarita--Schwinger projectors with the photon-vertex matrix 
element in Eq.~\eqref{final phenpart}. Representative samples of 
these structures, grouped by the form factor to which they are 
proportional, read:
\begin{align}
\Pi^{Had}_{\mu\nu}(p,q) &= -\frac{\lambda^2_{\mathrm{P_{bb}^*}}}{[(p+q)^2 - m^2_{\mathrm{P_{bb}^*}}][p^2 - m^2_{\mathrm{P_{bb}^*}}]} \nonumber\\
& \times \Bigg\{ F_1(q^2)\bigg[  
\frac{4(\varepsilon\!\cdot\! p)\, p_\mu p_\nu}{3m_{\mathrm{P_{bb}^*}}} 
- \frac{4(\varepsilon\!\cdot\! q)\, p_\mu p_\nu}{3m_{\mathrm{P_{bb}^*}}} 
+ \frac{4(\varepsilon\!\cdot\! p)\, p_\mu q_\nu}{3m_{\mathrm{P_{bb}^*}}} 
- \frac{4(\varepsilon\!\cdot\! q)\, p_\mu q_\nu}{3m_{\mathrm{P_{bb}^*}}} - \frac{2\, p_\mu p_\nu\, q^2\,\eslash}{3m_{\mathrm{P_{bb}^*}}^2}  + \cdots \bigg] \nonumber\\
&\quad + F_2(q^2)\bigg[  
- \frac{2\,\varepsilon_\nu\, p_\mu\, q^2}{3\, m_{\mathrm{P_{bb}^*}}}
- \frac{5\, p_\mu p_\nu\, q^2\,\eslash}{3\, m_{\mathrm{P_{bb}^*}}^2}
+ \frac{2(p\!\cdot\! q)\, p_\mu p_\nu\, q^2\,\eslash}{3\, m_{\mathrm{P_{bb}^*}}^4}
- \frac{p_\nu\, q^2 q_\mu\,\eslash}{m_{\mathrm{P_{bb}^*}}^2}
- \frac{2\, p_\mu\, q^2 q_\nu\,\eslash}{3\, m_{\mathrm{P_{bb}^*}}^2} + \cdots \bigg] \nonumber\\
&\quad + F_3(q^2)\bigg[ - \frac{(\varepsilon\!\cdot\! q)\, p_\mu p_\nu\, q^2}{6\, m_{\mathrm{P_{bb}^*}}^3}
- \frac{(\varepsilon\!\cdot\! q)\, p_\mu\, q^2 q_\nu}{6\, m_{\mathrm{P_{bb}^*}}^3}
+ \frac{p_\mu p_\nu\, q^2}{6\, m_{\mathrm{P_{bb}^*}}^2}\,\eslash
- \frac{p_\mu p_\nu\, (q^2)^2}{6\, m_{\mathrm{P_{bb}^*}}^4}\,\eslash
- \frac{p_\nu\, q^2 q_\mu}{4\, m_{\mathrm{P_{bb}^*}}^2}\,\eslash + \cdots \bigg] \nonumber\\
&\quad + F_4(q^2)\bigg[ 
- \frac{(p\!\cdot\! q)\, p_\mu p_\nu\, q^2}{6\, m_{\mathrm{P_{bb}^*}}^4}\,\eslash
+ \frac{(p\!\cdot\! q)^2\, p_\mu p_\nu\, q^2}{6\, m_{\mathrm{P_{bb}^*}}^6}\,\eslash
- \frac{p_\mu p_\nu\, (q^2)^2}{24\, m_{\mathrm{P_{bb}^*}}^4}\,\eslash
+ \frac{(p\!\cdot\! q)\, p_\mu p_\nu\, (q^2)^2}{6\, m_{\mathrm{P_{bb}^*}}^6}\,\eslash + \cdots \bigg]\Bigg\},
\label{full_pi_munu}
\end{align}
where the ellipses $(\cdots)$ denote additional Lorentz 
structures of the same form, systematically generated by the 
Dirac algebra and processed in the same manner. The complete, 
unabbreviated expression for $\Pi^{Had}_{\mu\nu}(p,q)$, 
organized term by term per form-factor sector $F_i(q^2)$, is 
collected in Appendix~\ref{appc1}.

In principle, Eqs.~\eqref{full_pi} and~\eqref{full_pi_munu} 
provide the hadronic representation of the correlation functions. 
However, before proceeding to the numerical analysis, two 
technical aspects must be addressed. First, not all Lorentz 
structures appearing in the spin-$3/2$ correlation function are 
linearly independent. Second, the interpolating current $J_\mu$ 
for the spin-$3/2$ states also couples to spin-$1/2$ pentaquark 
states, and these unwanted contributions must be eliminated. 
The coupling of $J_\mu$ to spin-$1/2$ states can be parametrized 
through
\begin{align}
\langle 0 | J_{\mu}(0) | B(p, s = 1/2) \rangle = (A p_{\mu} + B \gamma_{\mu}) u(p, s = 1/2), \label{eq:spin12coupling}
\end{align}
where $A$ and $B$ are constants. As evident from 
Eq.~\eqref{eq:spin12coupling}, spin-$1/2$ contamination appears 
in structures proportional to $p_\mu$ and $\gamma_\mu$. To 
eliminate these pollutions and isolate the independent Lorentz 
structures in the spin-$3/2$ case, we adopt the gamma-matrix 
ordering $\gamma_\mu \pslash \eslash \qslash \gamma_\nu$ 
following~\cite{Belyaev:1982cd,Belyaev:1993ss}. Within 
this ordering, terms beginning with $\gamma_\mu$, ending with 
$\gamma_\nu$, or proportional to $p_\mu$ or $p_\nu$ are 
discarded as spin-$1/2$ contamination. For the spin-$1/2$ case, 
no such contamination is present, and the ordering 
$\pslash \eslash \qslash$ is adopted to isolate the MDM 
contribution.

\paragraph*{Reduction of the spin-$1/2$ correlation function.} 
After applying these procedures to Eq.~\eqref{full_pi}, the 
spin-$1/2$ hadronic representation reduces to
\begin{align}
\label{edmn0511}
\Pi^{Had}(p,q)&= \frac{\lambda^2_{\mathrm{P_{bb}}}}{[p^{2}-m_{{\mathrm{P_{bb}}}}^{2}][(p+q)^{2}-m_{{\mathrm{P_{bb}}}}^{2}]} \bigg[ \Big(f_1 (q^2) + f_2(q^2) \Big)\Big(-2\,m_{\mathrm{P_{bb}}}^2\,\eslash + 2\,(\varepsilon\cdot p)\,\pslash  - m_{\mathrm{P_{bb}}}\,\eslash\,\pslash- m_{\mathrm{P_{bb}}}\,\eslash\,\qslash \nonumber\\
&+ m_{\mathrm{P_{bb}}}\,\pslash\,\eslash + \pslash\,\eslash\,\qslash\Big)
+ \cdots
  \bigg].
\end{align}
As an illustrative example of this reduction procedure, consider 
the coefficient of $\gamma_\mu$ in the bracket of 
Eq.~\eqref{full_pi}, multiplied by $\varepsilon^\mu$. The 
relevant Dirac structure is 
$\gamma_5\,(\pslash+m_{\mathrm{P_{bb}}})\,\eslash\,(\pslash+\qslash+m_{\mathrm{P_{bb}}})\,\gamma_5$. 
Using $\{\gamma_5,\gamma_\mu\}=0$ together with 
$\gamma_5^2=\mathbb{1}$, this expression reduces to 
$(\pslash-m_{\mathrm{P_{bb}}})\,\eslash\,(\pslash+\qslash-m_{\mathrm{P_{bb}}})$. 
Expanding the product and applying the identity 
$\pslash\,\eslash = 2(\varepsilon\cdot p)-\eslash\,\pslash$, 
one obtains the four independent Lorentz structures appearing 
in Eq.~\eqref{edmn0511}: 
$2(\varepsilon\cdot p)\pslash$, 
$-m_{\mathrm{P_{bb}}}\,\eslash\,\pslash$, 
$-m_{\mathrm{P_{bb}}}\,\eslash\,\qslash$, and 
$\pslash\,\eslash\,\qslash$. The remaining 
$f_2(q^2)\,(\varepsilon\cdot p)\,(m_{\mathrm{P_{bb}}}\,\qslash - \pslash\,\qslash)/m_{\mathrm{P_{bb}}}$ 
contribution in Eq.~\eqref{full_pi} is processed analogously 
and reduces to the same set of independent structures, 
generating the overall combination $f_1(q^2)+f_2(q^2)$ as the 
coefficient of $\pslash\,\eslash\,\qslash$.

\paragraph*{Reduction of the spin-$3/2$ correlation function.} 
For the spin-$3/2$ case, the reduction is organized in four 
successive stages, each of which is documented in detail in 
Appendix~\ref{appc}: 
(i)~substitution of the polarization 
sum~\eqref{eq:raritasum} into Eq.~\eqref{final phenpart} and 
imposition of the on-shell condition 
$p^{2}=m^{2}_{P^{*}_{bb}}$, yielding the unreduced expression 
in Appendix~\ref{appc1}; 
(ii)~imposition of the real-photon kinematic conditions 
$q^{2}=0$ and $\varepsilon\!\cdot\! q=0$, yielding the 
gauge-fixed expression in Appendix~\ref{appc2}; 
(iii)~application of the gamma-matrix ordering 
$\gamma_\mu \pslash \eslash \qslash \gamma_\nu$ and removal of 
the spin-$1/2$ contamination, yielding the expression in 
Appendix~\ref{appc3}; and 
(iv)~projection of the resulting expression onto the four 
linearly independent Lorentz structures 
$g_{\mu\nu}\,\pslash\,\eslash\,\qslash$, 
$g_{\mu\nu}\,\eslash\,\qslash$, 
$q_\mu q_\nu\,\eslash\,\qslash$, and 
$(\varepsilon\!\cdot\! p)\,q_\mu q_\nu\,\pslash\,\qslash$, 
documented in Appendix~\ref{appc4}. To make this procedure 
transparent and reproducible, we work out below the contribution 
of the $F_1(q^2)\,g^{\alpha\beta}\,\eslash$ piece of the vertex 
in Eq.~\eqref{final phenpart}, restricted to the leading 
$g_{\mu\alpha}\,g_{\beta\nu}$ contractions of the two 
Rarita--Schwinger projectors; the remaining projector 
contractions and form-factor sectors are processed identically, 
with the full term-by-term algebra collected in 
Appendix~\ref{appc}.

The contribution under consideration reads
\begin{align}
\Pi^{Had,\,\mathrm{lead}}_{\mu\nu}\Big|_{F_1}
&=
-\,\frac{\lambda^{2}_{P^{*}_{bb}}\,F_1(q^2)\,g_{\mu\nu}}
{[(p+q)^2-m^2_{P^{*}_{bb}}]\,[p^2-m^2_{P^{*}_{bb}}]}\,
\big(\pslash+m_{P^{*}_{bb}}\big)\,\eslash\,
\big(\pslash+\qslash+m_{P^{*}_{bb}}\big),
\label{eq:F1lead_start}
\end{align}
where the factor $g_{\mu\nu}$ arises from the contraction 
$g_{\mu\alpha}\,g^{\alpha\beta}\,g_{\beta\nu}=g_{\mu\nu}$, and 
the on-shell condition $p^{2}=m^{2}_{P^{*}_{bb}}$ has been used 
in the spinor numerators. Expanding the Dirac chain in 
Eq.~\eqref{eq:F1lead_start} and using the anticommutation 
identity $\pslash\,\eslash = 2(\varepsilon\!\cdot\! p) - \eslash\,\pslash$ 
together with $\pslash^{2}=p^{2}=m^{2}_{P^{*}_{bb}}$:
\begin{align}
\big(\pslash+m_{P^{*}_{bb}}\big)\,\eslash\,\big(\pslash+\qslash+m_{P^{*}_{bb}}\big)
&=
\big[\,2(\varepsilon\!\cdot\! p)-\eslash\,\pslash + m_{P^{*}_{bb}}\,\eslash\,\big]\big(\pslash+\qslash+m_{P^{*}_{bb}}\big)
\nonumber\\
&=
2(\varepsilon\!\cdot\! p)\,\pslash + 2(\varepsilon\!\cdot\! p)\,\qslash + 2m_{P^{*}_{bb}}(\varepsilon\!\cdot\! p)
\nonumber\\
&\quad
-\,\eslash\,\pslash\,\pslash - \eslash\,\pslash\,\qslash - m_{P^{*}_{bb}}\,\eslash\,\pslash
\nonumber\\
&\quad
+\,m_{P^{*}_{bb}}\,\eslash\,\pslash + m_{P^{*}_{bb}}\,\eslash\,\qslash + m^{2}_{P^{*}_{bb}}\,\eslash.
\label{eq:F1lead_expand}
\end{align}
Applying $\eslash\,\pslash\,\pslash=m^{2}_{P^{*}_{bb}}\,\eslash$ 
cancels the $+m^{2}_{P^{*}_{bb}}\,\eslash$ contribution, while 
the pair $-m_{P^{*}_{bb}}\,\eslash\,\pslash + m_{P^{*}_{bb}}\,\eslash\,\pslash$ 
cancels identically. Applying once more 
$\eslash\,\pslash\,\qslash = 2(\varepsilon\!\cdot\! p)\,\qslash - \pslash\,\eslash\,\qslash$ 
to the surviving $-\eslash\,\pslash\,\qslash$ term collapses 
Eq.~\eqref{eq:F1lead_expand} to
\begin{align}
\big(\pslash+m_{P^{*}_{bb}}\big)\,\eslash\,\big(\pslash+\qslash+m_{P^{*}_{bb}}\big)
&=
2(\varepsilon\!\cdot\! p)\,\pslash + 2m_{P^{*}_{bb}}(\varepsilon\!\cdot\! p) + \pslash\,\eslash\,\qslash + m_{P^{*}_{bb}}\,\eslash\,\qslash.
\label{eq:F1lead_collapse}
\end{align}
Inserting Eq.~\eqref{eq:F1lead_collapse} back into 
Eq.~\eqref{eq:F1lead_start} produces four of the 
$F_1(q^2)$-proportional Lorentz structures in 
Eq.~\eqref{final phenpart11}:
\begin{align}
\Pi^{Had,\,\mathrm{lead}}_{\mu\nu}\Big|_{F_1}
&=
-\,\frac{\lambda^{2}_{P^{*}_{bb}}\,F_1(q^2)}{[(p+q)^2-m^2_{P^{*}_{bb}}][p^2-m^2_{P^{*}_{bb}}]}
\Big[\,2(\varepsilon\!\cdot\! p)\,g_{\mu\nu}\,\pslash + 2m_{P^{*}_{bb}}(\varepsilon\!\cdot\! p)\,g_{\mu\nu}
\nonumber\\
&\quad
+\,g_{\mu\nu}\,\pslash\,\eslash\,\qslash + m_{P^{*}_{bb}}\,g_{\mu\nu}\,\eslash\,\qslash\Big],
\label{eq:F1lead_result}
\end{align}
each of which can be identified, in turn, with the corresponding 
$g_{\mu\nu}$-dependent contributions to the $F_1(q^2)$ bracket 
of Eq.~\eqref{final phenpart11}. The remaining pieces of the 
two Rarita--Schwinger projectors --- namely 
$-\tfrac{1}{3}\gamma_\mu\gamma_\alpha$, 
$-\tfrac{2 p_\mu p_\alpha}{3 m^{2}_{P^{*}_{bb}}}$, and 
$\tfrac{p_\mu\gamma_\alpha-p_\alpha\gamma_\mu}{3 m_{P^{*}_{bb}}}$ 
on the left, with their analogues on the right --- contribute 
additional Lorentz structures of the types 
$\gamma_\mu(\ldots)\gamma_\nu$, $p_\mu p_\nu$, $p_\mu q_\nu$, 
$q_\mu p_\nu$, $q_\mu q_\nu$, and $\varepsilon_\mu q_\nu$. 
After the gamma-matrix ordering 
$\gamma_\mu \pslash \eslash \qslash \gamma_\nu$ is applied and 
all terms beginning with $\gamma_\mu$, ending with $\gamma_\nu$, 
or proportional to $p_\mu$ or $p_\nu$ are discarded as 
spin-$1/2$ contamination, the surviving structures combine with 
those of Eq.~\eqref{eq:F1lead_result} to reproduce the full 
$F_1(q^2)$ sector of Eq.~\eqref{final phenpart11}. The 
$F_2(q^2)$, $F_3(q^2)$, and $F_4(q^2)$ sectors are processed 
along identical lines, with the additional vertex factor 
$q^{\alpha}q^{\beta}/(2m_{P^{*}_{bb}})^{2}$ accompanying 
$F_3(q^2)$ and $F_4(q^2)$ generating the compact 
$q_\mu q_\nu$-tensor families that appear in 
Eq.~\eqref{final phenpart11}. The complete intermediate algebra 
for all four form-factor sectors --- in both unreduced 
(Appendix~\ref{appc1}) and gauge-fixed reduced 
(Appendix~\ref{appc2}) forms --- as well as the explicit 
projection onto the independent Lorentz structures 
(Appendix~\ref{appc3}), is collected in Appendix~\ref{appc}.

Applying these manipulations systematically to every Lorentz 
structure generated in stage~(i), to all four form-factor 
sectors, and on the sixteen cross-products of the two 
Rarita--Schwinger projectors, the spin-$3/2$ hadronic 
correlation function takes the reduced form
\begin{align}
\Pi^{Had}_{\mu\nu}(p,q)&=-\frac{\lambda_{_{{\mathrm{P_{bb}^*}}}}^{2}}{[(p+q)^{2}-m_{_{{\mathrm{P_{bb}^*}}}}^{2}][p^{2}-m_{_{{\mathrm{P_{bb}^*}}}}^{2}]} 
\nonumber\\
&
\times \Bigg\{  F_{1}(q^2) \bigg[  \frac{4\,(\varepsilon\!\cdot\! p)\, q_\mu q_\nu}{m_{P_{bb}^*}}
-\,2\,m_{P_{bb}^*}\,\varepsilon_\mu\, q_\nu
- 2\,m_{P_{bb}^*}\,(\varepsilon\!\cdot\! p)\, g_{\mu\nu}
- m_{P_{bb}^*}^{2}\, g_{\mu\nu}\,\eslash
+ \frac{4\,(\varepsilon\!\cdot\! p)\, q_\mu q_\nu \pslash}{m_{P_{bb}^*}^{2}}
- 2\,(\varepsilon\!\cdot\! p)\, g_{\mu\nu}\,\pslash
\nonumber\\
&
+ 2\,\varepsilon_\mu\, q_\nu\,\qslash
+ \frac{2\, q_\mu q_\nu \eslash\,\qslash}{m_{P_{bb}^*}}
- m_{P_{bb}^*}\, g_{\mu\nu}\,\eslash\,\qslash
- \frac{2\, q_\mu q_\nu \pslash\,\eslash}{m_{P_{bb}^*}}
+ \frac{2\,\varepsilon_\mu\, q_\nu \pslash\,\qslash}{m_{P_{bb}^*}}
+ \frac{2\, q_\mu q_\nu \pslash\,\eslash\,\qslash}{m_{P_{bb}^*}^{2}}
- g_{\mu\nu}\,\pslash\,\eslash\,\qslash \bigg ]
\nonumber\\
&
%%%%%%%%%%%%%%%%%%%%%%%%%%%%%%%%%%%%%%%%%%%%%%%%%
+F_{2}(q^2) \bigg[  q_\mu q_\nu\,\eslash
-\,\frac{2\,(\varepsilon\!\cdot\! p)\, q_\mu q_\nu}{m_{P_{bb}^*}}
- \frac{2\,(\varepsilon\!\cdot\! p)\, q_\mu q_\nu}{m_{P_{bb}^*}^{2}}\,\pslash
- \varepsilon_\mu\, q_\nu\,\qslash 
+ \frac{2\,(\varepsilon\!\cdot\! p)\, q_\mu q_\nu \qslash}{m_{P_{bb}^*}^{2}}
- (\varepsilon\!\cdot\! p)\, g_{\mu\nu}\,\qslash
- \frac{q_\mu q_\nu \eslash\,\qslash}{m_{P_{bb}^*}}
\nonumber\\
&
+ m_{P_{bb}^*}\, g_{\mu\nu}\,\eslash\,\qslash
+ \frac{2\, q_\mu q_\nu \pslash\,\eslash}{m_{P_{bb}^*}}
- \frac{2\,\varepsilon_\mu\, q_\nu \,\pslash\,\qslash}{m_{P_{bb}^*}}
+ \frac{2\,(\varepsilon\!\cdot\! p)\, q_\mu q_\nu \,\pslash\,\qslash}{m_{P_{bb}^*}^{3}}
- \frac{(\varepsilon\!\cdot\! p)\, g_{\mu\nu} \,\pslash\,\qslash}{m_{P_{bb}^*}}
- \frac{2\, q_\mu q_\nu\,\pslash\,\eslash\,\qslash}{m_{P_{bb}^*}^{2}}
+ g_{\mu\nu}\,\pslash\,\eslash\,\qslash \bigg]
%%%%%%%%%%%%%%%%%%%%%%%%%%%%%%%%%%%%%%%%%%%%%%%%%%%
\nonumber\\
&
+ F_{3}(q^2) \bigg[  -\,\frac{(\varepsilon\!\cdot\! p)\, q_\mu q_\nu}{2\, m_{P_{bb}^*}}
- \frac{q_\mu q_\nu \eslash}{4}
- \frac{(\varepsilon\!\cdot\! p)\, q_\mu q_\nu \pslash }{2\, m_{P_{bb}^*}^{2}}
- \frac{(\varepsilon\!\cdot\! p)\, q_\mu q_\nu \qslash}{2\, m_{P_{bb}^*}^{2}}\  
- \frac{q_\mu q_\nu \eslash\,\qslash}{2\, m_{P_{bb}^*}} 
- \frac{(\varepsilon\!\cdot\! p)\, q_\mu q_\nu \pslash\,\qslash}{2\, m_{P_{bb}^*}^{3}}
- \frac{q_\mu q_\nu \pslash\,\eslash\,\qslash}{4\, m_{P_{bb}^*}^{2}} \bigg]
%%%%%%%%%%%%%%%%%%%%%%%%%%%%%%%%%%%%%%%%%%%%%%%%%%%%%%%%%%%%%%%%%%%%%
\nonumber\\
&
+ F_{4}(q^2) \bigg[   \frac{q_\mu q_\nu \,\eslash\,\qslash }{8\, m_{P_{bb}^*}}
-\,\frac{(\varepsilon\!\cdot\! p)\, q_\mu q_\nu \,\qslash}{4\, m_{P_{bb}^*}^{2}}
- \frac{(\varepsilon\!\cdot\! p)\, q_\mu q_\nu \,\pslash\,\qslash }{4\, m_{P_{bb}^*}^{3}}
+ \frac{q_\mu q_\nu \,\pslash\,\eslash\,\qslash }{4\, m_{P_{bb}^*}^{2}} \bigg]
\Bigg\}. \label{final phenpart11}
\end{align}
The compact structure of the $F_3(q^2)$ and $F_4(q^2)$ sectors 
in Eq.~\eqref{final phenpart11}, as compared to the $F_1(q^2)$ 
and $F_2(q^2)$ sectors, reflects the high density of 
$(\varepsilon\!\cdot\! q)$ and $q^2$ factors carried by the 
vertex kinematics 
$q^{\alpha}q^{\beta}/(2m_{P^{*}_{bb}})^{2}$ in 
Eq.~\eqref{matelpar}, which vanish under the real-photon 
conditions $q^2=0$ and $\varepsilon\!\cdot\! q=0$. A detailed 
term-by-term comparison of the unreduced and reduced expressions 
that exhibits this collapse is presented in 
Appendix~\ref{appc2}.

After isolating the relevant independent Lorentz structures and 
removing all unwanted pollutions, the final expressions for the 
correlation functions are given by:
\begin{align}
\label{edmn050}
\Pi^{Had}(p,q)&=\frac{\lambda^2_{\mathrm{P_{bb}}}}{[(p+q)^2-m^2_{\mathrm{P_{bb}}}][p^2-m^2_{\mathrm{P_{bb}}}]}
  \bigg[\Big(f_1(q^2)+f_2(q^2)\Big)\pslash\eslash\qslash
  + \cdots \bigg],\\
\Pi^{Had}_{\mu\nu}(p,q)&=\frac{\lambda_{_{{\mathrm{P_{bb}^*}}}}^{2}}{[(p+q)^{2}-m_{_{{\mathrm{P_{bb}^*}}}}^{2}][p^{2}-m_{_{{\mathrm{P_{bb}^*}}}}^{2}]} 
\bigg[  F_{1}(q^2)  \, g_{\mu\nu}\pslash\eslash\qslash 
-F_{2}(q^2) \, m_{{\mathrm{P_{bb}^*}}}g_{\mu\nu}\eslash\qslash 
+
\frac{F_{3}(q^2)}{2m_{{\mathrm{P_{bb}^*}}}}q_{\mu}q_{\nu}\eslash\qslash \nonumber\\
&
+
\frac{F_{4}(q^2)}{4m_{{\mathrm{P_{bb}^*}}}^3}(\varepsilon.p)q_{\mu}q_{\nu}\pslash\qslash 
+
\cdots
\bigg]. \label{final phenpart1}
\end{align}
The choice of Lorentz structures in Eq.~\eqref{final phenpart1} 
is dictated by the linear independence of the form factors 
$F_1(q^2),\ldots,F_4(q^2)$ in Eq.~\eqref{final phenpart11}. A 
direct inspection of Eq.~\eqref{final phenpart11} reveals that 
$F_3(q^2)$ and $F_4(q^2)$ contribute only to structures 
proportional to $q_\mu q_\nu$, namely 
$q_\mu q_\nu\,\eslash\,\qslash$ and 
$(\varepsilon\!\cdot\! p)\,q_\mu q_\nu\,\pslash\,\qslash$, 
which are kinematically orthogonal to the $g_{\mu\nu}$-class 
structures $g_{\mu\nu}\,\pslash\,\eslash\,\qslash$ and 
$g_{\mu\nu}\,\eslash\,\qslash$ carrying the $F_1(q^2)$ and 
$F_2(q^2)$ information. This orthogonality, made explicit in 
Appendix~\ref{appc3}, ensures that the magnetic dipole form 
factor $G_M(0)=F_1(0)+F_2(0)$ can be cleanly extracted from the 
chosen structures, without contamination from the 
higher-multipole form factors $F_3(q^2)$ and $F_4(q^2)$. In the 
derivation leading to Eqs.~\eqref{edmn08} and~\eqref{edmn09}, 
the Lorentz structures $\pslash\,\eslash\,\qslash$, 
$g_{\mu\nu}\,\pslash\,\eslash\,\qslash$, and 
$g_{\mu\nu}\,\eslash\,\qslash$ are therefore chosen to isolate 
$\big(f_1(0)+f_2(0)\big)$, $F_1(0)$, and $F_2(0)$, 
respectively. Among the $F_1(q^2)$- and $F_2(q^2)$-bearing 
independent structures, the preference for these particular 
ones stems from their higher mass dimension and the consequent 
enhancement of pole-dominance behavior and 
operator-product-expansion convergence, thereby improving the 
reliability of the extracted MDMs. Several alternative 
independent Lorentz structures were examined, and the resulting 
predictions were found to differ by less than $5\%$, confirming 
the robustness of the present analysis.

To evaluate the magnetic form factors \( F_M(q^2) \) and \( G_M(q^2) \), corresponding respectively to the \( \rm{P_{bb}} \) and \( \rm{P_{bb}^*} \) pentaquarks, it is essential to rewrite them in terms of the previously introduced form factors  \( f_i(q^2) \) and  \( F_i(q^2) \). The explicit relations are presented below \cite{Leinweber:1990dv, Weber:1978dh, Nozawa:1990gt}:
\begin{align}
\label{edmn07}
F_M(q^2) &= f_1(q^2) + f_2(q^2),\\
G_{M}(q^2) &= \left[ F_1(q^2) + F_2(q^2)\right] ( 1+ \frac{4}{5}
\eta ) -\frac{2}{5} \left[ F_3(q^2)  
+
F_4(q^2)\right] \eta \left( 1 + \eta \right),
\end{align}  
where $\eta
= -\frac{q^2}{4m^2_{{\mathrm{P_{bb}^*}}}}$.    
The previously derived expressions enable the calculation of the electromagnetic form factors for the doubly-bottom states. Since our focus is on a real photon with \( q^2 = 0 \), it is necessary to employ the specific form factor expressions relevant to the MDM. These are detailed in the following equations:
\begin{align}
\label{edmn08}
\mu_{\mathrm{P_{bb}}} &= \frac{ e}{2\, m_{\mathrm{P_{bb}}}} \,F_M( 0),~~~~~%
\\
\mu_{{\mathrm{P_{bb}^*}}}&=\frac{e}{2m_{{\mathrm{P_{bb}^*}}}}G_{M}(0),
\label{edmn09}
\end{align}
with $F_{M}(0) = f_1(0)+f_2(0)$, and $G_{M}(0)= F_1(0)+F_2(0)$.

For the two spin-$\tfrac{3}{2}$ pentaquarks, $B\Sigma_b^{*}$ and
$B^{*}\Sigma_b$, we additionally evaluate the electric quadrupole and
magnetic octupole moments, which are given by
$\mathcal{Q}_{\mathrm{P_{bb}^*}} =
\frac{e}{(2m_{\mathrm{P_{bb}^*}})^{2}}\,G_{E2}(0)$ and
$\mathcal{O}_{\mathrm{P_{bb}^*}} =
\frac{e}{(2m_{\mathrm{P_{bb}^*}})^{3}}\,G_{M3}(0)$, with
$G_{E2}(0)=F_1(0)-\tfrac{1}{2}F_3(0)$ and
$G_{M3}(0)=F_1(0)+F_2(0)-\tfrac{1}{2}\left[F_3(0)+F_4(0)\right]$. The
additional form factors $F_3$ and $F_4$ entering these higher multipoles
are projected out from the same correlation function through the Lorentz
structures listed in Appendix~\ref{appc}, and the corresponding sum
rules follow by the identical operator-product-expansion and
Borel-transformation procedure used above for the magnetic dipole
moment. The complete analytic construction of the electric-quadrupole
and magnetic-octupole sum rules for multiquark states within this QCD light-cone sum rule  
framework, including the explicit higher-multipole projections, has been
presented in detail in Refs.~\cite{Ozdem:2026gwt, Ozdem:2025jda}, to
which we refer the reader for the full formalism.

Within the QCD framework, the quark fields are first contracted using Wick’s theorem, after which the operator product expansion (OPE) is applied. This procedure yields the correlation function expressed in terms of light and heavy quark propagators and the photon’s distribution amplitudes. The outcome of these operations can be summarized as follows:
\begin{align}
\label{QCD1}
\Pi^{\rm{QCD}-\mathrm{ B \Sigma_b}}(p,q)&= \frac{i}{3}\varepsilon^{abc} \varepsilon^{a^{\prime}b^{\prime}c^{\prime}}\, \int d^4x \, e^{ip\cdot x} 
\nonumber\\
& 
 \langle 0\mid \Big\{  
 \mbox{Tr}\Big[\gamma_{\alpha} S_d^{bb^\prime}(x) \gamma_{\beta}  
  \widetilde S_{u}^{aa^\prime}(x)\Big]
 \mbox{Tr}\Big[\gamma_5  S_{b}^{dd^\prime}(x) \gamma_5 S_{d}^{d^\prime d}(-x) \Big]  
 \nonumber\\
&     
  +2 \mbox{Tr}\Big[\gamma_{\alpha} S_u^{bb^\prime}(x) \gamma_{\beta}  
  \widetilde S_{u}^{aa^\prime}(x)\Big]
  \mbox{Tr}\Big[\gamma_5  S_{b}^{dd^\prime}(x) \gamma_5 S_{u}^{d^\prime d}(-x) \Big]  
 \nonumber\\
& + 2 \mbox{Tr}\Big[\gamma_{\alpha} S_u^{ba^\prime}(x) \gamma_{\beta}  
  \widetilde S_{u}^{ab^\prime}(x)\Big] 
\mbox{Tr}\Big[\gamma_5  S_{b}^{dd^\prime}(x) \gamma_5 S_{u}^{d^\prime d}(-x) \Big]  
 \Big\} \Big(\gamma^{\alpha}\gamma_5 S_{b}^{cc^\prime}(x) \gamma_5  \gamma^{\beta}\Big)
\mid 0 \rangle _\gamma \,,\\
%\nonumber\\
%\end{align}
%\begin{align}
%%%%%%%%%%%%%%%%%%%%%%%%%%%%%%%%%%%%%%%%%%%%%%%%%%%%%%%%%%%%%%%%%%%%%%%%%%%%%%%%%%%%%%%%%%%%%%%%%%%%%%%%%%%%%%%%%%%%%%%%%%%%%%%%%%%%%%%%%%%%%%%%%%%%%
\Pi_{\mu\nu}^{\rm{QCD}-\mathrm{ B \Sigma_b^*}}(p,q)&= \frac{i}{3}\varepsilon^{abc} \varepsilon^{a^{\prime}b^{\prime}c^{\prime}}\, \int d^4x \, e^{ip\cdot x} 
\nonumber\\
& 
 \langle 0\mid \Big\{ 
 \mbox{Tr}\Big[\gamma_{\mu} S_d^{bb^\prime}(x) \gamma_{\nu}  
  \widetilde S_{u}^{aa^\prime}(x)\Big]
 \mbox{Tr}\Big[\gamma_5  S_{b}^{dd^\prime}(x)\gamma_5 S_{d}^{d^\prime d}(-x) \Big]  
 \nonumber\\
&
  +2 \mbox{Tr}\Big[\gamma_{\mu} S_u^{bb^\prime}(x) \gamma_{\nu}  
  \widetilde S_{u}^{aa^\prime}(x)\Big] 
  \mbox{Tr}\Big[\gamma_5  S_{b}^{dd^\prime}(x)\gamma_5 S_{u}^{d^\prime d}(-x) \Big]  
 \nonumber\\
&  + 2 \mbox{Tr}\Big[\gamma_{\mu} S_u^{ba^\prime}(x) \gamma_{\nu}  
  \widetilde S_{u}^{ab^\prime}(x)\Big]
  \mbox{Tr}\Big[\gamma_5  S_{b}^{dd^\prime}(x)\gamma_5 S_{u}^{d^\prime d}(-x) \Big]  
  \Big\}S_{b}^{cc^\prime}(x)
\mid 0 \rangle _\gamma \,,\label{QCD2}\\
%\nonumber\\
%%%%%%%%%%%%%%%%%%%%%%%%%%%%%%%%%%%%%%%%%%%%%%%%%%%%%%%%%%%%%%%%%%%%%%%%%%%%%%%%%%%%%%%%%%%%%%%%%%%%%%%%%%%%%%%%%%%%%%%%%%%%%%%%%%%%%%%%%%%%%%%%%%%%%%%%%%%%%%%%%%%%%%%%
%\end{align}
%\begin{align}
\Pi_{\mu\nu}^{\rm{QCD}-\mathrm{ B^* \Sigma_b}}(p,q)&= \frac{i}{3}\varepsilon^{abc} \varepsilon^{a^{\prime}b^{\prime}c^{\prime}}\, \int d^4x \, e^{ip\cdot x} 
\nonumber\\
& 
 \langle 0\mid \Big\{ 
 \mbox{Tr}\Big[\gamma_{\mu} S_d^{bb^\prime}(x) \gamma_{\nu}  
  \widetilde S_{u}^{aa^\prime}(x)\Big]
 \mbox{Tr}\Big[\gamma_{\alpha} S_{b}^{dd^\prime}(x) \gamma_{\beta}  S_{d}^{d^\prime d}(-x)\Big]  
 \nonumber\\
&     
+2 \mbox{Tr}\Big[\gamma_{\mu} S_u^{bb^\prime}(x) \gamma_{\nu}\widetilde S_{u}^{aa^\prime}(x)\Big]
\mbox{Tr}\Big[\gamma_{\alpha} S_{b}^{dd^\prime}(x) \gamma_{\beta}  S_{u}^{d^\prime d}(-x)\Big]  
 \nonumber\\
&
+2 \mbox{Tr}\Big[\gamma_{\mu} S_u^{ba^\prime}(x) \gamma_{\nu}  
  \widetilde S_{u}^{ab^\prime}(x)\Big]  
\mbox{Tr}\Big[\gamma_{\alpha} S_{b}^{dd^\prime}(x) \gamma_{\beta}  S_{u}^{d^\prime d}(-x)\Big]  
 \Big\} \Big(\gamma^{\alpha}\gamma_5 S_{b}^{cc^\prime}(x) \gamma_5  \gamma^{\beta}\Big)
\mid 0 \rangle _\gamma \,, \label{QCD3}
\end{align}
where   
$\widetilde{S}_{b(q)}^{ij}(x)=CS_{b(q)}^{ij\mathrm{T}}(x)C$. The light and bottom quark propagators, denoted by $S_q(x)$ and $S_b(x)$, respectively, can be expressed as follows~\cite{Balitsky:1987bk, Belyaev:1985wza}:
\begin{align}
\label{edmn13}
S_{q}(x)&= S_q^{free}(x) - \frac{\langle \bar qq \rangle }{12} \Big(1-i\frac{m_{q} \xslash}{4}   \Big)- \frac{ \langle \bar qq \rangle }{192}m_0^2 x^2  \Big(1-i\frac{m_{q} \xslash}{6}   \Big)
-\frac {i g_s }{16 \pi^2 x^2} \int_0^1 du \, G^{\mu \nu} (ux)
\bigg[\bar u \rlap/{x} 
\sigma_{\mu \nu} + u \sigma_{\mu \nu} \rlap/{x}
 \bigg],\\
%\nonumber\\
%\end{align}%
%and
%
%\begin{align}
S_{b}(x)&=S_b^{free}(x)
%\nonumber\\
%&
-i\frac{m_{b}\,g_{s} }{16\pi ^{2}}  \int_0^1 dv \,G^{\mu \nu}(vx)\bigg[ (\sigma _{\mu \nu }{\xslash}
 % \nonumber\\
%&
+{\xslash}\sigma _{\mu \nu }) 
    \frac{K_{1}\big( m_{b}\sqrt{-x^{2}}\big) }{\sqrt{-x^{2}}}
   %\nonumber\\
  %&
 +2\sigma_{\mu \nu }K_{0}\big( m_{b}\sqrt{-x^{2}}\big)\bigg],
 \label{edmn14}
\end{align}%
with  
\begin{align}
 S_q^{free}(x)&=\frac{1}{2 \pi x^2}\Big(i \frac{\xslash}{x^2}- \frac{m_q}{2}\Big),\\
 %\nonumber\\
 %\end{align}
 %\begin{align}
 S_b^{free}(x)&=\frac{m_{b}^{2}}{4 \pi^{2}} \bigg[ \frac{K_{1}\big(m_{b}\sqrt{-x^{2}}\big) }{\sqrt{-x^{2}}}
+i\frac{{\xslash}~K_{2}\big( m_{b}\sqrt{-x^{2}}\big)}
{(\sqrt{-x^{2}})^{2}}\bigg],
\end{align}
where %$m_0$ is defined through the quark-gluon mixed condensate $ m_0^2= \langle 0 \mid \bar  q\, g_s\, \sigma_{\mu\nu}\, G^{\mu\nu}\, q \mid 0 \rangle / \langle \bar qq \rangle $, 
$G^{\mu\nu}$ is the gluon field-strength tensor, and $K_n$'s being the modified second type Bessel functions.  %Here, we use the following integral representation  of the  modified second-type Bessel function,     
%\begin{equation}\label{b2}
%K_n(m_Q\sqrt{-x^2})=\frac{\Gamma(n+ 1/2)~2^n}{m_Q^n \,\sqrt{\pi}}\int_0^\infty dt~\cos(m_Qt)\frac{(\sqrt{-x^2})^n}{(t^2-x^2)^{n+1/2}}.
%\end{equation}

Within the QCD framework, the correlation function receives two main types of contributions. The first corresponds to perturbative effects, where the photon is emitted at short distances. The second arises from nonperturbative effects involving photon emission at long distances. To extract the short-distance contributions, the following relation is employed: 
\begin{align}
\label{free}
S^{free}(x) \rightarrow \int d^4z\, S^{free} (x-z)\,\rlap/{\!A}(z)\, S^{free} (z)\,,
\end{align}
and the four propagators that survive in Eqs.~(\ref{QCD1})-(\ref{QCD3}) are considered to be free. %This amounts to taking $\bar T_4^{\gamma} (\underline{\alpha}) = 0$ and $S_{\gamma} (\underline {\alpha}) = \delta(\alpha_{\bar q})\delta(\alpha_{q})$ as the light-cone distribution amplitude in the three particle distribution amplitudes (see Ref. \cite{Li:2020rcg}).
To reliably incorporate nonperturbative (long-distance) effects, one typically employs the light-cone operator product expansion in terms of photon distribution amplitudes, which encode the relevant hadronic matrix elements in the presence of an external electromagnetic field. The corresponding transformation is performed as follows:
 \begin{align}
\label{edmn21}
S_{\alpha\beta}^{ab}(x) \rightarrow -\frac{1}{4} \Big[\bar{q}^a(x) \Gamma_i q^b(0)\Big]\Big(\Gamma_i\Big)_{\alpha\beta},
\end{align}
and the four propagators that remain in Eqs.~(\ref{QCD1})-(\ref{QCD3}) are considered to be full propagators. Here, $\Gamma_i = \{\textbf{1}, \gamma_5, \gamma_\mu, i\gamma_5 \gamma_\mu, \sigma_{\mu\nu}/2\}$. Applying the steps outlined in Eq.~(\ref{edmn21}) introduces matrix elements such as $\langle \gamma(q) | \bar{q}(x) \Gamma_i G_{\alpha\beta} q(0) | 0 \rangle$ and $\langle \gamma(q) | \bar{q}(x) \Gamma_i q(0) | 0 \rangle$, which are defined in terms of photon distribution amplitudes (DAs) and encapsulate essential nonperturbative contributions~\cite{Ball:2002ps}. The QCD-side of the correlation functions is constructed using Eqs.~(\ref{QCD1})–(\ref{edmn21}), and subsequently transformed into momentum space via Fourier transformation. Following these systematic procedures, one arrives at the final QCD representation of the MDMs. 
Given the technical nature of these aspects and the fact that they follow standard procedures, we omit a more detailed exposition here. Readers seeking a more comprehensive discussion are referred to \cite{Ozdem:2024suc}, where the relevant methods are presented in detail.

The QCD light-cone sum rules for MDMs are derived by equating the 
correlation functions computed at both the hadronic and quark-gluon 
levels, in accordance with the principle of quark-hadron duality. 
To enhance the contribution of the ground state while suppressing 
those of higher resonances and the continuum, we employ Borel 
transformation and continuum subtraction techniques. These standard 
procedures improve the reliability of the extracted physical 
observables. The resulting expression for the MDMs is given by:
\begin{align}
\label{edmn15}
\mu_{P} = \frac{e^{m_{P}^2/\mathrm{M}^2}}{\tilde \lambda_{P}}\,
\mathcal{R}_i(\mathrm{M}^2,\mathrm{s}_0),
\end{align}
where  $ \tilde \lambda_{P} = m_{P}\,\lambda_{P}^2$, and $P \in \{B\Sigma_b,\, B\Sigma_b^*,\, B^*\Sigma_b\}$ denotes 
the doubly-bottom pentaquark state under consideration, and $i = 1, 
2, 3$ refers to the corresponding sum-rule function 
$\mathcal{R}_i(\mathrm{M}^2,\mathrm{s}_0)$. The functions 
$\mathcal{R}_i(\mathrm{M}^2,\mathrm{s}_0)$ encapsulate the 
contributions from various quark and gluon configurations to the 
sum rules. These functions play a crucial role in the determination 
of the MDMs, and their explicit expressions are provided in 
Appendix~\ref{appa}.
%
%The explicit form of the $\mathcal R_i(\rm{M^2},\rm{s_0})$ functions are given in the Appendix \ref{appa}.

\end{widetext}

\section{Numerical analysis}\label{numerical}

In this section, we present a numerical evaluation of the expressions derived for the MDMs of the doubly-bottom pentaquark states. This analysis necessitates the use of specific input parameters, which are listed in Table~\ref{inputparameter}. Additionally, as seen from the sum rules provided in Eqs.~(\ref{sumrules})–(\ref{sumrules10}), the photon DAs play a crucial role in the calculations. The explicit forms of these DAs are given in \ref{appb}. 
%adopted from Ref.~\cite{Ball:2002ps}.

%\begin{widetext}
 %
 \begin{table}[htb]
	\addtolength{\tabcolsep}{10pt}
	\caption{Input parameters in this work.}
	\label{inputparameter}
	%	\begin{center}
%	\begin{ruledtabular}
		%\scalebox{1.0}{
\begin{tabular}{l|cccccc}
               \hline\hline
               \\
               Inputs& Values\\
               \\
                                        \hline\hline
$m_b$&$ 4.78 \pm 0.06$ GeV \cite{Workman:2022ynf}
                        \\
$m_{ B \Sigma_b}$&$  11.07^{+0.08}_{-0.08}$ GeV \cite{Wang:2024brl}%&  
                        \\
$m_{ B \Sigma_b^*}$&$ 11.09^{+0.08}_{-0.08}$ GeV \cite{Wang:2024brl}%& 
                        \\
$m_{ B^* \Sigma_b} $&$  11.12^{+0.08}_{-0.08}$ GeV \cite{Wang:2024brl}%& 
                       \\
$m_0^{2} $ & $ 0.8 \pm 0.1 $  \,GeV$^2 $ \cite{Ioffe:2005ym}  
                      \\
$f_{3\gamma} $&$ -0.0039 $ GeV$^2 $  \cite{Ball:2002ps}  %&
                       \\
$\chi $&$ 2.85 \pm 0.5 $ GeV$^{-2}$ \cite{Rohrwild:2007yt} %&
                       \\
$\langle \bar qq\rangle $&$ (-0.255 \pm 0.008)^3 $ GeV$^3$ \cite{Ioffe:2005ym}%& 
                       \\
$ \langle g_s^2G^2\rangle  $&$ 0.80 \pm 0.04 $ GeV$^4$ \cite{Narison:2018nbv} %&
                       \\
                       $\lambda_{ B \Sigma_b}  $&$ (19.89^{+2.65}_{-2.57})\times 10^{-3}$ GeV$^6$ \cite{Wang:2024brl} %& 
                       \\
$ \lambda_{ B \Sigma_b^*}   $&$ (12.10^{+1.56}_{-1.51})\times 10^{-3} $ GeV$^6$ \cite{Wang:2024brl}%&
                       \\
$\lambda_{ B^* \Sigma_b}   $&$(22.63^{+2.92}_{-2.84})\times 10^{-3}$ GeV$^6$   \cite{Wang:2024brl}%& 
                       \\
                                      \hline\hline
 \end{tabular}
%}
%\end{center}
%\end{ruledtabular}
\end{table}

%\end{widetext}

In addition to the previously discussed parameters, two further inputs play an essential role in the QCD sum rule analysis: the Borel mass parameter $\rm{M^2}$ and the continuum threshold $\rm{s_0}$. The numerical evaluation is performed within the bounds defined by the standard criteria of the QCD sum rule approach, which guarantee the stability and reliability of the results. 
%The continuum threshold parameter $\rm{s_0}$ is not an arbitrary input; instead, it defines the energy boundary beyond which contributions from the continuum and higher resonant states become relevant in the correlation function. Although different techniques have been suggested in the literature to determine the working interval of this parameter, a widely accepted range is given by $ (m_{P_{bb}} + 0.5)^2~\mathrm{GeV}^2 \leq s_0 \leq (m_{P_{bb}} + 0.8)^2~\mathrm{GeV}^2$. 
%
The continuum threshold $\rm{s_0}$ is not a completely arbitrary 
quantity; rather, it characterizes the scale at which the excited 
states and the continuum begin to contribute significantly to the 
correlation function. Several methodologies have been proposed for 
its estimation. One technique involves varying the parameter within 
a reasonable range while monitoring the emergence of a $\mathrm{M^2}$ 
window in which the predictions cease to depend on 
$\mathrm{M^2}$~\cite{Ligeti:1993qd}. An alternative technique treats 
$\rm{s_0}$ as a function of $\mathrm{M^2}$ and fixes the functional 
form of this relationship by requiring that the mass result be 
insensitive to $\mathrm{M^2}$~\cite{Lucha:2009uy}. A further approach 
allows the continuum threshold to be evaluated 
independently~\cite{Chen:2015moa, Chen:2016otp}. Each of these 
perspectives carries its own advantages and drawbacks regarding the 
reliability of the analysis. Nevertheless, the most widely adopted 
prescription is to constrain $\rm{s_0}$ to the range 
$(\rm{m_H}+0.4)^2\, \rm{GeV }^2 \leq \rm{s_0} \leq 
(\rm{m_H}+0.8)^2 \, \rm{GeV }^2$, guided by the mass gap between the 
ground state (1S) and the first radial excitation (2S) of the state 
under study. In the absence of experimental data on the excited 
states of doubly-bottom pentaquarks, a comparable estimate can be 
obtained by inspecting analogous multiquark systems for which such 
mass gaps have been discussed. If one adopts the tetraquark 
assignments, the $X(3915)$ and $X(4500)$ may be identified as the 1S 
and 2S states with $\rm{J^{PC}}=0^{++}$~\cite{Lebed:2016yvr, 
Wang:2016gxp}; the $Z_c(3900)$ and $Z_c(4430)$ as the 1S and 2S 
states with $\rm{J^{PC}}=1^{+-}$~\cite{Maiani:2014aja, 
Nielsen:2014mva, Wang:2014vha, Agaev:2017tzv}; the $Z_c(4020)$ and 
$Z_c(4600)$ as the 1S and 2S states with 
$\rm{J^{PC}}=1^{+-}$~\cite{Chen:2019osl, Wang:2019hnw}; and the 
$X(4140)$ and $X(4685)$ as the 1S and 2S states with 
$\rm{J^{PC}}=1^{++}$~\cite{Wang:2021ghk}. Across these assignments, 
the 1S--2S mass splittings are estimated to lie in the range of 
approximately $0.5$ to $0.6\,\rm{GeV}$. It is therefore reasonable to 
expect a comparable splitting for the doubly-bottom pentaquark 
states. Accordingly, we set the continuum threshold for the states 
considered here to 
$(\rm{m_{P_{bb}}}+0.5)^2\, \rm{GeV }^2 \leq \rm{s_0} \leq 
(\rm{m_{P_{bb}}}+0.8)^2 \, \rm{GeV }^2$.

The working region for the Borel parameter $\rm{M^2}$ is fixed by ensuring two fundamental conditions: pole dominance (PC) and the convergence of the operator product expansion (CVG). These constraints are imposed through the following relations:
\begin{align}
 \mbox{PC} &=\frac{\mathcal R_i (\rm{M^2},\rm{s_0})}{\mathcal R_i (\rm{M^2},\infty)} > 35\%,\\
% \nonumber\\
%\end{align}
%\begin{align}
 \mbox{CVG} &=\frac{\mathcal R_i^{\mbox{DimN}} (\rm{M^2},\rm{s_0})}{\mathcal R_i (\rm{M^2},\rm{s_0})} < 5\%,
 \end{align}
 where $\mathcal R_i^{\mbox{DimN}} (\rm{M^2},\rm{s_0})$ is  the highest dimensional term in the OPE of $\mathcal R_i (\rm{M^2},\rm{s_0})$. 
 The results of the corresponding CVG analysis are presented in Table~\ref{parameter}, and the outcomes of the PC analysis are likewise provided in the same table. 
 As can be observed from the table, the contribution of the highest-dimensional term in the CVG analysis is rather small, indicating that it is under sufficient control.   
 To illustrate the results of the PC analysis, Figure~\ref{Msqfig} demonstrates that, within the adopted Borel window, the PC contribution remains well above the continuum, thereby confirming the ground-state dominance. The PC curves in Fig.~\ref{Msqfig} are obtained by averaging over the 
continuum threshold $\rm{s_0}$ within the interval quoted in 
Table~\ref{parameter}; that is, the PC is computed for a dense set of 
$\rm{s_0}$ values spanning this interval and the resulting values are 
averaged. The spread of the PC over the interval is only a few 
percent, so that the averaged curve is representative of the entire 
$\rm{s_0}$ range and the resulting Borel window is essentially 
unaffected by the choice of $\rm{s_0}$. This confirms that the 
ground-state dominance displayed in the figure is insensitive to 
$\rm{s_0}$ within its established interval.   For completeness, Fig.~\ref{Msqfig1} displays the dependence of the MDMs of the doubly bottom pentaquark states on the Borel mass parameter $\rm{M^2}$ and the continuum threshold $\rm{s_0}$.  As evident from the figure, the magnetic moment remains largely insensitive to variations in $\rm{M^2}$ within the chosen window, thereby confirming the stability and reliability of the theoretical predictions.  
 Although the magnetic moment exhibits a more pronounced sensitivity to the continuum threshold $\rm{s_0}$, the resulting variations remain well within the estimated theoretical uncertainties of the employed sum-rule approach. This indicates that, despite the inherent dependence on the choice of $\rm{s_0}$, the predictions are stable and reliable. Moreover, the observed behavior reflects the expected physical consistency of the method: variations in $\rm{s_0}$ primarily affect the continuum contribution, while the ground-state dominance, as ensured by the pole contribution analysis, remains intact. Consequently, the extracted magnetic moments can be regarded as robust estimates, and their dependence on $\rm{s_0}$ does not compromise the overall validity of the theoretical predictions.
 
\begin{widetext}

\begin{table}[htb!]
	\addtolength{\tabcolsep}{10pt}
	\caption{Working regions of $\rm{s_0}$ and $\rm{M^2}$ together with the PC and CVG for the MDMs of the doubly-bottom pentaquarks.}
	\label{parameter}
	%	\begin{center}
	\begin{ruledtabular}
		%\scalebox{1.0}{
\begin{tabular}{l|cccccc}
 %               \hline\hline
 \\
States & $\rm{s_0}$ (GeV$^2$) & $\rm{M^2}$ (GeV$^2$)&  ~~  PC ($\%$) ~~ & ~~  CVG ($\%$) & $\mu\, (\mu_N)$\\
 \\ 
                                        \hline\hline                  \\     
$ B \Sigma_b$ & $134.0 - 140.0$ & $5.0 - 7.0 $ & $76.65 - 40.71$ &  $<1.0$  &$~~2.40^{+0.56}_{-0.47}$
                        \\
                    \\
$ B \Sigma_b^*$& $134.0 - 140.0$ & $5.0 - 7.0 $ & $75.67 - 40.18$ &  $<1.0$  &$- 2.84^{+0.78}_{-0.59}$
                       \\
                      \\
$ B^* \Sigma_b$ & $135.0 - 141.0$ & $5.5 - 8.0 $ & $77.16 - 39.16$ &  $<1.0$  &$~~5.17^{+1.08}_{-0.94}$
                      \\
                       \\
  %                                     \hline\hline
 \end{tabular}
%}
%\end{center}
\end{ruledtabular}
\end{table}
%\begin{widetext}

\begin{figure}[htb!]
%\centering
\includegraphics[width=0.32\textwidth]{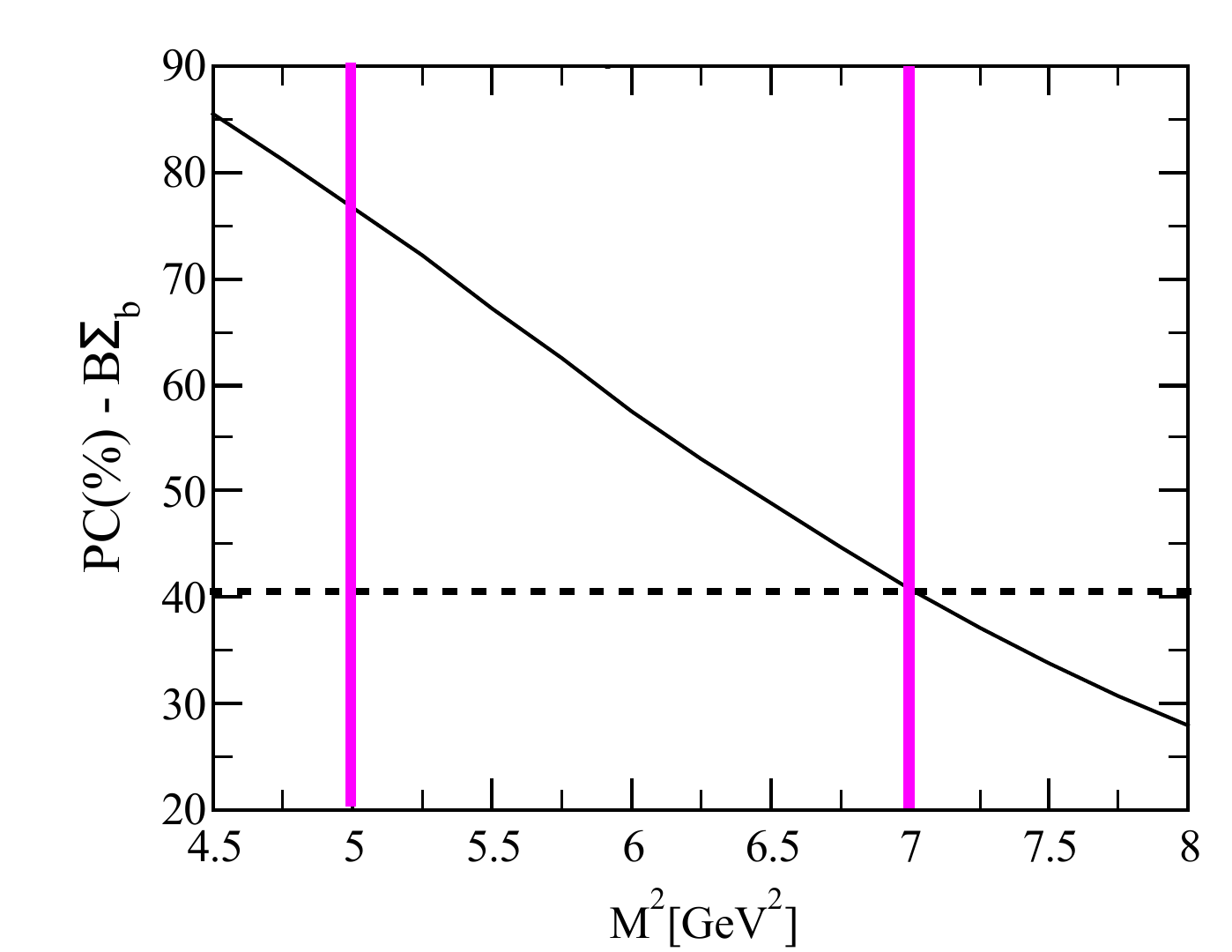}~~
\includegraphics[width=0.32\textwidth]{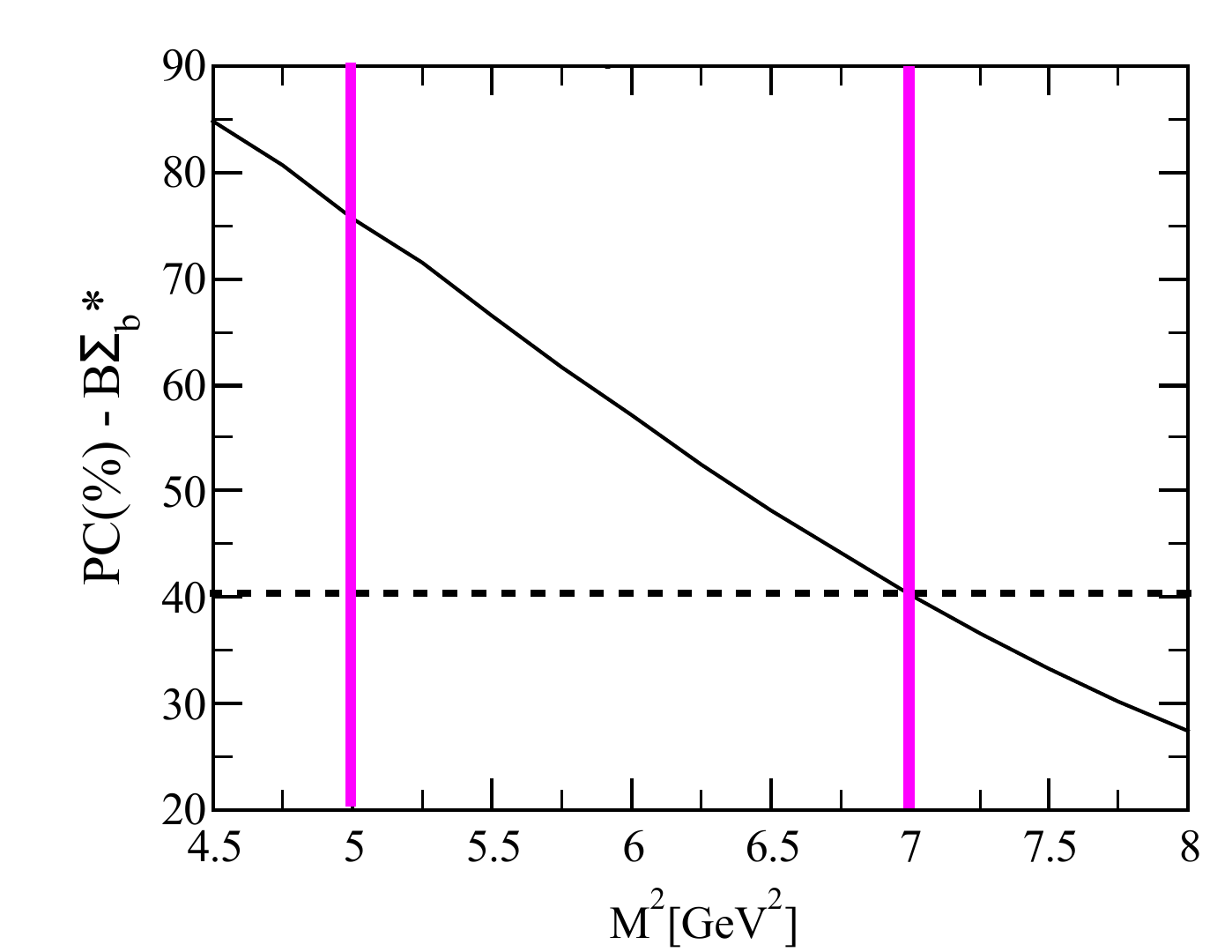}~~
\includegraphics[width=0.32\textwidth]{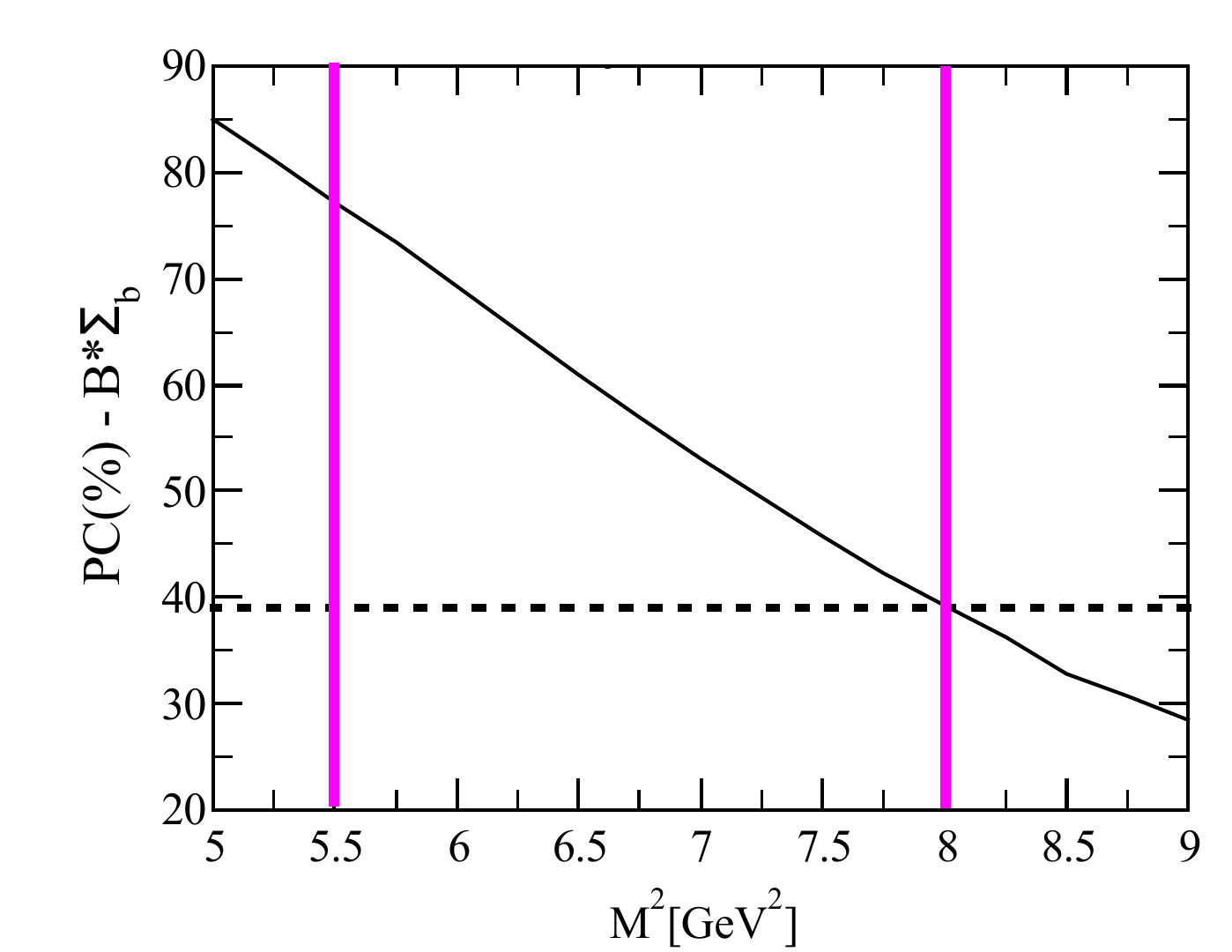}
 \caption{PC analysis of the MDM for doubly bottom pentaquarks as a function of $\rm{M^2}$. The region enclosed by the vertical lines represents the adopted Borel window, while the horizontal line marks the minimum value of the PC in this analysis. The PC results are obtained by averaging over the continuum threshold $\mathrm{s_0}$ within the interval specified in Table~\ref{parameter}.}
 \label{Msqfig}
  \end{figure}

\begin{figure}[htb!]
%\centering
\includegraphics[width=0.32\textwidth]{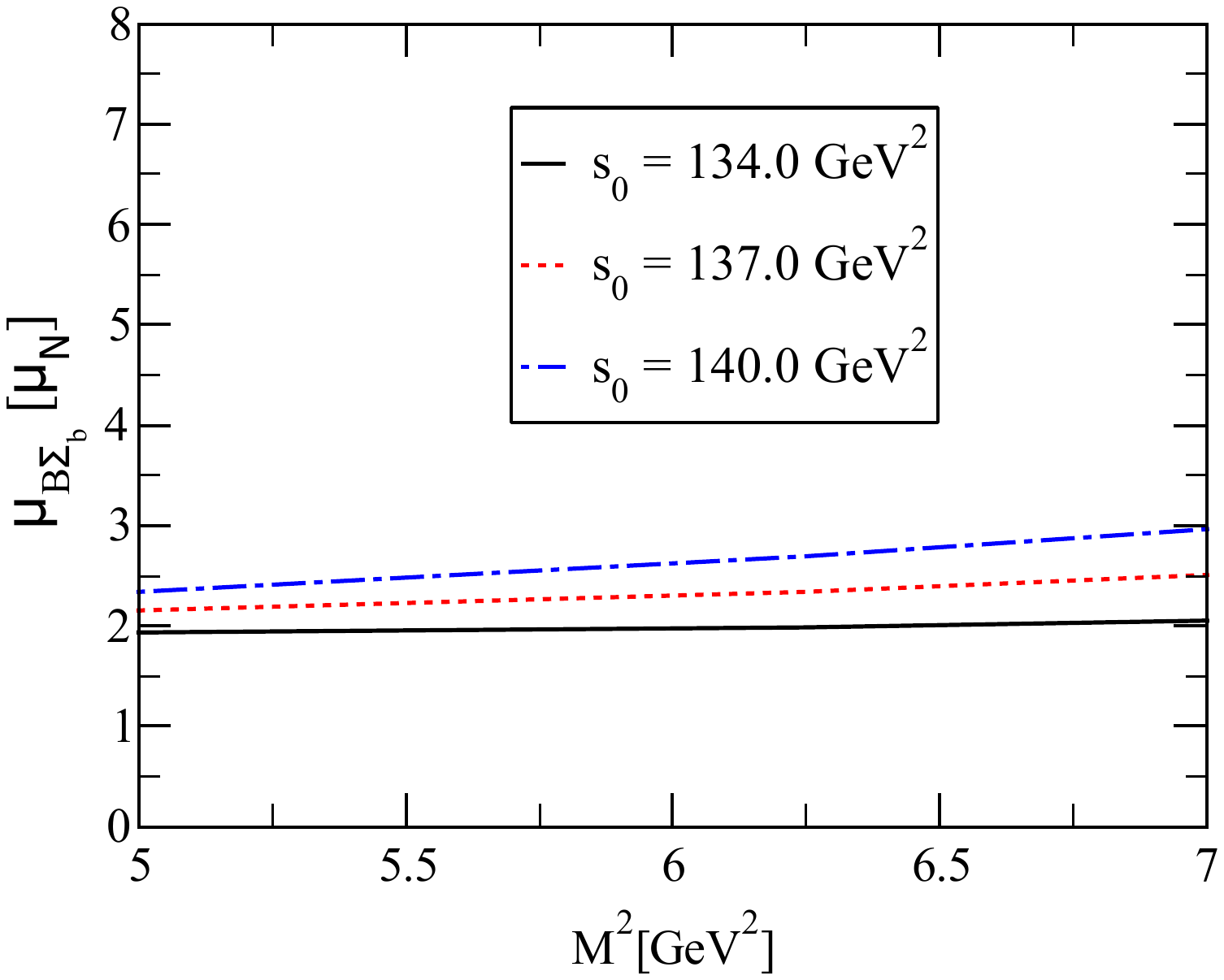}~~
\includegraphics[width=0.32\textwidth]{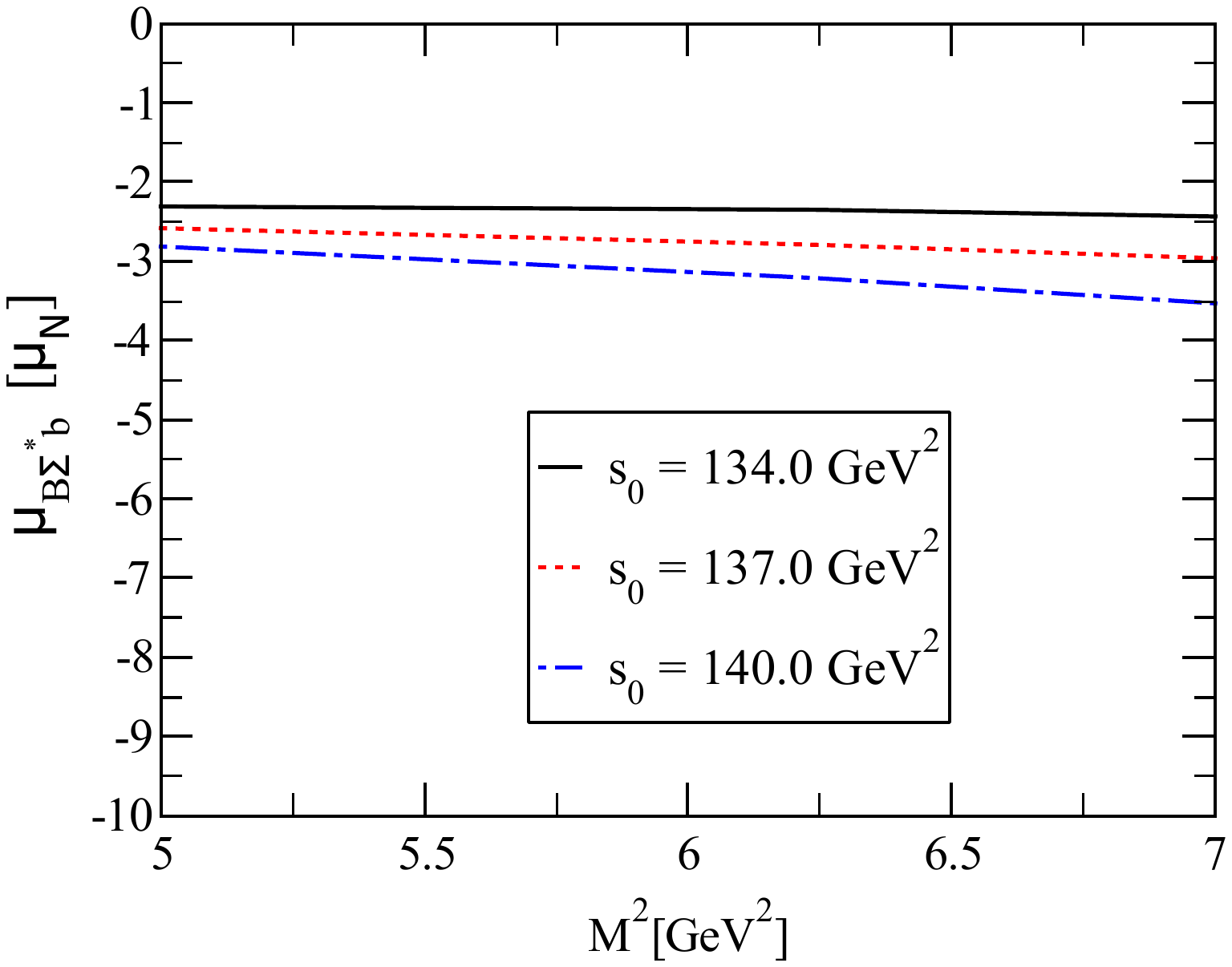}~~
\includegraphics[width=0.32\textwidth]{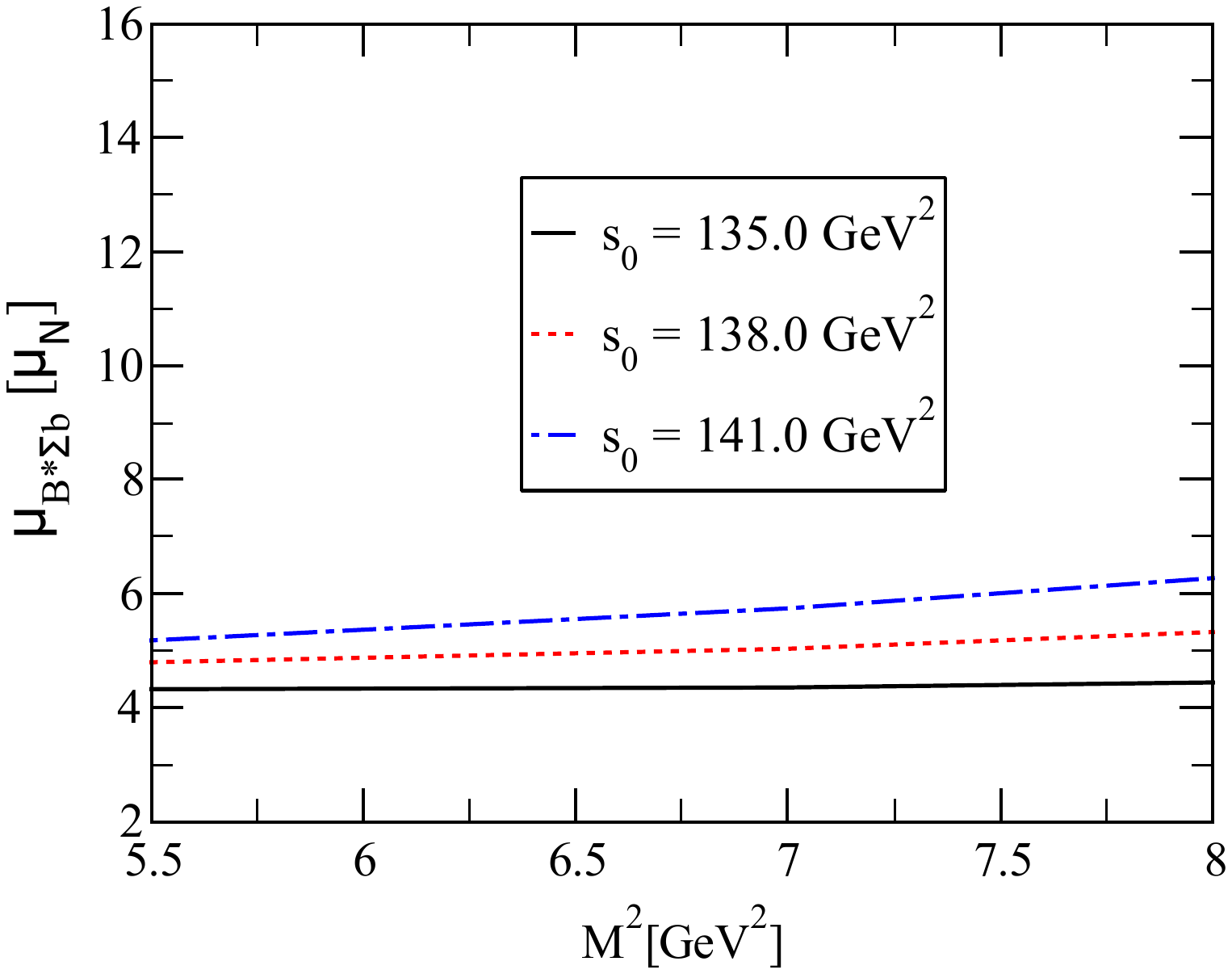}
 \caption{The MDMs of the doubly-bottom pentaquarks versus $\rm{M^2}$.}
 \label{Msqfig1}
  \end{figure}

\end{widetext}
Following the numerical analysis, the predicted values of the MDMs for the doubly-bottom pentaquark states with quantum numbers $\rm{J^P} =\frac{1}{2}^-$ and $\frac{3}{2}^-$ are summarized  in Table \ref{parameter}. %as follows: 
%\begin{align}
% & \mu_{B \Sigma_b}= 2.40^{+0.56}_{-0.47} ~\mu_N,\\
% \nonumber\\
% & \mu_{ B \Sigma_b^*}= - 2.84^{+0.78}_{-0.59} ~\mu_N,\\
% \nonumber\\
% & \mu_{B^* \Sigma_b}= 5.17^{+1.08}_{-0.94} ~\mu_N.
%\end{align}
%
The uncertainties in the obtained results primarily stem from the variations in the continuum threshold parameter $\rm{s_0}$ and the Borel mass parameter 
$\rm{M^2}$, as well as from the uncertainties associated with other input parameters, including those used in the DAs. We would like to point out that the contributions to the total uncertainty are approximately as follows: 15\% from the mass of the pentaquarks, 24\% from their residues, 30\% from the threshold parameter $\rm{s_0}$, 7\% from the Borel mass $\rm{M^2}$, 9\% from the photon DAs, and the remaining 15\% from other parameters.

%Several key points can be summarized from the numerical results presented in Table \ref{sonuc}:

Based on the obtained results, we summarize our findings as follows:
\begin{itemize}
  \item For the \( B \Sigma_b \) state, the dominant contribution to the magnetic moment originates from the perturbative part, which accounts for approximately 92\% of the total value. For the \( B \Sigma_b^* \) configuration, this fraction is about 84\%, while for the \( B^* \Sigma_b \) state, the perturbative contribution decreases to nearly 76\%. These results indicate that the perturbative component plays a leading role in all configurations, although its relative importance varies depending on the spin structure of the state.
  
  \item The significant variation in the MDMs among the \( B \Sigma_b \), \( B \Sigma_b^* \), and \( B^* \Sigma_b \) configurations highlights the crucial role of the internal spin structure and the dynamics of the constituent quarks. The observed differences, particularly in the sign and magnitude of the magnetic moments, reflect the sensitivity of these observables to the spin alignment between light and heavy quarks. In particular, the negative value obtained for the \( B \Sigma_b^* \) state indicates a destructive interference between spin components, whereas the enhanced magnetic moment in the \( B^* \Sigma_b \) configuration suggests a constructive spin alignment. These findings offer valuable insights into the electromagnetic structure and internal dynamics of doubly-bottom pentaquark systems.
  
  %\begin{widetext}
   
    \begin{table}[htb]
	\addtolength{\tabcolsep}{10pt}
	\caption{The contribution of light and heavy quarks to the MDMs of the $P_{bb}$ states (in units of $\mu_N$ ). These results have been obtained using the central values of the numerical input parameters.}
	\label{parameter2}
		\begin{center}
		%\begin{ruledtabular}
\begin{tabular}{l|ccccc}
	   \hline\hline
       \\
   States& $\mu_{u}$& $\mu_{d}$&  $\mu_{b}$& $\mu_{tot}$\\
   \\
\hline\hline
\\
$ B \Sigma_b$   & $ ~~2.54 $  & $ -0.14 $    &  $- $& $~~2.40$\\
\\
$B \Sigma_b^*$  & $ -0.275 $ & $ ~~0.015 $  &  $-2.58 $& $-2.84$\\
\\
$B^* \Sigma_b $ & $ ~~2.76 $  & $ -0.35$     & $ ~~2.76 $& $~~5.17$\\
\\
	   \hline\hline
	   %%%%%%%%%%%%%%%%%%%%%%%%
\end{tabular}
\end{center}
%\end{ruledtabular}
\end{table}

%  \end{widetext}
  
  \item The results presented in Table~\ref{parameter2} reveal the distinct roles of light and heavy quarks in shaping the magnetic moments of the \( P_{bb} \) pentaquark states. For the \( B \Sigma_b \) configuration, the dominant contribution originates from the up quark, yielding a sizable positive magnetic moment of \( \mu_{u} = 2.54~\mu_N \), while the down quark contributes a small negative correction (\( \mu_{d} = -0.14~\mu_N \)). The total magnetic moment, \( \mu_{\text{total}} = 2.40~\mu_N \), reflects primarily the influence of light quark dynamics, as the heavy \( b \)-quark contribution is negligible or effectively canceled in this state. 
  In contrast, the \( B \Sigma_b^* \) state exhibits a dominant negative contribution from the heavy \( b \)-quark (\( \mu_{b} = -2.58~\mu_N \)), which, combined with a relatively small net effect from the light quarks, results in a strongly negative total magnetic moment (\( \mu_{\text{total}} = -2.84~\mu_N \)). This indicates that spin alignment and internal structure in this state allow the heavy quark’s magnetic moment to play a significant role. 
  Finally, in the \( B^* \Sigma_b \) configuration, both light and heavy quarks contribute substantially and constructively. The up quark and \( b \)-quark yield \( \mu_u = 2.76~\mu_N \) and \( \mu_b = 2.76~\mu_N \), respectively, while the down quark slightly suppresses the total with \( \mu_d = -0.35~\mu_N \). The resulting total magnetic moment, \( \mu_{\text{total}} = 5.17~\mu_N \), is significantly enhanced, highlighting the cooperative effect between quark constituents in this configuration.  
  These observations underline the sensitivity of the magnetic moments to the internal spin-flavor configuration of the pentaquark states and demonstrate that both the composition and the spin alignment of the constituent quarks have a profound impact on the electromagnetic properties of such exotic hadrons.

 \item The MDMs predicted in this work for the
$B\Sigma_b$, $B\Sigma_b^{*}$, and $B^{*}\Sigma_b$ configurations are
compared with the results of~\cite{Ozdem:2022vip}, where compact
doubly-bottom pentaquark states were studied within the framework of
QCD light-cone sum rules. In that analysis, the spin-$\tfrac{1}{2}$
compact pentaquark states were found to possess negative magnetic
moments, whereas positive values were obtained for the spin-$\tfrac{3}{2}$
states. 
Our result for the $B\Sigma_b$ molecular configuration yields a
positive magnetic moment of about $2.40\,\mu_N$, which differs
markedly in both sign and magnitude from the compact
spin-$\tfrac{1}{2}$ predictions. This behavior indicates a strong
sensitivity of the MDM to the underlying
configuration of the state. In the spin-$\tfrac{3}{2}$ sector, the
$B^{*}\Sigma_b$ molecular state exhibits a magnetic moment of
approximately $5.17\,\mu_N$, showing a close numerical similarity to
the compact pentaquark result, while the $B\Sigma_b^{*}$ configuration
leads to an opposite-sign value. 
These observations suggest that, although certain spin-$\tfrac{3}{2}$
molecular configurations may reproduce the magnitude of the compact
pentaquark magnetic moments, the overall pattern of signs and
magnitudes is strongly model dependent. It must be emphasized, however,
that the sign of the magnetic moment alone cannot be regarded as an
unambiguous discriminator between the molecular and compact pictures.
Already among the states considered here, the two spin-$\tfrac{3}{2}$
molecular configurations carry opposite signs, so that the sign is
configuration dependent within the molecular scenario itself.
Furthermore, the spin-$\tfrac{1}{2}$ molecular partners associated with
the $B^{*}\Sigma_b$ and $B^{*}\Sigma_b^{*}$ systems could not be
included in the present analysis, for the reasons discussed above, and
their magnetic moments---whose signs cannot be anticipated---might
equally well turn out to be negative. Accordingly, the observation of a
negative spin-$\tfrac{1}{2}$ magnetic moment would not by itself
establish a compact internal structure. Rather than relying on a single
sign rule, we therefore regard the characteristic pattern of the
magnetic dipole and higher multipole moments across the accessible
spectrum, taken together with the masses and decay widths, as offering
complementary information for distinguishing molecular from compact
assignments in future experimental studies.

 \item The sensitivity of the MDMs to the spin alignment suggests that future experimental measurements could provide key insights into the internal structure of these pentaquark states. MDMs critically influence the electromagnetic transitions of pentaquark states, governing the rates and probabilities of photon emission between different spin-parity configurations. These moments directly affect the amplitudes of electromagnetic decay channels, thereby impacting the partial decay widths and lifetimes. Consequently, precise theoretical predictions of magnetic moments provide essential input for interpreting future experimental data on radiative decays. Moreover, when considered together with the masses and decay widths, differences in the magnetic moments can provide complementary information for distinguishing compact multiquark structures from hadronic molecular configurations, aiding in the clarification of the internal quark dynamics of exotic states.

 \item In addition to the MDMs, we have also evaluated the
electric quadrupole and magnetic octupole moments of the spin-$\tfrac{3}{2}$
molecular configurations $B\Sigma_b^{*}$ and $B^{*}\Sigma_b$, which
provide further insight into the spatial deformation and higher-order
spin--magnetic correlations of these states, following the definitions
and the sum-rule procedure given in Sec.~\ref{formalism}. For the $B\Sigma_b^{*}$
configuration, the electric quadrupole moment is found to be
$\mathcal{Q}_{B\Sigma_b^{*}} = (0.75 \pm 0.12)\times 10^{-1}\,\mathrm{fm}^2$,
indicating a small but nonvanishing deviation from spherical symmetry.
The positive sign of the quadrupole moment suggests a prolate-type
charge deformation, which is consistent with a molecular picture in
which the spin alignment of the constituents induces an elongated
charge distribution along the quantization axis. 
The corresponding magnetic octupole moment,
$\mathcal{O}_{B\Sigma_b^{*}} = (0.62 \pm 0.10)\times 10^{-3}\,\mathrm{fm}^3$,
is relatively suppressed, reflecting the weak contribution of
higher-order magnetic multipoles in this configuration. This behavior
indicates that, although the state exhibits a mild spatial deformation,
the spin--magnetic correlations beyond the dipole level remain limited,
as expected for a loosely bound molecular system dominated by a single
heavy-quark spin orientation. 
In contrast, the $B^{*}\Sigma_b$ molecular configuration displays a
markedly different multipole structure. The electric quadrupole moment
is obtained as $\mathcal{Q}_{B^{*}\Sigma_b} = -0.60 \pm 0.10 \,
\mathrm{fm}^2$, whose negative sign points to an oblate-type deformation
of the charge distribution. The considerably larger magnitude of this
quadrupole moment compared to the $B\Sigma_b^{*}$ case suggests a more
pronounced spatial anisotropy, arising from the interplay between the
spin-$1$ $B^{*}$ meson and the spin-$\tfrac{1}{2}$ $\Sigma_b$ baryon. 
Furthermore, the magnetic octupole moment of the $B^{*}\Sigma_b$ state
is found to be $\mathcal{O}_{B^{*}\Sigma_b} = (0.12 \pm 0.02)\times
10^{-1}\,\mathrm{fm}^3$, which is significantly enhanced relative to
that of the $B\Sigma_b^{*}$ configuration. This enhancement signals the
presence of stronger higher-order spin--magnetic correlations, driven
by the constructive coupling between the light- and heavy-quark sectors
in the $B^{*}\Sigma_b$ molecular system. Such a pattern is consistent
with the simultaneously large MDM obtained for this
state, indicating a coherent multipole hierarchy extending beyond the
dipole level. 
Overall, the distinct behaviors of the electric quadrupole and magnetic
octupole moments for the $B\Sigma_b^{*}$ and $B^{*}\Sigma_b$
configurations highlight the sensitivity of higher electromagnetic
multipoles to the internal spin arrangement and spatial structure of
doubly-bottom pentaquark molecules. These results reinforce the role of
electromagnetic multipole moments as powerful probes of hadronic
structure and suggest that measurements of quadrupole and octupole
observables, in addition to MDMs, could provide
valuable discrimination between different molecular configurations and
their compact counterparts.

\end{itemize}

\section{Summary}\label{summary}

In this study, we have investigated the MDMs of
doubly-bottom pentaquark states with spin-parities
$J^P=\tfrac{1}{2}^-$ and $\tfrac{3}{2}^-$, treating them as hadronic
molecules in the $B\Sigma_b$, $B\Sigma_b^{*}$, and $B^{*}\Sigma_b$
configurations. The analysis was performed within the framework of
QCD light-cone sum rules employing photon distribution amplitudes. 
Our results demonstrate that the MDMs are highly
sensitive to the internal spin structure and quark dynamics of the
states. In the $B\Sigma_b$ configuration, the magnetic moment is
dominated by the light-quark contributions, whereas in the
$B\Sigma_b^{*}$ case the heavy bottom-quark sector plays a leading
role. For the $B^{*}\Sigma_b$ molecular state, both light and heavy
sectors contribute constructively, resulting in a significantly
enhanced magnetic moment. Although perturbative contributions are
dominant in all configurations, their relative weight decreases from
approximately $92\%$ to $76\%$ as the spin structure becomes more
complex.  Moreover, the predicted electric quadrupole and magnetic octupole
moments of the spin-$\tfrac{3}{2}$ $B\Sigma_b^{*}$ and $B^{*}\Sigma_b$
configurations reveal distinct deformation patterns and higher-order
spin--magnetic correlations, further emphasizing the sensitivity of
electromagnetic multipole observables to the internal structure of
doubly-bottom pentaquark molecules.

A comparison with compact pentaquark predictions from \cite{Ozdem:2022vip} reveals differences in the magnetic dipole
moments that can be traced back to the distinct internal structures of
molecular and compact configurations. This suggests that magnetic
dipole moments, particularly when combined with the masses, decay
widths, and higher electromagnetic multipole moments, can provide
complementary information for distinguishing between these two
pictures. We caution, however, that the sign of the moment alone is not
a decisive criterion, since the spin-$\tfrac{1}{2}$ molecular partners
not treated here could also yield negative moments. 
These findings may have important experimental implications for
radiative decay channels and electromagnetic transitions, offering
valuable insight into the structure of exotic hadrons and providing
guidance for future searches at LHCb and Belle II.

%\clearpage

\appendix
 \begin{widetext}
\section{Explicit form of the $\mathcal R_i(\rm{M^2},\rm{s_0})$ functions} \label{appa}
This section presents the analytical outcomes of the MDMs analysis of the doubly-bottom pentaquark states. 

\subsection{The obtained sum rule for the MDM of the $B\Sigma_b$ state}
\begin{align}
\label{sumrules}
 \mathcal R_1 (\rm{M^2},\rm{s_0}) &= \frac{1}{2 ^{26}\times 3^2 \times 5^3\times 7^2 \pi^7} \Big[ 3 (e_d + 9 e_u) \big(149 I[0, 5] + 427 I[1, 4]\big) \Big] \nonumber\\
 %%%%%%%%%%%%%%%%%%%%%%%%%%%%%%%%%%%%%%%%%%%%%%
 &-\frac {m_b\langle g_s^2 G^2\rangle \langle \bar q q \rangle f_ {3\gamma } } {2^{20} \times 3^3 \pi^3} (e_d + 9 e_u) I_5[\psi^a] I[0, 0]\nonumber\\
    %%%%%%%%%%%%%%%%%%%%%%%%%%%%%%%%%%%%%%%%%%%%%%%%%%%%%%
 & -\frac { \langle \bar q q \rangle^2   } {2^{18} \times 3^2 \times 5^2 \pi^3}  \Big[ (e_d + 9 e_u)\Big ( 
      15  I[0, 2] h_\gamma[u_ 0] + 
       2 \chi I[0, 3] \varphi_\gamma[u_ 0]\Big)\Big]\nonumber\\
%   \end{align}
%\begin{align}
       %%%%%%%%%%%%%%%%%%%%%%%%%%%%%%%%%%%%%%%%%%%%%%%%%%%%%
       & + \frac {\langle g_s^2 G^2\rangle f_ {3 \gamma} } {2^{27}\times 3^5 \times 5 \pi^5} (e_d + 9 e_u)\Big [ 
   63 I_ 1[\mathcal V] + 80 (-9 I_5[\psi^a] + 4 \psi^a[u_ 0])\Big] I[
   0, 2]\nonumber\\
%   \end{align}
%\begin{align}
   %%%%%%%%%%%%%%%%%%%%%%%%%%%%%%%%%%%%%%%%%%%%%%%%%%%%%%
   & + \frac { m_b \langle \bar q q \rangle f_ {3 \gamma} } {2^{19}\times 3^4 \times 5 \pi^3} (e_d + 9 e_u)\Big[ 9 \big(I_ 1[\mathcal V] - 4 I_5[\psi^a]\big) I[0, 
      2] + 22 I[0, 2] \psi^a[u_ 0] \Big]\nonumber\\
      %%%%%%%%%%%%%%%%%%%%%%%%%%%%%%%%%%%%%%%%%%%%
      & + \frac { 
   f_ {3\gamma } } {2 ^{24}\times 3^2 \times 5^2 \times 7^2 \pi^5} (e_d + 
     9 e_u)\Big [ \big(5 I_ 1[\mathcal V] + 10 I_5[\psi^a] + 
      7 \psi^a[u_ 0]\big) I[0, 4]\Big].
\end{align}

\subsection{The obtained sum rule for the MDM of the $B\Sigma_b^*$ state}
\begin{align} \label{sumrules1}
 \mathcal R_2 (\rm{M^2},\rm{s_0}) &= F_1^{B\Sigma_b^*}(\rm{M^2},\rm{s_0})-\frac{1}{m_{B\Sigma_b^*}} F_2^{B\Sigma_b^*}(\rm{M^2},\rm{s_0}),
\end{align}
with 
\begin{align}
 F_1^{B\Sigma_b^*}(\rm{M^2},\rm{s_0})&= \frac{19 e_b }{2^{28} \times 3 \times 5^2 \times 7^2 \pi^7 } I[0, 5]
       %%%%%%%%%%%%%%%%%%%%%%%%%%%%%%%%%%%%%%%%%%%%
      \nonumber\\
      &- \frac{ m_b \langle g_s^2 G^2\rangle \langle \bar q q \rangle f_{3\gamma}}{2^{21} \times 3^6 \times 5 \pi^3} (e_d + 9 e_u)  \psi^a[u_0]I[0, 0] 
       %%%%%%%%%%%%%%%%%%%%%%%%%%%%%%%%%%%%%%%%%%%%
      \nonumber\\
      &+ \frac{  \langle \bar q q \rangle^2}{2^{21} \times 3^2 \pi^3} (e_d + 9 e_u) I_3[\mathcal S] I[0, 2]
       %%%%%%%%%%%%%%%%%%%%%%%%%%%%%%%%%%%%%%%%%%%%
      \nonumber\\
      &+\frac{ \langle g_s^2 G^2\rangle f_{3\gamma}}{2^{28}\times 3^4 \times 5  \pi^5} (e_d + 9 e_u)(26 I_1[\mathcal V] + \psi^a[u_0]) I[0, 2]
      %%%%%%%%%%%%%%%%%%%%%%%%%%%%%%%%%%%%%%%%%%%%
      \nonumber\\
       &+\frac{  m_b \langle \bar q q \rangle f_{3\gamma} }{2^{20}\times 3 ^2\times 5    \pi^3} (e_d + 9 e_u) I_1[\mathcal V] (m_0^2 I[0, 1] - I[0, 2])
%%%%%%%%%%%%%%%%%%%%%%%%%%%%%%%%%%%%%%%%%%%%
      \nonumber\\
       &-\frac{ f_{3\gamma} }{2^{26}\times 3 ^2\times 5    \pi^5} (e_d + 9 e_u) I_1[\mathcal V] I[0, 4],\\
%%%%%%%%%%%%%%%%%%%%%%%%%%%%%%%%%%%%%%%%%%%%%%%%%%%%%%%%%%%%%%%%%%%%%%%%%%%%%%%%%%%%%%%%%%%%%%%%%%%%%%%%%%%%%%%%%,
%\end{align}
%\begin{align}
%\nonumber\\
 F_2^{B\Sigma_b^*}(\rm{M^2},\rm{s_0})&=-\frac{223 e_b m_b }{2^{28} \times 3 \times 5^2 \times 7^2 \pi^7 } I[0, 5]
 %%%%%%%%%%%%%%%%%%%%%%%%%%%%%%%%%%%%%%%%%%%%
      \nonumber\\
      &+ \frac{m_b \langle g_s^2 G^2\rangle \langle \bar q q \rangle^2}{2^{24} \times 3^4 \pi^3} (e_d + 9 e_u)  I_1[\mathcal S] I[0, 0]
       %%%%%%%%%%%%%%%%%%%%%%%%%%%%%%%%%%%%%%%%%%%%
      \nonumber\\
      &+ \frac{ m_b^2\langle g_s^2 G^2\rangle \langle \bar q q \rangle f_{3\gamma}}{2^{21} \times 3^5 \pi^3} (e_d + 9 e_u) (3 I_5[\psi^a] + 2 \psi^a[u_0])I[0, 0] 
       %%%%%%%%%%%%%%%%%%%%%%%%%%%%%%%%%%%%%%%%%%%%
      \nonumber\\
%     \end{align}
%\begin{align}
    &- \frac{ m_b \langle \bar q q \rangle^2}{2^{19} \times 3 \times 5 \pi^3} (e_d + 9 e_u) I_3[\mathcal S] I[0, 2]
       %%%%%%%%%%%%%%%%%%%%%%%%%%%%%%%%%%%%%%%%%%%%
      \nonumber\\
            &+\frac{m_b \langle g_s^2 G^2\rangle f_{3\gamma}}{2^{30}\times 3^3 \times 5 \times  \pi^5} (e_d + 9 e_u)(85 I_1[\mathcal V] + 64 I_5[\psi^a]) I[0, 2]
      %%%%%%%%%%%%%%%%%%%%%%%%%%%%%%%%%%%%%%%%%%%%
      \nonumber\\
%      \end{align}
%\begin{align}
       &-\frac{  m_b^2 \langle \bar q q \rangle f_{3\gamma} }{2^{21}\times 3^2    \pi^3} (e_d + 9 e_u) I_1[\mathcal V] (m_0^2 I[0, 1] - I[0, 2])
%%%%%%%%%%%%%%%%%%%%%%%%%%%%%%%%%%%%%%%%%%%%
      \nonumber\\
       &+\frac{  m_b f_{3\gamma} }{2^{24}\times 3 \times 5^2    \pi^5} (e_d + 9 e_u) I_1[\mathcal V] I[0, 4].
\end{align}

\subsection{The obtained sum rule for the MDM of the $B^*\Sigma_b$ state}
\begin{align}\label{sumrules2}
 \mathcal R_3 (\rm{M^2},\rm{s_0}) &= F_1^{B^*\Sigma_b}(\rm{M^2},\rm{s_0})-\frac{1}{m_{B^*\Sigma_b}} F_2^{B^*\Sigma_b}(\rm{M^2},\rm{s_0}),
\end{align}
with 
\begin{align}
 F_1^{B^*\Sigma_b}(\rm{M^2},\rm{s_0})&= \frac{1 }{2^{27}\times 3 \times 5 ^2 \times 7^2   \pi^7 } (95 e_b - 78 (e_d + 4 e_u)) I[0, 5]
 %%%%%%%%%%%%%%%%%%%%%%%%%%%%%%%%%%%%%%%%%%%%%%%%%%%%%%%%%%%%%
 \nonumber\\
 & +
 \frac{ m_b \langle g_s^2 G^2\rangle \langle \bar q q \rangle }{2^{26} \times 3^6 \times 5 \pi^5}\Big[
 -248 (e_d + 4 e_u) \mathbb A[_ 0] I[0, 1] + 
 3348 (e_d + 4 e_u) I_ 4[\mathcal S] I[0, 1] - 
 4428 e_d I_ 4[\mathcal {\tilde S}] I[0, 1] \nonumber\\
 &+ 
 263 \chi e_d I[0, 2] \varphi_\gamma[u_ 0] + 
 1052 \chi e_u I[0, 2] \varphi_\gamma[u_ 0] + 
 704 (e_d + 9 e_u) f_ {3\gamma} \pi^2 I[0, 0] \psi^a[u_ 0]
 \Big]
 %%%%%%%%%%%%%%%%%%%%%%%%%%%%%%%%%%%%%%%%%%%%%%%%%%%%%%%%%%%%%
 \nonumber\\
  & +\frac{ \langle g_s^2 G^2\rangle f_ {3\gamma}}{2^{28}\times 3 ^4\times 5 \times 7   \pi^5} \Big[ 336 (e_d + 9 e_u) I_ 1[\mathcal V] - (e_d + 
    4 e_u) (28 I_ 2[\mathcal V] + 71 \psi^a[u_ 0])  \Big] I[0,2] 
    %%%%%%%%%%%%%%%%%%%%%%%%%%%%%%%%%%%%%%%%%%%%%%%%%%%%%%%%%%%%%
 \nonumber\\
  & +
 \frac{ m_b \langle \bar q q \rangle }{2^{24}\times 3 ^2\times 5 ^3 \times 7 \pi^5}\Big[
 -5600 (e_d + 
    9 e_u) f_ {3\gamma} \pi^2  I_ 1[\mathcal V] (m_ 0^2 I[0, 1] - 
    I[0, 2]) + 
 420 ((e_d + 4 e_u) (\mathbb A[u_ 0] \nonumber\\
%  \end{align}
%\begin{align}
 &- 8 I_ 4[\mathcal S]) + 
    8 e_d I_ 4[\mathcal {\tilde S}]) I[0, 3] - 
 149 \chi (e_d + 4 e_u) I[0, 4] \varphi_\gamma[u_ 0] \Big]
 %%%%%%%%%%%%%%%%%%%%%%%%%%%%%%%%%%%%%%%%%%%%%%%%%%%%%%%%%%%%%
 \nonumber\\
 & +\frac{  f_ {3\gamma}}{2^{26}\times 3 ^3\times 5 ^3 \times 7   \pi^5} (e_d + 4 e_u)  (1400 I_2[\mathcal V] + 2223  \psi^a[u_0])I[0, 4],%\\
\end{align}
\begin{align}
 F_2^{B^*\Sigma_b}(\rm{M^2},\rm{s_0})&= \frac{m_b }{2^{27}\times 3^4 \times 5 ^3 \times 7^2 \pi^7 } \Big[ -54 (e_d + 4 e_u) (88 I[0, 5] + 427 I[1, 4]) + 
 5 e_b (839 I[0, 5] + 19390 I[1, 4])\Big]
 %%%%%%%%%%%%%%%%%%%%%%%%%%%%%%%%%%%%%%%%%%%%%%%%%%%%%%%%%%%%%
 \nonumber\\
 & + \frac{m_b \langle g_s^2 G^2\rangle \langle \bar q q \rangle^2}{2^{24} \times 3^4 \pi^3} (e_d + 9 e_u) I_1[\mathcal S] I[0, 0]
       %%%%%%%%%%%%%%%%%%%%%%%%%%%%%%%%%%%%%%%%%%%%
      \nonumber\\
      &+ \frac{ m_b^2\langle g_s^2 G^2\rangle \langle \bar q q \rangle }{2^{21} \times 3^5 \pi^3} \Big[ \Big (124 (e_d + 4 e_u) \mathbb A[_ 0] - 
    45 \big ((e_d + 4 e_u) I_ 2[\mathcal S] - 
        e_d I_ 2[\mathcal {\tilde S}] - 
        8 (e_d + 4 e_u) I_ 4[\mathcal S] + 
        24 e_d I_ 4[\mathcal {\tilde S}]\nonumber\\
      & + 
        32 (e_d + 4 e_u) I_ 6[h_ {\gamma}]\big)\Big) I[0, 1] - 
 172 \chi (e_d + 4 e_u) I[0, 2] \varphi_\gamma[u_ 0] \Big]
       %%%%%%%%%%%%%%%%%%%%%%%%%%%%%%%%%%%%%%%%%%%%
      \nonumber\\
      &+ \frac{ m_b \langle \bar q q \rangle^2}{2^{18} \times 3 \times 5 \pi^3} (e_d + 9 e_u) \Big[  (m_0^2 I[0, 1] - I[0, 2])I_1[\mathcal S] + I_3[\mathcal S] I[0, 2]\Big]
       %%%%%%%%%%%%%%%%%%%%%%%%%%%%%%%%%%%%%%%%%%%%
      \nonumber\\
      &+\frac{m_b \langle g_s^2 G^2\rangle f_{3\gamma}}{2^{29}\times 3^4 \times 5 \times  \pi^5} \Big[  5 (e_d + 9 e_u) I_1[\mathcal V] + 164 (e_d + 4 e_u) I_2[\mathcal V] - 
 32 \big (5 (7 e_d + 34 e_u) I_5[\psi^a] 
 \nonumber\\
        &+ 4 (e_d + 4 e_u) \psi^a[u_0]\big)\Big] I[0,2]
      %%%%%%%%%%%%%%%%%%%%%%%%%%%%%%%%%%%%%%%%%%%%
      \nonumber\\
% \end{align}
% \begin{align}
       &+\frac{  m_b^2 \langle \bar q q \rangle }{2^{23}\times 3 ^5\times 5 ^2   \pi^5} \Big[  -9 \Big (19 (e_d + 4 e_u) \mathbb A[u_ 0] + 
    8 \big (-15 (e_d + 4 e_u) I_ 4[\mathcal S] + 
        15 e_d I_ 4[\mathcal {\tilde S}] + 
        46 (e_d + 4 e_u) 
        \nonumber\\
       & \times I_ 6[h_ {\gamma}]\big)\Big) I[0, 3] + 
 56 \chi (e_d + 4 e_u) I[0, 4] \varphi_\gamma[u_ 0] \Big]
%%%%%%%%%%%%%%%%%%%%%%%%%%%%%%%%%%%%%%%%%%%%
      \nonumber\\
       &-\frac{  m_b f_{3\gamma} }{2^{24}\times 3^3 \times 5^2    \pi^5} \Big[ 12 (e_d + 9 e_u) I_ 1[\mathcal V] + (e_d + 
    4 e_u) (8 I_ 2[\mathcal V] + 60 I_ 5[\psi^a] + 
       3 \psi^a[u_ 0])\Big] I[0, 4],
    %\nonumber\\
    \label{sumrules10}
\end{align}

\noindent where the symbols appearing in Eqs.~(\ref{sumrules})-(\ref{sumrules10}) are defined as follows: 
$e_u$, $e_d$, and $e_b$ denote the electric charges of the up, down, and bottom quarks, respectively, in units of the elementary charge $e$; 
$\langle \bar{q} q \rangle$ represents the light-quark condensate; 
$\langle g_s^2 G^2 \rangle$ is the gluon condensate; 
$f_{3\gamma}$ corresponds to a non-perturbative constant measuring the chirality-flipping soft part of a photon's structure; 
$\chi$ is the magnetic susceptibility of the quark condensate; 
and $m_0^2$ is defined through the mixed quark-gluon condensate as $\langle \bar{q} g_s \sigma_{\mu\nu} G^{\mu\nu} q \rangle = m_0^2 \langle \bar{q} q \rangle$.  
The explicit forms of the functions $I[n,m]$, $I_i[F]$, and the various photon distribution amplitudes ($\psi^a$, $\psi^v$, $h_\gamma$, $\varphi_\gamma$, $\mathbb{A}$, $\mathcal{S}$, $\tilde{\mathcal{S}}$, $\mathcal{A}$, $\mathcal{V}$, $\mathcal{T}_1$, $\mathcal{T}_2$, $\mathcal{T}_3$, $\mathcal{T}_4$) together with the integration measures and conventions, are provided in Appendix~\ref{appb}.  
The functions $I_i[\mathcal{F}]$ and $I[n,m]$ are defined as follows:
\begin{align}
 I[n,m]&= \int_{\mathcal M}^{\rm{s_0}} ds \, s^n (s-\mathcal M)^m\,e^{-s/\rm{M^2}},~
 \nonumber\\
 %  \end{align}
 %\begin{align}
 I_1[\mathcal{F}]&=\int D_{\alpha_i} \int_0^1 dv~ \mathcal{F}(\alpha_{\bar q},\alpha_q,\alpha_g) \delta'(\alpha_ q +\bar v \alpha_g-u_0),\nonumber\\
  I_2[\mathcal{F}]&=\int D_{\alpha_i} \int_0^1 dv~ \mathcal{F}(\alpha_{\bar q},\alpha_q,\alpha_g) \delta'(\alpha_{\bar q}+ v \alpha_g-u_0),\nonumber\\
    I_3[\mathcal{F}]&=\int D_{\alpha_i} \int_0^1 dv~ \mathcal{F}(\alpha_{\bar q},\alpha_q,\alpha_g) \delta(\alpha_ q +\bar v \alpha_g-u_0),\nonumber\\
   I_4[\mathcal{F}]&=\int D_{\alpha_i} \int_0^1 dv~ \mathcal{F}(\alpha_{\bar q},\alpha_q,\alpha_g) \delta(\alpha_{\bar q}+ v \alpha_g-u_0),\nonumber\\
% \end{align}
% \begin{align}
   I_5[\mathcal{F}]&=\int_0^1 du~ \mathcal{F}(u)\delta'(u-u_0),\nonumber\\
 I_6[\mathcal{F}]&=\int_0^1 du~ \mathcal{F}(u),
 \end{align}
 with $\mathcal M = 4\,m_b^2$,  and $\mathcal{F}$ being the relevant photon DAs. 

 The Borel transformations are carried out using the standard formulas given below. At the hadronic level, the transformation applied to the double-pole structure yields:
\begin{align}
 \mathcal{B}\left\{ \frac{1}{\big[ (p^2-m_i^2)((p+q)^2-m_f^2) \big]} \right\} \rightarrow e^{-m_i^2/\mathrm M_1^2 - m_f^2/\mathrm M_2^2}.
\end{align}
At the quark-gluon level, for the generic denominator structure appearing in the QCD side, one employs:
\begin{align}
 \mathcal{B}\left\{ \frac{1}{\big( m^2 - \bar u p^2 - u (p+q)^2 \big)^{\alpha}} \right\} \rightarrow (\mathrm M^2)^{(2-\alpha)} \, \delta(u-u_0) \, e^{-m^2/\mathrm M^2},
\end{align}
where the auxiliary Borel parameters are defined as:
$
 \mathrm M^2 = \frac{\mathrm M_1^2 \mathrm M_2^2}{ \mathrm M_1^2 + \mathrm M_2^2}$ and  
 $u_0 = \frac{\mathrm M_1^2}{\mathrm M_1^2 + \mathrm M_2^2}.
$

In these expressions, $\mathrm M_1^2$ and $\mathrm M_2^2$ denote the Borel parameters associated with the initial and final pentaquark channels, respectively. Given that the initial and final states are identical in the present analysis, it is natural to adopt the symmetric choice $\mathrm M_1^2 = \mathrm M_2^2 = 2 \mathrm M^2$, which leads to $u_0 = 1/2$. This selection ensures a balanced treatment of both sides of the correlation function and provides sufficient suppression of contributions from higher resonances and the continuum.

 \section{Photon Distribution Amplitudes}\label{appb}

In this appendix, we present the explicit forms of the non-perturbative matrix elements that appear in the QCD side of the light-cone sum rule calculations. These matrix elements, which involve photon DAs, are defined as $\langle \gamma(q) | \bar{q}(x) \Gamma_i q(0) | 0 \rangle$ and $\langle \gamma(q) | \bar{q}(x) \Gamma_i G_{\mu\nu} q(0) | 0 \rangle$, where $\Gamma_i$ denotes the relevant Dirac matrix structures. Following the conventions established in~\cite{Ball:2002ps}, these matrix elements can be expressed as:

\begin{eqnarray}
\langle \gamma(q) \vert  \bar q(x) \gamma_\mu q(0) \vert 0 \rangle
&=& e_q f_{3 \gamma} \left(\varepsilon_\mu - q_\mu \frac{\varepsilon
x}{q x} \right) \int_0^1 du e^{i \bar u q x} \psi^v(u), \\
\langle \gamma(q) \vert \bar q(x) \gamma_\mu \gamma_5 q(0) \vert 0
\rangle  &=& - \frac{1}{4} e_q f_{3 \gamma} \epsilon_{\mu \nu \alpha
\beta } \varepsilon^\nu q^\alpha x^\beta \int_0^1 du e^{i \bar u q
x} \psi^a(u), \\
\langle \gamma(q) \vert  \bar q(x) \sigma_{\mu \nu} q(0) \vert  0
\rangle  &=& -i e_q \langle \bar q q \rangle (\varepsilon_\mu q_\nu - \varepsilon_\nu
q_\mu) \int_0^1 du e^{i \bar u qx} \left(\chi \varphi_\gamma(u) +
\frac{x^2}{16} \mathbb{A} (u) \right) \nonumber \\ 
&&-\frac{i}{2(qx)}  e_q \bar qq \left[x_\nu \left(\varepsilon_\mu - q_\mu
\frac{\varepsilon x}{qx}\right) - x_\mu \left(\varepsilon_\nu -
q_\nu \frac{\varepsilon x}{q x}\right) \right] \int_0^1 du e^{i \bar
u q x} h_\gamma(u),  \\
\langle \gamma(q) | \bar q(x) g_s G_{\mu \nu} (v x) q(0) \vert 0
\rangle &=& -i e_q \langle \bar q q \rangle \left(\varepsilon_\mu q_\nu - \varepsilon_\nu
q_\mu \right) \int {\cal D}\alpha_i e^{i (\alpha_{\bar q} + v
\alpha_g) q x} {\cal S}(\alpha_i),  \\
\langle \gamma(q) | \bar q(x) g_s \tilde G_{\mu \nu}(v
x) i \gamma_5  q(0) \vert 0 \rangle &=& -i e_q \langle \bar q q \rangle \left(\varepsilon_\mu q_\nu -
\varepsilon_\nu q_\mu \right) \int {\cal D}\alpha_i e^{i
(\alpha_{\bar q} + v \alpha_g) q x} \tilde {\cal S}(\alpha_i),  \\
\langle \gamma(q) \vert \bar q(x) g_s \tilde G_{\mu \nu}(v x)
\gamma_\alpha \gamma_5 q(0) \vert 0 \rangle &=& e_q f_{3 \gamma}
q_\alpha (\varepsilon_\mu q_\nu - \varepsilon_\nu q_\mu) \int {\cal
D}\alpha_i e^{i (\alpha_{\bar q} + v \alpha_g) q x} {\cal
A}(\alpha_i), \\
\langle \gamma(q) \vert \bar q(x) g_s G_{\mu \nu}(v x) i
\gamma_\alpha q(0) \vert 0 \rangle &=& e_q f_{3 \gamma} q_\alpha
(\varepsilon_\mu q_\nu - \varepsilon_\nu q_\mu) \int {\cal
D}\alpha_i e^{i (\alpha_{\bar q} + v \alpha_g) q x} {\cal
V}(\alpha_i), \\
\langle \gamma(q) \vert \bar q(x)
\sigma_{\alpha \beta} g_s G_{\mu \nu}(v x) q(0) \vert 0 \rangle  &=&
e_q \langle \bar q q \rangle \left\{
        \left[\left(\varepsilon_\mu - q_\mu \frac{\varepsilon x}{q x}\right)\left(g_{\alpha \nu} -
        \frac{1}{qx} (q_\alpha x_\nu + q_\nu x_\alpha)\right) \right. \right. q_\beta \nonumber \\
 && -
         \left(\varepsilon_\mu - q_\mu \frac{\varepsilon x}{q x}\right)\left(g_{\beta \nu} -
        \frac{1}{qx} (q_\beta x_\nu + q_\nu x_\beta)\right) q_\alpha
\nonumber \\
&&-
         \left(\varepsilon_\nu - q_\nu \frac{\varepsilon x}{q x}\right)\left(g_{\alpha \mu} -
        \frac{1}{qx} (q_\alpha x_\mu + q_\mu x_\alpha)\right) q_\beta \nonumber \\
&&+
         \left. \left(\varepsilon_\nu - q_\nu \frac{\varepsilon x}{q.x}\right)\left( g_{\beta \mu} -
        \frac{1}{qx} (q_\beta x_\mu + q_\mu x_\beta)\right) q_\alpha \right]
   \int {\cal D}\alpha_i e^{i (\alpha_{\bar q} + v \alpha_g) qx} {\cal T}_1(\alpha_i) \nonumber \\
%\end{eqnarray}
%\begin{eqnarray}
 &&+
        \left[\left(\varepsilon_\alpha - q_\alpha \frac{\varepsilon x}{qx}\right)
        \left(g_{\mu \beta} - \frac{1}{qx}(q_\mu x_\beta + q_\beta x_\mu)\right) \right. q_\nu \nonumber \\
        &&
-
         \left(\varepsilon_\alpha - q_\alpha \frac{\varepsilon x}{qx}\right)
        \left(g_{\nu \beta} - \frac{1}{qx}(q_\nu x_\beta + q_\beta x_\nu)\right)  q_\mu \nonumber \\
          && -
         \left(\varepsilon_\beta - q_\beta \frac{\varepsilon x}{qx}\right)
        \left(g_{\mu \alpha} - \frac{1}{qx}(q_\mu x_\alpha + q_\alpha x_\mu)\right) q_\nu \nonumber \\
     %     \end{eqnarray}
    %\begin{eqnarray}
      &&+
         \left. \left(\varepsilon_\beta - q_\beta \frac{\varepsilon x}{qx}\right)
        \left(g_{\nu \alpha} - \frac{1}{qx}(q_\nu x_\alpha + q_\alpha x_\nu) \right) q_\mu
        \right]      
    \int {\cal D} \alpha_i e^{i (\alpha_{\bar q} + v \alpha_g) qx} {\cal T}_2(\alpha_i) \nonumber \\
&&+\frac{1}{qx} (q_\mu x_\nu - q_\nu x_\mu)
        (\varepsilon_\alpha q_\beta - \varepsilon_\beta q_\alpha)
    \int {\cal D} \alpha_i e^{i (\alpha_{\bar q} + v \alpha_g) qx} {\cal T}_3(\alpha_i) \nonumber \\ &&+
        \left. \frac{1}{qx} (q_\alpha x_\beta - q_\beta x_\alpha)
        (\varepsilon_\mu q_\nu - \varepsilon_\nu q_\mu)
    \int {\cal D} \alpha_i e^{i (\alpha_{\bar q} + v \alpha_g) qx} {\cal T}_4(\alpha_i)
                        \right\}.
\end{eqnarray}

In these expressions, the functions $\varphi_\gamma(u)$, $\psi^v(u)$, $\psi^a(u)$, ${\cal A}(\alpha_i)$, ${\cal V}(\alpha_i)$, $h_\gamma(u)$, $\mathbb{A}(u)$, ${\cal S}(\alpha_i)$, $\tilde {\cal S}(\alpha_i)$, ${\cal T}_1(\alpha_i)$, ${\cal T}_2(\alpha_i)$, ${\cal T}_3(\alpha_i)$, and ${\cal T}_4(\alpha_i)$ represent the photon DAs of different twists. Specifically, $\varphi_\gamma(u)$ corresponds to the leading twist-2 DA, while $\psi^v(u)$, $\psi^a(u)$, ${\cal A}(\alpha_i)$, and ${\cal V}(\alpha_i)$ are associated with twist-3 contributions. The remaining functions, namely $h_\gamma(u)$, $\mathbb{A}(u)$, ${\cal S}(\alpha_i)$, $\tilde {\cal S}(\alpha_i)$, ${\cal T}_1(\alpha_i)$, ${\cal T}_2(\alpha_i)$, ${\cal T}_3(\alpha_i)$, and ${\cal T}_4(\alpha_i)$, encode the twist-4 effects.

The integration measure ${\cal D} \alpha_i$ appearing in the multi-variable distributions is defined as:
\begin{eqnarray}
\int {\cal D} \alpha_i = \int_0^1 d \alpha_{\bar q} \int_0^1 d
\alpha_q \int_0^1 d \alpha_g \,\delta(1-\alpha_{\bar
q}-\alpha_q-\alpha_g). 
\end{eqnarray}

The explicit functional forms of the photon DAs employed in our analysis are given below:
\begin{eqnarray}
\varphi_\gamma(u) &=& 6 u \bar u \left( 1 + \varphi_2(\mu)
C_2^{\frac{3}{2}}(u - \bar u) \right), \nonumber \\
\psi^v(u) &=& 3 \left(3 (2 u - 1)^2 -1 \right)+\frac{3}{64} \left(15
w^V_\gamma - 5 w^A_\gamma\right)
                        \left(3 - 30 (2 u - 1)^2 + 35 (2 u -1)^4
                        \right), \nonumber \\
   %                         \end{eqnarray}
   %  \begin{eqnarray}
\psi^a(u) &=& \left(1- (2 u -1)^2\right)\left(5 (2 u -1)^2 -1\right)
\frac{5}{2}
    \left(1 + \frac{9}{16} w^V_\gamma - \frac{3}{16} w^A_\gamma
    \right), \nonumber \\
h_\gamma(u) &=& - 10 \left(1 + 2 \kappa^+\right) C_2^{\frac{1}{2}}(u
- \bar u), \nonumber \\
\mathbb{A}(u) &=& 40 u^2 \bar u^2 \left(3 \kappa - \kappa^+
+1\right)  +
        8 (\zeta_2^+ - 3 \zeta_2) \left[u \bar u (2 + 13 u \bar u) +  2 u^3 (10 -15 u + 6 u^2) \ln(u)\right. \nonumber \\ &&  \left.
                + 2 \bar u^3 (10 - 15 \bar u + 6 \bar u^2)
        \ln(\bar u) \right], \nonumber \\
{\cal A}(\alpha_i) &=& 360 \alpha_q \alpha_{\bar q} \alpha_g^2
        \left(1 + w^A_\gamma \frac{1}{2} (7 \alpha_g - 3)\right), \nonumber \\
{\cal V}(\alpha_i) &=& 540 w^V_\gamma (\alpha_q - \alpha_{\bar q})
\alpha_q \alpha_{\bar q}
                \alpha_g^2, \nonumber \\
{\cal T}_1(\alpha_i) &=& -120 (3 \zeta_2 + \zeta_2^+)(\alpha_{\bar
q} - \alpha_q)
        \alpha_{\bar q} \alpha_q \alpha_g, \nonumber \\
{\cal T}_2(\alpha_i) &=& 30 \alpha_g^2 (\alpha_{\bar q} - \alpha_q)
    \left((\kappa - \kappa^+) + (\zeta_1 - \zeta_1^+)(1 - 2\alpha_g) +
    \zeta_2 (3 - 4 \alpha_g)\right), \nonumber \\
{\cal T}_3(\alpha_i) &=& - 120 (3 \zeta_2 - \zeta_2^+)(\alpha_{\bar
q} -\alpha_q)
        \alpha_{\bar q} \alpha_q \alpha_g, \nonumber \\
{\cal T}_4(\alpha_i) &=& 30 \alpha_g^2 (\alpha_{\bar q} - \alpha_q)
    \left((\kappa + \kappa^+) + (\zeta_1 + \zeta_1^+)(1 - 2\alpha_g) +
    \zeta_2 (3 - 4 \alpha_g)\right), \nonumber \\
{\cal S}(\alpha_i) &=& 30\alpha_g^2\left[(\kappa +
\kappa^+)(1-\alpha_g)+(\zeta_1 + \zeta_1^+)(1 - \alpha_g)(1 -
2\alpha_g)+\zeta_2\left(3 (\alpha_{\bar q} - \alpha_q)^2-\alpha_g(1 - \alpha_g)\right)\right], \nonumber \\
\tilde {\cal S}(\alpha_i) &=&-30\alpha_g^2\left[(\kappa -\kappa^+)(1-\alpha_g)+(\zeta_1 - \zeta_1^+)(1 - \alpha_g)(1 -
2\alpha_g)+\zeta_2 \left(3 (\alpha_{\bar q} -\alpha_q)^2-\alpha_g(1 - \alpha_g)\right)\right].
\end{eqnarray}

The numerical values of the parameters entering these DAs are taken as $\varphi_2(1~\mathrm{GeV}) = 0$, $w^V_\gamma = 3.8 \pm 1.8$, $w^A_\gamma = -2.1 \pm 1.0$, $\kappa = 0.2$, $\kappa^+ = 0$, $\zeta_1 = 0.4$, and $\zeta_2 = 0.3$, following the conventions of~\cite{Ball:2002ps}.

%\section{Acknowledgments}
% \newpage
%\section{Full unreduced spin-3/2 hadronic correlation function}\label{appc}

\section{Step-by-step derivation of the spin-3/2 hadronic 
correlation function}
\label{appc}

This appendix documents, in explicit form, every intermediate 
expression that connects the hadronic representation of the 
correlation function in Eq.~\eqref{final phenpart} to the projected 
form in Eq.~\eqref{final phenpart11}. The derivation proceeds in 
four stages, each presented in a dedicated subsection:
\begin{enumerate}
\item[(i)] Application of the polarization sums of 
Eqs.~\eqref{eq:diracsum}--\eqref{eq:raritasum} and the on-shell 
substitution $p^{2}=m_{\mathrm{P_{bb}^{*}}}^{2}$, with no further 
simplification (Sec.~\ref{appc1}). The output is a Lorentz 
decomposition in which every term generated by the Dirac algebra 
appears explicitly.
\item[(ii)] Imposition of the real-photon kinematic conditions 
$q^{2}=0$ and $\varepsilon\!\cdot\! q=0$ on the unreduced 
expressions of Sec.~\ref{appc1} (Sec.~\ref{appc2}). The output is 
the gauge-fixed expression, which still contains Lorentz 
structures originating from the spin-$1/2$ pollution of the 
interpolating current $\mathrm{J}_{\mu}^{\mathrm{P_{bb}^{*}}}(x)$.
\item[(iii)] Application of the gamma-matrix ordering 
$\gamma_{\mu}\,\pslash\,\eslash\,\qslash\,\gamma_{\nu}$ to the 
gauge-fixed expressions of Sec.~\ref{appc2} and removal of all 
Lorentz structures beginning with $\gamma_{\mu}$, ending with 
$\gamma_{\nu}$, or proportional to $p_{\mu}$ or $p_{\nu}$, which 
are identified as spin-$1/2$ contamination 
(Sec.~\ref{appc3}). The output contains only those Lorentz 
structures that contribute to the magnetic dipole moment.
\item[(iv)] Projection of the resulting expressions onto the 
linearly independent Lorentz structures 
$g_{\mu\nu}\,\pslash\,\eslash\,\qslash$, 
$g_{\mu\nu}\,\eslash\,\qslash$, $q_{\mu}q_{\nu}\,\eslash\,\qslash$, 
and $(\varepsilon\!\cdot\! p)\,q_{\mu}q_{\nu}\,\pslash\,\qslash$ 
(Sec.~\ref{appc4}), yielding the coefficients of the form factors 
$F_{1}(q^{2})$, $F_{2}(q^{2})$, $F_{3}(q^{2})$, and $F_{4}(q^{2})$ 
entering Eq.~\eqref{final phenpart11}.
\end{enumerate}
The expressions in stages (i), (ii), and (iii) are organized by 
the form factor to which they are proportional, so that each 
$F_{i}(q^{2})$ contribution can be traced independently across the 
derivation. All quantities are expressed in terms of 
$m_{\mathrm{P_{bb}^{*}}}$, with the on-shell substitution 
$p^{2}=m_{\mathrm{P_{bb}^{*}}}^{2}$ applied throughout.

\subsection{Stage (i): unreduced form after polarization sums and 
on-shell substitution}
\label{appc1}

The hadronic correlation function 
$\Pi^{\mathrm{Had}}_{\mu\nu}(p,q)$ of Eq.~\eqref{final phenpart} 
contains the Rarita-Schwinger projectors $P_{\mu\alpha}(p)$ and 
$P_{\beta\nu}(p+q)$ on either side of the photon-vertex insertion. 
Substituting the explicit form of these projectors 
[Eq.~\eqref{eq:raritasum}], expanding the products, and applying 
the on-shell condition $p^{2}=m_{\mathrm{P_{bb}^{*}}}^{2}$, the 
correlation function decomposes into four sectors, each 
proportional to a single form factor $F_{i}(q^{2})$:
\begin{align}
\Pi^{\mathrm{Had}}_{\mu\nu}(p,q)=-
\frac{\lambda^{2}_{\mathrm{P_{bb}^{*}}}}
{[(p+q)^{2}-m^{2}_{\mathrm{P_{bb}^{*}}}]\,
[p^{2}-m^{2}_{\mathrm{P_{bb}^{*}}}]}\;
\sum_{i=1}^{4} F_{i}(q^{2})\,\mathcal{T}^{(i)}_{\mu\nu}(p,q,
\varepsilon),
\label{eq:appc_decomposition}
\end{align}
where $\mathcal{T}^{(i)}_{\mu\nu}$ denotes the tensor structure 
multiplying $F_{i}(q^{2})$. The four subsections below collect 
these tensor structures one by one, in their fully unreduced form. 
No gamma-matrix reordering, projection, or kinematic constraint 
beyond $p^{2}=m_{\mathrm{P_{bb}^{*}}}^{2}$ has been applied at this 
stage. The terms are grouped by the Dirac chain they contain (no 
chain, single slash, double slash, etc.) to facilitate the 
identification of the structures that survive the subsequent 
ordering and gauge-fixing steps.

\subsection{Lorentz structures proportional to $F_1(q^2)$}

\begin{align}
{F}_1(q^2) & \propto\,\Bigg\{ 
\frac{8m_{\mathrm{P_{bb}^*}}}{3}\,\varepsilon_\nu\, p_\mu 
+ \frac{4(\varepsilon\!\cdot\! p)\, p_\mu p_\nu}{3m_{\mathrm{P_{bb}^*}}} 
- \frac{4(\varepsilon\!\cdot\! q)\, p_\mu p_\nu}{3m_{\mathrm{P_{bb}^*}}} 
+ \frac{4(\varepsilon\!\cdot\! p)\, p_\mu q_\nu}{3m_{\mathrm{P_{bb}^*}}} 
- \frac{4(\varepsilon\!\cdot\! q)\, p_\mu q_\nu}{3m_{\mathrm{P_{bb}^*}}} - \frac{2\, p_\mu p_\nu\, q^2\,\eslash}{3m_{\mathrm{P_{bb}^*}}^2} - \frac{2\, p_\mu\, q^2\, q_\nu\,\eslash}{3m_{\mathrm{P_{bb}^*}}^2} \nonumber\\
&\quad 
- \frac{4}{3}\,\varepsilon_\nu\, p_\mu\,\pslash 
+ \frac{4(\varepsilon\!\cdot\! p)\, p_\mu p_\nu\,\pslash}{3m_{\mathrm{P_{bb}^*}}^2} 
- \frac{4(\varepsilon\!\cdot\! q)\, p_\mu p_\nu\,\pslash}{3m_{\mathrm{P_{bb}^*}}^2} 
- \frac{8(\varepsilon\!\cdot\! p)\,(p\!\cdot\! q)\, p_\mu p_\nu\,\pslash}{3m_{\mathrm{P_{bb}^*}}^4} 
+ \frac{4(\varepsilon\!\cdot\! p)\, p_\nu q_\mu\,\pslash}{m_{\mathrm{P_{bb}^*}}^2}
- \frac{4(\varepsilon\!\cdot\! q)\, p_\mu q_\nu\,\pslash}{3m_{\mathrm{P_{bb}^*}}^2} 
\nonumber\\
&\quad  
- \frac{8(\varepsilon\!\cdot\! p)\,(p\!\cdot\! q)\, p_\mu q_\nu\,\pslash}{3m_{\mathrm{P_{bb}^*}}^4} 
+ \frac{4(\varepsilon\!\cdot\! p)\, q_\mu q_\nu\,\pslash}{m_{\mathrm{P_{bb}^*}}^2} 
- 6(\varepsilon\!\cdot\! p)\, g_{\mu\nu}\,\pslash 
+ \frac{4(\varepsilon\!\cdot\! q)\, p_\mu p_\nu\,\qslash}{3m_{\mathrm{P_{bb}^*}}^2} 
+ \frac{4(\varepsilon\!\cdot\! q)\, p_\mu q_\nu\,\qslash}{3m_{\mathrm{P_{bb}^*}}^2} 
+ 2(\varepsilon\!\cdot\! p)\, p_\nu\,\gamma_\mu 
\nonumber\\
&\quad 
+ \frac{4(\varepsilon\!\cdot\! q)\, p_\nu}{3}\,\gamma_\mu 
+ 2(\varepsilon\!\cdot\! p)\, q_\nu\,\gamma_\mu 
+ \frac{4(\varepsilon\!\cdot\! q)\, q_\nu}{3}\,\gamma_\mu 
+ 2(\varepsilon\!\cdot\! p)\, p_\mu\,\gamma_\nu 
+ \frac{4\, p_\mu p_\nu}{3m_{\mathrm{P_{bb}^*}}}\,\eslash\,\pslash 
- \frac{4(p\!\cdot\! q)\, p_\mu p_\nu}{3m_{\mathrm{P_{bb}^*}}^3}\,\eslash\,\pslash 
\nonumber\\
&\quad 
+ \frac{2\, p_\nu q_\mu}{m_{\mathrm{P_{bb}^*}}}\,\eslash\,\pslash 
+ \frac{2\, p_\mu q_\nu}{3m_{\mathrm{P_{bb}^*}}}\,\eslash\,\pslash 
- \frac{4(p\!\cdot\! q)\, p_\mu q_\nu}{3m_{\mathrm{P_{bb}^*}}^3}\,\eslash\,\pslash 
+ \frac{2\, q_\mu q_\nu}{m_{\mathrm{P_{bb}^*}}}\,\eslash\,\pslash 
- 3m_{\mathrm{P_{bb}^*}}\, g_{\mu\nu}\,\eslash\,\pslash 
+ \frac{4\, p_\mu p_\nu}{m_{\mathrm{P_{bb}^*}}}\,\eslash\,\qslash 
- \frac{4(p\!\cdot\! q)\, p_\mu p_\nu}{3m_{\mathrm{P_{bb}^*}}^3}\,\eslash\,\qslash 
\nonumber\\
&\quad 
+ \frac{2\, p_\nu q_\mu}{m_{\mathrm{P_{bb}^*}}}\,\eslash\,\qslash 
+ \frac{2\, p_\mu q_\nu}{m_{\mathrm{P_{bb}^*}}}\,\eslash\,\qslash 
- \frac{4(p\!\cdot\! q)\, p_\mu q_\nu}{3m_{\mathrm{P_{bb}^*}}^3}\,\eslash\,\qslash 
+ \frac{2\, q_\mu q_\nu}{m_{\mathrm{P_{bb}^*}}}\,\eslash\,\qslash 
- 3m_{\mathrm{P_{bb}^*}}\, g_{\mu\nu}\,\eslash\,\qslash 
- \frac{2m_{\mathrm{P_{bb}^*}}\, p_\mu}{3}\,\eslash\,\gamma_\nu 
- \frac{p_\mu\, q^2}{3m_{\mathrm{P_{bb}^*}}}\,\eslash\,\gamma_\nu 
\nonumber\\
&\quad 
+ \frac{4\, p_\mu p_\nu}{3m_{\mathrm{P_{bb}^*}}}\,\pslash\,\eslash  
- \frac{2\, p_\mu p_\nu\, q^2}{3m_{\mathrm{P_{bb}^*}}^3}\,\pslash\,\eslash 
+ \frac{2\, p_\nu q_\mu}{m_{\mathrm{P_{bb}^*}}}\,\pslash\,\eslash 
+ \frac{2\, p_\mu q_\nu}{3m_{\mathrm{P_{bb}^*}}}\,\pslash\,\eslash 
- \frac{2\, p_\mu\, q^2 q_\nu}{3m_{\mathrm{P_{bb}^*}}^3}\,\pslash\,\eslash 
+ \frac{2\, q_\mu q_\nu}{m_{\mathrm{P_{bb}^*}}}\,\pslash\,\eslash 
- 3m_{\mathrm{P_{bb}^*}}\, g_{\mu\nu}\,\pslash\,\eslash 
\nonumber\\
&\quad - \frac{4m_{\mathrm{P_{bb}^*}}}{3}\,\varepsilon_\nu\,\pslash\,\gamma_\mu 
- \frac{2(\varepsilon\!\cdot\! p)\, p_\nu}{3m_{\mathrm{P_{bb}^*}}}\,\pslash\,\gamma_\mu 
+ \frac{4(\varepsilon\!\cdot\! q)\, p_\nu}{3m_{\mathrm{P_{bb}^*}}}\,\pslash\,\gamma_\mu 
- \frac{4(\varepsilon\!\cdot\! p)\,(p\!\cdot\! q)\, p_\nu}{3m_{\mathrm{P_{bb}^*}}^3}\,\pslash\,\gamma_\mu 
- \frac{4(\varepsilon\!\cdot\! p)\, q_\nu}{3m_{\mathrm{P_{bb}^*}}}\,\pslash\,\gamma_\mu 
\nonumber\\
&\quad + \frac{4(\varepsilon\!\cdot\! q)\, q_\nu}{3m_{\mathrm{P_{bb}^*}}}\,\pslash\,\gamma_\mu 
- \frac{4(\varepsilon\!\cdot\! p)\,(p\!\cdot\! q)\, q_\nu}{3m_{\mathrm{P_{bb}^*}}^3}\,\pslash\,\gamma_\mu 
- \frac{2(\varepsilon\!\cdot\! p)\, p_\mu}{3m_{\mathrm{P_{bb}^*}}}\,\pslash\,\gamma_\nu 
- \frac{4(\varepsilon\!\cdot\! p)\,(p\!\cdot\! q)\, p_\mu}{3m_{\mathrm{P_{bb}^*}}^3}\,\pslash\,\gamma_\nu 
\nonumber\\
&\quad + \frac{2(\varepsilon\!\cdot\! p)\, q_\mu}{m_{\mathrm{P_{bb}^*}}}\,\pslash\,\gamma_\nu 
+ \frac{4\, p_\mu p_\nu}{3m_{\mathrm{P_{bb}^*}}}\,\qslash\,\eslash 
+ \frac{4\, p_\mu q_\nu}{3m_{\mathrm{P_{bb}^*}}}\,\qslash\,\eslash 
+ \frac{2(\varepsilon\!\cdot\! q)\, p_\mu}{3m_{\mathrm{P_{bb}^*}}}\,\qslash\,\gamma_\nu 
+ 2m_{\mathrm{P_{bb}^*}}\, p_\nu\,\gamma_\mu\,\eslash 
\nonumber%\\
\end{align}

\begin{align}
&\quad - \frac{4(p\!\cdot\! q)\, p_\nu}{3m_{\mathrm{P_{bb}^*}}}\,\gamma_\mu\,\eslash 
+ \frac{2\, p_\nu\, q^2}{3m_{\mathrm{P_{bb}^*}}}\,\gamma_\mu\,\eslash 
- \frac{4(p\!\cdot\! q)\, q_\nu}{3m_{\mathrm{P_{bb}^*}}}\,\gamma_\mu\,\eslash 
+ \frac{2\, q^2 q_\nu}{3m_{\mathrm{P_{bb}^*}}}\,\gamma_\mu\,\eslash 
- \frac{4(\varepsilon\!\cdot\! p)\, p_\nu}{3m_{\mathrm{P_{bb}^*}}}\,\gamma_\mu\,\pslash 
- \frac{4(\varepsilon\!\cdot\! p)\, q_\nu}{3m_{\mathrm{P_{bb}^*}}}\,\gamma_\mu\,\pslash 
\nonumber\\
&\quad 
- \frac{4(\varepsilon\!\cdot\! q)\, p_\nu}{3m_{\mathrm{P_{bb}^*}}}\,\gamma_\mu\,\qslash 
- \frac{4(\varepsilon\!\cdot\! q)\, q_\nu}{3m_{\mathrm{P_{bb}^*}}}\,\gamma_\mu\,\qslash 
- 2m_{\mathrm{P_{bb}^*}}\, p_\mu\,\gamma_\nu\,\eslash 
- p_\nu\,\eslash\,\pslash\,\gamma_\mu 
- q_\nu\,\eslash\,\pslash\,\gamma_\mu 
+ \frac{p_\mu}{3}\,\eslash\,\pslash\,\gamma_\nu 
\nonumber\\
&\quad 
- \frac{2(p\!\cdot\! q)\, p_\mu}{3m_{\mathrm{P_{bb}^*}}^2}\,\eslash\,\pslash\,\gamma_\nu 
+ q_\mu\,\eslash\,\pslash\,\gamma_\nu 
+ \frac{2\, p_\mu p_\nu}{3m_{\mathrm{P_{bb}^*}}^2}\,\eslash\,\qslash\,\pslash 
+ \frac{2\, p_\mu q_\nu}{3m_{\mathrm{P_{bb}^*}}^2}\,\eslash\,\qslash\,\pslash 
- p_\nu\,\eslash\,\qslash\,\gamma_\mu 
- q_\nu\,\eslash\,\qslash\,\gamma_\mu 
\nonumber\\
&\quad + \frac{4\, p_\mu}{3}\,\eslash\,\qslash\,\gamma_\nu 
- \frac{2(p\!\cdot\! q)\, p_\mu}{3m_{\mathrm{P_{bb}^*}}^2}\,\eslash\,\qslash\,\gamma_\nu 
+ q_\mu\,\eslash\,\qslash\,\gamma_\nu 
+ \frac{4\, p_\mu p_\nu}{3m_{\mathrm{P_{bb}^*}}^2}\,\pslash\,\eslash\,\qslash 
- \frac{4(p\!\cdot\! q)\, p_\mu p_\nu}{3m_{\mathrm{P_{bb}^*}}^4}\,\pslash\,\eslash\,\qslash 
+ \frac{2\, p_\nu q_\mu}{m_{\mathrm{P_{bb}^*}}^2}\,\pslash\,\eslash\,\qslash 
\nonumber\\
&\quad 
+ \frac{2\, p_\mu q_\nu}{3m_{\mathrm{P_{bb}^*}}^2}\,\pslash\,\eslash\,\qslash 
- \frac{4(p\!\cdot\! q)\, p_\mu q_\nu}{3m_{\mathrm{P_{bb}^*}}^4}\,\pslash\,\eslash\,\qslash 
+ \frac{2\, q_\mu q_\nu}{m_{\mathrm{P_{bb}^*}}^2}\,\pslash\,\eslash\,\qslash 
- 3\, g_{\mu\nu}\,\pslash\,\eslash\,\qslash 
- p_\nu\,\pslash\,\eslash\,\gamma_\mu 
- q_\nu\,\pslash\,\eslash\,\gamma_\mu 
\nonumber\\
&\quad 
+ p_\mu\,\pslash\,\eslash\,\gamma_\nu 
+ \frac{2(p\!\cdot\! q)\, p_\mu}{3m_{\mathrm{P_{bb}^*}}^2}\,\pslash\,\eslash\,\gamma_\nu 
- \frac{p_\mu\, q^2}{3m_{\mathrm{P_{bb}^*}}^2}\,\pslash\,\eslash\,\gamma_\nu 
+ q_\mu\,\pslash\,\eslash\,\gamma_\nu 
- \frac{2\, p_\mu p_\nu}{3m_{\mathrm{P_{bb}^*}}^2}\,\pslash\,\qslash\,\eslash 
- \frac{2\, p_\mu q_\nu}{3m_{\mathrm{P_{bb}^*}}^2}\,\pslash\,\qslash\,\eslash 
\nonumber\\
%\end{align}
%\begin{align}
&\quad 
- \frac{2(\varepsilon\!\cdot\! p)\, p_\mu}{3m_{\mathrm{P_{bb}^*}}^2}\,\pslash\,\qslash\,\gamma_\nu 
+ \frac{2(\varepsilon\!\cdot\! q)\, p_\mu}{3m_{\mathrm{P_{bb}^*}}^2}\,\pslash\,\qslash\,\gamma_\nu 
- \frac{p_\nu}{3}\,\gamma_\mu\,\pslash\,\eslash 
+ \frac{4(p\!\cdot\! q)\, p_\nu}{3m_{\mathrm{P_{bb}^*}}^2}\,\gamma_\mu\,\pslash\,\eslash 
+ \frac{2\, p_\nu\, q^2}{3m_{\mathrm{P_{bb}^*}}^2}\,\gamma_\mu\,\pslash\,\eslash 
\nonumber\\
&\quad 
+ \frac{4(p\!\cdot\! q)\, q_\nu}{3m_{\mathrm{P_{bb}^*}}^2}\,\gamma_\mu\,\pslash\,\eslash 
+ \frac{2\, q^2 q_\nu}{3m_{\mathrm{P_{bb}^*}}^2}\,\gamma_\mu\,\pslash\,\eslash 
- \frac{2(\varepsilon\!\cdot\! p)\, p_\nu}{m_{\mathrm{P_{bb}^*}}^2}\,\gamma_\mu\,\pslash\,\qslash 
- \frac{4(\varepsilon\!\cdot\! q)\, p_\nu}{3m_{\mathrm{P_{bb}^*}}^2}\,\gamma_\mu\,\pslash\,\qslash 
- \frac{2(\varepsilon\!\cdot\! p)\, q_\nu}{m_{\mathrm{P_{bb}^*}}^2}\,\gamma_\mu\,\pslash\,\qslash 
\nonumber\\
&\quad 
- \frac{4(\varepsilon\!\cdot\! q)\, q_\nu}{3m_{\mathrm{P_{bb}^*}}^2}\,\gamma_\mu\,\pslash\,\qslash 
+ 2(\varepsilon\!\cdot\! p)\,\gamma_\mu\,\pslash\,\gamma_\nu 
- \frac{2(\varepsilon\!\cdot\! p)\,(p\!\cdot\! q)}{3m_{\mathrm{P_{bb}^*}}^2}\,\gamma_\mu\,\pslash\,\gamma_\nu 
+ p_\mu\,\gamma_\nu\,\pslash\,\eslash 
+ 2(\varepsilon\!\cdot\! p)\,\gamma_\mu\,\gamma_\nu\,\pslash 
\nonumber\\
&\quad 
+ \frac{2\, p_\mu p_\nu}{3m_{\mathrm{P_{bb}^*}}^2}\,\qslash\,\eslash\,\pslash 
+ \frac{2\, p_\mu q_\nu}{3m_{\mathrm{P_{bb}^*}}^2}\,\qslash\,\eslash\,\pslash 
+ \frac{2\, p_\mu}{3}\,\qslash\,\eslash\,\gamma_\nu 
+ \frac{p_\nu}{3}\,\gamma_\mu\,\eslash\,\pslash 
- \frac{2(p\!\cdot\! q)\, p_\nu}{3m_{\mathrm{P_{bb}^*}}^2}\,\gamma_\mu\,\eslash\,\pslash 
- \frac{2(p\!\cdot\! q)\, q_\nu}{3m_{\mathrm{P_{bb}^*}}^2}\,\gamma_\mu\,\eslash\,\pslash 
\nonumber\\
&\quad + \frac{2\, p_\nu}{3}\,\gamma_\mu\,\eslash\,\qslash 
- \frac{2(p\!\cdot\! q)\, p_\nu}{3m_{\mathrm{P_{bb}^*}}^2}\,\gamma_\mu\,\eslash\,\qslash 
- \frac{q_\nu}{3}\,\gamma_\mu\,\eslash\,\qslash 
- \frac{2(p\!\cdot\! q)\, q_\nu}{3m_{\mathrm{P_{bb}^*}}^2}\,\gamma_\mu\,\eslash\,\qslash 
- \frac{2(p\!\cdot\! q)}{3}\,\gamma_\mu\,\eslash\,\gamma_\nu 
+ \frac{q^2}{3}\,\gamma_\mu\,\eslash\,\gamma_\nu 
\nonumber\\
&\quad - \frac{2(\varepsilon\!\cdot\! p)}{3}\,\gamma_\mu\,\pslash\,\gamma_\nu 
- \frac{2(\varepsilon\!\cdot\! q)}{3}\,\gamma_\mu\,\qslash\,\gamma_\nu 
- p_\mu\,\gamma_\nu\,\eslash\,\pslash 
- p_\mu\,\gamma_\nu\,\eslash\,\qslash 
- \frac{2\, p_\mu}{3}\,\gamma_\nu\,\pslash\,\eslash 
- \frac{2\, p_\mu}{3}\,\gamma_\nu\,\qslash\,\eslash 
\nonumber\\
&\quad 
+ m_{\mathrm{P_{bb}^*}}\,\eslash\,\pslash\,\gamma_\nu\,\gamma_\mu 
- \frac{2\, p_\mu}{3m_{\mathrm{P_{bb}^*}}}\,\eslash\,\qslash\,\gamma_\nu\,\pslash 
+ m_{\mathrm{P_{bb}^*}}\,\eslash\,\qslash\,\gamma_\nu\,\gamma_\mu 
- \frac{p_\nu}{m_{\mathrm{P_{bb}^*}}}\,\pslash\,\eslash\,\qslash\,\gamma_\mu 
- \frac{q_\nu}{m_{\mathrm{P_{bb}^*}}}\,\pslash\,\eslash\,\qslash\,\gamma_\mu 
+ \frac{2\, p_\mu}{3m_{\mathrm{P_{bb}^*}}}\,\pslash\,\eslash\,\qslash\,\gamma_\nu \nonumber\\
&\quad 
- \frac{2(p\!\cdot\! q)\, p_\mu}{3m_{\mathrm{P_{bb}^*}}^3}\,\pslash\,\eslash\,\qslash\,\gamma_\nu 
+ \frac{q_\mu}{m_{\mathrm{P_{bb}^*}}}\,\pslash\,\eslash\,\qslash\,\gamma_\nu 
+ m\,\pslash\,\eslash\,\gamma_\nu\,\gamma_\mu 
- \frac{p_\mu}{3m_{\mathrm{P_{bb}^*}}}\,\pslash\,\qslash\,\eslash\,\gamma_\nu 
- \frac{p_\nu}{3m_{\mathrm{P_{bb}^*}}}\,\pslash\,\gamma_\mu\,\eslash\,\qslash 
- \frac{2(p\!\cdot\! q)\, p_\nu}{3m_{\mathrm{P_{bb}^*}}^3}\,\pslash\,\gamma_\mu\,\eslash\,\qslash 
\nonumber\\
&\quad 
- \frac{2\, q_\nu}{3m_{\mathrm{P_{bb}^*}}}\,\pslash\,\gamma_\mu\,\eslash\,\qslash 
- \frac{2(p\!\cdot\! q)\, q_\nu}{3m_{\mathrm{P_{bb}^*}}^3}\,\pslash\,\gamma_\mu\,\eslash\,\qslash 
+ \frac{m_{\mathrm{P_{bb}^*}}}{3}\,\pslash\,\gamma_\mu\,\eslash\,\gamma_\nu 
+ \frac{(p\!\cdot\! q)}{m_{\mathrm{P_{bb}^*}}}\,\pslash\,\gamma_\mu\,\eslash\,\gamma_\nu 
+ \frac{q^2}{3m_{\mathrm{P_{bb}^*}}}\,\pslash\,\gamma_\mu\,\eslash\,\gamma_\nu 
\nonumber\\
&\quad 
- \frac{2\, p_\nu}{3m_{\mathrm{P_{bb}^*}}}\,\pslash\,\gamma_\mu\,\qslash\,\eslash 
- \frac{2\, q_\nu}{3m_{\mathrm{P_{bb}^*}}}\,\pslash\,\gamma_\mu\,\qslash\,\eslash 
- \frac{4(\varepsilon\!\cdot\! p)}{3m_{\mathrm{P_{bb}^*}}}\,\pslash\,\gamma_\mu\,\qslash\,\gamma_\nu 
- \frac{2(\varepsilon\!\cdot\! q)}{3m_{\mathrm{P_{bb}^*}}}\,\pslash\,\gamma_\mu\,\qslash\,\gamma_\nu 
+m_{\mathrm{P_{bb}^*}}\,\pslash\,\gamma_\mu\,\gamma_\nu\,\eslash 
\nonumber\\
&\quad 
- \frac{p_\mu}{m_{\mathrm{P_{bb}^*}}}\,\pslash\,\gamma_\nu\,\eslash\,\qslash 
- \frac{2\, p_\mu}{3m_{\mathrm{P_{bb}^*}}}\,\pslash\,\gamma_\nu\,\qslash\,\eslash 
+ \frac{p_\mu}{3m_{\mathrm{P_{bb}^*}}}\,\qslash\,\eslash\,\pslash\,\gamma_\nu 
- \frac{m_{\mathrm{P_{bb}^*}}}{3}\,\gamma_\mu\,\eslash\,\pslash\,\gamma_\nu 
- \frac{(p\!\cdot\! q)}{3m_{\mathrm{P_{bb}^*}}}\,\gamma_\mu\,\eslash\,\pslash\,\gamma_\nu \nonumber\\
&\quad 
+ \frac{p_\nu}{3m_{\mathrm{P_{bb}^*}}}\,\gamma_\mu\,\eslash\,\qslash\,\pslash 
+ \frac{q_\nu}{3m_{\mathrm{P_{bb}^*}}}\,\gamma_\mu\,\eslash\,\qslash\,\pslash 
- \frac{m_{\mathrm{P_{bb}^*}}}{3}\,\gamma_\mu\,\eslash\,\qslash\,\gamma_\nu 
- \frac{(p\!\cdot\! q)}{3m_{\mathrm{P_{bb}^*}}}\,\gamma_\mu\,\eslash\,\qslash\,\gamma_\nu 
- \frac{2m_{\mathrm{P_{bb}^*}}}{3}\,\gamma_\mu\,\eslash\,\gamma_\nu\,\pslash 
\nonumber\\
&\quad 
- \frac{2\, p_\nu}{3m_{\mathrm{P_{bb}^*}}}\,\gamma_\mu\,\pslash\,\eslash\,\qslash 
- \frac{2\, q_\nu}{3m_{\mathrm{P_{bb}^*}}}\,\gamma_\mu\,\pslash\,\eslash\,\qslash 
- \frac{2\, p_\nu}{3m_{\mathrm{P_{bb}^*}}}\,\gamma_\mu\,\qslash\,\eslash\,\pslash 
- \frac{2\, q_\nu}{3m_{\mathrm{P_{bb}^*}}}\,\gamma_\mu\,\qslash\,\eslash\,\pslash 
+ m_{\mathrm{P_{bb}^*}}\,\gamma_\mu\,\gamma_\nu\,\eslash\,\pslash 
+ m_{\mathrm{P_{bb}^*}}\,\gamma_\mu\,\gamma_\nu\,\eslash\,\qslash 
\nonumber\\
&\quad 
+ \frac{2m_{\mathrm{P_{bb}^*}}}{3}\,\gamma_\mu\,\gamma_\nu\,\pslash\,\eslash 
+ \frac{2m_{\mathrm{P_{bb}^*}}}{3}\,\gamma_\mu\,\gamma_\nu\,\qslash\,\eslash 
+ \pslash\,\eslash\,\qslash\,\gamma_\nu\,\gamma_\mu 
- \frac{1}{3}\,\pslash\,\gamma_\mu\,\eslash\,\qslash\,\gamma_\nu 
- \frac{(p\!\cdot\! q)}{3m_{\mathrm{P_{bb}^*}}^2}\,\pslash\,\gamma_\mu\,\eslash\,\qslash\,\gamma_\nu 
- \frac{1}{3}\,\pslash\,\gamma_\mu\,\qslash\,\eslash\,\gamma_\nu \nonumber\\
&\quad 
+ \pslash\,\gamma_\mu\,\gamma_\nu\,\eslash\,\qslash 
+ \frac{2}{3}\,\pslash\,\gamma_\mu\,\gamma_\nu\,\qslash\,\eslash 
- \frac{1}{3}\,\gamma_\mu\,\eslash\,\qslash\,\gamma_\nu\,\pslash 
- \frac{1}{3}\,\gamma_\mu\,\pslash\,\eslash\,\qslash\,\gamma_\nu 
- \frac{1}{3}\,\gamma_\mu\,\qslash\,\eslash\,\pslash\,\gamma_\nu 
\Bigg\}.
\label{F1_appendix}
\end{align}

\subsection{Lorentz structures proportional to $F_2(q^2)$}

\begin{align}
   {F}_2 &\propto\,\Bigg\{ 
- \frac{2\,\varepsilon_\nu\, p_\mu\, q^2}{3\, m_{\mathrm{P_{bb}^*}}}
- \frac{5\, p_\mu p_\nu\, q^2\,\eslash}{3\, m_{\mathrm{P_{bb}^*}}^2}
+ \frac{2(p\!\cdot\! q)\, p_\mu p_\nu\, q^2\,\eslash}{3\, m_{\mathrm{P_{bb}^*}}^4}
- \frac{p_\nu\, q^2 q_\mu\,\eslash}{m_{\mathrm{P_{bb}^*}}^2}
- \frac{2\, p_\mu\, q^2 q_\nu\,\eslash}{3\, m_{\mathrm{P_{bb}^*}}^2} \nonumber\\
&\quad
+ \frac{2(p\!\cdot\! q)\, p_\mu\, q^2 q_\nu\,\eslash}{3\, m_{\mathrm{P_{bb}^*}}^4}
- \frac{q^2 q_\mu q_\nu\,\eslash}{m_{\mathrm{P_{bb}^*}}^2}
+ \frac{3\, q^2 g_{\mu\nu}\,\eslash}{2}
- \frac{4(p\!\cdot\! q)\,\varepsilon_\nu\, p_\mu\,\pslash}{3\, m_{\mathrm{P_{bb}^*}}^2}
+ \frac{2(\varepsilon\!\cdot\! q)\, p_\mu p_\nu\,\pslash}{m_{\mathrm{P_{bb}^*}}^2}
- \frac{2\,\varepsilon_\nu\, p_\mu\, q^2\,\pslash}{3\, m_{\mathrm{P_{bb}^*}}^2} \nonumber\\
&\quad
+ \frac{4(\varepsilon\!\cdot\! p)\, p_\mu q_\nu\,\pslash}{3\, m_{\mathrm{P_{bb}^*}}^2}
+ \frac{2(\varepsilon\!\cdot\! q)\, p_\mu q_\nu\,\pslash}{m_{\mathrm{P_{bb}^*}}^2}
- \frac{4(\varepsilon\!\cdot\! q)\, p_\nu}{3}\,\gamma_\mu
+ \frac{2\,\varepsilon_\nu\, q^2}{3}\,\gamma_\mu
- \frac{4(\varepsilon\!\cdot\! q)\, q_\nu}{3}\,\gamma_\mu
- \frac{14\, p_\mu p_\nu}{3\, m_{\mathrm{P_{bb}^*}}}\,\eslash\,\qslash 
\nonumber\\
&\quad
+ \frac{4(p\!\cdot\! q)\, p_\mu p_\nu}{3\, m_{\mathrm{P_{bb}^*}}^3}\,\eslash\,\qslash
- \frac{2\, p_\nu q_\mu}{m_{\mathrm{P_{bb}^*}}}\,\eslash\,\qslash
- \frac{8\, p_\mu q_\nu}{3\, m_{\mathrm{P_{bb}^*}}}\,\eslash\,\qslash
+ \frac{4(p\!\cdot\! q)\, p_\mu q_\nu}{3\, m_{\mathrm{P_{bb}^*}}^3}\,\eslash\,\qslash
- \frac{2\, q_\mu q_\nu}{m_{\mathrm{P_{bb}^*}}}\,\eslash\,\qslash
+ 3\, m_{\mathrm{P_{bb}^*}}\, g_{\mu\nu}\,\eslash\,\qslash 
\nonumber\\
%\end{align}
%\begin{align}
&\quad
+ \frac{p_\nu\, q^2}{2\, m_{\mathrm{P_{bb}^*}}}\,\eslash\,\gamma_\mu
+ \frac{q^2 q_\nu}{2\, m_{\mathrm{P_{bb}^*}}}\,\eslash\,\gamma_\mu
+ \frac{(p\!\cdot\! q)\, p_\mu}{3\, m_{\mathrm{P_{bb}^*}}}\,\eslash\,\gamma_\nu
+ \frac{(p\!\cdot\! q)\, p_\mu\, q^2}{3\, m_{\mathrm{P_{bb}^*}}^3}\,\eslash\,\gamma_\nu
- \frac{q^2 q_\mu}{2\, m_{\mathrm{P_{bb}^*}}}\,\eslash\,\gamma_\nu \nonumber\\ 
&\quad
- \frac{4(p\!\cdot\! q)\, p_\mu p_\nu}{3\, m_{\mathrm{P_{bb}^*}}^3}\,\pslash\,\eslash
+ \frac{4(p\!\cdot\! q)^2\, p_\mu p_\nu}{3\, m_{\mathrm{P_{bb}^*}}^5}\,\pslash\,\eslash
+ \frac{2\, p_\mu p_\nu\, q^2}{3\, m_{\mathrm{P_{bb}^*}}^3}\,\pslash\,\eslash
+ \frac{2(p\!\cdot\! q)\, p_\mu p_\nu\, q^2}{3\, m_{\mathrm{P_{bb}^*}}^5}\,\pslash\,\eslash
- \frac{2(p\!\cdot\! q)\, p_\nu q_\mu}{m_{\mathrm{P_{bb}^*}}^3}\,\pslash\,\eslash \nonumber\\
&\quad
- \frac{p_\nu\, q^2 q_\mu}{m_{\mathrm{P_{bb}^*}}^3}\,\pslash\,\eslash
- \frac{2(p\!\cdot\! q)\, p_\mu q_\nu}{3\, m_{\mathrm{P_{bb}^*}}^3}\,\pslash\,\eslash
+ \frac{4(p\!\cdot\! q)^2\, p_\mu q_\nu}{3\, m_{\mathrm{P_{bb}^*}}^5}\,\pslash\,\eslash
+ \frac{p_\mu\, q^2 q_\nu}{m_{\mathrm{P_{bb}^*}}^3}\,\pslash\,\eslash
+ \frac{2(p\!\cdot\! q)\, p_\mu\, q^2 q_\nu}{3\, m_{\mathrm{P_{bb}^*}}^5}\,\pslash\,\eslash \nonumber\\
&\quad
- \frac{2(p\!\cdot\! q)\, q_\mu q_\nu}{m_{\mathrm{P_{bb}^*}}^3}\,\pslash\,\eslash
- \frac{q^2 q_\mu q_\nu}{m_{\mathrm{P_{bb}^*}}^3}\,\pslash\,\eslash
+ \frac{3(p\!\cdot\! q)\, g_{\mu\nu}}{m_{\mathrm{P_{bb}^*}}}\,\pslash\,\eslash
+ \frac{3\, q^2 g_{\mu\nu}}{2\, m_{\mathrm{P_{bb}^*}}}\,\pslash\,\eslash
+ \frac{4(\varepsilon\!\cdot\! p)\, p_\mu p_\nu}{3\, m_{\mathrm{P_{bb}^*}}^3}\,\pslash\,\qslash \nonumber\\
&\quad
- \frac{2(\varepsilon\!\cdot\! q)\, p_\mu p_\nu}{m_{\mathrm{P_{bb}^*}}^3}\,\pslash\,\qslash
- \frac{4(\varepsilon\!\cdot\! p)\,(p\!\cdot\! q)\, p_\mu p_\nu}{3\, m_{\mathrm{P_{bb}^*}}^5}\,\pslash\,\qslash
+ \frac{2(\varepsilon\!\cdot\! p)\, p_\nu q_\mu}{m_{\mathrm{P_{bb}^*}}^3}\,\pslash\,\qslash
+ \frac{2(\varepsilon\!\cdot\! p)\, p_\mu q_\nu}{3\, m_{\mathrm{P_{bb}^*}}^3}\,\pslash\,\qslash \nonumber\\
&\quad
- \frac{2(\varepsilon\!\cdot\! q)\, p_\mu q_\nu}{m_{\mathrm{P_{bb}^*}}^3}\,\pslash\,\qslash
- \frac{4(\varepsilon\!\cdot\! p)\,(p\!\cdot\! q)\, p_\mu q_\nu}{3\, m_{\mathrm{P_{bb}^*}}^5}\,\pslash\,\qslash
+ \frac{2(\varepsilon\!\cdot\! p)\, q_\mu q_\nu}{m_{\mathrm{P_{bb}^*}}^3}\,\pslash\,\qslash
- \frac{3(\varepsilon\!\cdot\! p)\, g_{\mu\nu}}{m_{\mathrm{P_{bb}^*}}}\,\pslash\,\qslash \nonumber\\
&\quad
+ \frac{4(p\!\cdot\! q)\,\varepsilon_\nu}{3\, m_{\mathrm{P_{bb}^*}}}\,\pslash\,\gamma_\mu
- \frac{2(\varepsilon\!\cdot\! q)\, p_\nu}{3\, m_{\mathrm{P_{bb}^*}}}\,\pslash\,\gamma_\mu
+ \frac{2\,\varepsilon_\nu\, q^2}{3\, m_{\mathrm{P_{bb}^*}}}\,\pslash\,\gamma_\mu
- \frac{(\varepsilon\!\cdot\! p)\, p_\nu\, q^2}{m_{\mathrm{P_{bb}^*}}^3}\,\pslash\,\gamma_\mu
- \frac{4(\varepsilon\!\cdot\! p)\, q_\nu}{3\, m_{\mathrm{P_{bb}^*}}}\,\pslash\,\gamma_\mu 
\nonumber\\
&\quad
- \frac{2(\varepsilon\!\cdot\! q)\, q_\nu}{3\, m_{\mathrm{P_{bb}^*}}}\,\pslash\,\gamma_\mu
- \frac{(\varepsilon\!\cdot\! p)\, q^2 q_\nu}{m_{\mathrm{P_{bb}^*}}^3}\,\pslash\,\gamma_\mu
- \frac{(\varepsilon\!\cdot\! p)\, p_\mu\, q^2}{3\, m_{\mathrm{P_{bb}^*}}^3}\,\pslash\,\gamma_\nu
- \frac{p_\mu p_\nu\, q^2}{3\, m_{\mathrm{P_{bb}^*}}^3}\,\qslash\,\eslash
- \frac{p_\mu\, q^2 q_\nu}{3\, m_{\mathrm{P_{bb}^*}}^3}\,\qslash\,\eslash \nonumber\\
&\quad
- \frac{2(\varepsilon\!\cdot\! q)\, p_\mu p_\nu}{3\, m_{\mathrm{P_{bb}^*}}^3}\,\qslash\,\pslash
- \frac{2(\varepsilon\!\cdot\! q)\, p_\mu q_\nu}{3\, m_{\mathrm{P_{bb}^*}}^3}\,\qslash\,\pslash
- \frac{(\varepsilon\!\cdot\! p)\, p_\mu}{3\, m_{\mathrm{P_{bb}^*}}}\,\qslash\,\gamma_\nu
- \frac{(p\!\cdot\! q)\, p_\nu}{m_{\mathrm{P_{bb}^*}}}\,\gamma_\mu\,\eslash
- \frac{p_\nu\, q^2}{m_{\mathrm{P_{bb}^*}}}\,\gamma_\mu\,\eslash \nonumber\\
&\quad
+ \frac{(p\!\cdot\! q)\, p_\nu\, q^2}{3\, m_{\mathrm{P_{bb}^*}}^3}\,\gamma_\mu\,\eslash
- \frac{(p\!\cdot\! q)\, q_\nu}{m_{\mathrm{P_{bb}^*}}}\,\gamma_\mu\,\eslash
- \frac{q^2 q_\nu}{2\, m_{\mathrm{P_{bb}^*}}}\,\gamma_\mu\,\eslash
+ \frac{(p\!\cdot\! q)\, q^2 q_\nu}{3\, m_{\mathrm{P_{bb}^*}}^3}\,\gamma_\mu\,\eslash
+ \frac{(\varepsilon\!\cdot\! p)\, p_\nu}{m_{\mathrm{P_{bb}^*}}}\,\gamma_\mu\,\qslash \nonumber\\
&\quad
+ \frac{4(\varepsilon\!\cdot\! q)\, p_\nu}{3\, m_{\mathrm{P_{bb}^*}}}\,\gamma_\mu\,\qslash
+ \frac{(\varepsilon\!\cdot\! p)\, q_\nu}{m_{\mathrm{P_{bb}^*}}}\,\gamma_\mu\,\qslash
+ \frac{4(\varepsilon\!\cdot\! q)\, q_\nu}{3\, m_{\mathrm{P_{bb}^*}}}\,\gamma_\mu\,\qslash
+ \frac{p_\mu\, q^2}{2\, m_{\mathrm{P_{bb}^*}}}\,\gamma_\nu\,\eslash
+ \frac{p_\mu\, q^2}{6\, m_{\mathrm{P_{bb}^*}}^2}\,\eslash\,\pslash\,\gamma_\nu \nonumber\\
&\quad
- \frac{4\, p_\mu p_\nu}{3\, m_{\mathrm{P_{bb}^*}}^2}\,\eslash\,\qslash\,\pslash
+ \frac{2(p\!\cdot\! q)\, p_\mu p_\nu}{3\, m_{\mathrm{P_{bb}^*}}^4}\,\eslash\,\qslash\,\pslash
- \frac{p_\nu q_\mu}{m_{\mathrm{P_{bb}^*}}^2}\,\eslash\,\qslash\,\pslash
- \frac{p_\mu q_\nu}{m_{\mathrm{P_{bb}^*}}^2}\,\eslash\,\qslash\,\pslash
+ \frac{2(p\!\cdot\! q)\, p_\mu q_\nu}{3\, m_{\mathrm{P_{bb}^*}}^4}\,\eslash\,\qslash\,\pslash \nonumber\\
&\quad
- \frac{q_\mu q_\nu}{m_{\mathrm{P_{bb}^*}}^2}\,\eslash\,\qslash\,\pslash
+ \frac{3\, g_{\mu\nu}}{2}\,\eslash\,\qslash\,\pslash
+ p_\nu\,\eslash\,\qslash\,\gamma_\mu
+ q_\nu\,\eslash\,\qslash\,\gamma_\mu
- \frac{5\, p_\mu}{3}\,\eslash\,\qslash\,\gamma_\nu
+ \frac{2(p\!\cdot\! q)\, p_\mu}{3\, m_{\mathrm{P_{bb}^*}}^2}\,\eslash\,\qslash\,\gamma_\nu \nonumber\\
&\quad
- q_\mu\,\eslash\,\qslash\,\gamma_\nu
+ \frac{p_\mu\, q^2}{3\, m_{\mathrm{P_{bb}^*}}^2}\,\eslash\,\gamma_\nu\,\pslash
- \frac{q^2}{2}\,\eslash\,\gamma_\nu\,\gamma_\mu
- \frac{p_\mu p_\nu}{m_{\mathrm{P_{bb}^*}}^2}\,\pslash\,\eslash\,\qslash
+ \frac{2(p\!\cdot\! q)\, p_\mu p_\nu}{3\, m_{\mathrm{P_{bb}^*}}^4}\,\pslash\,\eslash\,\qslash
- \frac{p_\nu q_\mu}{m_{\mathrm{P_{bb}^*}}^2}\,\pslash\,\eslash\,\qslash \nonumber\\
&\quad
- \frac{2\, p_\mu q_\nu}{3\, m_{\mathrm{P_{bb}^*}}^2}\,\pslash\,\eslash\,\qslash
+ \frac{2(p\!\cdot\! q)\, p_\mu q_\nu}{3\, m_{\mathrm{P_{bb}^*}}^4}\,\pslash\,\eslash\,\qslash
- \frac{q_\mu q_\nu}{m_{\mathrm{P_{bb}^*}}^2}\,\pslash\,\eslash\,\qslash
+ \frac{3\, g_{\mu\nu}}{2}\,\pslash\,\eslash\,\qslash
+ \frac{(p\!\cdot\! q)\, p_\nu}{m_{\mathrm{P_{bb}^*}}^2}\,\pslash\,\eslash\,\gamma_\mu \nonumber\\
&\quad
+ \frac{p_\nu\, q^2}{2\, m_{\mathrm{P_{bb}^*}}^2}\,\pslash\,\eslash\,\gamma_\mu
+ \frac{(p\!\cdot\! q)\, q_\nu}{m_{\mathrm{P_{bb}^*}}^2}\,\pslash\,\eslash\,\gamma_\mu
+ \frac{q^2 q_\nu}{2\, m_{\mathrm{P_{bb}^*}}^2}\,\pslash\,\eslash\,\gamma_\mu
+ \frac{2(p\!\cdot\! q)^2\, p_\mu}{3\, m_{\mathrm{P_{bb}^*}}^4}\,\pslash\,\eslash\,\gamma_\nu
+ \frac{5\, p_\mu\, q^2}{6\, m_{\mathrm{P_{bb}^*}}^2}\,\pslash\,\eslash\,\gamma_\nu 
\nonumber%\\
\end{align}

\begin{align}
&\quad
+ \frac{(p\!\cdot\! q)\, p_\mu\, q^2}{3\, m_{\mathrm{P_{bb}^*}}^4}\,\pslash\,\eslash\,\gamma_\nu
- \frac{(p\!\cdot\! q)\, q_\mu}{m_{\mathrm{P_{bb}^*}}^2}\,\pslash\,\eslash\,\gamma_\nu
- \frac{q^2 q_\mu}{2\, m_{\mathrm{P_{bb}^*}}^2}\,\pslash\,\eslash\,\gamma_\nu
- \frac{p_\mu p_\nu}{3\, m_{\mathrm{P_{bb}^*}}^2}\,\pslash\,\qslash\,\eslash
- \frac{2(p\!\cdot\! q)\, p_\mu p_\nu}{3\, m_{\mathrm{P_{bb}^*}}^4}\,\pslash\,\qslash\,\eslash \nonumber\\
&\quad
- \frac{p_\mu p_\nu\, q^2}{3\, m_{\mathrm{P_{bb}^*}}^4}\,\pslash\,\qslash\,\eslash
- \frac{p_\mu q_\nu}{3\, m_{\mathrm{P_{bb}^*}}^2}\,\pslash\,\qslash\,\eslash
- \frac{2(p\!\cdot\! q)\, p_\mu q_\nu}{3\, m_{\mathrm{P_{bb}^*}}^4}\,\pslash\,\qslash\,\eslash
- \frac{p_\mu\, q^2 q_\nu}{3\, m_{\mathrm{P_{bb}^*}}^4}\,\pslash\,\qslash\,\eslash
- \frac{(\varepsilon\!\cdot\! p)\, p_\nu}{m_{\mathrm{P_{bb}^*}}^2}\,\pslash\,\qslash\,\gamma_\mu \nonumber\\
&\quad
- \frac{(\varepsilon\!\cdot\! p)\, q_\nu}{m_{\mathrm{P_{bb}^*}}^2}\,\pslash\,\qslash\,\gamma_\mu
- \frac{(\varepsilon\!\cdot\! q)\, p_\mu}{m_{\mathrm{P_{bb}^*}}^2}\,\pslash\,\qslash\,\gamma_\nu
- \frac{2(\varepsilon\!\cdot\! p)\,(p\!\cdot\! q)\, p_\mu}{3\, m_{\mathrm{P_{bb}^*}}^4}\,\pslash\,\qslash\,\gamma_\nu
+ \frac{(\varepsilon\!\cdot\! p)\, q_\mu}{m_{\mathrm{P_{bb}^*}}^2}\,\pslash\,\qslash\,\gamma_\nu \nonumber\\
&\quad
+ \frac{2(p\!\cdot\! q)^2\, p_\nu}{3\, m_{\mathrm{P_{bb}^*}}^4}\,\pslash\,\gamma_\mu\,\eslash
+ \frac{p_\nu\, q^2}{6\, m_{\mathrm{P_{bb}^*}}^2}\,\pslash\,\gamma_\mu\,\eslash
+ \frac{(p\!\cdot\! q)\, p_\nu\, q^2}{3\, m_{\mathrm{P_{bb}^*}}^4}\,\pslash\,\gamma_\mu\,\eslash
+ \frac{(p\!\cdot\! q)\, q_\nu}{3\, m_{\mathrm{P_{bb}^*}}^2}\,\pslash\,\gamma_\mu\,\eslash
+ \frac{2(p\!\cdot\! q)^2\, q_\nu}{3\, m_{\mathrm{P_{bb}^*}}^4}\,\pslash\,\gamma_\mu\,\eslash \nonumber\\
&\quad
+ \frac{q^2 q_\nu}{3\, m_{\mathrm{P_{bb}^*}}^2}\,\pslash\,\gamma_\mu\,\eslash
+ \frac{(p\!\cdot\! q)\, q^2 q_\nu}{3\, m_{\mathrm{P_{bb}^*}}^4}\,\pslash\,\gamma_\mu\,\eslash
+ \frac{2(\varepsilon\!\cdot\! q)\, p_\nu}{3\, m_{\mathrm{P_{bb}^*}}^2}\,\pslash\,\gamma_\mu\,\qslash
- \frac{2(\varepsilon\!\cdot\! p)\,(p\!\cdot\! q)\, p_\nu}{3\, m_{\mathrm{P_{bb}^*}}^4}\,\pslash\,\gamma_\mu\,\qslash \nonumber\\
&\quad
- \frac{(\varepsilon\!\cdot\! p)\, q_\nu}{3\, m_{\mathrm{P_{bb}^*}}^2}\,\pslash\,\gamma_\mu\,\qslash
+ \frac{2(\varepsilon\!\cdot\! q)\, q_\nu}{3\, m_{\mathrm{P_{bb}^*}}^2}\,\pslash\,\gamma_\mu\,\qslash
- \frac{2(\varepsilon\!\cdot\! p)\,(p\!\cdot\! q)\, q_\nu}{3\, m_{\mathrm{P_{bb}^*}}^4}\,\pslash\,\gamma_\mu\,\qslash
- \frac{2(\varepsilon\!\cdot\! p)\, q^2}{3\, m_{\mathrm{P_{bb}^*}}^2}\,\pslash\,\gamma_\mu\,\gamma_\nu \nonumber\\
%\end{align}
%\begin{align}
&\quad
+ \frac{(p\!\cdot\! q)\, p_\mu}{m_{\mathrm{P_{bb}^*}}^2}\,\pslash\,\gamma_\nu\,\eslash
+ \frac{p_\mu\, q^2}{2\, m_{\mathrm{P_{bb}^*}}^2}\,\pslash\,\gamma_\nu\,\eslash
- \frac{(\varepsilon\!\cdot\! p)\, p_\mu}{m_{\mathrm{P_{bb}^*}}^2}\,\pslash\,\gamma_\nu\,\qslash
- \frac{p_\mu\, q^2}{6\, m_{\mathrm{P_{bb}^*}}^2}\,\qslash\,\eslash\,\gamma_\nu \nonumber\\
&\quad
- \frac{(\varepsilon\!\cdot\! q)\, p_\mu}{3\, m_{\mathrm{P_{bb}^*}}^2}\,\qslash\,\pslash\,\gamma_\nu
- \frac{p_\nu\, q^2}{2\, m_{\mathrm{P_{bb}^*}}^2}\,\gamma_\mu\,\eslash\,\pslash
- \frac{q^2 q_\nu}{2\, m_{\mathrm{P_{bb}^*}}^2}\,\gamma_\mu\,\eslash\,\pslash
+ \frac{p_\nu}{3}\,\gamma_\mu\,\eslash\,\qslash
+ \frac{q_\nu}{3}\,\gamma_\mu\,\eslash\,\qslash \nonumber\\
&\quad
- \frac{2(p\!\cdot\! q)}{3}\,\gamma_\mu\,\eslash\,\gamma_\nu
- \frac{q^2}{2}\,\gamma_\mu\,\eslash\,\gamma_\nu
+ \frac{(p\!\cdot\! q)\, q^2}{6\, m_{\mathrm{P_{bb}^*}}^2}\,\gamma_\mu\,\eslash\,\gamma_\nu
+ \frac{2(p\!\cdot\! q)\, p_\nu}{3\, m_{\mathrm{P_{bb}^*}}^2}\,\gamma_\mu\,\pslash\,\eslash
+ \frac{p_\nu\, q^2}{3\, m_{\mathrm{P_{bb}^*}}^2}\,\gamma_\mu\,\pslash\,\eslash \nonumber\\
&\quad
+ \frac{2(p\!\cdot\! q)\, q_\nu}{3\, m_{\mathrm{P_{bb}^*}}^2}\,\gamma_\mu\,\pslash\,\eslash
+ \frac{q^2 q_\nu}{3\, m_{\mathrm{P_{bb}^*}}^2}\,\gamma_\mu\,\pslash\,\eslash
- \frac{2(\varepsilon\!\cdot\! p)\, p_\nu}{3\, m_{\mathrm{P_{bb}^*}}^2}\,\gamma_\mu\,\pslash\,\qslash
- \frac{2(\varepsilon\!\cdot\! p)\, q_\nu}{3\, m_{\mathrm{P_{bb}^*}}^2}\,\gamma_\mu\,\pslash\,\qslash
\nonumber\\
&\quad
+ \frac{p_\nu\, q^2}{3\, m_{\mathrm{P_{bb}^*}}^2}\,\gamma_\mu\,\qslash\,\eslash
+ \frac{q^2 q_\nu}{3\, m_{\mathrm{P_{bb}^*}}^2}\,\gamma_\mu\,\qslash\,\eslash
+ \frac{2(\varepsilon\!\cdot\! q)\, p_\nu}{3\, m_{\mathrm{P_{bb}^*}}^2}\,\gamma_\mu\,\qslash\,\pslash
+ \frac{2(\varepsilon\!\cdot\! q)\, q_\nu}{3\, m_{\mathrm{P_{bb}^*}}^2}\,\gamma_\mu\,\qslash\,\pslash
+ \frac{2(\varepsilon\!\cdot\! p)}{3}\,\gamma_\mu\,\qslash\,\gamma_\nu \nonumber\\
&\quad
+ \frac{2(\varepsilon\!\cdot\! q)}{3}\,\gamma_\mu\,\qslash\,\gamma_\nu
- \frac{q^2}{2}\,\gamma_\mu\,\gamma_\nu\,\eslash
+ \frac{p_\mu\, q^2}{2\, m_{\mathrm{P_{bb}^*}}}\,\gamma_\nu\,\eslash
+ \frac{p_\nu}{2\, m_{\mathrm{P_{bb}^*}}}\,\eslash\,\qslash\,\pslash\,\gamma_\mu
+ \frac{q_\nu}{2\, m_{\mathrm{P_{bb}^*}}}\,\eslash\,\qslash\,\pslash\,\gamma_\mu \nonumber\\
&\quad
- \frac{p_\mu}{6\, m_{\mathrm{P_{bb}^*}}}\,\eslash\,\qslash\,\pslash\,\gamma_\nu
+ \frac{(p\!\cdot\! q)\, p_\mu}{3\, m_{\mathrm{P_{bb}^*}}^3}\,\eslash\,\qslash\,\pslash\,\gamma_\nu
- \frac{q_\mu}{2\, m_{\mathrm{P_{bb}^*}}}\,\eslash\,\qslash\,\pslash\,\gamma_\nu
+ \frac{2\, p_\mu}{3\, m_{\mathrm{P_{bb}^*}}}\,\eslash\,\qslash\,\gamma_\nu\,\pslash
- m_{\mathrm{P_{bb}^*}}\,\eslash\,\qslash\,\gamma_\nu\,\gamma_\mu \nonumber\\
&\quad
+ \frac{p_\nu}{2\, m_{\mathrm{P_{bb}^*}}}\,\pslash\,\eslash\,\qslash\,\gamma_\mu
+ \frac{q_\nu}{2\, m_{\mathrm{P_{bb}^*}}}\,\pslash\,\eslash\,\qslash\,\gamma_\mu
- \frac{5\, p_\mu}{6\, m_{\mathrm{P_{bb}^*}}}\,\pslash\,\eslash\,\qslash\,\gamma_\nu
+ \frac{(p\!\cdot\! q)\, p_\mu}{3\, m_{\mathrm{P_{bb}^*}}^3}\,\pslash\,\eslash\,\qslash\,\gamma_\nu
- \frac{q_\mu}{2\, m_{\mathrm{P_{bb}^*}}}\,\pslash\,\eslash\,\qslash\,\gamma_\nu \nonumber\\
&\quad
- \frac{(p\!\cdot\! q)}{m_{\mathrm{P_{bb}^*}}}\,\pslash\,\eslash\,\gamma_\nu\,\gamma_\mu
- \frac{q^2}{2\, m_{\mathrm{P_{bb}^*}}}\,\pslash\,\eslash\,\gamma_\nu\,\gamma_\mu
- \frac{p_\mu}{3\, m_{\mathrm{P_{bb}^*}}}\,\pslash\,\qslash\,\eslash\,\gamma_\nu
- \frac{(p\!\cdot\! q)\, p_\mu}{3\, m_{\mathrm{P_{bb}^*}}^3}\,\pslash\,\qslash\,\eslash\,\gamma_\nu
- \frac{p_\mu\, q^2}{6\, m_{\mathrm{P_{bb}^*}}^3}\,\pslash\,\qslash\,\eslash\,\gamma_\nu \nonumber\\
&\quad
+ \frac{(\varepsilon\!\cdot\! p)}{m_{\mathrm{P_{bb}^*}}}\,\pslash\,\qslash\,\gamma_\nu\,\gamma_\mu
+ \frac{p_\nu}{2\, m_{\mathrm{P_{bb}^*}}}\,\pslash\,\gamma_\mu\,\eslash\,\qslash
+ \frac{(p\!\cdot\! q)\, p_\nu}{m_{\mathrm{P_{bb}^*}}^3}\,\pslash\,\gamma_\mu\,\eslash\,\qslash
+ \frac{q_\nu}{m_{\mathrm{P_{bb}^*}}}\,\pslash\,\gamma_\mu\,\eslash\,\qslash
+ \frac{(p\!\cdot\! q)\, q_\nu}{m_{\mathrm{P_{bb}^*}}^3}\,\pslash\,\gamma_\mu\,\eslash\,\qslash \nonumber\\
&\quad
- \frac{2(p\!\cdot\! q)}{3\, m_{\mathrm{P_{bb}^*}}}\,\pslash\,\gamma_\mu\,\eslash\,\gamma_\nu
+ \frac{(p\!\cdot\! q)^2}{3\, m_{\mathrm{P_{bb}^*}}^3}\,\pslash\,\gamma_\mu\,\eslash\,\gamma_\nu
+ \frac{(p\!\cdot\! q)\, q^2}{6\, m_{\mathrm{P_{bb}^*}}^3}\,\pslash\,\gamma_\mu\,\eslash\,\gamma_\nu
+ \frac{2(p\!\cdot\! q)\, p_\nu}{3\, m_{\mathrm{P_{bb}^*}}^3}\,\pslash\,\gamma_\mu\,\qslash\,\eslash \nonumber\\
&\quad
+ \frac{p_\nu\, q^2}{3\, m_{\mathrm{P_{bb}^*}}^3}\,\pslash\,\gamma_\mu\,\qslash\,\eslash
+ \frac{2(p\!\cdot\! q)\, q_\nu}{3\, m_{\mathrm{P_{bb}^*}}^3}\,\pslash\,\gamma_\mu\,\qslash\,\eslash
+ \frac{q^2 q_\nu}{3\, m_{\mathrm{P_{bb}^*}}^3}\,\pslash\,\gamma_\mu\,\qslash\,\eslash
+ \frac{2(\varepsilon\!\cdot\! p)}{3\, m_{\mathrm{P_{bb}^*}}}\,\pslash\,\gamma_\mu\,\qslash\,\gamma_\nu \nonumber\\
&\quad
+ \frac{(\varepsilon\!\cdot\! q)}{3\, m_{\mathrm{P_{bb}^*}}}\,\pslash\,\gamma_\mu\,\qslash\,\gamma_\nu
- \frac{(\varepsilon\!\cdot\! p)\,(p\!\cdot\! q)}{3\, m_{\mathrm{P_{bb}^*}}^3}\,\pslash\,\gamma_\mu\,\qslash\,\gamma_\nu
- \frac{(p\!\cdot\! q)}{m_{\mathrm{P_{bb}^*}}}\,\pslash\,\gamma_\mu\,\gamma_\nu\,\eslash
- \frac{q^2}{2\, m_{\mathrm{P_{bb}^*}}}\,\pslash\,\gamma_\mu\,\gamma_\nu\,\eslash \nonumber\\
&\quad
+ \frac{(\varepsilon\!\cdot\! p)}{m_{\mathrm{P_{bb}^*}}}\,\pslash\,\gamma_\mu\,\gamma_\nu\,\qslash
+ \frac{3\, p_\mu}{2\, m_{\mathrm{P_{bb}^*}}}\,\pslash\,\gamma_\nu\,\eslash\,\qslash
+ \frac{p_\mu}{m_{\mathrm{P_{bb}^*}}}\,\pslash\,\gamma_\nu\,\qslash\,\eslash
- \frac{q^2}{6\, m_{\mathrm{P_{bb}^*}}}\,\gamma_\mu\,\eslash\,\pslash\,\gamma_\nu
+ \frac{p_\nu}{6\, m_{\mathrm{P_{bb}^*}}}\,\gamma_\mu\,\eslash\,\qslash\,\pslash \nonumber\\
&\quad
+ \frac{(p\!\cdot\! q)\, p_\nu}{3\, m_{\mathrm{P_{bb}^*}}^3}\,\gamma_\mu\,\eslash\,\qslash\,\pslash
+ \frac{q_\nu}{3\, m_{\mathrm{P_{bb}^*}}}\,\gamma_\mu\,\eslash\,\qslash\,\pslash
+ \frac{(p\!\cdot\! q)\, q_\nu}{3\, m_{\mathrm{P_{bb}^*}}^3}\,\gamma_\mu\,\eslash\,\qslash\,\pslash
+ \frac{m_{\mathrm{P_{bb}^*}}}{3}\,\gamma_\mu\,\eslash\,\qslash\,\gamma_\nu
+ \frac{q^2}{6\, m_{\mathrm{P_{bb}^*}}}\,\gamma_\mu\,\eslash\,\gamma_\nu\,\pslash \nonumber%\\
\end{align}

\begin{align}
&\quad
+ \frac{2\, p_\nu}{3\, m_{\mathrm{P_{bb}^*}}}\,\gamma_\mu\,\pslash\,\eslash\,\qslash
+ \frac{2\, q_\nu}{3\, m_{\mathrm{P_{bb}^*}}}\,\gamma_\mu\,\pslash\,\eslash\,\qslash
+ \frac{(p\!\cdot\! q)}{3\, m_{\mathrm{P_{bb}^*}}}\,\gamma_\mu\,\pslash\,\eslash\,\gamma_\nu
+ \frac{q^2}{6\, m_{\mathrm{P_{bb}^*}}}\,\gamma_\mu\,\pslash\,\eslash\,\gamma_\nu
+ \frac{p_\nu}{3\, m_{\mathrm{P_{bb}^*}}}\,\gamma_\mu\,\pslash\,\qslash\,\eslash \nonumber\\
&\quad
+ \frac{q_\nu}{3\, m_{\mathrm{P_{bb}^*}}}\,\gamma_\mu\,\pslash\,\qslash\,\eslash
- \frac{(\varepsilon\!\cdot\! p)}{3\, m_{\mathrm{P_{bb}^*}}}\,\gamma_\mu\,\pslash\,\qslash\,\gamma_\nu
+ \frac{q^2}{6\, m_{\mathrm{P_{bb}^*}}}\,\gamma_\mu\,\qslash\,\eslash\,\gamma_\nu
+ \frac{(\varepsilon\!\cdot\! q)}{3\, m_{\mathrm{P_{bb}^*}}}\,\gamma_\mu\,\qslash\,\pslash\,\gamma_\nu \nonumber\\
&\quad
- m_{\mathrm{P_{bb}^*}}\,\gamma_\mu\,\gamma_\nu\,\eslash\,\qslash
- \frac{2\, m_{\mathrm{P_{bb}^*}}}{3}\,\gamma_\mu\,\gamma_\nu\,\qslash\,\eslash
+ \frac{p_\mu}{6\, m_{\mathrm{P_{bb}^*}}}\,\gamma_\nu\,\eslash\,\qslash\,\pslash
- \frac{1}{2}\,\eslash\,\qslash\,\pslash\,\gamma_\nu\,\gamma_\mu
- \frac{1}{2}\,\pslash\,\eslash\,\qslash\,\gamma_\nu\,\gamma_\mu \nonumber\\
&\quad
+ \frac{1}{2}\,\pslash\,\gamma_\mu\,\eslash\,\qslash\,\gamma_\nu
+ \frac{(p\!\cdot\! q)}{2\, m_{\mathrm{P_{bb}^*}}^2}\,\pslash\,\gamma_\mu\,\eslash\,\qslash\,\gamma_\nu
+ \frac{(p\!\cdot\! q)}{3\, m_{\mathrm{P_{bb}^*}}^2}\,\pslash\,\gamma_\mu\,\qslash\,\eslash\,\gamma_\nu
+ \frac{q^2}{6\, m_{\mathrm{P_{bb}^*}}^2}\,\pslash\,\gamma_\mu\,\qslash\,\eslash\,\gamma_\nu \nonumber\\
&\quad
- \frac{1}{2}\,\pslash\,\gamma_\mu\,\gamma_\nu\,\eslash\,\qslash
- \frac{1}{3}\,\pslash\,\gamma_\mu\,\gamma_\nu\,\qslash\,\eslash
+ \frac{1}{6}\,\gamma_\mu\,\eslash\,\qslash\,\pslash\,\gamma_\nu
+ \frac{(p\!\cdot\! q)}{6\, m_{\mathrm{P_{bb}^*}}^2}\,\gamma_\mu\,\eslash\,\qslash\,\pslash\,\gamma_\nu
+ \frac{1}{3}\,\gamma_\mu\,\pslash\,\eslash\,\qslash\,\gamma_\nu
+ \frac{1}{3}\,\gamma_\mu\,\pslash\,\qslash\,\eslash\,\gamma_\nu
- \frac{1}{6}\,\gamma_\mu\,\gamma_\nu\,\eslash\,\qslash\,\pslash
\Bigg\}.
\label{F2_appendix}
\end{align}

\subsection{Lorentz structures proportional to $F_3(q^2)$}

\begin{align}
{F}_3 & \propto \,\Bigg\{ 
- \frac{(\varepsilon\!\cdot\! q)\, p_\mu p_\nu\, q^2}{6\, m_{\mathrm{P_{bb}^*}}^3}
- \frac{(\varepsilon\!\cdot\! q)\, p_\mu\, q^2 q_\nu}{6\, m_{\mathrm{P_{bb}^*}}^3}
+ \frac{p_\mu p_\nu\, q^2}{6\, m_{\mathrm{P_{bb}^*}}^2}\,\eslash
- \frac{p_\mu p_\nu\, (q^2)^2}{6\, m_{\mathrm{P_{bb}^*}}^4}\,\eslash
- \frac{p_\nu\, q^2 q_\mu}{4\, m_{\mathrm{P_{bb}^*}}^2}\,\eslash \nonumber\\
&\quad 
+ \frac{5\, p_\mu\, q^2 q_\nu}{12\, m_{\mathrm{P_{bb}^*}}^2}\,\eslash
- \frac{p_\mu\, (q^2)^2 q_\nu}{6\, m_{\mathrm{P_{bb}^*}}^4}\,\eslash
- \frac{q^2 q_\mu q_\nu}{4\, m_{\mathrm{P_{bb}^*}}^2}\,\eslash
- \frac{2(\varepsilon\!\cdot\! p)(p\!\cdot\! q)^2\, p_\mu p_\nu}{3\, m_{\mathrm{P_{bb}^*}}^6}\,\pslash
- \frac{(\varepsilon\!\cdot\! p)\, p_\mu p_\nu\, q^2}{6\, m_{\mathrm{P_{bb}^*}}^4}\,\pslash \nonumber\\
&\quad 
- \frac{(\varepsilon\!\cdot\! q)\, p_\mu p_\nu\, q^2}{6\, m_{\mathrm{P_{bb}^*}}^4}\,\pslash
- \frac{2(\varepsilon\!\cdot\! p)(p\!\cdot\! q)\, p_\mu p_\nu\, q^2}{3\, m_{\mathrm{P_{bb}^*}}^6}\,\pslash
+ \frac{(\varepsilon\!\cdot\! p)(p\!\cdot\! q)\, p_\nu q_\mu}{m_{\mathrm{P_{bb}^*}}^4}\,\pslash
+ \frac{(\varepsilon\!\cdot\! p)\, p_\nu\, q^2 q_\mu}{m_{\mathrm{P_{bb}^*}}^4}\,\pslash \nonumber\\
&\quad 
+ \frac{(\varepsilon\!\cdot\! p)(p\!\cdot\! q)\, p_\mu q_\nu}{m_{\mathrm{P_{bb}^*}}^4}\,\pslash
- \frac{2(\varepsilon\!\cdot\! p)(p\!\cdot\! q)^2\, p_\mu q_\nu}{3\, m_{\mathrm{P_{bb}^*}}^6}\,\pslash
- \frac{(\varepsilon\!\cdot\! p)\, p_\mu\, q^2 q_\nu}{6\, m_{\mathrm{P_{bb}^*}}^4}\,\pslash
- \frac{(\varepsilon\!\cdot\! q)\, p_\mu\, q^2 q_\nu}{6\, m_{\mathrm{P_{bb}^*}}^4}\,\pslash \nonumber\\
&\quad 
- \frac{2(\varepsilon\!\cdot\! p)(p\!\cdot\! q)\, p_\mu\, q^2 q_\nu}{3\, m_{\mathrm{P_{bb}^*}}^6}\,\pslash
- \frac{3(\varepsilon\!\cdot\! p)\, q_\mu q_\nu}{2\, m_{\mathrm{P_{bb}^*}}^2}\,\pslash
+ \frac{(\varepsilon\!\cdot\! p)(p\!\cdot\! q)\, q_\mu q_\nu}{m_{\mathrm{P_{bb}^*}}^4}\,\pslash
+ \frac{(\varepsilon\!\cdot\! p)\, q^2 q_\mu q_\nu}{m_{\mathrm{P_{bb}^*}}^4}\,\pslash \nonumber\\
&\quad 
- \frac{(\varepsilon\!\cdot\! q)\, p_\mu p_\nu}{3\, m_{\mathrm{P_{bb}^*}}^2}\,\qslash
+ \frac{(\varepsilon\!\cdot\! q)(p\!\cdot\! q)\, p_\mu p_\nu}{3\, m_{\mathrm{P_{bb}^*}}^4}\,\qslash
+ \frac{(\varepsilon\!\cdot\! q)\, p_\mu p_\nu\, q^2}{3\, m_{\mathrm{P_{bb}^*}}^4}\,\qslash
- \frac{(\varepsilon\!\cdot\! q)\, p_\mu q_\nu}{2\, m_{\mathrm{P_{bb}^*}}^2}\,\qslash
+ \frac{(\varepsilon\!\cdot\! q)(p\!\cdot\! q)\, p_\mu q_\nu}{3\, m_{\mathrm{P_{bb}^*}}^4}\,\qslash \nonumber\\
&\quad 
+ \frac{(\varepsilon\!\cdot\! q)\, p_\mu\, q^2 q_\nu}{3\, m_{\mathrm{P_{bb}^*}}^4}\,\qslash
+ \frac{(\varepsilon\!\cdot\! q)\, p_\nu\, q^2}{6\, m_{\mathrm{P_{bb}^*}}^2}\,\gamma_\mu
+ \frac{(\varepsilon\!\cdot\! q)\, q^2 q_\nu}{6\, m_{\mathrm{P_{bb}^*}}^2}\,\gamma_\mu
- \frac{(\varepsilon\!\cdot\! q)\, p_\mu\, q^2}{6\, m_{\mathrm{P_{bb}^*}}^2}\,\gamma_\nu
- \frac{(p\!\cdot\! q)^2\, p_\mu p_\nu}{3\, m_{\mathrm{P_{bb}^*}}^5}\,\eslash\,\pslash \nonumber\\
&\quad 
- \frac{p_\mu p_\nu\, q^2}{12\, m_{\mathrm{P_{bb}^*}}^3}\,\eslash\,\pslash
- \frac{(p\!\cdot\! q)\, p_\mu p_\nu\, q^2}{3\, m_{\mathrm{P_{bb}^*}}^5}\,\eslash\,\pslash
+ \frac{(p\!\cdot\! q)\, p_\nu q_\mu}{2\, m_{\mathrm{P_{bb}^*}}^3}\,\eslash\,\pslash
+ \frac{p_\nu\, q^2 q_\mu}{2\, m_{\mathrm{P_{bb}^*}}^3}\,\eslash\,\pslash
+ \frac{(p\!\cdot\! q)\, p_\mu q_\nu}{2\, m_{\mathrm{P_{bb}^*}}^3}\,\eslash\,\pslash \nonumber\\
&\quad 
- \frac{(p\!\cdot\! q)^2\, p_\mu q_\nu}{3\, m_{\mathrm{P_{bb}^*}}^5}\,\eslash\,\pslash
- \frac{p_\mu\, q^2 q_\nu}{12\, m_{\mathrm{P_{bb}^*}}^3}\,\eslash\,\pslash
- \frac{(p\!\cdot\! q)\, p_\mu\, q^2 q_\nu}{3\, m_{\mathrm{P_{bb}^*}}^5}\,\eslash\,\pslash
- \frac{3\, q_\mu q_\nu}{4\, m_{\mathrm{P_{bb}^*}}}\,\eslash\,\pslash
+ \frac{(p\!\cdot\! q)\, q_\mu q_\nu}{2\, m_{\mathrm{P_{bb}^*}}^3}\,\eslash\,\pslash \nonumber\\
&\quad 
+ \frac{q^2 q_\mu q_\nu}{2\, m_{\mathrm{P_{bb}^*}}^3}\,\eslash\,\pslash
- \frac{(p\!\cdot\! q)^2\, p_\mu p_\nu}{3\, m_{\mathrm{P_{bb}^*}}^5}\,\eslash\,\qslash
+ \frac{p_\mu p_\nu\, q^2}{12\, m_{\mathrm{P_{bb}^*}}^3}\,\eslash\,\qslash
- \frac{(p\!\cdot\! q)\, p_\mu p_\nu\, q^2}{3\, m_{\mathrm{P_{bb}^*}}^5}\,\eslash\,\qslash
+ \frac{(p\!\cdot\! q)\, p_\nu q_\mu}{2\, m_{\mathrm{P_{bb}^*}}^3}\,\eslash\,\qslash \nonumber\\
&\quad 
+ \frac{p_\nu\, q^2 q_\mu}{2\, m_{\mathrm{P_{bb}^*}}^3}\,\eslash\,\qslash
+ \frac{(p\!\cdot\! q)\, p_\mu q_\nu}{6\, m_{\mathrm{P_{bb}^*}}^3}\,\eslash\,\qslash
- \frac{(p\!\cdot\! q)^2\, p_\mu q_\nu}{3\, m_{\mathrm{P_{bb}^*}}^5}\,\eslash\,\qslash
+ \frac{p_\mu\, q^2 q_\nu}{12\, m_{\mathrm{P_{bb}^*}}^3}\,\eslash\,\qslash
- \frac{(p\!\cdot\! q)\, p_\mu\, q^2 q_\nu}{3\, m_{\mathrm{P_{bb}^*}}^5}\,\eslash\,\qslash \nonumber\\
&\quad 
- \frac{q_\mu q_\nu}{4\, m_{\mathrm{P_{bb}^*}}}\,\eslash\,\qslash
+ \frac{(p\!\cdot\! q)\, q_\mu q_\nu}{2\, m_{\mathrm{P_{bb}^*}}^3}\,\eslash\,\qslash
+ \frac{q^2 q_\mu q_\nu}{2\, m_{\mathrm{P_{bb}^*}}^3}\,\eslash\,\qslash
+ \frac{p_\mu\, q^2}{6\, m_{\mathrm{P_{bb}^*}}}\,\eslash\,\gamma_\nu
+ \frac{(p\!\cdot\! q)\, p_\mu\, q^2}{12\, m_{\mathrm{P_{bb}^*}}^3}\,\eslash\,\gamma_\nu
- \frac{p_\mu\, (q^2)^2}{12\, m_{\mathrm{P_{bb}^*}}^3}\,\eslash\,\gamma_\nu \nonumber\\
&\quad 
- \frac{q^2 q_\mu}{4\, m_{\mathrm{P_{bb}^*}}}\,\eslash\,\gamma_\nu
- \frac{(p\!\cdot\! q)^2\, p_\mu p_\nu}{3\, m_{\mathrm{P_{bb}^*}}^5}\,\pslash\,\eslash
- \frac{p_\mu p_\nu\, q^2}{12\, m_{\mathrm{P_{bb}^*}}^3}\,\pslash\,\eslash
- \frac{(p\!\cdot\! q)\, p_\mu p_\nu\, q^2}{3\, m_{\mathrm{P_{bb}^*}}^5}\,\pslash\,\eslash
- \frac{p_\mu p_\nu\, (q^2)^2}{6\, m_{\mathrm{P_{bb}^*}}^5}\,\pslash\,\eslash 
\nonumber%\\
\end{align}

\begin{align}
&\quad 
+ \frac{(p\!\cdot\! q)\, p_\nu q_\mu}{2\, m_{\mathrm{P_{bb}^*}}^3}\,\pslash\,\eslash
+ \frac{p_\nu\, q^2 q_\mu}{4\, m_{\mathrm{P_{bb}^*}}^3}\,\pslash\,\eslash
+ \frac{(p\!\cdot\! q)\, p_\mu q_\nu}{2\, m_{\mathrm{P_{bb}^*}}^3}\,\pslash\,\eslash
- \frac{(p\!\cdot\! q)^2\, p_\mu q_\nu}{3\, m_{\mathrm{P_{bb}^*}}^5}\,\pslash\,\eslash
+ \frac{p_\mu\, q^2 q_\nu}{6\, m_{\mathrm{P_{bb}^*}}^3}\,\pslash\,\eslash 
\nonumber\\
&\quad 
- \frac{(p\!\cdot\! q)\, p_\mu\, q^2 q_\nu}{3\, m_{\mathrm{P_{bb}^*}}^5}\,\pslash\,\eslash
- \frac{p_\mu\, (q^2)^2 q_\nu}{6\, m_{\mathrm{P_{bb}^*}}^5}\,\pslash\,\eslash
- \frac{3\, q_\mu q_\nu}{4\, m_{\mathrm{P_{bb}^*}}}\,\pslash\,\eslash
+ \frac{(p\!\cdot\! q)\, q_\mu q_\nu}{2\, m_{\mathrm{P_{bb}^*}}^3}\,\pslash\,\eslash
+ \frac{q^2 q_\mu q_\nu}{4\, m_{\mathrm{P_{bb}^*}}^3}\,\pslash\,\eslash \nonumber\\
&\quad 
+ \frac{(\varepsilon\!\cdot\! q)\, p_\mu p_\nu}{6\, m_{\mathrm{P_{bb}^*}}^3}\,\pslash\,\qslash
+ \frac{2(\varepsilon\!\cdot\! p)(p\!\cdot\! q)\, p_\mu p_\nu}{3\, m_{\mathrm{P_{bb}^*}}^5}\,\pslash\,\qslash
+ \frac{(\varepsilon\!\cdot\! q)(p\!\cdot\! q)\, p_\mu p_\nu}{3\, m_{\mathrm{P_{bb}^*}}^5}\,\pslash\,\qslash
+ \frac{(\varepsilon\!\cdot\! p)\, p_\mu p_\nu\, q^2}{3\, m_{\mathrm{P_{bb}^*}}^5}\,\pslash\,\qslash \nonumber\\
&\quad 
+ \frac{(\varepsilon\!\cdot\! q)\, p_\mu p_\nu\, q^2}{3\, m_{\mathrm{P_{bb}^*}}^5}\,\pslash\,\qslash
- \frac{(\varepsilon\!\cdot\! p)\, p_\nu q_\mu}{2\, m_{\mathrm{P_{bb}^*}}^3}\,\pslash\,\qslash
- \frac{(\varepsilon\!\cdot\! p)\, p_\mu q_\nu}{6\, m_{\mathrm{P_{bb}^*}}^3}\,\pslash\,\qslash
+ \frac{2(\varepsilon\!\cdot\! p)(p\!\cdot\! q)\, p_\mu q_\nu}{3\, m_{\mathrm{P_{bb}^*}}^5}\,\pslash\,\qslash \nonumber\\
&\quad 
+ \frac{(\varepsilon\!\cdot\! q)(p\!\cdot\! q)\, p_\mu q_\nu}{3\, m_{\mathrm{P_{bb}^*}}^5}\,\pslash\,\qslash
+ \frac{(\varepsilon\!\cdot\! p)\, p_\mu\, q^2 q_\nu}{3\, m_{\mathrm{P_{bb}^*}}^5}\,\pslash\,\qslash
+ \frac{(\varepsilon\!\cdot\! q)\, p_\mu\, q^2 q_\nu}{3\, m_{\mathrm{P_{bb}^*}}^5}\,\pslash\,\qslash
- \frac{(\varepsilon\!\cdot\! p)\, q_\mu q_\nu}{2\, m_{\mathrm{P_{bb}^*}}^3}\,\pslash\,\qslash \nonumber\\
&\quad 
- \frac{(\varepsilon\!\cdot\! p)(p\!\cdot\! q)^2\, p_\nu}{3\, m_{\mathrm{P_{bb}^*}}^5}\,\pslash\,\gamma_\mu
+ \frac{(\varepsilon\!\cdot\! p)\, p_\nu\, q^2}{6\, m_{\mathrm{P_{bb}^*}}^3}\,\pslash\,\gamma_\mu
+ \frac{(\varepsilon\!\cdot\! q)\, p_\nu\, q^2}{6\, m_{\mathrm{P_{bb}^*}}^3}\,\pslash\,\gamma_\mu
- \frac{(\varepsilon\!\cdot\! p)(p\!\cdot\! q)\, p_\nu\, q^2}{3\, m_{\mathrm{P_{bb}^*}}^5}\,\pslash\,\gamma_\mu \nonumber\\
&\quad 
+ \frac{(\varepsilon\!\cdot\! p)(p\!\cdot\! q)\, q_\nu}{2\, m_{\mathrm{P_{bb}^*}}^3}\,\pslash\,\gamma_\mu
- \frac{(\varepsilon\!\cdot\! p)(p\!\cdot\! q)^2\, q_\nu}{3\, m_{\mathrm{P_{bb}^*}}^5}\,\pslash\,\gamma_\mu
+ \frac{(\varepsilon\!\cdot\! p)\, q^2 q_\nu}{6\, m_{\mathrm{P_{bb}^*}}^3}\,\pslash\,\gamma_\mu
+ \frac{(\varepsilon\!\cdot\! q)\, q^2 q_\nu}{6\, m_{\mathrm{P_{bb}^*}}^3}\,\pslash\,\gamma_\mu \nonumber\\
&\quad 
- \frac{(\varepsilon\!\cdot\! p)(p\!\cdot\! q)\, q^2 q_\nu}{3\, m_{\mathrm{P_{bb}^*}}^5}\,\pslash\,\gamma_\mu
- \frac{(\varepsilon\!\cdot\! p)(p\!\cdot\! q)^2\, p_\mu}{3\, m_{\mathrm{P_{bb}^*}}^5}\,\pslash\,\gamma_\nu
- \frac{(\varepsilon\!\cdot\! p)\, p_\mu\, q^2}{6\, m_{\mathrm{P_{bb}^*}}^3}\,\pslash\,\gamma_\nu
- \frac{(\varepsilon\!\cdot\! q)\, p_\mu\, q^2}{6\, m_{\mathrm{P_{bb}^*}}^3}\,\pslash\,\gamma_\nu \nonumber\\
%\end{align}
%\begin{align}
&\quad 
- \frac{(\varepsilon\!\cdot\! p)(p\!\cdot\! q)\, p_\mu\, q^2}{3\, m_{\mathrm{P_{bb}^*}}^5}\,\pslash\,\gamma_\nu
+ \frac{(\varepsilon\!\cdot\! p)(p\!\cdot\! q)\, q_\mu}{2\, m_{\mathrm{P_{bb}^*}}^3}\,\pslash\,\gamma_\nu
+ \frac{(\varepsilon\!\cdot\! p)\, q^2 q_\mu}{2\, m_{\mathrm{P_{bb}^*}}^3}\,\pslash\,\gamma_\nu
+ \frac{(p\!\cdot\! q)\, p_\mu p_\nu}{6\, m_{\mathrm{P_{bb}^*}}^3}\,\qslash\,\eslash \nonumber\\
&\quad 
+ \frac{p_\mu p_\nu\, q^2}{3\, m_{\mathrm{P_{bb}^*}}^3}\,\qslash\,\eslash
- \frac{p_\mu q_\nu}{6\, m_{\mathrm{P_{bb}^*}}}\,\qslash\,\eslash
+ \frac{(p\!\cdot\! q)\, p_\mu q_\nu}{6\, m_{\mathrm{P_{bb}^*}}^3}\,\qslash\,\eslash
+ \frac{p_\mu\, q^2 q_\nu}{3\, m_{\mathrm{P_{bb}^*}}^3}\,\qslash\,\eslash
+ \frac{(\varepsilon\!\cdot\! q)\, p_\mu p_\nu}{6\, m_{\mathrm{P_{bb}^*}}^3}\,\qslash\,\pslash \nonumber\\
&\quad 
+ \frac{(\varepsilon\!\cdot\! q)\, p_\mu q_\nu}{6\, m_{\mathrm{P_{bb}^*}}^3}\,\qslash\,\pslash
- \frac{(\varepsilon\!\cdot\! q)\, p_\mu}{3\, m_{\mathrm{P_{bb}^*}}}\,\qslash\,\gamma_\nu
+ \frac{(\varepsilon\!\cdot\! q)(p\!\cdot\! q)\, p_\mu}{6\, m_{\mathrm{P_{bb}^*}}^3}\,\qslash\,\gamma_\nu
+ \frac{(\varepsilon\!\cdot\! q)\, p_\mu\, q^2}{6\, m_{\mathrm{P_{bb}^*}}^3}\,\qslash\,\gamma_\nu \nonumber\\
&\quad 
- \frac{(p\!\cdot\! q)^2\, p_\nu}{3\, m_{\mathrm{P_{bb}^*}}^3}\,\gamma_\mu\,\eslash
- \frac{(p\!\cdot\! q)\, p_\nu\, q^2}{12\, m_{\mathrm{P_{bb}^*}}^3}\,\gamma_\mu\,\eslash
+ \frac{p_\nu\, (q^2)^2}{6\, m_{\mathrm{P_{bb}^*}}^3}\,\gamma_\mu\,\eslash
+ \frac{(p\!\cdot\! q)\, q_\nu}{2\, m_{\mathrm{P_{bb}^*}}}\,\gamma_\mu\,\eslash
- \frac{(p\!\cdot\! q)^2\, q_\nu}{3\, m_{\mathrm{P_{bb}^*}}^3}\,\gamma_\mu\,\eslash \nonumber\\
&\quad 
- \frac{q^2 q_\nu}{4\, m_{\mathrm{P_{bb}^*}}}\,\gamma_\mu\,\eslash
- \frac{(p\!\cdot\! q)\, q^2 q_\nu}{12\, m_{\mathrm{P_{bb}^*}}^3}\,\gamma_\mu\,\eslash
+ \frac{(q^2)^2 q_\nu}{6\, m_{\mathrm{P_{bb}^*}}^3}\,\gamma_\mu\,\eslash
- \frac{(\varepsilon\!\cdot\! q)(p\!\cdot\! q)\, p_\nu}{3\, m_{\mathrm{P_{bb}^*}}^3}\,\gamma_\mu\,\qslash
- \frac{(\varepsilon\!\cdot\! q)\, p_\nu\, q^2}{3\, m_{\mathrm{P_{bb}^*}}^3}\,\gamma_\mu\,\qslash \nonumber\\
&\quad 
+ \frac{(\varepsilon\!\cdot\! q)\, q_\nu}{6\, m_{\mathrm{P_{bb}^*}}}\,\gamma_\mu\,\qslash
- \frac{(\varepsilon\!\cdot\! q)(p\!\cdot\! q)\, q_\nu}{3\, m_{\mathrm{P_{bb}^*}}^3}\,\gamma_\mu\,\qslash
- \frac{(\varepsilon\!\cdot\! q)\, q^2 q_\nu}{3\, m_{\mathrm{P_{bb}^*}}^3}\,\gamma_\mu\,\qslash
+ \frac{(\varepsilon\!\cdot\! q)\, q^2}{6\, m_{\mathrm{P_{bb}^*}}}\,\gamma_\mu\,\gamma_\nu \nonumber\\
&\quad 
+ \frac{(p\!\cdot\! q)\, p_\mu p_\nu}{6\, m_{\mathrm{P_{bb}^*}}^4}\,\eslash\,\pslash\,\qslash
- \frac{p_\nu q_\mu}{4\, m_{\mathrm{P_{bb}^*}}^2}\,\eslash\,\pslash\,\qslash
+ \frac{(p\!\cdot\! q)\, p_\mu q_\nu}{6\, m_{\mathrm{P_{bb}^*}}^4}\,\eslash\,\pslash\,\qslash
- \frac{q_\mu q_\nu}{4\, m_{\mathrm{P_{bb}^*}}^2}\,\eslash\,\pslash\,\qslash \nonumber\\
&\quad 
- \frac{(p\!\cdot\! q)^2\, p_\mu}{6\, m_{\mathrm{P_{bb}^*}}^4}\,\eslash\,\pslash\,\gamma_\nu
- \frac{p_\mu\, q^2}{12\, m_{\mathrm{P_{bb}^*}}^2}\,\eslash\,\pslash\,\gamma_\nu
- \frac{(p\!\cdot\! q)\, p_\mu\, q^2}{6\, m_{\mathrm{P_{bb}^*}}^4}\,\eslash\,\pslash\,\gamma_\nu
+ \frac{(p\!\cdot\! q)\, q_\mu}{4\, m_{\mathrm{P_{bb}^*}}^2}\,\eslash\,\pslash\,\gamma_\nu
+ \frac{q^2 q_\mu}{4\, m_{\mathrm{P_{bb}^*}}^2}\,\eslash\,\pslash\,\gamma_\nu \nonumber\\
&\quad 
- \frac{(p\!\cdot\! q)^2\, p_\mu}{6\, m_{\mathrm{P_{bb}^*}}^4}\,\eslash\,\qslash\,\gamma_\nu
- \frac{(p\!\cdot\! q)\, p_\mu\, q^2}{6\, m_{\mathrm{P_{bb}^*}}^4}\,\eslash\,\qslash\,\gamma_\nu
+ \frac{(p\!\cdot\! q)\, q_\mu}{4\, m_{\mathrm{P_{bb}^*}}^2}\,\eslash\,\qslash\,\gamma_\nu
+ \frac{q^2 q_\mu}{4\, m_{\mathrm{P_{bb}^*}}^2}\,\eslash\,\qslash\,\gamma_\nu
- \frac{p_\mu\, q^2}{12\, m_{\mathrm{P_{bb}^*}}^2}\,\eslash\,\gamma_\nu\,\qslash \nonumber\\
&\quad 
+ \frac{(p\!\cdot\! q)\, p_\mu p_\nu}{6\, m_{\mathrm{P_{bb}^*}}^4}\,\pslash\,\eslash\,\qslash
- \frac{(p\!\cdot\! q)^2\, p_\mu p_\nu}{3\, m_{\mathrm{P_{bb}^*}}^6}\,\pslash\,\eslash\,\qslash
+ \frac{p_\mu p_\nu\, q^2}{12\, m_{\mathrm{P_{bb}^*}}^4}\,\pslash\,\eslash\,\qslash
- \frac{(p\!\cdot\! q)\, p_\mu p_\nu\, q^2}{3\, m_{\mathrm{P_{bb}^*}}^6}\,\pslash\,\eslash\,\qslash \nonumber\\
&\quad 
- \frac{p_\nu q_\mu}{4\, m_{\mathrm{P_{bb}^*}}^2}\,\pslash\,\eslash\,\qslash
+ \frac{(p\!\cdot\! q)\, p_\nu q_\mu}{2\, m_{\mathrm{P_{bb}^*}}^4}\,\pslash\,\eslash\,\qslash
+ \frac{p_\nu\, q^2 q_\mu}{2\, m_{\mathrm{P_{bb}^*}}^4}\,\pslash\,\eslash\,\qslash
+ \frac{(p\!\cdot\! q)\, p_\mu q_\nu}{3\, m_{\mathrm{P_{bb}^*}}^4}\,\pslash\,\eslash\,\qslash \nonumber\\
&\quad 
- \frac{(p\!\cdot\! q)^2\, p_\mu q_\nu}{3\, m_{\mathrm{P_{bb}^*}}^6}\,\pslash\,\eslash\,\qslash
+ \frac{p_\mu\, q^2 q_\nu}{12\, m_{\mathrm{P_{bb}^*}}^4}\,\pslash\,\eslash\,\qslash
- \frac{(p\!\cdot\! q)\, p_\mu\, q^2 q_\nu}{3\, m_{\mathrm{P_{bb}^*}}^6}\,\pslash\,\eslash\,\qslash
- \frac{q_\mu q_\nu}{2\, m_{\mathrm{P_{bb}^*}}^2}\,\pslash\,\eslash\,\qslash
+ \frac{(p\!\cdot\! q)\, q_\mu q_\nu}{2\, m_{\mathrm{P_{bb}^*}}^4}\,\pslash\,\eslash\,\qslash \nonumber\\
&\quad 
+ \frac{q^2 q_\mu q_\nu}{2\, m_{\mathrm{P_{bb}^*}}^4}\,\pslash\,\eslash\,\qslash
- \frac{(p\!\cdot\! q)^2\, p_\mu}{6\, m_{\mathrm{P_{bb}^*}}^4}\,\pslash\,\eslash\,\gamma_\nu
- \frac{p_\mu\, q^2}{12\, m_{\mathrm{P_{bb}^*}}^2}\,\pslash\,\eslash\,\gamma_\nu
- \frac{(p\!\cdot\! q)\, p_\mu\, q^2}{12\, m_{\mathrm{P_{bb}^*}}^4}\,\pslash\,\eslash\,\gamma_\nu 
\nonumber%\\
\end{align}

\begin{align}
&\quad 
- \frac{p_\mu\, (q^2)^2}{12\, m_{\mathrm{P_{bb}^*}}^4}\,\pslash\,\eslash\,\gamma_\nu
+ \frac{(p\!\cdot\! q)\, q_\mu}{4\, m_{\mathrm{P_{bb}^*}}^2}\,\pslash\,\eslash\,\gamma_\nu
- \frac{(p\!\cdot\! q)\, p_\mu p_\nu}{6\, m_{\mathrm{P_{bb}^*}}^4}\,\pslash\,\qslash\,\eslash
- \frac{p_\mu p_\nu\, q^2}{6\, m_{\mathrm{P_{bb}^*}}^4}\,\pslash\,\qslash\,\eslash
+ \frac{p_\mu q_\nu}{12\, m_{\mathrm{P_{bb}^*}}^2}\,\pslash\,\qslash\,\eslash \nonumber\\
&\quad 
- \frac{(p\!\cdot\! q)\, p_\mu q_\nu}{6\, m_{\mathrm{P_{bb}^*}}^4}\,\pslash\,\qslash\,\eslash
- \frac{p_\mu\, q^2 q_\nu}{6\, m_{\mathrm{P_{bb}^*}}^4}\,\pslash\,\qslash\,\eslash
+ \frac{(\varepsilon\!\cdot\! q)\, p_\mu}{6\, m_{\mathrm{P_{bb}^*}}^2}\,\pslash\,\qslash\,\gamma_\nu
+ \frac{(\varepsilon\!\cdot\! p)(p\!\cdot\! q)\, p_\mu}{6\, m_{\mathrm{P_{bb}^*}}^4}\,\pslash\,\qslash\,\gamma_\nu \nonumber\\
&\quad 
+ \frac{(\varepsilon\!\cdot\! q)(p\!\cdot\! q)\, p_\mu}{6\, m_{\mathrm{P_{bb}^*}}^4}\,\pslash\,\qslash\,\gamma_\nu
+ \frac{(\varepsilon\!\cdot\! p)\, p_\mu\, q^2}{6\, m_{\mathrm{P_{bb}^*}}^4}\,\pslash\,\qslash\,\gamma_\nu
+ \frac{(\varepsilon\!\cdot\! q)\, p_\mu\, q^2}{6\, m_{\mathrm{P_{bb}^*}}^4}\,\pslash\,\qslash\,\gamma_\nu
+ \frac{(p\!\cdot\! q)^2\, p_\nu}{6\, m_{\mathrm{P_{bb}^*}}^4}\,\pslash\,\gamma_\mu\,\eslash \nonumber\\
&\quad 
- \frac{p_\nu\, q^2}{12\, m_{\mathrm{P_{bb}^*}}^2}\,\pslash\,\gamma_\mu\,\eslash
+ \frac{5(p\!\cdot\! q)\, p_\nu\, q^2}{12\, m_{\mathrm{P_{bb}^*}}^4}\,\pslash\,\gamma_\mu\,\eslash
+ \frac{p_\nu\, (q^2)^2}{6\, m_{\mathrm{P_{bb}^*}}^4}\,\pslash\,\gamma_\mu\,\eslash
- \frac{(p\!\cdot\! q)\, q_\nu}{4\, m_{\mathrm{P_{bb}^*}}^2}\,\pslash\,\gamma_\mu\,\eslash \nonumber\\
&\quad 
+ \frac{(p\!\cdot\! q)^2\, q_\nu}{6\, m_{\mathrm{P_{bb}^*}}^4}\,\pslash\,\gamma_\mu\,\eslash
- \frac{q^2 q_\nu}{3\, m_{\mathrm{P_{bb}^*}}^2}\,\pslash\,\gamma_\mu\,\eslash
+ \frac{5(p\!\cdot\! q)\, q^2 q_\nu}{12\, m_{\mathrm{P_{bb}^*}}^4}\,\pslash\,\gamma_\mu\,\eslash
+ \frac{(q^2)^2 q_\nu}{6\, m_{\mathrm{P_{bb}^*}}^4}\,\pslash\,\gamma_\mu\,\eslash \nonumber\\
&\quad 
+ \frac{(\varepsilon\!\cdot\! q)\, p_\nu}{6\, m_{\mathrm{P_{bb}^*}}^2}\,\pslash\,\gamma_\mu\,\qslash
- \frac{(\varepsilon\!\cdot\! p)(p\!\cdot\! q)\, p_\nu}{6\, m_{\mathrm{P_{bb}^*}}^4}\,\pslash\,\gamma_\mu\,\qslash
- \frac{(\varepsilon\!\cdot\! q)(p\!\cdot\! q)\, p_\nu}{3\, m_{\mathrm{P_{bb}^*}}^4}\,\pslash\,\gamma_\mu\,\qslash
- \frac{(\varepsilon\!\cdot\! p)\, p_\nu\, q^2}{3\, m_{\mathrm{P_{bb}^*}}^4}\,\pslash\,\gamma_\mu\,\qslash \nonumber\\
&\quad 
- \frac{(\varepsilon\!\cdot\! q)\, p_\nu\, q^2}{3\, m_{\mathrm{P_{bb}^*}}^4}\,\pslash\,\gamma_\mu\,\qslash
+ \frac{(\varepsilon\!\cdot\! p)\, q_\nu}{6\, m_{\mathrm{P_{bb}^*}}^2}\,\pslash\,\gamma_\mu\,\qslash
+ \frac{(\varepsilon\!\cdot\! q)\, q_\nu}{3\, m_{\mathrm{P_{bb}^*}}^2}\,\pslash\,\gamma_\mu\,\qslash
- \frac{(\varepsilon\!\cdot\! p)(p\!\cdot\! q)\, q_\nu}{6\, m_{\mathrm{P_{bb}^*}}^4}\,\pslash\,\gamma_\mu\,\qslash \nonumber\\
&\quad 
- \frac{(\varepsilon\!\cdot\! q)(p\!\cdot\! q)\, q_\nu}{3\, m_{\mathrm{P_{bb}^*}}^4}\,\pslash\,\gamma_\mu\,\qslash
- \frac{(\varepsilon\!\cdot\! p)\, q^2 q_\nu}{3\, m_{\mathrm{P_{bb}^*}}^4}\,\pslash\,\gamma_\mu\,\qslash
- \frac{(\varepsilon\!\cdot\! q)\, q^2 q_\nu}{3\, m_{\mathrm{P_{bb}^*}}^4}\,\pslash\,\gamma_\mu\,\qslash
- \frac{(\varepsilon\!\cdot\! p)(p\!\cdot\! q)^2}{6\, m_{\mathrm{P_{bb}^*}}^4}\,\pslash\,\gamma_\mu\,\gamma_\nu \nonumber\\
&\quad 
+ \frac{(\varepsilon\!\cdot\! p)\, q^2}{6\, m_{\mathrm{P_{bb}^*}}^2}\,\pslash\,\gamma_\mu\,\gamma_\nu
+ \frac{(\varepsilon\!\cdot\! q)\, q^2}{6\, m_{\mathrm{P_{bb}^*}}^2}\,\pslash\,\gamma_\mu\,\gamma_\nu
- \frac{(\varepsilon\!\cdot\! p)(p\!\cdot\! q)\, q^2}{6\, m_{\mathrm{P_{bb}^*}}^4}\,\pslash\,\gamma_\mu\,\gamma_\nu
- \frac{(\varepsilon\!\cdot\! p)(p\!\cdot\! q)\, p_\mu}{3\, m_{\mathrm{P_{bb}^*}}^4}\,\pslash\,\gamma_\nu\,\qslash \nonumber\\
%\end{align}
%\begin{align}
&\quad 
+ \frac{(\varepsilon\!\cdot\! p)\, q_\mu}{2\, m_{\mathrm{P_{bb}^*}}^2}\,\pslash\,\gamma_\nu\,\qslash
+ \frac{(p\!\cdot\! q)\, p_\mu p_\nu}{6\, m_{\mathrm{P_{bb}^*}}^4}\,\qslash\,\eslash\,\pslash
+ \frac{p_\mu p_\nu\, q^2}{6\, m_{\mathrm{P_{bb}^*}}^4}\,\qslash\,\eslash\,\pslash
- \frac{p_\mu q_\nu}{12\, m_{\mathrm{P_{bb}^*}}^2}\,\qslash\,\eslash\,\pslash \nonumber\\
&\quad 
+ \frac{(p\!\cdot\! q)\, p_\mu q_\nu}{6\, m_{\mathrm{P_{bb}^*}}^4}\,\qslash\,\eslash\,\pslash
+ \frac{p_\mu\, q^2 q_\nu}{6\, m_{\mathrm{P_{bb}^*}}^4}\,\qslash\,\eslash\,\pslash
+ \frac{p_\mu\, q^2}{6\, m_{\mathrm{P_{bb}^*}}^2}\,\qslash\,\eslash\,\gamma_\nu
+ \frac{(\varepsilon\!\cdot\! q)\, p_\mu}{6\, m_{\mathrm{P_{bb}^*}}^2}\,\qslash\,\pslash\,\gamma_\nu \nonumber\\
&\quad 
- \frac{(p\!\cdot\! q)^2\, p_\nu}{6\, m_{\mathrm{P_{bb}^*}}^4}\,\gamma_\mu\,\eslash\,\pslash
+ \frac{p_\nu\, q^2}{12\, m_{\mathrm{P_{bb}^*}}^2}\,\gamma_\mu\,\eslash\,\pslash
- \frac{(p\!\cdot\! q)\, p_\nu\, q^2}{6\, m_{\mathrm{P_{bb}^*}}^4}\,\gamma_\mu\,\eslash\,\pslash
+ \frac{(p\!\cdot\! q)\, q_\nu}{4\, m_{\mathrm{P_{bb}^*}}^2}\,\gamma_\mu\,\eslash\,\pslash \nonumber\\
&\quad 
- \frac{(p\!\cdot\! q)^2\, q_\nu}{6\, m_{\mathrm{P_{bb}^*}}^4}\,\gamma_\mu\,\eslash\,\pslash
+ \frac{q^2 q_\nu}{12\, m_{\mathrm{P_{bb}^*}}^2}\,\gamma_\mu\,\eslash\,\pslash
- \frac{(p\!\cdot\! q)\, q^2 q_\nu}{6\, m_{\mathrm{P_{bb}^*}}^4}\,\gamma_\mu\,\eslash\,\pslash
+ \frac{(p\!\cdot\! q)\, p_\nu}{6\, m_{\mathrm{P_{bb}^*}}^2}\,\gamma_\mu\,\eslash\,\qslash
- \frac{(p\!\cdot\! q)^2\, p_\nu}{6\, m_{\mathrm{P_{bb}^*}}^4}\,\gamma_\mu\,\eslash\,\qslash \nonumber\\
&\quad 
- \frac{p_\nu\, q^2}{12\, m_{\mathrm{P_{bb}^*}}^2}\,\gamma_\mu\,\eslash\,\qslash
- \frac{(p\!\cdot\! q)\, p_\nu\, q^2}{6\, m_{\mathrm{P_{bb}^*}}^4}\,\gamma_\mu\,\eslash\,\qslash
+ \frac{(p\!\cdot\! q)\, q_\nu}{4\, m_{\mathrm{P_{bb}^*}}^2}\,\gamma_\mu\,\eslash\,\qslash
- \frac{(p\!\cdot\! q)^2\, q_\nu}{6\, m_{\mathrm{P_{bb}^*}}^4}\,\gamma_\mu\,\eslash\,\qslash \nonumber\\
&\quad 
- \frac{q^2 q_\nu}{12\, m_{\mathrm{P_{bb}^*}}^2}\,\gamma_\mu\,\eslash\,\qslash
- \frac{(p\!\cdot\! q)\, q^2 q_\nu}{6\, m_{\mathrm{P_{bb}^*}}^4}\,\gamma_\mu\,\eslash\,\qslash
- \frac{(p\!\cdot\! q)^2}{6\, m_{\mathrm{P_{bb}^*}}^2}\,\gamma_\mu\,\eslash\,\gamma_\nu
+ \frac{(q^2)^2}{12\, m_{\mathrm{P_{bb}^*}}^2}\,\gamma_\mu\,\eslash\,\gamma_\nu \nonumber\\
&\quad 
+ \frac{(p\!\cdot\! q)\, p_\nu}{6\, m_{\mathrm{P_{bb}^*}}^2}\,\gamma_\mu\,\qslash\,\eslash
+ \frac{(p\!\cdot\! q)\, q_\nu}{6\, m_{\mathrm{P_{bb}^*}}^2}\,\gamma_\mu\,\qslash\,\eslash
- \frac{(\varepsilon\!\cdot\! q)\, p_\nu}{6\, m_{\mathrm{P_{bb}^*}}^2}\,\gamma_\mu\,\qslash\,\pslash
- \frac{(\varepsilon\!\cdot\! q)\, q_\nu}{6\, m_{\mathrm{P_{bb}^*}}^2}\,\gamma_\mu\,\qslash\,\pslash \nonumber\\
&\quad 
- \frac{(\varepsilon\!\cdot\! q)(p\!\cdot\! q)}{6\, m_{\mathrm{P_{bb}^*}}^2}\,\gamma_\mu\,\qslash\,\gamma_\nu
- \frac{(\varepsilon\!\cdot\! q)\, q^2}{6\, m_{\mathrm{P_{bb}^*}}^2}\,\gamma_\mu\,\qslash\,\gamma_\nu
- \frac{(p\!\cdot\! q)\, p_\mu}{6\, m_{\mathrm{P_{bb}^*}}^3}\,\eslash\,\pslash\,\gamma_\nu\,\qslash
+ \frac{q_\mu}{4\, m_{\mathrm{P_{bb}^*}}}\,\eslash\,\pslash\,\gamma_\nu\,\qslash \nonumber\\
&\quad 
- \frac{(p\!\cdot\! q)^2\, p_\mu}{6\, m_{\mathrm{P_{bb}^*}}^5}\,\pslash\,\eslash\,\qslash\,\gamma_\nu
- \frac{(p\!\cdot\! q)\, p_\mu\, q^2}{6\, m_{\mathrm{P_{bb}^*}}^5}\,\pslash\,\eslash\,\qslash\,\gamma_\nu
+ \frac{(p\!\cdot\! q)\, q_\mu}{4\, m_{\mathrm{P_{bb}^*}}^3}\,\pslash\,\eslash\,\qslash\,\gamma_\nu
+ \frac{q^2 q_\mu}{4\, m_{\mathrm{P_{bb}^*}}^3}\,\pslash\,\eslash\,\qslash\,\gamma_\nu \nonumber\\
&\quad 
- \frac{(p\!\cdot\! q)\, p_\mu}{6\, m_{\mathrm{P_{bb}^*}}^3}\,\pslash\,\eslash\,\gamma_\nu\,\qslash
- \frac{p_\mu\, q^2}{12\, m_{\mathrm{P_{bb}^*}}^3}\,\pslash\,\eslash\,\gamma_\nu\,\qslash
+ \frac{q_\mu}{4\, m_{\mathrm{P_{bb}^*}}}\,\pslash\,\eslash\,\gamma_\nu\,\qslash
- \frac{(p\!\cdot\! q)\, p_\mu}{12\, m_{\mathrm{P_{bb}^*}}^3}\,\pslash\,\qslash\,\eslash\,\gamma_\nu \nonumber\\
&\quad 
- \frac{p_\mu\, q^2}{12\, m_{\mathrm{P_{bb}^*}}^3}\,\pslash\,\qslash\,\eslash\,\gamma_\nu
- \frac{(p\!\cdot\! q)\, p_\nu}{12\, m_{\mathrm{P_{bb}^*}}^3}\,\pslash\,\gamma_\mu\,\eslash\,\qslash
- \frac{(p\!\cdot\! q)^2\, p_\nu}{6\, m_{\mathrm{P_{bb}^*}}^5}\,\pslash\,\gamma_\mu\,\eslash\,\qslash
- \frac{p_\nu\, q^2}{12\, m_{\mathrm{P_{bb}^*}}^3}\,\pslash\,\gamma_\mu\,\eslash\,\qslash \nonumber\\
&\quad 
- \frac{(p\!\cdot\! q)\, p_\nu\, q^2}{6\, m_{\mathrm{P_{bb}^*}}^5}\,\pslash\,\gamma_\mu\,\eslash\,\qslash
- \frac{(p\!\cdot\! q)^2\, q_\nu}{6\, m_{\mathrm{P_{bb}^*}}^5}\,\pslash\,\gamma_\mu\,\eslash\,\qslash
- \frac{q^2 q_\nu}{12\, m_{\mathrm{P_{bb}^*}}^3}\,\pslash\,\gamma_\mu\,\eslash\,\qslash
- \frac{(p\!\cdot\! q)\, q^2 q_\nu}{6\, m_{\mathrm{P_{bb}^*}}^5}\,\pslash\,\gamma_\mu\,\eslash\,\qslash \nonumber%\\
\end{align}

\begin{align}
&\quad 
+ \frac{(p\!\cdot\! q)^2}{12\, m_{\mathrm{P_{bb}^*}}^3}\,\pslash\,\gamma_\mu\,\eslash\,\gamma_\nu
- \frac{q^2}{12\, m_{\mathrm{P_{bb}^*}}}\,\pslash\,\gamma_\mu\,\eslash\,\gamma_\nu
+ \frac{(p\!\cdot\! q)\, q^2}{4\, m_{\mathrm{P_{bb}^*}}^3}\,\pslash\,\gamma_\mu\,\eslash\,\gamma_\nu
+ \frac{(q^2)^2}{12\, m_{\mathrm{P_{bb}^*}}^3}\,\pslash\,\gamma_\mu\,\eslash\,\gamma_\nu \nonumber\\
&\quad 
- \frac{(p\!\cdot\! q)\, p_\nu}{6\, m_{\mathrm{P_{bb}^*}}^3}\,\pslash\,\gamma_\mu\,\qslash\,\eslash
- \frac{p_\nu\, q^2}{6\, m_{\mathrm{P_{bb}^*}}^3}\,\pslash\,\gamma_\mu\,\qslash\,\eslash
+ \frac{q_\nu}{12\, m_{\mathrm{P_{bb}^*}}}\,\pslash\,\gamma_\mu\,\qslash\,\eslash
- \frac{(p\!\cdot\! q)\, q_\nu}{6\, m_{\mathrm{P_{bb}^*}}^3}\,\pslash\,\gamma_\mu\,\qslash\,\eslash \nonumber\\
&\quad 
- \frac{q^2 q_\nu}{6\, m_{\mathrm{P_{bb}^*}}^3}\,\pslash\,\gamma_\mu\,\qslash\,\eslash
+ \frac{(\varepsilon\!\cdot\! q)}{6\, m_{\mathrm{P_{bb}^*}}}\,\pslash\,\gamma_\mu\,\qslash\,\gamma_\nu
- \frac{(\varepsilon\!\cdot\! p)(p\!\cdot\! q)}{6\, m_{\mathrm{P_{bb}^*}}^3}\,\pslash\,\gamma_\mu\,\qslash\,\gamma_\nu
- \frac{(\varepsilon\!\cdot\! q)(p\!\cdot\! q)}{6\, m_{\mathrm{P_{bb}^*}}^3}\,\pslash\,\gamma_\mu\,\qslash\,\gamma_\nu \nonumber\\
&\quad 
- \frac{(\varepsilon\!\cdot\! p)\, q^2}{6\, m_{\mathrm{P_{bb}^*}}^3}\,\pslash\,\gamma_\mu\,\qslash\,\gamma_\nu
- \frac{(\varepsilon\!\cdot\! q)\, q^2}{6\, m_{\mathrm{P_{bb}^*}}^3}\,\pslash\,\gamma_\mu\,\qslash\,\gamma_\nu
- \frac{(\varepsilon\!\cdot\! p)(p\!\cdot\! q)}{6\, m_{\mathrm{P_{bb}^*}}^3}\,\pslash\,\gamma_\mu\,\gamma_\nu\,\qslash
+ \frac{(p\!\cdot\! q)\, p_\mu}{12\, m_{\mathrm{P_{bb}^*}}^3}\,\qslash\,\eslash\,\pslash\,\gamma_\nu \nonumber\\
&\quad 
+ \frac{p_\mu\, q^2}{12\, m_{\mathrm{P_{bb}^*}}^3}\,\qslash\,\eslash\,\pslash\,\gamma_\nu
+ \frac{(p\!\cdot\! q)\, p_\nu}{12\, m_{\mathrm{P_{bb}^*}}^3}\,\gamma_\mu\,\eslash\,\pslash\,\qslash
+ \frac{(p\!\cdot\! q)\, q_\nu}{12\, m_{\mathrm{P_{bb}^*}}^3}\,\gamma_\mu\,\eslash\,\pslash\,\qslash
- \frac{(p\!\cdot\! q)^2}{12\, m_{\mathrm{P_{bb}^*}}^3}\,\gamma_\mu\,\eslash\,\pslash\,\gamma_\nu \nonumber\\
&\quad 
+ \frac{q^2}{12\, m_{\mathrm{P_{bb}^*}}}\,\gamma_\mu\,\eslash\,\pslash\,\gamma_\nu
- \frac{(p\!\cdot\! q)\, q^2}{12\, m_{\mathrm{P_{bb}^*}}^3}\,\gamma_\mu\,\eslash\,\pslash\,\gamma_\nu
- \frac{(p\!\cdot\! q)^2}{12\, m_{\mathrm{P_{bb}^*}}^3}\,\gamma_\mu\,\eslash\,\qslash\,\gamma_\nu
- \frac{(p\!\cdot\! q)\, q^2}{12\, m_{\mathrm{P_{bb}^*}}^3}\,\gamma_\mu\,\eslash\,\qslash\,\gamma_\nu \nonumber\\
&\quad 
- \frac{(p\!\cdot\! q)}{6\, m_{\mathrm{P_{bb}^*}}}\,\gamma_\mu\,\eslash\,\gamma_\nu\,\qslash
+ \frac{q^2}{12\, m_{\mathrm{P_{bb}^*}}}\,\gamma_\mu\,\eslash\,\gamma_\nu\,\qslash
- \frac{(p\!\cdot\! q)\, p_\nu}{6\, m_{\mathrm{P_{bb}^*}}^3}\,\gamma_\mu\,\qslash\,\eslash\,\pslash
- \frac{p_\nu\, q^2}{6\, m_{\mathrm{P_{bb}^*}}^3}\,\gamma_\mu\,\qslash\,\eslash\,\pslash \nonumber\\
&\quad 
+ \frac{q_\nu}{12\, m_{\mathrm{P_{bb}^*}}}\,\gamma_\mu\,\qslash\,\eslash\,\pslash
- \frac{(p\!\cdot\! q)\, q_\nu}{6\, m_{\mathrm{P_{bb}^*}}^3}\,\gamma_\mu\,\qslash\,\eslash\,\pslash
- \frac{q^2 q_\nu}{6\, m_{\mathrm{P_{bb}^*}}^3}\,\gamma_\mu\,\qslash\,\eslash\,\pslash
+ \frac{(p\!\cdot\! q)}{6\, m_{\mathrm{P_{bb}^*}}}\,\gamma_\mu\,\qslash\,\eslash\,\gamma_\nu
- \frac{(\varepsilon\!\cdot\! q)}{6\, m_{\mathrm{P_{bb}^*}}}\,\gamma_\mu\,\qslash\,\pslash\,\gamma_\nu \nonumber\\
&\quad 
- \frac{(p\!\cdot\! q)}{12\, m_{\mathrm{P_{bb}^*}}^2}\,\pslash\,\gamma_\mu\,\eslash\,\qslash\,\gamma_\nu
- \frac{q^2}{12\, m_{\mathrm{P_{bb}^*}}^2}\,\pslash\,\gamma_\mu\,\eslash\,\qslash\,\gamma_\nu
+ \frac{(p\!\cdot\! q)}{12\, m_{\mathrm{P_{bb}^*}}^2}\,\pslash\,\gamma_\mu\,\eslash\,\gamma_\nu\,\qslash
+ \frac{q^2}{12\, m_{\mathrm{P_{bb}^*}}^2}\,\pslash\,\gamma_\mu\,\eslash\,\gamma_\nu\,\qslash \nonumber\\
&\quad 
- \frac{(p\!\cdot\! q)}{12\, m_{\mathrm{P_{bb}^*}}^2}\,\pslash\,\gamma_\mu\,\qslash\,\eslash\,\gamma_\nu
- \frac{q^2}{12\, m_{\mathrm{P_{bb}^*}}^2}\,\pslash\,\gamma_\mu\,\qslash\,\eslash\,\gamma_\nu
- \frac{(p\!\cdot\! q)}{12\, m_{\mathrm{P_{bb}^*}}^2}\,\gamma_\mu\,\eslash\,\pslash\,\gamma_\nu\,\qslash
- \frac{(p\!\cdot\! q)}{12\, m_{\mathrm{P_{bb}^*}}^2}\,\gamma_\mu\,\qslash\,\eslash\,\pslash\,\gamma_\nu 
- \frac{q^2}{12\, m_{\mathrm{P_{bb}^*}}^2}\,\gamma_\mu\,\qslash\,\eslash\,\pslash\,\gamma_\nu
\Bigg\}.
\label{F3_appendix}
\end{align}

\subsection{Lorentz structures proportional to $F_4(q^2)$}

\begin{align}
F_4(q^2) & \propto  
\,\Bigg\{ 
- \frac{(p\!\cdot\! q)\, p_\mu p_\nu\, q^2}{6\, m_{\mathrm{P_{bb}^*}}^4}\,\eslash
+ \frac{(p\!\cdot\! q)^2\, p_\mu p_\nu\, q^2}{6\, m_{\mathrm{P_{bb}^*}}^6}\,\eslash
- \frac{p_\mu p_\nu\, (q^2)^2}{24\, m_{\mathrm{P_{bb}^*}}^4}\,\eslash
+ \frac{(p\!\cdot\! q)\, p_\mu p_\nu\, (q^2)^2}{6\, m_{\mathrm{P_{bb}^*}}^6}\,\eslash \nonumber\\
&\quad
+ \frac{p_\nu\, q^2 q_\mu}{4\, m_{\mathrm{P_{bb}^*}}^2}\,\eslash
- \frac{(p\!\cdot\! q)\, p_\nu\, q^2 q_\mu}{4\, m_{\mathrm{P_{bb}^*}}^4}\,\eslash
- \frac{p_\nu\, (q^2)^2 q_\mu}{4\, m_{\mathrm{P_{bb}^*}}^4}\,\eslash
- \frac{5(p\!\cdot\! q)\, p_\mu\, q^2 q_\nu}{12\, m_{\mathrm{P_{bb}^*}}^4}\,\eslash
+ \frac{(p\!\cdot\! q)^2\, p_\mu\, q^2 q_\nu}{6\, m_{\mathrm{P_{bb}^*}}^6}\,\eslash \nonumber\\
&\quad
- \frac{p_\mu\, (q^2)^2 q_\nu}{24\, m_{\mathrm{P_{bb}^*}}^4}\,\eslash
+ \frac{(p\!\cdot\! q)\, p_\mu\, (q^2)^2 q_\nu}{6\, m_{\mathrm{P_{bb}^*}}^6}\,\eslash
+ \frac{5\, q^2 q_\mu q_\nu}{8\, m_{\mathrm{P_{bb}^*}}^2}\,\eslash
- \frac{(p\!\cdot\! q)\, q^2 q_\mu q_\nu}{4\, m_{\mathrm{P_{bb}^*}}^4}\,\eslash
- \frac{(q^2)^2 q_\mu q_\nu}{4\, m_{\mathrm{P_{bb}^*}}^4}\,\eslash \nonumber\\
&\quad
+ \frac{(\varepsilon\!\cdot\! q)\, p_\mu p_\nu\, q^2}{6\, m_{\mathrm{P_{bb}^*}}^4}\,\pslash
+ \frac{(\varepsilon\!\cdot\! p)(p\!\cdot\! q)\, p_\mu p_\nu\, q^2}{3\, m_{\mathrm{P_{bb}^*}}^6}\,\pslash
+ \frac{(\varepsilon\!\cdot\! p)\, p_\mu p_\nu\, (q^2)^2}{6\, m_{\mathrm{P_{bb}^*}}^6}\,\pslash
- \frac{(\varepsilon\!\cdot\! p)\, p_\nu\, q^2 q_\mu}{4\, m_{\mathrm{P_{bb}^*}}^4}\,\pslash \nonumber\\
&\quad
- \frac{(\varepsilon\!\cdot\! p)\, p_\mu\, q^2 q_\nu}{4\, m_{\mathrm{P_{bb}^*}}^4}\,\pslash
+ \frac{(\varepsilon\!\cdot\! q)\, p_\mu\, q^2 q_\nu}{6\, m_{\mathrm{P_{bb}^*}}^4}\,\pslash
+ \frac{(\varepsilon\!\cdot\! p)(p\!\cdot\! q)\, p_\mu\, q^2 q_\nu}{3\, m_{\mathrm{P_{bb}^*}}^6}\,\pslash
+ \frac{(\varepsilon\!\cdot\! p)\, p_\mu\, (q^2)^2 q_\nu}{6\, m_{\mathrm{P_{bb}^*}}^6}\,\pslash \nonumber\\
&\quad
- \frac{(\varepsilon\!\cdot\! p)\, q^2 q_\mu q_\nu}{4\, m_{\mathrm{P_{bb}^*}}^4}\,\pslash
+ \frac{(\varepsilon\!\cdot\! q)(p\!\cdot\! q)\, p_\mu p_\nu}{6\, m_{\mathrm{P_{bb}^*}}^4}\,\qslash
+ \frac{(\varepsilon\!\cdot\! q)\, p_\mu p_\nu\, q^2}{12\, m_{\mathrm{P_{bb}^*}}^4}\,\qslash
+ \frac{(\varepsilon\!\cdot\! q)(p\!\cdot\! q)\, p_\mu q_\nu}{6\, m_{\mathrm{P_{bb}^*}}^4}\,\qslash 
\nonumber%\\
\end{align}

\begin{align}
&\quad
+ \frac{(\varepsilon\!\cdot\! q)\, p_\mu\, q^2 q_\nu}{12\, m_{\mathrm{P_{bb}^*}}^4}\,\qslash
- \frac{(\varepsilon\!\cdot\! q)\, p_\nu\, q^2}{6\, m_{\mathrm{P_{bb}^*}}^2}\,\gamma_\mu
- \frac{(\varepsilon\!\cdot\! q)\, q^2 q_\nu}{6\, m_{\mathrm{P_{bb}^*}}^2}\,\gamma_\mu
+ \frac{(p\!\cdot\! q)\, p_\mu p_\nu\, q^2}{6\, m_{\mathrm{P_{bb}^*}}^5}\,\eslash\,\pslash \nonumber\\
&\quad
+ \frac{p_\mu p_\nu\, (q^2)^2}{12\, m_{\mathrm{P_{bb}^*}}^5}\,\eslash\,\pslash
- \frac{p_\nu\, q^2 q_\mu}{8\, m_{\mathrm{P_{bb}^*}}^3}\,\eslash\,\pslash
- \frac{p_\mu\, q^2 q_\nu}{8\, m_{\mathrm{P_{bb}^*}}^3}\,\eslash\,\pslash
+ \frac{(p\!\cdot\! q)\, p_\mu\, q^2 q_\nu}{6\, m_{\mathrm{P_{bb}^*}}^5}\,\eslash\,\pslash
+ \frac{p_\mu\, (q^2)^2 q_\nu}{12\, m_{\mathrm{P_{bb}^*}}^5}\,\eslash\,\pslash \nonumber\\
&\quad
- \frac{q^2 q_\mu q_\nu}{8\, m_{\mathrm{P_{bb}^*}}^3}\,\eslash\,\pslash
+ \frac{(p\!\cdot\! q)^2\, p_\mu p_\nu}{6\, m_{\mathrm{P_{bb}^*}}^5}\,\eslash\,\qslash
+ \frac{(p\!\cdot\! q)\, p_\mu p_\nu\, q^2}{4\, m_{\mathrm{P_{bb}^*}}^5}\,\eslash\,\qslash
- \frac{(p\!\cdot\! q)\, p_\nu q_\mu}{4\, m_{\mathrm{P_{bb}^*}}^3}\,\eslash\,\qslash
- \frac{3\, p_\nu\, q^2 q_\mu}{8\, m_{\mathrm{P_{bb}^*}}^3}\,\eslash\,\qslash \nonumber\\
&\quad
- \frac{(p\!\cdot\! q)\, p_\mu q_\nu}{6\, m_{\mathrm{P_{bb}^*}}^3}\,\eslash\,\qslash
+ \frac{(p\!\cdot\! q)^2\, p_\mu q_\nu}{6\, m_{\mathrm{P_{bb}^*}}^5}\,\eslash\,\qslash
+ \frac{(p\!\cdot\! q)\, p_\mu\, q^2 q_\nu}{4\, m_{\mathrm{P_{bb}^*}}^5}\,\eslash\,\qslash
+ \frac{q_\mu q_\nu}{4\, m_{\mathrm{P_{bb}^*}}}\,\eslash\,\qslash
- \frac{(p\!\cdot\! q)\, q_\mu q_\nu}{4\, m_{\mathrm{P_{bb}^*}}^3}\,\eslash\,\qslash \nonumber\\
&\quad
- \frac{3\, q^2 q_\mu q_\nu}{8\, m_{\mathrm{P_{bb}^*}}^3}\,\eslash\,\qslash
- \frac{(p\!\cdot\! q)\, p_\mu\, q^2}{6\, m_{\mathrm{P_{bb}^*}}^3}\,\eslash\,\gamma_\nu
+ \frac{(p\!\cdot\! q)^2\, p_\mu\, q^2}{12\, m_{\mathrm{P_{bb}^*}}^5}\,\eslash\,\gamma_\nu
- \frac{p_\mu\, (q^2)^2}{24\, m_{\mathrm{P_{bb}^*}}^3}\,\eslash\,\gamma_\nu
+ \frac{(p\!\cdot\! q)\, p_\mu\, (q^2)^2}{12\, m_{\mathrm{P_{bb}^*}}^5}\,\eslash\,\gamma_\nu \nonumber\\
&\quad
+ \frac{q^2 q_\mu}{4\, m_{\mathrm{P_{bb}^*}}}\,\eslash\,\gamma_\nu
- \frac{(p\!\cdot\! q)\, q^2 q_\mu}{8\, m_{\mathrm{P_{bb}^*}}^3}\,\eslash\,\gamma_\nu
- \frac{(q^2)^2 q_\mu}{8\, m_{\mathrm{P_{bb}^*}}^3}\,\eslash\,\gamma_\nu
+ \frac{(p\!\cdot\! q)^3\, p_\mu p_\nu}{3\, m_{\mathrm{P_{bb}^*}}^7}\,\pslash\,\eslash \nonumber\\
&\quad
+ \frac{(p\!\cdot\! q)\, p_\mu p_\nu\, q^2}{12\, m_{\mathrm{P_{bb}^*}}^5}\,\pslash\,\eslash
+ \frac{(p\!\cdot\! q)^2\, p_\mu p_\nu\, q^2}{2\, m_{\mathrm{P_{bb}^*}}^7}\,\pslash\,\eslash
+ \frac{5\, p_\mu p_\nu\, (q^2)^2}{24\, m_{\mathrm{P_{bb}^*}}^5}\,\pslash\,\eslash
+ \frac{(p\!\cdot\! q)\, p_\mu p_\nu\, (q^2)^2}{6\, m_{\mathrm{P_{bb}^*}}^7}\,\pslash\,\eslash \nonumber\\
&\quad
- \frac{(p\!\cdot\! q)^2\, p_\nu q_\mu}{2\, m_{\mathrm{P_{bb}^*}}^5}\,\pslash\,\eslash
+ \frac{p_\nu\, q^2 q_\mu}{8\, m_{\mathrm{P_{bb}^*}}^3}\,\pslash\,\eslash
- \frac{3(p\!\cdot\! q)\, p_\nu\, q^2 q_\mu}{4\, m_{\mathrm{P_{bb}^*}}^5}\,\pslash\,\eslash
- \frac{p_\nu\, (q^2)^2 q_\mu}{4\, m_{\mathrm{P_{bb}^*}}^5}\,\pslash\,\eslash
- \frac{(p\!\cdot\! q)^2\, p_\mu q_\nu}{2\, m_{\mathrm{P_{bb}^*}}^5}\,\pslash\,\eslash \nonumber\\
&\quad
+ \frac{(p\!\cdot\! q)^3\, p_\mu q_\nu}{3\, m_{\mathrm{P_{bb}^*}}^7}\,\pslash\,\eslash
- \frac{3\, p_\mu\, q^2 q_\nu}{8\, m_{\mathrm{P_{bb}^*}}^3}\,\pslash\,\eslash
- \frac{(p\!\cdot\! q)\, p_\mu\, q^2 q_\nu}{6\, m_{\mathrm{P_{bb}^*}}^5}\,\pslash\,\eslash
+ \frac{(p\!\cdot\! q)^2\, p_\mu\, q^2 q_\nu}{2\, m_{\mathrm{P_{bb}^*}}^7}\,\pslash\,\eslash
+ \frac{5\, p_\mu\, (q^2)^2 q_\nu}{24\, m_{\mathrm{P_{bb}^*}}^5}\,\pslash\,\eslash \nonumber\\
&\quad
+ \frac{(p\!\cdot\! q)\, p_\mu\, (q^2)^2 q_\nu}{6\, m_{\mathrm{P_{bb}^*}}^7}\,\pslash\,\eslash
+ \frac{3(p\!\cdot\! q)\, q_\mu q_\nu}{4\, m_{\mathrm{P_{bb}^*}}^3}\,\pslash\,\eslash
- \frac{(p\!\cdot\! q)^2\, q_\mu q_\nu}{2\, m_{\mathrm{P_{bb}^*}}^5}\,\pslash\,\eslash
+ \frac{q^2 q_\mu q_\nu}{2\, m_{\mathrm{P_{bb}^*}}^3}\,\pslash\,\eslash
- \frac{3(p\!\cdot\! q)\, q^2 q_\mu q_\nu}{4\, m_{\mathrm{P_{bb}^*}}^5}\,\pslash\,\eslash \nonumber\\
&\quad
- \frac{(q^2)^2 q_\mu q_\nu}{4\, m_{\mathrm{P_{bb}^*}}^5}\,\pslash\,\eslash
- \frac{(\varepsilon\!\cdot\! q)(p\!\cdot\! q)\, p_\mu p_\nu}{3\, m_{\mathrm{P_{bb}^*}}^5}\,\pslash\,\qslash
- \frac{(\varepsilon\!\cdot\! p)(p\!\cdot\! q)^2\, p_\mu p_\nu}{3\, m_{\mathrm{P_{bb}^*}}^7}\,\pslash\,\qslash
- \frac{(\varepsilon\!\cdot\! p)\, p_\mu p_\nu\, q^2}{12\, m_{\mathrm{P_{bb}^*}}^5}\,\pslash\,\qslash \nonumber\\
&\quad
- \frac{5(\varepsilon\!\cdot\! q)\, p_\mu p_\nu\, q^2}{12\, m_{\mathrm{P_{bb}^*}}^5}\,\pslash\,\qslash
- \frac{(\varepsilon\!\cdot\! p)(p\!\cdot\! q)\, p_\mu p_\nu\, q^2}{3\, m_{\mathrm{P_{bb}^*}}^7}\,\pslash\,\qslash
+ \frac{(\varepsilon\!\cdot\! p)(p\!\cdot\! q)\, p_\nu q_\mu}{2\, m_{\mathrm{P_{bb}^*}}^5}\,\pslash\,\qslash
+ \frac{(\varepsilon\!\cdot\! p)\, p_\nu\, q^2 q_\mu}{2\, m_{\mathrm{P_{bb}^*}}^5}\,\pslash\,\qslash \nonumber\\
&\quad
+ \frac{(\varepsilon\!\cdot\! q)\, p_\mu q_\nu}{4\, m_{\mathrm{P_{bb}^*}}^3}\,\pslash\,\qslash
+ \frac{(\varepsilon\!\cdot\! p)(p\!\cdot\! q)\, p_\mu q_\nu}{6\, m_{\mathrm{P_{bb}^*}}^5}\,\pslash\,\qslash
- \frac{(\varepsilon\!\cdot\! q)(p\!\cdot\! q)\, p_\mu q_\nu}{3\, m_{\mathrm{P_{bb}^*}}^5}\,\pslash\,\qslash
- \frac{(\varepsilon\!\cdot\! p)(p\!\cdot\! q)^2\, p_\mu q_\nu}{3\, m_{\mathrm{P_{bb}^*}}^7}\,\pslash\,\qslash \nonumber\\
&\quad
- \frac{(\varepsilon\!\cdot\! p)\, p_\mu\, q^2 q_\nu}{12\, m_{\mathrm{P_{bb}^*}}^5}\,\pslash\,\qslash
- \frac{5(\varepsilon\!\cdot\! q)\, p_\mu\, q^2 q_\nu}{12\, m_{\mathrm{P_{bb}^*}}^5}\,\pslash\,\qslash
- \frac{(\varepsilon\!\cdot\! p)(p\!\cdot\! q)\, p_\mu\, q^2 q_\nu}{3\, m_{\mathrm{P_{bb}^*}}^7}\,\pslash\,\qslash
- \frac{(\varepsilon\!\cdot\! p)\, q_\mu q_\nu}{4\, m_{\mathrm{P_{bb}^*}}^3}\,\pslash\,\qslash \nonumber\\
%\end{align}
%\begin{align}
&\quad
+ \frac{(\varepsilon\!\cdot\! p)(p\!\cdot\! q)\, q_\mu q_\nu}{2\, m_{\mathrm{P_{bb}^*}}^5}\,\pslash\,\qslash
+ \frac{(\varepsilon\!\cdot\! p)\, q^2 q_\mu q_\nu}{2\, m_{\mathrm{P_{bb}^*}}^5}\,\pslash\,\qslash
- \frac{(\varepsilon\!\cdot\! q)\, p_\nu\, q^2}{12\, m_{\mathrm{P_{bb}^*}}^3}\,\pslash\,\gamma_\mu
- \frac{(\varepsilon\!\cdot\! p)(p\!\cdot\! q)\, p_\nu\, q^2}{12\, m_{\mathrm{P_{bb}^*}}^5}\,\pslash\,\gamma_\mu \nonumber\\
&\quad
- \frac{(\varepsilon\!\cdot\! p)\, p_\nu\, (q^2)^2}{6\, m_{\mathrm{P_{bb}^*}}^5}\,\pslash\,\gamma_\mu
+ \frac{(\varepsilon\!\cdot\! p)\, q^2 q_\nu}{4\, m_{\mathrm{P_{bb}^*}}^3}\,\pslash\,\gamma_\mu
- \frac{(\varepsilon\!\cdot\! q)\, q^2 q_\nu}{12\, m_{\mathrm{P_{bb}^*}}^3}\,\pslash\,\gamma_\mu
- \frac{(\varepsilon\!\cdot\! p)(p\!\cdot\! q)\, q^2 q_\nu}{12\, m_{\mathrm{P_{bb}^*}}^5}\,\pslash\,\gamma_\mu \nonumber\\
&\quad
- \frac{(\varepsilon\!\cdot\! p)\, (q^2)^2 q_\nu}{6\, m_{\mathrm{P_{bb}^*}}^5}\,\pslash\,\gamma_\mu
+ \frac{(\varepsilon\!\cdot\! q)\, p_\mu\, q^2}{6\, m_{\mathrm{P_{bb}^*}}^3}\,\pslash\,\gamma_\nu
+ \frac{(\varepsilon\!\cdot\! p)(p\!\cdot\! q)\, p_\mu\, q^2}{4\, m_{\mathrm{P_{bb}^*}}^5}\,\pslash\,\gamma_\nu
+ \frac{(\varepsilon\!\cdot\! p)\, p_\mu\, (q^2)^2}{12\, m_{\mathrm{P_{bb}^*}}^5}\,\pslash\,\gamma_\nu \nonumber\\
&\quad
- \frac{(\varepsilon\!\cdot\! p)\, q^2 q_\mu}{4\, m_{\mathrm{P_{bb}^*}}^3}\,\pslash\,\gamma_\nu
- \frac{(p\!\cdot\! q)\, p_\mu p_\nu\, q^2}{12\, m_{\mathrm{P_{bb}^*}}^5}\,\qslash\,\eslash
- \frac{p_\mu p_\nu\, (q^2)^2}{12\, m_{\mathrm{P_{bb}^*}}^5}\,\qslash\,\eslash
+ \frac{p_\mu\, q^2 q_\nu}{24\, m_{\mathrm{P_{bb}^*}}^3}\,\qslash\,\eslash \nonumber\\
&\quad
- \frac{(p\!\cdot\! q)\, p_\mu\, q^2 q_\nu}{12\, m_{\mathrm{P_{bb}^*}}^5}\,\qslash\,\eslash
- \frac{p_\mu\, (q^2)^2 q_\nu}{12\, m_{\mathrm{P_{bb}^*}}^5}\,\qslash\,\eslash
- \frac{(\varepsilon\!\cdot\! q)(p\!\cdot\! q)\, p_\mu p_\nu}{6\, m_{\mathrm{P_{bb}^*}}^5}\,\qslash\,\pslash
- \frac{(\varepsilon\!\cdot\! q)\, p_\mu p_\nu\, q^2}{6\, m_{\mathrm{P_{bb}^*}}^5}\,\qslash\,\pslash \nonumber\\
&\quad
+ \frac{(\varepsilon\!\cdot\! q)\, p_\mu q_\nu}{12\, m_{\mathrm{P_{bb}^*}}^3}\,\qslash\,\pslash
- \frac{(\varepsilon\!\cdot\! q)(p\!\cdot\! q)\, p_\mu q_\nu}{6\, m_{\mathrm{P_{bb}^*}}^5}\,\qslash\,\pslash
- \frac{(\varepsilon\!\cdot\! q)\, p_\mu\, q^2 q_\nu}{6\, m_{\mathrm{P_{bb}^*}}^5}\,\qslash\,\pslash
+ \frac{(\varepsilon\!\cdot\! q)(p\!\cdot\! q)\, p_\mu}{6\, m_{\mathrm{P_{bb}^*}}^3}\,\qslash\,\gamma_\nu 
\nonumber%\\
\end{align}

\begin{align}
&\quad
+ \frac{(\varepsilon\!\cdot\! q)\, p_\mu\, q^2}{12\, m_{\mathrm{P_{bb}^*}}^3}\,\qslash\,\gamma_\nu
- \frac{(p\!\cdot\! q)\, p_\nu\, q^2}{6\, m_{\mathrm{P_{bb}^*}}^3}\,\gamma_\mu\,\eslash
+ \frac{(p\!\cdot\! q)^2\, p_\nu\, q^2}{12\, m_{\mathrm{P_{bb}^*}}^5}\,\gamma_\mu\,\eslash
- \frac{p_\nu\, (q^2)^2}{8\, m_{\mathrm{P_{bb}^*}}^3}\,\gamma_\mu\,\eslash \nonumber\\
&\quad
+ \frac{(p\!\cdot\! q)\, p_\nu\, (q^2)^2}{12\, m_{\mathrm{P_{bb}^*}}^5}\,\gamma_\mu\,\eslash
+ \frac{q^2 q_\nu}{4\, m_{\mathrm{P_{bb}^*}}}\,\gamma_\mu\,\eslash
- \frac{7(p\!\cdot\! q)\, q^2 q_\nu}{24\, m_{\mathrm{P_{bb}^*}}^3}\,\gamma_\mu\,\eslash
+ \frac{(p\!\cdot\! q)^2\, q^2 q_\nu}{12\, m_{\mathrm{P_{bb}^*}}^5}\,\gamma_\mu\,\eslash \nonumber\\
&\quad
- \frac{(q^2)^2 q_\nu}{8\, m_{\mathrm{P_{bb}^*}}^3}\,\gamma_\mu\,\eslash
+ \frac{(p\!\cdot\! q)\, (q^2)^2 q_\nu}{12\, m_{\mathrm{P_{bb}^*}}^5}\,\gamma_\mu\,\eslash
+ \frac{(\varepsilon\!\cdot\! q)\, p_\nu\, q^2}{12\, m_{\mathrm{P_{bb}^*}}^3}\,\gamma_\mu\,\pslash
+ \frac{(\varepsilon\!\cdot\! q)\, q^2 q_\nu}{12\, m_{\mathrm{P_{bb}^*}}^3}\,\gamma_\mu\,\pslash \nonumber\\
&\quad
+ \frac{(\varepsilon\!\cdot\! q)(p\!\cdot\! q)\, p_\nu}{6\, m_{\mathrm{P_{bb}^*}}^3}\,\gamma_\mu\,\qslash
+ \frac{(\varepsilon\!\cdot\! q)\, p_\nu\, q^2}{4\, m_{\mathrm{P_{bb}^*}}^3}\,\gamma_\mu\,\qslash
- \frac{(\varepsilon\!\cdot\! q)\, q_\nu}{6\, m_{\mathrm{P_{bb}^*}}}\,\gamma_\mu\,\qslash
+ \frac{(\varepsilon\!\cdot\! q)(p\!\cdot\! q)\, q_\nu}{6\, m_{\mathrm{P_{bb}^*}}^3}\,\gamma_\mu\,\qslash
+ \frac{(\varepsilon\!\cdot\! q)\, q^2 q_\nu}{4\, m_{\mathrm{P_{bb}^*}}^3}\,\gamma_\mu\,\qslash \nonumber\\
&\quad
- \frac{(\varepsilon\!\cdot\! q)\, q^2}{6\, m_{\mathrm{P_{bb}^*}}}\,\gamma_\mu\,\gamma_\nu
- \frac{p_\mu p_\nu\, q^2}{24\, m_{\mathrm{P_{bb}^*}}^4}\,\eslash\,\pslash\,\qslash
- \frac{p_\mu\, q^2 q_\nu}{24\, m_{\mathrm{P_{bb}^*}}^4}\,\eslash\,\pslash\,\qslash
+ \frac{(p\!\cdot\! q)\, p_\mu\, q^2}{8\, m_{\mathrm{P_{bb}^*}}^4}\,\eslash\,\pslash\,\gamma_\nu \nonumber\\
&\quad
+ \frac{p_\mu\, (q^2)^2}{24\, m_{\mathrm{P_{bb}^*}}^4}\,\eslash\,\pslash\,\gamma_\nu
- \frac{q^2 q_\mu}{8\, m_{\mathrm{P_{bb}^*}}^2}\,\eslash\,\pslash\,\gamma_\nu
+ \frac{(p\!\cdot\! q)^2\, p_\mu p_\nu}{6\, m_{\mathrm{P_{bb}^*}}^6}\,\eslash\,\qslash\,\pslash
+ \frac{(p\!\cdot\! q)\, p_\mu p_\nu\, q^2}{6\, m_{\mathrm{P_{bb}^*}}^6}\,\eslash\,\qslash\,\pslash \nonumber\\
&\quad
- \frac{(p\!\cdot\! q)\, p_\nu q_\mu}{4\, m_{\mathrm{P_{bb}^*}}^4}\,\eslash\,\qslash\,\pslash
- \frac{p_\nu\, q^2 q_\mu}{4\, m_{\mathrm{P_{bb}^*}}^4}\,\eslash\,\qslash\,\pslash
- \frac{(p\!\cdot\! q)\, p_\mu q_\nu}{12\, m_{\mathrm{P_{bb}^*}}^4}\,\eslash\,\qslash\,\pslash
+ \frac{(p\!\cdot\! q)^2\, p_\mu q_\nu}{6\, m_{\mathrm{P_{bb}^*}}^6}\,\eslash\,\qslash\,\pslash \nonumber\\
&\quad
+ \frac{(p\!\cdot\! q)\, p_\mu\, q^2 q_\nu}{6\, m_{\mathrm{P_{bb}^*}}^6}\,\eslash\,\qslash\,\pslash
+ \frac{q_\mu q_\nu}{8\, m_{\mathrm{P_{bb}^*}}^2}\,\eslash\,\qslash\,\pslash
- \frac{(p\!\cdot\! q)\, q_\mu q_\nu}{4\, m_{\mathrm{P_{bb}^*}}^4}\,\eslash\,\qslash\,\pslash
- \frac{q^2 q_\mu q_\nu}{4\, m_{\mathrm{P_{bb}^*}}^4}\,\eslash\,\qslash\,\pslash \nonumber\\
&\quad
+ \frac{(p\!\cdot\! q)\, p_\mu\, q^2}{6\, m_{\mathrm{P_{bb}^*}}^4}\,\eslash\,\qslash\,\gamma_\nu
- \frac{q^2 q_\mu}{4\, m_{\mathrm{P_{bb}^*}}^2}\,\eslash\,\qslash\,\gamma_\nu
+ \frac{(p\!\cdot\! q)\, p_\mu\, q^2}{12\, m_{\mathrm{P_{bb}^*}}^4}\,\eslash\,\gamma_\nu\,\qslash
- \frac{q^2 q_\mu}{8\, m_{\mathrm{P_{bb}^*}}^2}\,\eslash\,\gamma_\nu\,\qslash \nonumber\\
&\quad
- \frac{p_\mu p_\nu\, q^2}{8\, m_{\mathrm{P_{bb}^*}}^4}\,\pslash\,\eslash\,\qslash
+ \frac{(p\!\cdot\! q)\, p_\mu p_\nu\, q^2}{12\, m_{\mathrm{P_{bb}^*}}^6}\,\pslash\,\eslash\,\qslash
- \frac{p_\nu\, q^2 q_\mu}{8\, m_{\mathrm{P_{bb}^*}}^4}\,\pslash\,\eslash\,\qslash
- \frac{(p\!\cdot\! q)\, p_\mu q_\nu}{12\, m_{\mathrm{P_{bb}^*}}^4}\,\pslash\,\eslash\,\qslash \nonumber\\
&\quad
- \frac{p_\mu\, q^2 q_\nu}{8\, m_{\mathrm{P_{bb}^*}}^4}\,\pslash\,\eslash\,\qslash
+ \frac{(p\!\cdot\! q)\, p_\mu\, q^2 q_\nu}{12\, m_{\mathrm{P_{bb}^*}}^6}\,\pslash\,\eslash\,\qslash
+ \frac{q_\mu q_\nu}{8\, m_{\mathrm{P_{bb}^*}}^2}\,\pslash\,\eslash\,\qslash
- \frac{q^2 q_\mu q_\nu}{8\, m_{\mathrm{P_{bb}^*}}^4}\,\pslash\,\eslash\,\qslash \nonumber\\
&\quad
+ \frac{(p\!\cdot\! q)^3\, p_\mu}{6\, m_{\mathrm{P_{bb}^*}}^6}\,\pslash\,\eslash\,\gamma_\nu
- \frac{(p\!\cdot\! q)\, p_\mu\, q^2}{24\, m_{\mathrm{P_{bb}^*}}^4}\,\pslash\,\eslash\,\gamma_\nu
+ \frac{(p\!\cdot\! q)^2\, p_\mu\, q^2}{4\, m_{\mathrm{P_{bb}^*}}^6}\,\pslash\,\eslash\,\gamma_\nu
+ \frac{p_\mu\, (q^2)^2}{12\, m_{\mathrm{P_{bb}^*}}^4}\,\pslash\,\eslash\,\gamma_\nu \nonumber\\
&\quad
+ \frac{(p\!\cdot\! q)\, p_\mu\, (q^2)^2}{12\, m_{\mathrm{P_{bb}^*}}^6}\,\pslash\,\eslash\,\gamma_\nu
- \frac{(p\!\cdot\! q)^2\, q_\mu}{4\, m_{\mathrm{P_{bb}^*}}^4}\,\pslash\,\eslash\,\gamma_\nu
+ \frac{q^2 q_\mu}{8\, m_{\mathrm{P_{bb}^*}}^2}\,\pslash\,\eslash\,\gamma_\nu
- \frac{3(p\!\cdot\! q)\, q^2 q_\mu}{8\, m_{\mathrm{P_{bb}^*}}^4}\,\pslash\,\eslash\,\gamma_\nu \nonumber\\
&\quad
- \frac{(q^2)^2 q_\mu}{8\, m_{\mathrm{P_{bb}^*}}^4}\,\pslash\,\eslash\,\gamma_\nu
- \frac{(p\!\cdot\! q)^2\, p_\mu p_\nu}{6\, m_{\mathrm{P_{bb}^*}}^6}\,\pslash\,\qslash\,\eslash
- \frac{(p\!\cdot\! q)\, p_\mu p_\nu\, q^2}{4\, m_{\mathrm{P_{bb}^*}}^6}\,\pslash\,\qslash\,\eslash
- \frac{p_\mu p_\nu\, (q^2)^2}{12\, m_{\mathrm{P_{bb}^*}}^6}\,\pslash\,\qslash\,\eslash \nonumber\\
&\quad
+ \frac{(p\!\cdot\! q)\, p_\mu q_\nu}{12\, m_{\mathrm{P_{bb}^*}}^4}\,\pslash\,\qslash\,\eslash
- \frac{(p\!\cdot\! q)^2\, p_\mu q_\nu}{6\, m_{\mathrm{P_{bb}^*}}^6}\,\pslash\,\qslash\,\eslash
+ \frac{p_\mu\, q^2 q_\nu}{24\, m_{\mathrm{P_{bb}^*}}^4}\,\pslash\,\qslash\,\eslash
- \frac{(p\!\cdot\! q)\, p_\mu\, q^2 q_\nu}{4\, m_{\mathrm{P_{bb}^*}}^6}\,\pslash\,\qslash\,\eslash \nonumber\\
&\quad
- \frac{p_\mu\, (q^2)^2 q_\nu}{12\, m_{\mathrm{P_{bb}^*}}^6}\,\pslash\,\qslash\,\eslash
- \frac{(\varepsilon\!\cdot\! q)(p\!\cdot\! q)\, p_\mu}{12\, m_{\mathrm{P_{bb}^*}}^4}\,\pslash\,\qslash\,\gamma_\nu
- \frac{(\varepsilon\!\cdot\! p)(p\!\cdot\! q)^2\, p_\mu}{6\, m_{\mathrm{P_{bb}^*}}^6}\,\pslash\,\qslash\,\gamma_\nu
- \frac{(\varepsilon\!\cdot\! q)\, p_\mu\, q^2}{6\, m_{\mathrm{P_{bb}^*}}^4}\,\pslash\,\qslash\,\gamma_\nu \nonumber\\
%\end{align}
%\begin{align}
&\quad
- \frac{(\varepsilon\!\cdot\! p)(p\!\cdot\! q)\, p_\mu\, q^2}{6\, m_{\mathrm{P_{bb}^*}}^6}\,\pslash\,\qslash\,\gamma_\nu
+ \frac{(\varepsilon\!\cdot\! p)(p\!\cdot\! q)\, q_\mu}{4\, m_{\mathrm{P_{bb}^*}}^4}\,\pslash\,\qslash\,\gamma_\nu
+ \frac{(\varepsilon\!\cdot\! p)\, q^2 q_\mu}{4\, m_{\mathrm{P_{bb}^*}}^4}\,\pslash\,\qslash\,\gamma_\nu \nonumber\\
&\quad
+ \frac{(p\!\cdot\! q)^3\, p_\nu}{6\, m_{\mathrm{P_{bb}^*}}^6}\,\pslash\,\gamma_\mu\,\eslash
- \frac{(p\!\cdot\! q)\, p_\nu\, q^2}{8\, m_{\mathrm{P_{bb}^*}}^4}\,\pslash\,\gamma_\mu\,\eslash
+ \frac{(p\!\cdot\! q)^2\, p_\nu\, q^2}{4\, m_{\mathrm{P_{bb}^*}}^6}\,\pslash\,\gamma_\mu\,\eslash
- \frac{p_\nu\, (q^2)^2}{24\, m_{\mathrm{P_{bb}^*}}^4}\,\pslash\,\gamma_\mu\,\eslash \nonumber\\
&\quad
+ \frac{(p\!\cdot\! q)\, p_\nu\, (q^2)^2}{12\, m_{\mathrm{P_{bb}^*}}^6}\,\pslash\,\gamma_\mu\,\eslash
- \frac{(p\!\cdot\! q)^2\, q_\nu}{4\, m_{\mathrm{P_{bb}^*}}^4}\,\pslash\,\gamma_\mu\,\eslash
+ \frac{(p\!\cdot\! q)^3\, q_\nu}{6\, m_{\mathrm{P_{bb}^*}}^6}\,\pslash\,\gamma_\mu\,\eslash
+ \frac{q^2 q_\nu}{8\, m_{\mathrm{P_{bb}^*}}^2}\,\pslash\,\gamma_\mu\,\eslash
- \frac{(p\!\cdot\! q)\, q^2 q_\nu}{4\, m_{\mathrm{P_{bb}^*}}^4}\,\pslash\,\gamma_\mu\,\eslash \nonumber\\
&\quad
+ \frac{(p\!\cdot\! q)^2\, q^2 q_\nu}{4\, m_{\mathrm{P_{bb}^*}}^6}\,\pslash\,\gamma_\mu\,\eslash
- \frac{(q^2)^2 q_\nu}{24\, m_{\mathrm{P_{bb}^*}}^4}\,\pslash\,\gamma_\mu\,\eslash
+ \frac{(p\!\cdot\! q)\, (q^2)^2 q_\nu}{12\, m_{\mathrm{P_{bb}^*}}^6}\,\pslash\,\gamma_\mu\,\eslash
- \frac{(\varepsilon\!\cdot\! p)(p\!\cdot\! q)^2\, p_\nu}{6\, m_{\mathrm{P_{bb}^*}}^6}\,\pslash\,\gamma_\mu\,\qslash \nonumber\\
&\quad
+ \frac{(\varepsilon\!\cdot\! p)\, p_\nu\, q^2}{12\, m_{\mathrm{P_{bb}^*}}^4}\,\pslash\,\gamma_\mu\,\qslash
+ \frac{(\varepsilon\!\cdot\! q)\, p_\nu\, q^2}{12\, m_{\mathrm{P_{bb}^*}}^4}\,\pslash\,\gamma_\mu\,\qslash
- \frac{(\varepsilon\!\cdot\! p)(p\!\cdot\! q)\, p_\nu\, q^2}{6\, m_{\mathrm{P_{bb}^*}}^6}\,\pslash\,\gamma_\mu\,\qslash
- \frac{(\varepsilon\!\cdot\! q)\, q_\nu}{12\, m_{\mathrm{P_{bb}^*}}^2}\,\pslash\,\gamma_\mu\,\qslash 
\nonumber%\\
\end{align}

\begin{align}
&\quad
+ \frac{(\varepsilon\!\cdot\! p)(p\!\cdot\! q)\, q_\nu}{12\, m_{\mathrm{P_{bb}^*}}^4}\,\pslash\,\gamma_\mu\,\qslash
- \frac{(\varepsilon\!\cdot\! p)(p\!\cdot\! q)^2\, q_\nu}{6\, m_{\mathrm{P_{bb}^*}}^6}\,\pslash\,\gamma_\mu\,\qslash
+ \frac{(\varepsilon\!\cdot\! p)\, q^2 q_\nu}{12\, m_{\mathrm{P_{bb}^*}}^4}\,\pslash\,\gamma_\mu\,\qslash
+ \frac{(\varepsilon\!\cdot\! q)\, q^2 q_\nu}{12\, m_{\mathrm{P_{bb}^*}}^4}\,\pslash\,\gamma_\mu\,\qslash \nonumber\\
&\quad
- \frac{(\varepsilon\!\cdot\! p)(p\!\cdot\! q)\, q^2 q_\nu}{6\, m_{\mathrm{P_{bb}^*}}^6}\,\pslash\,\gamma_\mu\,\qslash
- \frac{(\varepsilon\!\cdot\! q)\, q^2}{12\, m_{\mathrm{P_{bb}^*}}^2}\,\pslash\,\gamma_\mu\,\gamma_\nu
- \frac{(\varepsilon\!\cdot\! p)\, (q^2)^2}{12\, m_{\mathrm{P_{bb}^*}}^4}\,\pslash\,\gamma_\mu\,\gamma_\nu
+ \frac{(\varepsilon\!\cdot\! p)\, p_\mu\, q^2}{12\, m_{\mathrm{P_{bb}^*}}^4}\,\pslash\,\gamma_\nu\,\qslash \nonumber\\
&\quad
- \frac{(p\!\cdot\! q)\, p_\mu\, q^2}{24\, m_{\mathrm{P_{bb}^*}}^4}\,\qslash\,\eslash\,\gamma_\nu
- \frac{p_\mu\, (q^2)^2}{24\, m_{\mathrm{P_{bb}^*}}^4}\,\qslash\,\eslash\,\gamma_\nu
- \frac{(\varepsilon\!\cdot\! q)(p\!\cdot\! q)\, p_\mu}{12\, m_{\mathrm{P_{bb}^*}}^4}\,\qslash\,\pslash\,\gamma_\nu
- \frac{(\varepsilon\!\cdot\! q)\, p_\mu\, q^2}{12\, m_{\mathrm{P_{bb}^*}}^4}\,\qslash\,\pslash\,\gamma_\nu \nonumber\\
&\quad
- \frac{(p\!\cdot\! q)\, p_\nu\, q^2}{24\, m_{\mathrm{P_{bb}^*}}^4}\,\gamma_\mu\,\eslash\,\pslash
- \frac{p_\nu\, (q^2)^2}{12\, m_{\mathrm{P_{bb}^*}}^4}\,\gamma_\mu\,\eslash\,\pslash
+ \frac{q^2 q_\nu}{8\, m_{\mathrm{P_{bb}^*}}^2}\,\gamma_\mu\,\eslash\,\pslash
- \frac{(p\!\cdot\! q)\, q^2 q_\nu}{24\, m_{\mathrm{P_{bb}^*}}^4}\,\gamma_\mu\,\eslash\,\pslash \nonumber\\
&\quad
- \frac{(q^2)^2 q_\nu}{12\, m_{\mathrm{P_{bb}^*}}^4}\,\gamma_\mu\,\eslash\,\pslash
- \frac{(p\!\cdot\! q)^2\, p_\nu}{12\, m_{\mathrm{P_{bb}^*}}^4}\,\gamma_\mu\,\eslash\,\qslash
+ \frac{p_\nu\, q^2}{12\, m_{\mathrm{P_{bb}^*}}^2}\,\gamma_\mu\,\eslash\,\qslash
- \frac{(p\!\cdot\! q)\, p_\nu\, q^2}{24\, m_{\mathrm{P_{bb}^*}}^4}\,\gamma_\mu\,\eslash\,\qslash \nonumber\\
&\quad
- \frac{(p\!\cdot\! q)^2\, q_\nu}{12\, m_{\mathrm{P_{bb}^*}}^4}\,\gamma_\mu\,\eslash\,\qslash
+ \frac{q^2 q_\nu}{12\, m_{\mathrm{P_{bb}^*}}^2}\,\gamma_\mu\,\eslash\,\qslash
- \frac{(p\!\cdot\! q)\, q^2 q_\nu}{24\, m_{\mathrm{P_{bb}^*}}^4}\,\gamma_\mu\,\eslash\,\qslash
- \frac{(p\!\cdot\! q)\, q^2}{12\, m_{\mathrm{P_{bb}^*}}^2}\,\gamma_\mu\,\eslash\,\gamma_\nu \nonumber\\
&\quad
+ \frac{(p\!\cdot\! q)^2\, q^2}{24\, m_{\mathrm{P_{bb}^*}}^4}\,\gamma_\mu\,\eslash\,\gamma_\nu
- \frac{(q^2)^2}{24\, m_{\mathrm{P_{bb}^*}}^2}\,\gamma_\mu\,\eslash\,\gamma_\nu
+ \frac{(p\!\cdot\! q)\, (q^2)^2}{24\, m_{\mathrm{P_{bb}^*}}^4}\,\gamma_\mu\,\eslash\,\gamma_\nu
+ \frac{(\varepsilon\!\cdot\! q)\, q^2}{12\, m_{\mathrm{P_{bb}^*}}^2}\,\gamma_\mu\,\pslash\,\gamma_\nu \nonumber\\
&\quad
+ \frac{(p\!\cdot\! q)\, p_\nu\, q^2}{12\, m_{\mathrm{P_{bb}^*}}^4}\,\gamma_\mu\,\qslash\,\eslash
+ \frac{p_\nu\, (q^2)^2}{12\, m_{\mathrm{P_{bb}^*}}^4}\,\gamma_\mu\,\qslash\,\eslash
- \frac{q^2 q_\nu}{24\, m_{\mathrm{P_{bb}^*}}^2}\,\gamma_\mu\,\qslash\,\eslash
+ \frac{(p\!\cdot\! q)\, q^2 q_\nu}{12\, m_{\mathrm{P_{bb}^*}}^4}\,\gamma_\mu\,\qslash\,\eslash \nonumber\\
&\quad
+ \frac{(q^2)^2 q_\nu}{12\, m_{\mathrm{P_{bb}^*}}^4}\,\gamma_\mu\,\qslash\,\eslash
+ \frac{(\varepsilon\!\cdot\! q)(p\!\cdot\! q)\, p_\nu}{6\, m_{\mathrm{P_{bb}^*}}^4}\,\gamma_\mu\,\qslash\,\pslash
+ \frac{(\varepsilon\!\cdot\! q)\, p_\nu\, q^2}{6\, m_{\mathrm{P_{bb}^*}}^4}\,\gamma_\mu\,\qslash\,\pslash
- \frac{(\varepsilon\!\cdot\! q)\, q_\nu}{12\, m_{\mathrm{P_{bb}^*}}^2}\,\gamma_\mu\,\qslash\,\pslash \nonumber\\
&\quad
+ \frac{(\varepsilon\!\cdot\! q)(p\!\cdot\! q)\, q_\nu}{6\, m_{\mathrm{P_{bb}^*}}^4}\,\gamma_\mu\,\qslash\,\pslash
+ \frac{(\varepsilon\!\cdot\! q)\, q^2 q_\nu}{6\, m_{\mathrm{P_{bb}^*}}^4}\,\gamma_\mu\,\qslash\,\pslash
+ \frac{(\varepsilon\!\cdot\! q)\, q^2}{12\, m_{\mathrm{P_{bb}^*}}^2}\,\gamma_\mu\,\qslash\,\gamma_\nu
+ \frac{p_\mu\, q^2}{24\, m_{\mathrm{P_{bb}^*}}^3}\,\eslash\,\pslash\,\gamma_\nu\,\qslash \nonumber\\
&\quad
+ \frac{(p\!\cdot\! q)^2\, p_\mu}{12\, m_{\mathrm{P_{bb}^*}}^5}\,\eslash\,\qslash\,\pslash\,\gamma_\nu
+ \frac{(p\!\cdot\! q)\, p_\mu\, q^2}{12\, m_{\mathrm{P_{bb}^*}}^5}\,\eslash\,\qslash\,\pslash\,\gamma_\nu
- \frac{(p\!\cdot\! q)\, q_\mu}{8\, m_{\mathrm{P_{bb}^*}}^3}\,\eslash\,\qslash\,\pslash\,\gamma_\nu
- \frac{q^2 q_\mu}{8\, m_{\mathrm{P_{bb}^*}}^3}\,\eslash\,\qslash\,\pslash\,\gamma_\nu \nonumber\\
&\quad
+ \frac{(p\!\cdot\! q)^2\, p_\mu}{12\, m_{\mathrm{P_{bb}^*}}^5}\,\pslash\,\eslash\,\qslash\,\gamma_\nu
+ \frac{(p\!\cdot\! q)\, p_\mu\, q^2}{12\, m_{\mathrm{P_{bb}^*}}^5}\,\pslash\,\eslash\,\qslash\,\gamma_\nu
- \frac{(p\!\cdot\! q)\, q_\mu}{8\, m_{\mathrm{P_{bb}^*}}^3}\,\pslash\,\eslash\,\qslash\,\gamma_\nu
- \frac{q^2 q_\mu}{8\, m_{\mathrm{P_{bb}^*}}^3}\,\pslash\,\eslash\,\qslash\,\gamma_\nu \nonumber\\
&\quad
+ \frac{(p\!\cdot\! q)^2\, p_\mu}{6\, m_{\mathrm{P_{bb}^*}}^5}\,\pslash\,\eslash\,\gamma_\nu\,\qslash
+ \frac{p_\mu\, q^2}{8\, m_{\mathrm{P_{bb}^*}}^3}\,\pslash\,\eslash\,\gamma_\nu\,\qslash
+ \frac{(p\!\cdot\! q)\, p_\mu\, q^2}{12\, m_{\mathrm{P_{bb}^*}}^5}\,\pslash\,\eslash\,\gamma_\nu\,\qslash
- \frac{(p\!\cdot\! q)\, q_\mu}{4\, m_{\mathrm{P_{bb}^*}}^3}\,\pslash\,\eslash\,\gamma_\nu\,\qslash \nonumber\\
&\quad
- \frac{q^2 q_\mu}{8\, m_{\mathrm{P_{bb}^*}}^3}\,\pslash\,\eslash\,\gamma_\nu\,\qslash
- \frac{(p\!\cdot\! q)^2\, p_\mu}{12\, m_{\mathrm{P_{bb}^*}}^5}\,\pslash\,\qslash\,\eslash\,\gamma_\nu
- \frac{(p\!\cdot\! q)\, p_\mu\, q^2}{8\, m_{\mathrm{P_{bb}^*}}^5}\,\pslash\,\qslash\,\eslash\,\gamma_\nu
- \frac{p_\mu\, (q^2)^2}{24\, m_{\mathrm{P_{bb}^*}}^5}\,\pslash\,\qslash\,\eslash\,\gamma_\nu \nonumber\\
&\quad
+ \frac{(p\!\cdot\! q)^2\, p_\nu}{6\, m_{\mathrm{P_{bb}^*}}^5}\,\pslash\,\gamma_\mu\,\eslash\,\qslash
+ \frac{p_\nu\, q^2}{24\, m_{\mathrm{P_{bb}^*}}^3}\,\pslash\,\gamma_\mu\,\eslash\,\qslash
+ \frac{5(p\!\cdot\! q)\, p_\nu\, q^2}{24\, m_{\mathrm{P_{bb}^*}}^5}\,\pslash\,\gamma_\mu\,\eslash\,\qslash
- \frac{(p\!\cdot\! q)\, q_\nu}{8\, m_{\mathrm{P_{bb}^*}}^3}\,\pslash\,\gamma_\mu\,\eslash\,\qslash \nonumber\\
&\quad
+ \frac{(p\!\cdot\! q)^2\, q_\nu}{6\, m_{\mathrm{P_{bb}^*}}^5}\,\pslash\,\gamma_\mu\,\eslash\,\qslash
+ \frac{q^2 q_\nu}{24\, m_{\mathrm{P_{bb}^*}}^3}\,\pslash\,\gamma_\mu\,\eslash\,\qslash
+ \frac{5(p\!\cdot\! q)\, q^2 q_\nu}{24\, m_{\mathrm{P_{bb}^*}}^5}\,\pslash\,\gamma_\mu\,\eslash\,\qslash
+ \frac{(p\!\cdot\! q)^3}{12\, m_{\mathrm{P_{bb}^*}}^5}\,\pslash\,\gamma_\mu\,\eslash\,\gamma_\nu \nonumber\\
&\quad
- \frac{(p\!\cdot\! q)\, q^2}{12\, m_{\mathrm{P_{bb}^*}}^3}\,\pslash\,\gamma_\mu\,\eslash\,\gamma_\nu
+ \frac{(p\!\cdot\! q)^2\, q^2}{8\, m_{\mathrm{P_{bb}^*}}^5}\,\pslash\,\gamma_\mu\,\eslash\,\gamma_\nu
+ \frac{(p\!\cdot\! q)\, (q^2)^2}{24\, m_{\mathrm{P_{bb}^*}}^5}\,\pslash\,\gamma_\mu\,\eslash\,\gamma_\nu
+ \frac{(p\!\cdot\! q)^2\, p_\nu}{6\, m_{\mathrm{P_{bb}^*}}^5}\,\pslash\,\gamma_\mu\,\qslash\,\eslash \nonumber
\end{align}

\begin{align}
&\quad
+ \frac{(p\!\cdot\! q)\, p_\nu\, q^2}{4\, m_{\mathrm{P_{bb}^*}}^5}\,\pslash\,\gamma_\mu\,\qslash\,\eslash
+ \frac{p_\nu\, (q^2)^2}{12\, m_{\mathrm{P_{bb}^*}}^5}\,\pslash\,\gamma_\mu\,\qslash\,\eslash
- \frac{(p\!\cdot\! q)\, q_\nu}{12\, m_{\mathrm{P_{bb}^*}}^3}\,\pslash\,\gamma_\mu\,\qslash\,\eslash
+ \frac{(p\!\cdot\! q)^2\, q_\nu}{6\, m_{\mathrm{P_{bb}^*}}^5}\,\pslash\,\gamma_\mu\,\qslash\,\eslash \nonumber\\
&\quad
- \frac{q^2 q_\nu}{24\, m_{\mathrm{P_{bb}^*}}^3}\,\pslash\,\gamma_\mu\,\qslash\,\eslash
+ \frac{(p\!\cdot\! q)\, q^2 q_\nu}{4\, m_{\mathrm{P_{bb}^*}}^5}\,\pslash\,\gamma_\mu\,\qslash\,\eslash
+ \frac{(q^2)^2 q_\nu}{12\, m_{\mathrm{P_{bb}^*}}^5}\,\pslash\,\gamma_\mu\,\qslash\,\eslash
- \frac{(\varepsilon\!\cdot\! q)(p\!\cdot\! q)}{12\, m_{\mathrm{P_{bb}^*}}^3}\,\pslash\,\gamma_\mu\,\qslash\,\gamma_\nu \nonumber\\
&\quad
- \frac{(\varepsilon\!\cdot\! p)(p\!\cdot\! q)^2}{12\, m_{\mathrm{P_{bb}^*}}^5}\,\pslash\,\gamma_\mu\,\qslash\,\gamma_\nu
- \frac{(\varepsilon\!\cdot\! p)(p\!\cdot\! q)\, q^2}{12\, m_{\mathrm{P_{bb}^*}}^5}\,\pslash\,\gamma_\mu\,\qslash\,\gamma_\nu
- \frac{(\varepsilon\!\cdot\! p)\, q^2}{12\, m_{\mathrm{P_{bb}^*}}^3}\,\pslash\,\gamma_\mu\,\gamma_\nu\,\qslash
+ \frac{p_\nu\, q^2}{24\, m_{\mathrm{P_{bb}^*}}^3}\,\gamma_\mu\,\eslash\,\pslash\,\qslash \nonumber\\
&\quad
+ \frac{q^2 q_\nu}{24\, m_{\mathrm{P_{bb}^*}}^3}\,\gamma_\mu\,\eslash\,\pslash\,\qslash
- \frac{(q^2)^2}{24\, m_{\mathrm{P_{bb}^*}}^3}\,\gamma_\mu\,\eslash\,\pslash\,\gamma_\nu
+ \frac{(p\!\cdot\! q)^2\, p_\nu}{12\, m_{\mathrm{P_{bb}^*}}^5}\,\gamma_\mu\,\eslash\,\qslash\,\pslash
+ \frac{(p\!\cdot\! q)\, p_\nu\, q^2}{12\, m_{\mathrm{P_{bb}^*}}^5}\,\gamma_\mu\,\eslash\,\qslash\,\pslash \nonumber\\
&\quad
- \frac{(p\!\cdot\! q)\, q_\nu}{24\, m_{\mathrm{P_{bb}^*}}^3}\,\gamma_\mu\,\eslash\,\qslash\,\pslash
+ \frac{(p\!\cdot\! q)^2\, q_\nu}{12\, m_{\mathrm{P_{bb}^*}}^5}\,\gamma_\mu\,\eslash\,\qslash\,\pslash
+ \frac{(p\!\cdot\! q)\, q^2 q_\nu}{12\, m_{\mathrm{P_{bb}^*}}^5}\,\gamma_\mu\,\eslash\,\qslash\,\pslash
- \frac{(p\!\cdot\! q)^2}{12\, m_{\mathrm{P_{bb}^*}}^3}\,\gamma_\mu\,\eslash\,\qslash\,\gamma_\nu \nonumber\\
&\quad
- \frac{q^2}{12\, m_{\mathrm{P_{bb}^*}}}\,\gamma_\mu\,\eslash\,\qslash\,\gamma_\nu
+ \frac{(p\!\cdot\! q)\, q^2}{24\, m_{\mathrm{P_{bb}^*}}^3}\,\gamma_\mu\,\eslash\,\qslash\,\gamma_\nu
+ \frac{(p\!\cdot\! q)\, q^2}{24\, m_{\mathrm{P_{bb}^*}}^3}\,\gamma_\mu\,\qslash\,\eslash\,\gamma_\nu
+ \frac{(q^2)^2}{24\, m_{\mathrm{P_{bb}^*}}^3}\,\gamma_\mu\,\qslash\,\eslash\,\gamma_\nu \nonumber\\
&\quad
+ \frac{(\varepsilon\!\cdot\! q)(p\!\cdot\! q)}{12\, m_{\mathrm{P_{bb}^*}}^3}\,\gamma_\mu\,\qslash\,\pslash\,\gamma_\nu
+ \frac{(\varepsilon\!\cdot\! q)\, q^2}{12\, m_{\mathrm{P_{bb}^*}}^3}\,\gamma_\mu\,\qslash\,\pslash\,\gamma_\nu
+ \frac{(p\!\cdot\! q)^2}{8\, m_{\mathrm{P_{bb}^*}}^4}\,\pslash\,\gamma_\mu\,\eslash\,\qslash\,\gamma_\nu
+ \frac{(p\!\cdot\! q)\, q^2}{8\, m_{\mathrm{P_{bb}^*}}^4}\,\pslash\,\gamma_\mu\,\eslash\,\qslash\,\gamma_\nu \nonumber\\
&\quad
+ \frac{(p\!\cdot\! q)^2}{12\, m_{\mathrm{P_{bb}^*}}^4}\,\pslash\,\gamma_\mu\,\eslash\,\gamma_\nu\,\qslash
- \frac{q^2}{24\, m_{\mathrm{P_{bb}^*}}^2}\,\pslash\,\gamma_\mu\,\eslash\,\gamma_\nu\,\qslash
+ \frac{(p\!\cdot\! q)\, q^2}{24\, m_{\mathrm{P_{bb}^*}}^4}\,\pslash\,\gamma_\mu\,\eslash\,\gamma_\nu\,\qslash
+ \frac{(p\!\cdot\! q)^2}{12\, m_{\mathrm{P_{bb}^*}}^4}\,\pslash\,\gamma_\mu\,\qslash\,\eslash\,\gamma_\nu \nonumber\\
&\quad
+ \frac{(p\!\cdot\! q)\, q^2}{8\, m_{\mathrm{P_{bb}^*}}^4}\,\pslash\,\gamma_\mu\,\qslash\,\eslash\,\gamma_\nu
+ \frac{(q^2)^2}{24\, m_{\mathrm{P_{bb}^*}}^4}\,\pslash\,\gamma_\mu\,\qslash\,\eslash\,\gamma_\nu
- \frac{q^2}{24\, m_{\mathrm{P_{bb}^*}}^2}\,\gamma_\mu\,\eslash\,\pslash\,\gamma_\nu\,\qslash
+ \frac{(p\!\cdot\! q)^2}{24\, m_{\mathrm{P_{bb}^*}}^4}\,\gamma_\mu\,\eslash\,\qslash\,\pslash\,\gamma_\nu \nonumber\\
&\quad
+ \frac{(p\!\cdot\! q)\, q^2}{24\, m_{\mathrm{P_{bb}^*}}^4}\,\gamma_\mu\,\eslash\,\qslash\,\pslash\,\gamma_\nu
\Bigg\}.
\label{F4_appendix}
\end{align}

\subsection{Stage (ii): imposition of the real-photon kinematics}
\label{appc2}

In this stage, the real-photon kinematic conditions 
\begin{align}
q^{2}=0, \qquad \varepsilon\!\cdot\! q=0
\end{align}
are imposed on the unreduced expressions of Sec.~\ref{appc1}. 
All Lorentz structures proportional to either $q^{2}$ or 
$(\varepsilon\!\cdot\! q)$ are thereby eliminated. The 
gamma-matrix ordering 
$\gamma_{\mu}\,\pslash\,\eslash\,\qslash\,\gamma_{\nu}$ has 
\textit{not yet} been applied at this stage, so the resulting 
gauge-fixed expressions still contain Lorentz structures with 
$\gamma_{\mu}$ at the leftmost position, $\gamma_{\nu}$ at the 
rightmost position, or proportional to $p_{\mu}$ or $p_{\nu}$. 
These structures originate from the spin-$1/2$ pollution of the 
interpolating current $\mathrm{J}_{\mu}^{\mathrm{P_{bb}^{*}}}(x)$, 
as parametrized in Eq.~\eqref{eq:spin12coupling}, and will be 
removed in Sec.~\ref{appc3}.

The effect of the present stage is most pronounced in the 
$F_{3}(q^{2})$ and $F_{4}(q^{2})$ sectors, whose unreduced 
structures carry a high density of $q^{2}$ and 
$(\varepsilon\!\cdot\! q)$ factors inherited from the vertex 
kinematics $q^{\alpha}q^{\beta}/(2m_{\mathrm{P_{bb}^{*}}})^{2}$ 
of Eq.~\eqref{matelpar}. The gauge-fixed expressions, organized 
by form factor, read as follows.

\subsubsection{Gauge-fixed tensor structure proportional to  
$F_{1}(q^{2})$ prior to spin-1/2 contamination removal)} \label{F1_appendix11}

\begin{align}
\label{F1_appendix1}
{F}_1 & \propto\,\Bigg\{ 
\frac{8(\varepsilon\!\cdot\! p)\, p_\mu p_\nu}{3\, m_{\mathrm{P_{bb}^*}}}
- \frac{16(\varepsilon\!\cdot\! p)(p\!\cdot\! q)\, p_\mu p_\nu}{3\, m_{\mathrm{P_{bb}^*}}^3}
+ \frac{4(\varepsilon\!\cdot\! p)\, p_\nu q_\mu}{m_{\mathrm{P_{bb}^*}}}
- \frac{16(\varepsilon\!\cdot\! p)(p\!\cdot\! q)\, p_\mu q_\nu}{3\, m_{\mathrm{P_{bb}^*}}^3} \nonumber\\
&\quad
+ \frac{4(\varepsilon\!\cdot\! p)\, q_\mu q_\nu}{m_{\mathrm{P_{bb}^*}}}
- 2\, m_{\mathrm{P_{bb}^*}}\,(\varepsilon\!\cdot\! p)\, g_{\mu\nu}
+ \frac{8(p\!\cdot\! q)\, p_\mu p_\nu}{3\, m_{\mathrm{P_{bb}^*}}^2}\,\eslash
- 2\, p_\nu q_\mu\,\eslash
- \frac{2\, p_\mu q_\nu}{3}\,\eslash
+ \frac{8(p\!\cdot\! q)\, p_\mu q_\nu}{3\, m_{\mathrm{P_{bb}^*}}^2}\,\eslash \nonumber\\
&\quad
- 2\, q_\mu q_\nu\,\eslash
+ \frac{4(\varepsilon\!\cdot\! p)\, p_\mu p_\nu}{3\, m_{\mathrm{P_{bb}^*}}^2}\,\pslash
- \frac{8(\varepsilon\!\cdot\! p)(p\!\cdot\! q)\, p_\mu p_\nu}{3\, m_{\mathrm{P_{bb}^*}}^4}\,\pslash
+ \frac{4(\varepsilon\!\cdot\! p)\, p_\nu q_\mu}{m_{\mathrm{P_{bb}^*}}^2}\,\pslash
- \frac{8(\varepsilon\!\cdot\! p)(p\!\cdot\! q)\, p_\mu q_\nu}{3\, m_{\mathrm{P_{bb}^*}}^4}\,\pslash \nonumber\\
&\quad
+ \frac{4(\varepsilon\!\cdot\! p)\, q_\mu q_\nu}{m_{\mathrm{P_{bb}^*}}^2}\,\pslash
- 2(\varepsilon\!\cdot\! p)\, g_{\mu\nu}\,\pslash
+ \frac{2\,\varepsilon_\nu\, p_\mu}{3}\,\qslash
+ 2\,\varepsilon_\mu\, p_\nu\,\qslash
- \frac{4(\varepsilon\!\cdot\! p)\, p_\mu p_\nu}{m_{\mathrm{P_{bb}^*}}^2}\,\qslash
+ 2\,\varepsilon_\mu\, q_\nu\,\qslash \nonumber\\
&\quad
- \frac{4(\varepsilon\!\cdot\! p)\, p_\mu q_\nu}{m_{\mathrm{P_{bb}^*}}^2}\,\qslash
+ \frac{2(\varepsilon\!\cdot\! p)\, p_\nu}{3}\,\gamma_\mu
- \frac{4(\varepsilon\!\cdot\! p)(p\!\cdot\! q)\, p_\nu}{3\, m_{\mathrm{P_{bb}^*}}^2}\,\gamma_\mu
- \frac{4(\varepsilon\!\cdot\! p)(p\!\cdot\! q)\, q_\nu}{3\, m_{\mathrm{P_{bb}^*}}^2}\,\gamma_\mu
+ \frac{2(\varepsilon\!\cdot\! p)\, p_\mu}{3}\,\gamma_\nu \nonumber\\
&\quad
- \frac{8(\varepsilon\!\cdot\! p)(p\!\cdot\! q)\, p_\mu}{3\, m_{\mathrm{P_{bb}^*}}^2}\,\gamma_\nu
+ 2(\varepsilon\!\cdot\! p)\, q_\mu\,\gamma_\nu
- \frac{8(p\!\cdot\! q)\, p_\mu p_\nu}{3\, m_{\mathrm{P_{bb}^*}}^3}\,\eslash\,\qslash
+ \frac{2\, p_\nu q_\mu}{m_{\mathrm{P_{bb}^*}}}\,\eslash\,\qslash
- \frac{4\, p_\mu q_\nu}{3\, m_{\mathrm{P_{bb}^*}}}\,\eslash\,\qslash \nonumber\\
&\quad
- \frac{8(p\!\cdot\! q)\, p_\mu q_\nu}{3\, m_{\mathrm{P_{bb}^*}}^3}\,\eslash\,\qslash
+ \frac{2\, q_\mu q_\nu}{m_{\mathrm{P_{bb}^*}}}\,\eslash\,\qslash
- m_{\mathrm{P_{bb}^*}}\, g_{\mu\nu}\,\eslash\,\qslash
+ \frac{8(p\!\cdot\! q)\, p_\mu}{3\, m_{\mathrm{P_{bb}^*}}}\,\eslash\,\gamma_\nu
- 2\, m_{\mathrm{P_{bb}^*}}\, q_\mu\,\eslash\,\gamma_\nu \nonumber\\
&\quad
+ \frac{4(p\!\cdot\! q)\, p_\mu p_\nu}{3\, m_{\mathrm{P_{bb}^*}}^3}\,\pslash\,\eslash
- \frac{2\, p_\nu q_\mu}{m_{\mathrm{P_{bb}^*}}}\,\pslash\,\eslash
+ \frac{2\, p_\mu q_\nu}{3\, m_{\mathrm{P_{bb}^*}}}\,\pslash\,\eslash
+ \frac{4(p\!\cdot\! q)\, p_\mu q_\nu}{3\, m_{\mathrm{P_{bb}^*}}^3}\,\pslash\,\eslash
- \frac{2\, q_\mu q_\nu}{m_{\mathrm{P_{bb}^*}}}\,\pslash\,\eslash \nonumber\\
&\quad
- \frac{2\,\varepsilon_\nu\, p_\mu}{3\, m_{\mathrm{P_{bb}^*}}}\,\pslash\,\qslash
+ \frac{2\,\varepsilon_\mu\, p_\nu}{m_{\mathrm{P_{bb}^*}}}\,\pslash\,\qslash
+ \frac{2\,\varepsilon_\mu\, q_\nu}{m_{\mathrm{P_{bb}^*}}}\,\pslash\,\qslash
- \frac{2(\varepsilon\!\cdot\! p)\, p_\mu}{3\, m_{\mathrm{P_{bb}^*}}}\,\pslash\,\gamma_\nu
- \frac{4(\varepsilon\!\cdot\! p)(p\!\cdot\! q)\, p_\mu}{3\, m_{\mathrm{P_{bb}^*}}^3}\,\pslash\,\gamma_\nu \nonumber\\
&\quad
+ \frac{2(\varepsilon\!\cdot\! p)\, q_\mu}{m_{\mathrm{P_{bb}^*}}}\,\pslash\,\gamma_\nu
+ 2\, m_{\mathrm{P_{bb}^*}}\,\varepsilon_\mu\,\qslash\,\gamma_\nu
- \frac{10(\varepsilon\!\cdot\! p)\, p_\mu}{3\, m_{\mathrm{P_{bb}^*}}}\,\qslash\,\gamma_\nu
+ \frac{2(p\!\cdot\! q)\, p_\nu}{3\, m_{\mathrm{P_{bb}^*}}}\,\gamma_\mu\,\eslash
- \frac{2\, m_{\mathrm{P_{bb}^*}}\, q_\nu}{3}\,\gamma_\mu\,\eslash \nonumber\\
&\quad
+ \frac{2(p\!\cdot\! q)\, q_\nu}{3\, m_{\mathrm{P_{bb}^*}}}\,\gamma_\mu\,\eslash
- \frac{2(\varepsilon\!\cdot\! p)\, p_\nu}{3\, m_{\mathrm{P_{bb}^*}}}\,\gamma_\mu\,\pslash
+ \frac{4(\varepsilon\!\cdot\! p)(p\!\cdot\! q)\, p_\nu}{3\, m_{\mathrm{P_{bb}^*}}^3}\,\gamma_\mu\,\pslash
+ \frac{4(\varepsilon\!\cdot\! p)(p\!\cdot\! q)\, q_\nu}{3\, m_{\mathrm{P_{bb}^*}}^3}\,\gamma_\mu\,\pslash \nonumber\\
&\quad
+ \frac{2\, m_{\mathrm{P_{bb}^*}}\,\varepsilon_\nu}{3}\,\gamma_\mu\,\qslash
- \frac{2(\varepsilon\!\cdot\! p)\, p_\nu}{m_{\mathrm{P_{bb}^*}}}\,\gamma_\mu\,\qslash
- \frac{2(\varepsilon\!\cdot\! p)\, q_\nu}{m_{\mathrm{P_{bb}^*}}}\,\gamma_\mu\,\qslash
+ \frac{2\, m_{\mathrm{P_{bb}^*}}\,(\varepsilon\!\cdot\! p)}{3}\,\gamma_\mu\,\gamma_\nu
- \frac{2(\varepsilon\!\cdot\! p)(p\!\cdot\! q)}{3\, m_{\mathrm{P_{bb}^*}}}\,\gamma_\mu\,\gamma_\nu \nonumber\\
&\quad
- p_\mu\,\eslash\,\qslash\,\gamma_\nu
- \frac{4(p\!\cdot\! q)\, p_\mu}{3\, m_{\mathrm{P_{bb}^*}}^2}\,\eslash\,\qslash\,\gamma_\nu
+ q_\mu\,\eslash\,\qslash\,\gamma_\nu
+ \frac{2\, p_\mu p_\nu}{m_{\mathrm{P_{bb}^*}}^2}\,\pslash\,\eslash\,\qslash
- \frac{4(p\!\cdot\! q)\, p_\mu p_\nu}{3\, m_{\mathrm{P_{bb}^*}}^4}\,\pslash\,\eslash\,\qslash \nonumber\\
&\quad
+ \frac{2\, p_\nu q_\mu}{m_{\mathrm{P_{bb}^*}}^2}\,\pslash\,\eslash\,\qslash
+ \frac{4\, p_\mu q_\nu}{3\, m_{\mathrm{P_{bb}^*}}^2}\,\pslash\,\eslash\,\qslash
- \frac{4(p\!\cdot\! q)\, p_\mu q_\nu}{3\, m_{\mathrm{P_{bb}^*}}^4}\,\pslash\,\eslash\,\qslash
+ \frac{2\, q_\mu q_\nu}{m_{\mathrm{P_{bb}^*}}^2}\,\pslash\,\eslash\,\qslash
- g_{\mu\nu}\,\pslash\,\eslash\,\qslash \nonumber\\
&\quad
+ \frac{4(p\!\cdot\! q)\, p_\mu}{3\, m_{\mathrm{P_{bb}^*}}^2}\,\pslash\,\eslash\,\gamma_\nu
- 2\, q_\mu\,\pslash\,\eslash\,\gamma_\nu
+ 2\,\varepsilon_\mu\,\pslash\,\qslash\,\gamma_\nu
- \frac{2(\varepsilon\!\cdot\! p)\, p_\mu}{3\, m_{\mathrm{P_{bb}^*}}^2}\,\pslash\,\qslash\,\gamma_\nu
- p_\nu\,\gamma_\mu\,\eslash\,\qslash \nonumber\\
&\quad
- \frac{2(p\!\cdot\! q)\, p_\nu}{3\, m_{\mathrm{P_{bb}^*}}^2}\,\gamma_\mu\,\eslash\,\qslash
- \frac{4\, q_\nu}{3}\,\gamma_\mu\,\eslash\,\qslash
- \frac{2(p\!\cdot\! q)\, q_\nu}{3\, m_{\mathrm{P_{bb}^*}}^2}\,\gamma_\mu\,\eslash\,\qslash
+ \frac{2(p\!\cdot\! q)}{3}\,\gamma_\mu\,\eslash\,\gamma_\nu
- \frac{2(p\!\cdot\! q)\, p_\nu}{3\, m_{\mathrm{P_{bb}^*}}^2}\,\gamma_\mu\,\pslash\,\eslash \nonumber\\
&\quad
+ \frac{2\, q_\nu}{3}\,\gamma_\mu\,\pslash\,\eslash
- \frac{2(p\!\cdot\! q)\, q_\nu}{3\, m_{\mathrm{P_{bb}^*}}^2}\,\gamma_\mu\,\pslash\,\eslash
- \frac{2\,\varepsilon_\nu}{3}\,\gamma_\mu\,\pslash\,\qslash
+ \frac{2(\varepsilon\!\cdot\! p)\, p_\nu}{m_{\mathrm{P_{bb}^*}}^2}\,\gamma_\mu\,\pslash\,\qslash
+ \frac{2(\varepsilon\!\cdot\! p)\, q_\nu}{m_{\mathrm{P_{bb}^*}}^2}\,\gamma_\mu\,\pslash\,\qslash \nonumber\\
&\quad
- \frac{2(\varepsilon\!\cdot\! p)}{3}\,\gamma_\mu\,\pslash\,\gamma_\nu
+ \frac{2(\varepsilon\!\cdot\! p)(p\!\cdot\! q)}{3\, m_{\mathrm{P_{bb}^*}}^2}\,\gamma_\mu\,\pslash\,\gamma_\nu
- \frac{4(\varepsilon\!\cdot\! p)}{3}\,\gamma_\mu\,\qslash\,\gamma_\nu
+ \frac{p_\mu}{m_{\mathrm{P_{bb}^*}}}\,\pslash\,\eslash\,\qslash\,\gamma_\nu
- \frac{2(p\!\cdot\! q)\, p_\mu}{3\, m_{\mathrm{P_{bb}^*}}^3}\,\pslash\,\eslash\,\qslash\,\gamma_\nu \nonumber\\
&\quad
+ \frac{q_\mu}{m_{\mathrm{P_{bb}^*}}}\,\pslash\,\eslash\,\qslash\,\gamma_\nu
- m_{\mathrm{P_{bb}^*}}\,\gamma_\mu\,\eslash\,\qslash\,\gamma_\nu
- \frac{(p\!\cdot\! q)}{3\, m_{\mathrm{P_{bb}^*}}}\,\gamma_\mu\,\eslash\,\qslash\,\gamma_\nu
+ \frac{p_\nu}{m_{\mathrm{P_{bb}^*}}}\,\gamma_\mu\,\pslash\,\eslash\,\qslash
+ \frac{2(p\!\cdot\! q)\, p_\nu}{3\, m_{\mathrm{P_{bb}^*}}^3}\,\gamma_\mu\,\pslash\,\eslash\,\qslash \nonumber\\
&\quad
+ \frac{4\, q_\nu}{3\, m_{\mathrm{P_{bb}^*}}}\,\gamma_\mu\,\pslash\,\eslash\,\qslash
+ \frac{2(p\!\cdot\! q)\, q_\nu}{3\, m_{\mathrm{P_{bb}^*}}^3}\,\gamma_\mu\,\pslash\,\eslash\,\qslash
- \frac{2(p\!\cdot\! q)}{3\, m_{\mathrm{P_{bb}^*}}}\,\gamma_\mu\,\pslash\,\eslash\,\gamma_\nu
+ \frac{4(\varepsilon\!\cdot\! p)}{3\, m_{\mathrm{P_{bb}^*}}}\,\gamma_\mu\,\pslash\,\qslash\,\gamma_\nu \nonumber\\
&\quad
+ \gamma_\mu\,\pslash\,\eslash\,\qslash\,\gamma_\nu
+ \frac{(p\!\cdot\! q)}{3\, m_{\mathrm{P_{bb}^*}}^2}\,\gamma_\mu\,\pslash\,\eslash\,\qslash\,\gamma_\nu
\Bigg\}.
\end{align}

\subsubsection{Gauge-fixed tensor structure proportional to  
$F_{2}(q^{2})$ prior to spin-1/2 contamination removal)} \label{F2_appendix11}
\begin{align}
{F}_2 & \propto\,\Bigg\{ 
- \frac{2\,\varepsilon_\nu\,(p\!\cdot\! q)\, p_\mu}{3\, m_{\mathrm{P_{bb}^*}}}
+ \frac{2\,\varepsilon_\mu\,(p\!\cdot\! q)\, p_\nu}{m_{\mathrm{P_{bb}^*}}}
- \frac{2(\varepsilon\!\cdot\! p)\, p_\nu q_\mu}{m_{\mathrm{P_{bb}^*}}}
+ \frac{2\,\varepsilon_\mu\,(p\!\cdot\! q)\, q_\nu}{m_{\mathrm{P_{bb}^*}}}
+ \frac{2(\varepsilon\!\cdot\! p)\, p_\mu q_\nu}{3\, m_{\mathrm{P_{bb}^*}}}
- \frac{2(\varepsilon\!\cdot\! p)\, q_\mu q_\nu}{m_{\mathrm{P_{bb}^*}}} \nonumber\\
&\quad
- \frac{14(p\!\cdot\! q)\, p_\mu p_\nu}{3\, m_{\mathrm{P_{bb}^*}}^2}\,\eslash
+ \frac{8(p\!\cdot\! q)^2\, p_\mu p_\nu}{3\, m_{\mathrm{P_{bb}^*}}^4}\,\eslash
+ 2\, p_\nu q_\mu\,\eslash
- \frac{2(p\!\cdot\! q)\, p_\nu q_\mu}{m_{\mathrm{P_{bb}^*}}^2}\,\eslash
+ \frac{2\, p_\mu q_\nu}{3}\,\eslash
- \frac{10(p\!\cdot\! q)\, p_\mu q_\nu}{3\, m_{\mathrm{P_{bb}^*}}^2}\,\eslash \nonumber\\
&\quad
+ \frac{8(p\!\cdot\! q)^2\, p_\mu q_\nu}{3\, m_{\mathrm{P_{bb}^*}}^4}\,\eslash
+ 2\, q_\mu q_\nu\,\eslash
- \frac{2(p\!\cdot\! q)\, q_\mu q_\nu}{m_{\mathrm{P_{bb}^*}}^2}\,\eslash
+ (p\!\cdot\! q)\, g_{\mu\nu}\,\eslash
+ \frac{2\,\varepsilon_\nu\,(p\!\cdot\! q)\, p_\mu}{3\, m_{\mathrm{P_{bb}^*}}^2}\,\pslash
+ \frac{2\,\varepsilon_\mu\,(p\!\cdot\! q)\, p_\nu}{m_{\mathrm{P_{bb}^*}}^2}\,\pslash \nonumber\\
&\quad
- \frac{2(\varepsilon\!\cdot\! p)\, p_\nu q_\mu}{m_{\mathrm{P_{bb}^*}}^2}\,\pslash
+ \frac{2\,\varepsilon_\mu\,(p\!\cdot\! q)\, q_\nu}{m_{\mathrm{P_{bb}^*}}^2}\,\pslash
- \frac{2(\varepsilon\!\cdot\! p)\, p_\mu q_\nu}{3\, m_{\mathrm{P_{bb}^*}}^2}\,\pslash
- \frac{2(\varepsilon\!\cdot\! p)\, q_\mu q_\nu}{m_{\mathrm{P_{bb}^*}}^2}\,\pslash
- \frac{2\,\varepsilon_\nu\, p_\mu}{3}\,\qslash
- 2\,\varepsilon_\mu\, p_\nu\,\qslash \nonumber\\
&\quad
+ \frac{14(\varepsilon\!\cdot\! p)\, p_\mu p_\nu}{3\, m_{\mathrm{P_{bb}^*}}^2}\,\qslash
- \frac{8(\varepsilon\!\cdot\! p)(p\!\cdot\! q)\, p_\mu p_\nu}{3\, m_{\mathrm{P_{bb}^*}}^4}\,\qslash
+ \frac{2(\varepsilon\!\cdot\! p)\, p_\nu q_\mu}{m_{\mathrm{P_{bb}^*}}^2}\,\qslash
- 2\,\varepsilon_\mu\, q_\nu\,\qslash
+ \frac{10(\varepsilon\!\cdot\! p)\, p_\mu q_\nu}{3\, m_{\mathrm{P_{bb}^*}}^2}\,\qslash \nonumber\\
&\quad
- \frac{8(\varepsilon\!\cdot\! p)(p\!\cdot\! q)\, p_\mu q_\nu}{3\, m_{\mathrm{P_{bb}^*}}^4}\,\qslash
+ \frac{2(\varepsilon\!\cdot\! p)\, q_\mu q_\nu}{m_{\mathrm{P_{bb}^*}}^2}\,\qslash
- (\varepsilon\!\cdot\! p)\, g_{\mu\nu}\,\qslash
- \frac{2\,\varepsilon_\nu\,(p\!\cdot\! q)}{3}\,\gamma_\mu
+ \frac{2(\varepsilon\!\cdot\! p)\, q_\nu}{3}\,\gamma_\mu
+ 2\,\varepsilon_\mu\,(p\!\cdot\! q)\,\gamma_\nu \nonumber\\
&\quad
- 2(\varepsilon\!\cdot\! p)\, q_\mu\,\gamma_\nu
+ \frac{2(p\!\cdot\! q)\, p_\mu p_\nu}{m_{\mathrm{P_{bb}^*}}^3}\,\eslash\,\qslash
- \frac{2\, p_\nu q_\mu}{m_{\mathrm{P_{bb}^*}}}\,\eslash\,\qslash
+ \frac{4\, p_\mu q_\nu}{3\, m_{\mathrm{P_{bb}^*}}}\,\eslash\,\qslash
+ \frac{2(p\!\cdot\! q)\, p_\mu q_\nu}{m_{\mathrm{P_{bb}^*}}^3}\,\eslash\,\qslash \nonumber\\
&\quad
- \frac{2\, q_\mu q_\nu}{m_{\mathrm{P_{bb}^*}}}\,\eslash\,\qslash
+ m_{\mathrm{P_{bb}^*}}\, g_{\mu\nu}\,\eslash\,\qslash
- \frac{3(p\!\cdot\! q)\, p_\mu}{m_{\mathrm{P_{bb}^*}}}\,\eslash\,\gamma_\nu
+ \frac{4(p\!\cdot\! q)^2\, p_\mu}{3\, m_{\mathrm{P_{bb}^*}}^3}\,\eslash\,\gamma_\nu
+ 2\, m_{\mathrm{P_{bb}^*}}\, q_\mu\,\eslash\,\gamma_\nu
- \frac{(p\!\cdot\! q)\, q_\mu}{m_{\mathrm{P_{bb}^*}}}\,\eslash\,\gamma_\nu \nonumber\\
&\quad
- \frac{4(p\!\cdot\! q)\, p_\mu p_\nu}{3\, m_{\mathrm{P_{bb}^*}}^3}\,\pslash\,\eslash
+ \frac{4(p\!\cdot\! q)^2\, p_\mu p_\nu}{3\, m_{\mathrm{P_{bb}^*}}^5}\,\pslash\,\eslash
+ \frac{2\, p_\nu q_\mu}{m_{\mathrm{P_{bb}^*}}}\,\pslash\,\eslash
- \frac{2(p\!\cdot\! q)\, p_\nu q_\mu}{m_{\mathrm{P_{bb}^*}}^3}\,\pslash\,\eslash
- \frac{2\, p_\mu q_\nu}{3\, m_{\mathrm{P_{bb}^*}}}\,\pslash\,\eslash \nonumber\\
&\quad
- \frac{2(p\!\cdot\! q)\, p_\mu q_\nu}{3\, m_{\mathrm{P_{bb}^*}}^3}\,\pslash\,\eslash
+ \frac{4(p\!\cdot\! q)^2\, p_\mu q_\nu}{3\, m_{\mathrm{P_{bb}^*}}^5}\,\pslash\,\eslash
+ \frac{2\, q_\mu q_\nu}{m_{\mathrm{P_{bb}^*}}}\,\pslash\,\eslash
- \frac{2(p\!\cdot\! q)\, q_\mu q_\nu}{m_{\mathrm{P_{bb}^*}}^3}\,\pslash\,\eslash
+ \frac{(p\!\cdot\! q)\, g_{\mu\nu}}{m_{\mathrm{P_{bb}^*}}}\,\pslash\,\eslash \nonumber\\
&\quad
+ \frac{2\,\varepsilon_\nu\, p_\mu}{3\, m_{\mathrm{P_{bb}^*}}}\,\pslash\,\qslash
- \frac{2\,\varepsilon_\mu\, p_\nu}{m_{\mathrm{P_{bb}^*}}}\,\pslash\,\qslash
+ \frac{4(\varepsilon\!\cdot\! p)\, p_\mu p_\nu}{3\, m_{\mathrm{P_{bb}^*}}^3}\,\pslash\,\qslash
- \frac{4(\varepsilon\!\cdot\! p)(p\!\cdot\! q)\, p_\mu p_\nu}{3\, m_{\mathrm{P_{bb}^*}}^5}\,\pslash\,\qslash
+ \frac{2(\varepsilon\!\cdot\! p)\, p_\nu q_\mu}{m_{\mathrm{P_{bb}^*}}^3}\,\pslash\,\qslash \nonumber\\
&\quad
- \frac{2\,\varepsilon_\mu\, q_\nu}{m_{\mathrm{P_{bb}^*}}}\,\pslash\,\qslash
+ \frac{2(\varepsilon\!\cdot\! p)\, p_\mu q_\nu}{3\, m_{\mathrm{P_{bb}^*}}^3}\,\pslash\,\qslash
- \frac{4(\varepsilon\!\cdot\! p)(p\!\cdot\! q)\, p_\mu q_\nu}{3\, m_{\mathrm{P_{bb}^*}}^5}\,\pslash\,\qslash
+ \frac{2(\varepsilon\!\cdot\! p)\, q_\mu q_\nu}{m_{\mathrm{P_{bb}^*}}^3}\,\pslash\,\qslash
- \frac{(\varepsilon\!\cdot\! p)\, g_{\mu\nu}}{m_{\mathrm{P_{bb}^*}}}\,\pslash\,\qslash \nonumber\\
&\quad
+ \frac{2\,\varepsilon_\mu\,(p\!\cdot\! q)}{m_{\mathrm{P_{bb}^*}}}\,\pslash\,\gamma_\nu
- \frac{2(\varepsilon\!\cdot\! p)\, q_\mu}{m_{\mathrm{P_{bb}^*}}}\,\pslash\,\gamma_\nu
- 2\, m_{\mathrm{P_{bb}^*}}\,\varepsilon_\mu\,\qslash\,\gamma_\nu
+ \frac{3(\varepsilon\!\cdot\! p)\, p_\mu}{m_{\mathrm{P_{bb}^*}}}\,\qslash\,\gamma_\nu
- \frac{4(\varepsilon\!\cdot\! p)(p\!\cdot\! q)\, p_\mu}{3\, m_{\mathrm{P_{bb}^*}}^3}\,\qslash\,\gamma_\nu \nonumber\\
&\quad
+ \frac{(\varepsilon\!\cdot\! p)\, q_\mu}{m_{\mathrm{P_{bb}^*}}}\,\qslash\,\gamma_\nu
- \frac{5(p\!\cdot\! q)\, p_\nu}{3\, m_{\mathrm{P_{bb}^*}}}\,\gamma_\mu\,\eslash
+ \frac{2(p\!\cdot\! q)^2\, p_\nu}{3\, m_{\mathrm{P_{bb}^*}}^3}\,\gamma_\mu\,\eslash
+ \frac{2\, m_{\mathrm{P_{bb}^*}}\, q_\nu}{3}\,\gamma_\mu\,\eslash
- \frac{4(p\!\cdot\! q)\, q_\nu}{3\, m_{\mathrm{P_{bb}^*}}}\,\gamma_\mu\,\eslash
+ \frac{2(p\!\cdot\! q)^2\, q_\nu}{3\, m_{\mathrm{P_{bb}^*}}^3}\,\gamma_\mu\,\eslash \nonumber\\
&\quad
+ \frac{2\,\varepsilon_\nu\,(p\!\cdot\! q)}{3\, m_{\mathrm{P_{bb}^*}}}\,\gamma_\mu\,\pslash
- \frac{2(\varepsilon\!\cdot\! p)\, q_\nu}{3\, m_{\mathrm{P_{bb}^*}}}\,\gamma_\mu\,\pslash
- \frac{2\, m_{\mathrm{P_{bb}^*}}\,\varepsilon_\nu}{3}\,\gamma_\mu\,\qslash
+ \frac{5(\varepsilon\!\cdot\! p)\, p_\nu}{3\, m_{\mathrm{P_{bb}^*}}}\,\gamma_\mu\,\qslash
- \frac{2(\varepsilon\!\cdot\! p)(p\!\cdot\! q)\, p_\nu}{3\, m_{\mathrm{P_{bb}^*}}^3}\,\gamma_\mu\,\qslash \nonumber\\
&\quad
+ \frac{4(\varepsilon\!\cdot\! p)\, q_\nu}{3\, m_{\mathrm{P_{bb}^*}}}\,\gamma_\mu\,\qslash
- \frac{2(\varepsilon\!\cdot\! p)(p\!\cdot\! q)\, q_\nu}{3\, m_{\mathrm{P_{bb}^*}}^3}\,\gamma_\mu\,\qslash
+ p_\mu\,\eslash\,\qslash\,\gamma_\nu
+ \frac{(p\!\cdot\! q)\, p_\mu}{m_{\mathrm{P_{bb}^*}}^2}\,\eslash\,\qslash\,\gamma_\nu
- q_\mu\,\eslash\,\qslash\,\gamma_\nu \nonumber\\
&\quad
- \frac{2\, p_\mu p_\nu}{m_{\mathrm{P_{bb}^*}}^2}\,\pslash\,\eslash\,\qslash
+ \frac{2(p\!\cdot\! q)\, p_\mu p_\nu}{m_{\mathrm{P_{bb}^*}}^4}\,\pslash\,\eslash\,\qslash
- \frac{2\, p_\nu q_\mu}{m_{\mathrm{P_{bb}^*}}^2}\,\pslash\,\eslash\,\qslash
- \frac{4\, p_\mu q_\nu}{3\, m_{\mathrm{P_{bb}^*}}^2}\,\pslash\,\eslash\,\qslash \nonumber\\
&\quad
+ \frac{2(p\!\cdot\! q)\, p_\mu q_\nu}{m_{\mathrm{P_{bb}^*}}^4}\,\pslash\,\eslash\,\qslash
- \frac{2\, q_\mu q_\nu}{m_{\mathrm{P_{bb}^*}}^2}\,\pslash\,\eslash\,\qslash
+ g_{\mu\nu}\,\pslash\,\eslash\,\qslash
- \frac{(p\!\cdot\! q)\, p_\mu}{m_{\mathrm{P_{bb}^*}}^2}\,\pslash\,\eslash\,\gamma_\nu
+ \frac{2(p\!\cdot\! q)^2\, p_\mu}{3\, m_{\mathrm{P_{bb}^*}}^4}\,\pslash\,\eslash\,\gamma_\nu \nonumber\\
&\quad
+ 2\, q_\mu\,\pslash\,\eslash\,\gamma_\nu
- \frac{(p\!\cdot\! q)\, q_\mu}{m_{\mathrm{P_{bb}^*}}^2}\,\pslash\,\eslash\,\gamma_\nu
- 2\,\varepsilon_\mu\,\pslash\,\qslash\,\gamma_\nu
+ \frac{(\varepsilon\!\cdot\! p)\, p_\mu}{m_{\mathrm{P_{bb}^*}}^2}\,\pslash\,\qslash\,\gamma_\nu
- \frac{2(\varepsilon\!\cdot\! p)(p\!\cdot\! q)\, p_\mu}{3\, m_{\mathrm{P_{bb}^*}}^4}\,\pslash\,\qslash\,\gamma_\nu \nonumber\\
&\quad
+ \frac{(\varepsilon\!\cdot\! p)\, q_\mu}{m_{\mathrm{P_{bb}^*}}^2}\,\pslash\,\qslash\,\gamma_\nu
+ p_\nu\,\gamma_\mu\,\eslash\,\qslash
+ \frac{4\, q_\nu}{3}\,\gamma_\mu\,\eslash\,\qslash
- (p\!\cdot\! q)\,\gamma_\mu\,\eslash\,\gamma_\nu
+ \frac{(p\!\cdot\! q)^2}{3\, m_{\mathrm{P_{bb}^*}}^2}\,\gamma_\mu\,\eslash\,\gamma_\nu
+ \frac{5(p\!\cdot\! q)\, p_\nu}{3\, m_{\mathrm{P_{bb}^*}}^2}\,\gamma_\mu\,\pslash\,\eslash \nonumber\\
&\quad
- \frac{2(p\!\cdot\! q)^2\, p_\nu}{3\, m_{\mathrm{P_{bb}^*}}^4}\,\gamma_\mu\,\pslash\,\eslash
- \frac{2\, q_\nu}{3}\,\gamma_\mu\,\pslash\,\eslash
+ \frac{4(p\!\cdot\! q)\, q_\nu}{3\, m_{\mathrm{P_{bb}^*}}^2}\,\gamma_\mu\,\pslash\,\eslash
- \frac{2(p\!\cdot\! q)^2\, q_\nu}{3\, m_{\mathrm{P_{bb}^*}}^4}\,\gamma_\mu\,\pslash\,\eslash
+ \frac{2\,\varepsilon_\nu}{3}\,\gamma_\mu\,\pslash\,\qslash \nonumber%\\
\end{align}

\begin{align}
  &\quad
- \frac{5(\varepsilon\!\cdot\! p)\, p_\nu}{3\, m_{\mathrm{P_{bb}^*}}^2}\,\gamma_\mu\,\pslash\,\qslash
+ \frac{2(\varepsilon\!\cdot\! p)(p\!\cdot\! q)\, p_\nu}{3\, m_{\mathrm{P_{bb}^*}}^4}\,\gamma_\mu\,\pslash\,\qslash
- \frac{4(\varepsilon\!\cdot\! p)\, q_\nu}{3\, m_{\mathrm{P_{bb}^*}}^2}\,\gamma_\mu\,\pslash\,\qslash
+ \frac{2(\varepsilon\!\cdot\! p)(p\!\cdot\! q)\, q_\nu}{3\, m_{\mathrm{P_{bb}^*}}^4}\,\gamma_\mu\,\pslash\,\qslash \nonumber\\
&\quad
+ (\varepsilon\!\cdot\! p)\,\gamma_\mu\,\qslash\,\gamma_\nu
- \frac{(\varepsilon\!\cdot\! p)(p\!\cdot\! q)}{3\, m_{\mathrm{P_{bb}^*}}^2}\,\gamma_\mu\,\qslash\,\gamma_\nu
- \frac{p_\mu}{m_{\mathrm{P_{bb}^*}}}\,\pslash\,\eslash\,\qslash\,\gamma_\nu
+ \frac{(p\!\cdot\! q)\, p_\mu}{m_{\mathrm{P_{bb}^*}}^3}\,\pslash\,\eslash\,\qslash\,\gamma_\nu
- \frac{q_\mu}{m_{\mathrm{P_{bb}^*}}}\,\pslash\,\eslash\,\qslash\,\gamma_\nu \nonumber\\
&\quad
+ m_{\mathrm{P_{bb}^*}}\,\gamma_\mu\,\eslash\,\qslash\,\gamma_\nu
- \frac{p_\nu}{m_{\mathrm{P_{bb}^*}}}\,\gamma_\mu\,\pslash\,\eslash\,\qslash
+ \frac{(p\!\cdot\! q)\, p_\nu}{m_{\mathrm{P_{bb}^*}}^3}\,\gamma_\mu\,\pslash\,\eslash\,\qslash
- \frac{4\, q_\nu}{3\, m_{\mathrm{P_{bb}^*}}}\,\gamma_\mu\,\pslash\,\eslash\,\qslash
+ \frac{(p\!\cdot\! q)}{m_{\mathrm{P_{bb}^*}}}\,\gamma_\mu\,\pslash\,\eslash\,\gamma_\nu \nonumber\\
&\quad
- \frac{(p\!\cdot\! q)^2}{3\, m_{\mathrm{P_{bb}^*}}^3}\,\gamma_\mu\,\pslash\,\eslash\,\gamma_\nu
- \frac{(\varepsilon\!\cdot\! p)}{m_{\mathrm{P_{bb}^*}}}\,\gamma_\mu\,\pslash\,\qslash\,\gamma_\nu
+ \frac{(\varepsilon\!\cdot\! p)(p\!\cdot\! q)}{3\, m_{\mathrm{P_{bb}^*}}^3}\,\gamma_\mu\,\pslash\,\qslash\,\gamma_\nu
- \gamma_\mu\,\pslash\,\eslash\,\qslash\,\gamma_\nu
\Bigg\}.
\label{F2_appendix1}  
\end{align}

\subsubsection{Gauge-fixed tensor structure proportional to  
$F_{3}(q^{2})$ prior to spin-1/2 contamination removal)}  \label{F3_appendix11}
\begin{align}
\label{F3_appendix1}
{F}_3 & \propto\,\Bigg\{ 
- \frac{4(\varepsilon\!\cdot\! p)(p\!\cdot\! q)^2\, p_\mu p_\nu}{3\, m_{\mathrm{P_{bb}^*}}^5}
+ \frac{(\varepsilon\!\cdot\! p)(p\!\cdot\! q)\, p_\nu q_\mu}{m_{\mathrm{P_{bb}^*}}^3}
+ \frac{2(\varepsilon\!\cdot\! p)(p\!\cdot\! q)\, p_\mu q_\nu}{3\, m_{\mathrm{P_{bb}^*}}^3}
- \frac{4(\varepsilon\!\cdot\! p)(p\!\cdot\! q)^2\, p_\mu q_\nu}{3\, m_{\mathrm{P_{bb}^*}}^5} \nonumber\\
&\quad
- \frac{(\varepsilon\!\cdot\! p)\, q_\mu q_\nu}{2\, m_{\mathrm{P_{bb}^*}}}
+ \frac{(\varepsilon\!\cdot\! p)(p\!\cdot\! q)\, q_\mu q_\nu}{m_{\mathrm{P_{bb}^*}}^3}
- \frac{2(\varepsilon\!\cdot\! p)(p\!\cdot\! q)^2\, p_\mu p_\nu}{3\, m_{\mathrm{P_{bb}^*}}^6}\,\pslash
+ \frac{(\varepsilon\!\cdot\! p)(p\!\cdot\! q)\, p_\nu q_\mu}{m_{\mathrm{P_{bb}^*}}^4}\,\pslash \nonumber\\
&\quad
+ \frac{(\varepsilon\!\cdot\! p)(p\!\cdot\! q)\, p_\mu q_\nu}{3\, m_{\mathrm{P_{bb}^*}}^4}\,\pslash
- \frac{2(\varepsilon\!\cdot\! p)(p\!\cdot\! q)^2\, p_\mu q_\nu}{3\, m_{\mathrm{P_{bb}^*}}^6}\,\pslash
- \frac{(\varepsilon\!\cdot\! p)\, q_\mu q_\nu}{2\, m_{\mathrm{P_{bb}^*}}^2}\,\pslash
+ \frac{(\varepsilon\!\cdot\! p)(p\!\cdot\! q)\, q_\mu q_\nu}{m_{\mathrm{P_{bb}^*}}^4}\,\pslash \nonumber\\
&\quad
+ \frac{(\varepsilon\!\cdot\! p)(p\!\cdot\! q)\, p_\mu p_\nu}{3\, m_{\mathrm{P_{bb}^*}}^4}\,\qslash
- \frac{(\varepsilon\!\cdot\! p)\, p_\nu q_\mu}{2\, m_{\mathrm{P_{bb}^*}}^2}\,\qslash
+ \frac{(\varepsilon\!\cdot\! p)\, p_\mu q_\nu}{6\, m_{\mathrm{P_{bb}^*}}^2}\,\qslash
+ \frac{(\varepsilon\!\cdot\! p)(p\!\cdot\! q)\, p_\mu q_\nu}{3\, m_{\mathrm{P_{bb}^*}}^4}\,\qslash
- \frac{(\varepsilon\!\cdot\! p)\, q_\mu q_\nu}{2\, m_{\mathrm{P_{bb}^*}}^2}\,\qslash \nonumber\\
&\quad
- \frac{(\varepsilon\!\cdot\! p)(p\!\cdot\! q)^2\, p_\nu}{3\, m_{\mathrm{P_{bb}^*}}^4}\,\gamma_\mu
+ \frac{(\varepsilon\!\cdot\! p)(p\!\cdot\! q)\, q_\nu}{6\, m_{\mathrm{P_{bb}^*}}^2}\,\gamma_\mu
- \frac{(\varepsilon\!\cdot\! p)(p\!\cdot\! q)^2\, q_\nu}{3\, m_{\mathrm{P_{bb}^*}}^4}\,\gamma_\mu
- \frac{2(\varepsilon\!\cdot\! p)(p\!\cdot\! q)^2\, p_\mu}{3\, m_{\mathrm{P_{bb}^*}}^4}\,\gamma_\nu
+ \frac{(\varepsilon\!\cdot\! p)(p\!\cdot\! q)\, q_\mu}{2\, m_{\mathrm{P_{bb}^*}}^2}\,\gamma_\nu \nonumber\\
&\quad
- \frac{2(p\!\cdot\! q)^2\, p_\mu p_\nu}{3\, m_{\mathrm{P_{bb}^*}}^5}\,\eslash\,\qslash
+ \frac{(p\!\cdot\! q)\, p_\nu q_\mu}{2\, m_{\mathrm{P_{bb}^*}}^3}\,\eslash\,\qslash
+ \frac{(p\!\cdot\! q)\, p_\mu q_\nu}{3\, m_{\mathrm{P_{bb}^*}}^3}\,\eslash\,\qslash
- \frac{2(p\!\cdot\! q)^2\, p_\mu q_\nu}{3\, m_{\mathrm{P_{bb}^*}}^5}\,\eslash\,\qslash \nonumber\\
&\quad
- \frac{q_\mu q_\nu}{4\, m_{\mathrm{P_{bb}^*}}}\,\eslash\,\qslash
+ \frac{(p\!\cdot\! q)\, q_\mu q_\nu}{2\, m_{\mathrm{P_{bb}^*}}^3}\,\eslash\,\qslash
+ \frac{5(p\!\cdot\! q)\, p_\mu}{12\, m_{\mathrm{P_{bb}^*}}^3}\,\eslash\,\gamma_\nu
- \frac{q_\mu}{4\, m_{\mathrm{P_{bb}^*}}}\,\eslash\,\gamma_\nu
+ \frac{2(\varepsilon\!\cdot\! p)(p\!\cdot\! q)\, p_\mu p_\nu}{3\, m_{\mathrm{P_{bb}^*}}^5}\,\pslash\,\qslash \nonumber\\
&\quad
- \frac{(\varepsilon\!\cdot\! p)\, p_\nu q_\mu}{2\, m_{\mathrm{P_{bb}^*}}^3}\,\pslash\,\qslash
- \frac{(\varepsilon\!\cdot\! p)\, p_\mu q_\nu}{6\, m_{\mathrm{P_{bb}^*}}^3}\,\pslash\,\qslash
+ \frac{2(\varepsilon\!\cdot\! p)(p\!\cdot\! q)\, p_\mu q_\nu}{3\, m_{\mathrm{P_{bb}^*}}^5}\,\pslash\,\qslash
- \frac{(\varepsilon\!\cdot\! p)\, q_\mu q_\nu}{2\, m_{\mathrm{P_{bb}^*}}^3}\,\pslash\,\qslash \nonumber\\
&\quad
- \frac{(\varepsilon\!\cdot\! p)(p\!\cdot\! q)^2\, p_\nu}{3\, m_{\mathrm{P_{bb}^*}}^5}\,\pslash\,\gamma_\mu
+ \frac{(\varepsilon\!\cdot\! p)(p\!\cdot\! q)\, q_\nu}{2\, m_{\mathrm{P_{bb}^*}}^3}\,\pslash\,\gamma_\mu
- \frac{(\varepsilon\!\cdot\! p)(p\!\cdot\! q)^2\, q_\nu}{3\, m_{\mathrm{P_{bb}^*}}^5}\,\pslash\,\gamma_\mu
- \frac{(\varepsilon\!\cdot\! p)(p\!\cdot\! q)^2\, p_\mu}{3\, m_{\mathrm{P_{bb}^*}}^5}\,\pslash\,\gamma_\nu \nonumber\\
&\quad
+ \frac{(\varepsilon\!\cdot\! p)(p\!\cdot\! q)\, q_\mu}{2\, m_{\mathrm{P_{bb}^*}}^3}\,\pslash\,\gamma_\nu
+ \frac{(\varepsilon\!\cdot\! p)(p\!\cdot\! q)\, p_\mu}{2\, m_{\mathrm{P_{bb}^*}}^3}\,\qslash\,\gamma_\nu
- \frac{(\varepsilon\!\cdot\! p)\, q_\mu}{2\, m_{\mathrm{P_{bb}^*}}}\,\qslash\,\gamma_\nu
- \frac{(p\!\cdot\! q)^2\, p_\nu}{3\, m_{\mathrm{P_{bb}^*}}^3}\,\gamma_\mu\,\eslash
- \frac{(p\!\cdot\! q)^2\, q_\nu}{3\, m_{\mathrm{P_{bb}^*}}^3}\,\gamma_\mu\,\eslash \nonumber\\
&\quad
+ \frac{(\varepsilon\!\cdot\! p)(p\!\cdot\! q)^2\, p_\nu}{3\, m_{\mathrm{P_{bb}^*}}^5}\,\gamma_\mu\,\pslash
+ \frac{(\varepsilon\!\cdot\! p)(p\!\cdot\! q)\, q_\nu}{6\, m_{\mathrm{P_{bb}^*}}^3}\,\gamma_\mu\,\pslash
+ \frac{(\varepsilon\!\cdot\! p)(p\!\cdot\! q)^2\, q_\nu}{3\, m_{\mathrm{P_{bb}^*}}^5}\,\gamma_\mu\,\pslash
- \frac{(\varepsilon\!\cdot\! p)(p\!\cdot\! q)\, p_\nu}{6\, m_{\mathrm{P_{bb}^*}}^3}\,\gamma_\mu\,\qslash \nonumber\\
&\quad
- \frac{(\varepsilon\!\cdot\! p)(p\!\cdot\! q)\, q_\nu}{6\, m_{\mathrm{P_{bb}^*}}^3}\,\gamma_\mu\,\qslash
- \frac{(\varepsilon\!\cdot\! p)(p\!\cdot\! q)^2}{6\, m_{\mathrm{P_{bb}^*}}^3}\,\gamma_\mu\,\gamma_\nu
- \frac{(p\!\cdot\! q)^2\, p_\mu}{3\, m_{\mathrm{P_{bb}^*}}^4}\,\eslash\,\qslash\,\gamma_\nu
+ \frac{(p\!\cdot\! q)\, q_\mu}{4\, m_{\mathrm{P_{bb}^*}}^2}\,\eslash\,\qslash\,\gamma_\nu \nonumber\\
&\quad
- \frac{(p\!\cdot\! q)^2\, p_\mu p_\nu}{3\, m_{\mathrm{P_{bb}^*}}^6}\,\pslash\,\eslash\,\qslash
+ \frac{(p\!\cdot\! q)\, p_\nu q_\mu}{2\, m_{\mathrm{P_{bb}^*}}^4}\,\pslash\,\eslash\,\qslash
+ \frac{(p\!\cdot\! q)\, p_\mu q_\nu}{6\, m_{\mathrm{P_{bb}^*}}^4}\,\pslash\,\eslash\,\qslash
- \frac{(p\!\cdot\! q)^2\, p_\mu q_\nu}{3\, m_{\mathrm{P_{bb}^*}}^6}\,\pslash\,\eslash\,\qslash \nonumber%\\
\end{align}

\begin{align}
&\quad
- \frac{q_\mu q_\nu}{4\, m_{\mathrm{P_{bb}^*}}^2}\,\pslash\,\eslash\,\qslash
+ \frac{(p\!\cdot\! q)\, q_\mu q_\nu}{2\, m_{\mathrm{P_{bb}^*}}^4}\,\pslash\,\eslash\,\qslash
+ \frac{(\varepsilon\!\cdot\! p)(p\!\cdot\! q)\, p_\mu}{2\, m_{\mathrm{P_{bb}^*}}^4}\,\pslash\,\qslash\,\gamma_\nu
- \frac{(\varepsilon\!\cdot\! p)\, q_\mu}{2\, m_{\mathrm{P_{bb}^*}}^2}\,\pslash\,\qslash\,\gamma_\nu \nonumber\\
&\quad
- \frac{(p\!\cdot\! q)^2\, p_\nu}{6\, m_{\mathrm{P_{bb}^*}}^4}\,\gamma_\mu\,\eslash\,\qslash
+ \frac{(p\!\cdot\! q)\, q_\nu}{12\, m_{\mathrm{P_{bb}^*}}^2}\,\gamma_\mu\,\eslash\,\qslash
- \frac{(p\!\cdot\! q)^2\, q_\nu}{6\, m_{\mathrm{P_{bb}^*}}^4}\,\gamma_\mu\,\eslash\,\qslash
+ \frac{(p\!\cdot\! q)^2}{6\, m_{\mathrm{P_{bb}^*}}^2}\,\gamma_\mu\,\eslash\,\gamma_\nu \nonumber\\
&\quad
+ \frac{(\varepsilon\!\cdot\! p)(p\!\cdot\! q)\, p_\nu}{6\, m_{\mathrm{P_{bb}^*}}^4}\,\gamma_\mu\,\pslash\,\qslash
+ \frac{(\varepsilon\!\cdot\! p)(p\!\cdot\! q)\, q_\nu}{6\, m_{\mathrm{P_{bb}^*}}^4}\,\gamma_\mu\,\pslash\,\qslash
+ \frac{(\varepsilon\!\cdot\! p)(p\!\cdot\! q)^2}{6\, m_{\mathrm{P_{bb}^*}}^4}\,\gamma_\mu\,\pslash\,\gamma_\nu
- \frac{(p\!\cdot\! q)^2\, p_\mu}{6\, m_{\mathrm{P_{bb}^*}}^5}\,\pslash\,\eslash\,\qslash\,\gamma_\nu \nonumber\\
&\quad
+ \frac{(p\!\cdot\! q)\, q_\mu}{4\, m_{\mathrm{P_{bb}^*}}^3}\,\pslash\,\eslash\,\qslash\,\gamma_\nu
+ \frac{(p\!\cdot\! q)^2}{12\, m_{\mathrm{P_{bb}^*}}^4}\,\pslash\,\eslash\,\qslash\,\gamma_\nu
+ \frac{(p\!\cdot\! q)^2\, p_\nu}{6\, m_{\mathrm{P_{bb}^*}}^5}\,\gamma_\mu\,\pslash\,\eslash\,\qslash
+ \frac{(p\!\cdot\! q)^2\, q_\nu}{6\, m_{\mathrm{P_{bb}^*}}^5}\,\gamma_\mu\,\pslash\,\eslash\,\qslash \nonumber\\
&\quad
- \frac{(p\!\cdot\! q)^2}{12\, m_{\mathrm{P_{bb}^*}}^3}\,\gamma_\mu\,\pslash\,\eslash\,\gamma_\nu
+ \frac{(p\!\cdot\! q)^2}{12\, m_{\mathrm{P_{bb}^*}}^4}\,\gamma_\mu\,\pslash\,\eslash\,\qslash\,\gamma_\nu
\Bigg\}.
\end{align}

\subsubsection{Gauge-fixed tensor structure proportional to  
$F_{4}(q^{2})$ prior to spin-1/2 contamination removal)}  \label{F4_appendix11}
\begin{align}
{F}_4 & \propto\,\Bigg\{ 
\frac{(\varepsilon\!\cdot\! q)(p\!\cdot\! q)^2\, p_\mu p_\nu}{3\, m_{\mathrm{P_{bb}^*}}^5}
- \frac{(\varepsilon\!\cdot\! p)\, p_\nu q_\mu\, q^2}{4\, m_{\mathrm{P_{bb}^*}}^3}
- \frac{(\varepsilon\!\cdot\! q)(p\!\cdot\! q)\, p_\mu q_\nu}{6\, m_{\mathrm{P_{bb}^*}}^3}
+ \frac{(\varepsilon\!\cdot\! q)(p\!\cdot\! q)^2\, p_\mu q_\nu}{3\, m_{\mathrm{P_{bb}^*}}^5} \nonumber\\
&\quad
- \frac{(\varepsilon\!\cdot\! p)\, q_\mu q_\nu\, q^2}{4\, m_{\mathrm{P_{bb}^*}}^3}
+ \frac{2(p\!\cdot\! q)^3\, p_\mu p_\nu}{3\, m_{\mathrm{P_{bb}^*}}^6}\,\eslash
- \frac{(p\!\cdot\! q)^2\, p_\nu q_\mu}{2\, m_{\mathrm{P_{bb}^*}}^4}\,\eslash
- \frac{(p\!\cdot\! q)^2\, p_\mu q_\nu}{3\, m_{\mathrm{P_{bb}^*}}^4}\,\eslash
+ \frac{2(p\!\cdot\! q)^3\, p_\mu q_\nu}{3\, m_{\mathrm{P_{bb}^*}}^6}\,\eslash \nonumber\\
&\quad
+ \frac{(p\!\cdot\! q)\, q_\mu q_\nu}{4\, m_{\mathrm{P_{bb}^*}}^2}\,\eslash
- \frac{(p\!\cdot\! q)^2\, q_\mu q_\nu}{2\, m_{\mathrm{P_{bb}^*}}^4}\,\eslash
- \frac{(\varepsilon\!\cdot\! q)(p\!\cdot\! q)^2\, p_\mu p_\nu}{3\, m_{\mathrm{P_{bb}^*}}^6}\,\pslash
- \frac{(\varepsilon\!\cdot\! p)\, p_\nu q_\mu\, q^2}{4\, m_{\mathrm{P_{bb}^*}}^4}\,\pslash \nonumber\\
&\quad
+ \frac{(\varepsilon\!\cdot\! q)(p\!\cdot\! q)\, p_\mu q_\nu}{6\, m_{\mathrm{P_{bb}^*}}^4}\,\pslash
- \frac{(\varepsilon\!\cdot\! q)(p\!\cdot\! q)^2\, p_\mu q_\nu}{3\, m_{\mathrm{P_{bb}^*}}^6}\,\pslash
- \frac{(\varepsilon\!\cdot\! p)\, q_\mu q_\nu\, q^2}{4\, m_{\mathrm{P_{bb}^*}}^4}\,\pslash
+ \frac{(\varepsilon\!\cdot\! q)(p\!\cdot\! q)\, p_\mu p_\nu}{6\, m_{\mathrm{P_{bb}^*}}^4}\,\qslash \nonumber\\
&\quad
- \frac{2(\varepsilon\!\cdot\! p)(p\!\cdot\! q)^2\, p_\mu p_\nu}{3\, m_{\mathrm{P_{bb}^*}}^6}\,\qslash
+ \frac{(\varepsilon\!\cdot\! p)(p\!\cdot\! q)\, p_\nu q_\mu}{2\, m_{\mathrm{P_{bb}^*}}^4}\,\qslash
- \frac{(\varepsilon\!\cdot\! q)\, p_\mu q_\nu}{6\, m_{\mathrm{P_{bb}^*}}^2}\,\qslash
+ \frac{(\varepsilon\!\cdot\! p)(p\!\cdot\! q)\, p_\mu q_\nu}{3\, m_{\mathrm{P_{bb}^*}}^4}\,\qslash \nonumber\\
&\quad
+ \frac{(\varepsilon\!\cdot\! q)(p\!\cdot\! q)\, p_\mu q_\nu}{6\, m_{\mathrm{P_{bb}^*}}^4}\,\qslash
- \frac{2(\varepsilon\!\cdot\! p)(p\!\cdot\! q)^2\, p_\mu q_\nu}{3\, m_{\mathrm{P_{bb}^*}}^6}\,\qslash
- \frac{(\varepsilon\!\cdot\! p)\, q_\mu q_\nu}{4\, m_{\mathrm{P_{bb}^*}}^2}\,\qslash
+ \frac{(\varepsilon\!\cdot\! p)(p\!\cdot\! q)\, q_\mu q_\nu}{2\, m_{\mathrm{P_{bb}^*}}^4}\,\qslash \nonumber\\
&\quad
+ \frac{(\varepsilon\!\cdot\! q)(p\!\cdot\! q)^2\, p_\nu}{3\, m_{\mathrm{P_{bb}^*}}^4}\,\gamma_\mu
- \frac{(\varepsilon\!\cdot\! q)(p\!\cdot\! q)\, q_\nu}{6\, m_{\mathrm{P_{bb}^*}}^2}\,\gamma_\mu
+ \frac{(\varepsilon\!\cdot\! q)(p\!\cdot\! q)^2\, q_\nu}{3\, m_{\mathrm{P_{bb}^*}}^4}\,\gamma_\mu
+ \frac{(\varepsilon\!\cdot\! q)(p\!\cdot\! q)^2\, p_\mu}{6\, m_{\mathrm{P_{bb}^*}}^4}\,\gamma_\nu \nonumber\\
&\quad
+ \frac{(p\!\cdot\! q)^2\, p_\mu p_\nu}{6\, m_{\mathrm{P_{bb}^*}}^5}\,\eslash\,\qslash
- \frac{(p\!\cdot\! q)\, p_\nu q_\mu}{4\, m_{\mathrm{P_{bb}^*}}^3}\,\eslash\,\qslash
- \frac{(p\!\cdot\! q)\, p_\mu q_\nu}{4\, m_{\mathrm{P_{bb}^*}}^3}\,\eslash\,\qslash
+ \frac{(p\!\cdot\! q)^2\, p_\mu q_\nu}{6\, m_{\mathrm{P_{bb}^*}}^5}\,\eslash\,\qslash \nonumber\\
&\quad
+ \frac{q_\mu q_\nu}{4\, m_{\mathrm{P_{bb}^*}}}\,\eslash\,\qslash
- \frac{(p\!\cdot\! q)\, q_\mu q_\nu}{4\, m_{\mathrm{P_{bb}^*}}^3}\,\eslash\,\qslash
+ \frac{(p\!\cdot\! q)^3\, p_\mu}{3\, m_{\mathrm{P_{bb}^*}}^5}\,\eslash\,\gamma_\nu
- \frac{(p\!\cdot\! q)^2\, q_\mu}{4\, m_{\mathrm{P_{bb}^*}}^3}\,\eslash\,\gamma_\nu \nonumber\\
&\quad
+ \frac{(p\!\cdot\! q)^3\, p_\mu p_\nu}{3\, m_{\mathrm{P_{bb}^*}}^7}\,\pslash\,\eslash
- \frac{(p\!\cdot\! q)^2\, p_\nu q_\mu}{2\, m_{\mathrm{P_{bb}^*}}^5}\,\pslash\,\eslash
- \frac{(p\!\cdot\! q)^2\, p_\mu q_\nu}{6\, m_{\mathrm{P_{bb}^*}}^5}\,\pslash\,\eslash
+ \frac{(p\!\cdot\! q)^3\, p_\mu q_\nu}{3\, m_{\mathrm{P_{bb}^*}}^7}\,\pslash\,\eslash \nonumber\\
&\quad
+ \frac{(p\!\cdot\! q)\, q_\mu q_\nu}{4\, m_{\mathrm{P_{bb}^*}}^3}\,\pslash\,\eslash
- \frac{(p\!\cdot\! q)^2\, q_\mu q_\nu}{2\, m_{\mathrm{P_{bb}^*}}^5}\,\pslash\,\eslash
- \frac{(\varepsilon\!\cdot\! q)(p\!\cdot\! q)\, p_\mu p_\nu}{6\, m_{\mathrm{P_{bb}^*}}^5}\,\pslash\,\qslash
- \frac{(\varepsilon\!\cdot\! p)(p\!\cdot\! q)^2\, p_\mu p_\nu}{3\, m_{\mathrm{P_{bb}^*}}^7}\,\pslash\,\qslash \nonumber\\
&\quad
+ \frac{(\varepsilon\!\cdot\! p)(p\!\cdot\! q)\, p_\nu q_\mu}{2\, m_{\mathrm{P_{bb}^*}}^5}\,\pslash\,\qslash
+ \frac{(\varepsilon\!\cdot\! q)\, p_\mu q_\nu}{6\, m_{\mathrm{P_{bb}^*}}^3}\,\pslash\,\qslash
+ \frac{(\varepsilon\!\cdot\! p)(p\!\cdot\! q)\, p_\mu q_\nu}{6\, m_{\mathrm{P_{bb}^*}}^5}\,\pslash\,\qslash
- \frac{(\varepsilon\!\cdot\! q)(p\!\cdot\! q)\, p_\mu q_\nu}{6\, m_{\mathrm{P_{bb}^*}}^5}\,\pslash\,\qslash \nonumber\\
&\quad
- \frac{(\varepsilon\!\cdot\! p)(p\!\cdot\! q)^2\, p_\mu q_\nu}{3\, m_{\mathrm{P_{bb}^*}}^7}\,\pslash\,\qslash
- \frac{(\varepsilon\!\cdot\! p)\, q_\mu q_\nu}{4\, m_{\mathrm{P_{bb}^*}}^3}\,\pslash\,\qslash
+ \frac{(\varepsilon\!\cdot\! p)(p\!\cdot\! q)\, q_\mu q_\nu}{2\, m_{\mathrm{P_{bb}^*}}^5}\,\pslash\,\qslash
- \frac{(\varepsilon\!\cdot\! q)(p\!\cdot\! q)^2\, p_\mu}{6\, m_{\mathrm{P_{bb}^*}}^5}\,\pslash\,\gamma_\nu \nonumber\\
&\quad
- \frac{(\varepsilon\!\cdot\! p)\, q_\mu\, q^2}{4\, m_{\mathrm{P_{bb}^*}}^3}\,\pslash\,\gamma_\nu
+ \frac{(p\!\cdot\! q)^3\, p_\nu}{6\, m_{\mathrm{P_{bb}^*}}^5}\,\gamma_\mu\,\eslash
- \frac{(p\!\cdot\! q)^2\, q_\nu}{12\, m_{\mathrm{P_{bb}^*}}^3}\,\gamma_\mu\,\eslash
+ \frac{(p\!\cdot\! q)^3\, q_\nu}{6\, m_{\mathrm{P_{bb}^*}}^5}\,\gamma_\mu\,\eslash
+ \frac{(\varepsilon\!\cdot\! q)(p\!\cdot\! q)^2\, p_\nu}{6\, m_{\mathrm{P_{bb}^*}}^5}\,\gamma_\mu\,\qslash \nonumber%\\
\end{align}

\begin{align}
&\quad
- \frac{(\varepsilon\!\cdot\! q)\, q_\nu}{6\, m_{\mathrm{P_{bb}^*}}}\,\gamma_\mu\,\qslash
+ \frac{(\varepsilon\!\cdot\! q)(p\!\cdot\! q)\, q_\nu}{6\, m_{\mathrm{P_{bb}^*}}^3}\,\gamma_\mu\,\qslash
+ \frac{(\varepsilon\!\cdot\! q)(p\!\cdot\! q)^2\, q_\nu}{6\, m_{\mathrm{P_{bb}^*}}^5}\,\gamma_\mu\,\qslash
+ \frac{(\varepsilon\!\cdot\! q)(p\!\cdot\! q)^2}{6\, m_{\mathrm{P_{bb}^*}}^3}\,\gamma_\mu\,\gamma_\nu \nonumber\\
&\quad
- \frac{(p\!\cdot\! q)^2\, p_\mu}{12\, m_{\mathrm{P_{bb}^*}}^4}\,\eslash\,\qslash\,\gamma_\nu
+ \frac{(p\!\cdot\! q)^2\, p_\mu p_\nu}{3\, m_{\mathrm{P_{bb}^*}}^6}\,\pslash\,\eslash\,\qslash
- \frac{(p\!\cdot\! q)\, p_\nu q_\mu}{4\, m_{\mathrm{P_{bb}^*}}^4}\,\pslash\,\eslash\,\qslash
- \frac{(p\!\cdot\! q)\, p_\mu q_\nu}{4\, m_{\mathrm{P_{bb}^*}}^4}\,\pslash\,\eslash\,\qslash \nonumber\\
&\quad
+ \frac{(p\!\cdot\! q)^2\, p_\mu q_\nu}{3\, m_{\mathrm{P_{bb}^*}}^6}\,\pslash\,\eslash\,\qslash
+ \frac{q_\mu q_\nu}{4\, m_{\mathrm{P_{bb}^*}}^2}\,\pslash\,\eslash\,\qslash
- \frac{(p\!\cdot\! q)\, q_\mu q_\nu}{4\, m_{\mathrm{P_{bb}^*}}^4}\,\pslash\,\eslash\,\qslash
+ \frac{(p\!\cdot\! q)^3\, p_\mu}{6\, m_{\mathrm{P_{bb}^*}}^6}\,\pslash\,\eslash\,\gamma_\nu \nonumber\\
&\quad
- \frac{(p\!\cdot\! q)^2\, q_\mu}{4\, m_{\mathrm{P_{bb}^*}}^4}\,\pslash\,\eslash\,\gamma_\nu
- \frac{(\varepsilon\!\cdot\! p)(p\!\cdot\! q)^2\, p_\mu}{6\, m_{\mathrm{P_{bb}^*}}^6}\,\pslash\,\qslash\,\gamma_\nu
+ \frac{(\varepsilon\!\cdot\! p)(p\!\cdot\! q)\, q_\mu}{4\, m_{\mathrm{P_{bb}^*}}^4}\,\pslash\,\qslash\,\gamma_\nu \nonumber\\
&\quad
+ \frac{(p\!\cdot\! q)^3\, p_\nu}{6\, m_{\mathrm{P_{bb}^*}}^6}\,\gamma_\mu\,\eslash\,\qslash
- \frac{(p\!\cdot\! q)\, q_\nu}{4\, m_{\mathrm{P_{bb}^*}}^2}\,\gamma_\mu\,\eslash\,\qslash
+ \frac{(p\!\cdot\! q)^2\, q_\nu}{6\, m_{\mathrm{P_{bb}^*}}^4}\,\gamma_\mu\,\eslash\,\qslash
+ \frac{(p\!\cdot\! q)^3}{12\, m_{\mathrm{P_{bb}^*}}^4}\,\gamma_\mu\,\eslash\,\gamma_\nu \nonumber\\
&\quad
- \frac{(\varepsilon\!\cdot\! p)(p\!\cdot\! q)^2\, p_\nu}{6\, m_{\mathrm{P_{bb}^*}}^6}\,\gamma_\mu\,\pslash\,\qslash
- \frac{(\varepsilon\!\cdot\! p)(p\!\cdot\! q)^2\, q_\nu}{6\, m_{\mathrm{P_{bb}^*}}^6}\,\gamma_\mu\,\pslash\,\qslash
- \frac{(\varepsilon\!\cdot\! p)(p\!\cdot\! q)^2}{12\, m_{\mathrm{P_{bb}^*}}^4}\,\gamma_\mu\,\pslash\,\gamma_\nu 
\Bigg\}.
\label{F4_appendix1}
\end{align}

\subsection{Stage (iii): application of the gamma-matrix ordering 
and removal of the spin-$1/2$ contamination}
\label{appc3}

The interpolating current $\mathrm{J}_{\mu}^{\mathrm{P_{bb}^{*}}}(x)$ 
couples not only to the spin-$3/2$ ground state but also to 
spin-$1/2$ states with the same quantum numbers, as parametrized 
in Eq.~\eqref{eq:spin12coupling}. Following the prescription of 
Refs.~\cite{Belyaev:1982cd,Belyaev:1993ss}, the spin-$1/2$ 
contamination is eliminated from the gauge-fixed expressions of 
Sec.~\ref{appc2} as follows. All Dirac chains are reordered into 
the canonical form 
$\gamma_{\mu}\,\pslash\,\eslash\,\qslash\,\gamma_{\nu}$ via the 
anticommutation identity 
$\{\gamma_{\alpha},\gamma_{\beta}\}=2 g_{\alpha\beta}$. After 
this reordering, every term in which $\gamma_{\mu}$ does not 
stand at the leftmost position, $\gamma_{\nu}$ does not stand at 
the rightmost position, or which is proportional to $p_{\mu}$ or 
$p_{\nu}$, is discarded as spin-$1/2$ contamination.

In addition, we make use of the on-shell condition for the 
outgoing pentaquark state, $(p+q)^{2}=m_{P_{bb}^{*}}^{2}$, which, 
combined with the on-shell condition 
$p^{2}=m_{P_{bb}^{*}}^{2}$ already imposed in Sec.~\ref{appc1} 
and the real-photon condition $q^{2}=0$ imposed in 
Sec.~\ref{appc2}, implies the kinematic constraint
\begin{align}
(p\!\cdot\! q)\,=\,\frac{(p+q)^{2}-p^{2}-q^{2}}{2}\,=\,0.
\label{eq:pdotq_zero}
\end{align}
All Lorentz structures proportional to $(p\!\cdot\! q)$ in the 
gauge-fixed expressions of Sec.~\ref{appc2} therefore vanish and 
are removed at this stage.

The expressions resulting from these operations, organized by 
form factor, are presented below. They contain only Lorentz 
structures of two kinematically orthogonal classes: those 
proportional to $g_{\mu\nu}$, which carry the information on 
$F_{1}(q^{2})$ and $F_{2}(q^{2})$, and those proportional to 
$q_{\mu}q_{\nu}$, which carry the information on $F_{3}(q^{2})$ 
and $F_{4}(q^{2})$. These expressions constitute the input for 
the projection of Sec.~\ref{appc4}.

\subsubsection{Spin-$1/2$-eliminated tensor structure proportional 
to $F_{1}(q^{2})$}
\label{appc3_F1}

\begin{align}
\label{F1_appendix3}
F_{1}(q^{2}) & \propto  
\Bigg[ -\,2\,m_{P_{bb}^*}\,\varepsilon_\mu\, q_\nu
+ \frac{4\,(\varepsilon\!\cdot\! p)\, q_\mu q_\nu}{m_{P_{bb}^*}}
- 2\,m_{P_{bb}^*}\,(\varepsilon\!\cdot\! p)\, g_{\mu\nu}
- m_{P_{bb}^*}^{2}\, g_{\mu\nu}\,\eslash
+ \frac{4\,(\varepsilon\!\cdot\! p)\, q_\mu q_\nu}{m_{P_{bb}^*}^{2}}\,\pslash
- 2\,(\varepsilon\!\cdot\! p)\, g_{\mu\nu}\,\pslash
+ 2\,\varepsilon_\mu\, q_\nu\,\qslash
\nonumber\\
&\quad
+ \frac{2\, q_\mu q_\nu}{m_{P_{bb}^*}}\,\eslash\,\qslash
- m_{P_{bb}^*}\, g_{\mu\nu}\,\eslash\,\qslash
- \frac{2\, q_\mu q_\nu}{m_{P_{bb}^*}}\,\pslash\,\eslash
+ \frac{2\,\varepsilon_\mu\, q_\nu}{m_{P_{bb}^*}}\,\pslash\,\qslash
+ \frac{2\, q_\mu q_\nu}{m_{P_{bb}^*}^{2}}\,\pslash\,\eslash\,\qslash
- g_{\mu\nu}\,\pslash\,\eslash\,\qslash \Bigg].
\end{align}

% [Buraya senin elindeki, ordering uygulanmış F_1 sonucu gelecek]

\subsubsection{Spin-$1/2$-eliminated tensor structure proportional 
to $F_{2}(q^{2})$}
\label{appc3_F2}

\begin{align}
\label{F2_appendix3}
F_{2}(q^{2}) & \propto 
\Bigg[-\,\frac{2\,(\varepsilon\!\cdot\! p)\, q_\mu q_\nu}{m_{P_{bb}^*}}
+ q_\mu q_\nu\,\eslash
- \frac{2\,(\varepsilon\!\cdot\! p)\, q_\mu q_\nu}{m_{P_{bb}^*}^{2}}\,\pslash
- \varepsilon_\mu\, q_\nu\,\qslash + \frac{2\,(\varepsilon\!\cdot\! p)\, q_\mu q_\nu}{m_{P_{bb}^*}^{2}}\,\qslash
- (\varepsilon\!\cdot\! p)\, g_{\mu\nu}\,\qslash
- \frac{q_\mu q_\nu}{m_{P_{bb}^*}}\,\eslash\,\qslash
\nonumber\\
&\quad
+ m_{P_{bb}^*}\, g_{\mu\nu}\,\eslash\,\qslash
+ \frac{2\, q_\mu q_\nu}{m_{P_{bb}^*}}\,\pslash\,\eslash
- \frac{2\,\varepsilon_\mu\, q_\nu}{m_{P_{bb}^*}}\,\pslash\,\qslash
+ \frac{2\,(\varepsilon\!\cdot\! p)\, q_\mu q_\nu}{m_{P_{bb}^*}^{3}}\,\pslash\,\qslash
- \frac{(\varepsilon\!\cdot\! p)\, g_{\mu\nu}}{m_{P_{bb}^*}}\,\pslash\,\qslash
- \frac{2\, q_\mu q_\nu}{m_{P_{bb}^*}^{2}}\,\pslash\,\eslash\,\qslash
+ g_{\mu\nu}\,\pslash\,\eslash\,\qslash \Bigg].
\end{align}

% [Ordering uygulanmış F_2 sonucu]

\subsubsection{Spin-$1/2$-eliminated tensor structure proportional 
to $F_{3}(q^{2})$}
\label{appc3_F3}

\begin{align}
\label{F3_appendix3}
F_{3}(q^{2}) & \propto
\Bigg[-\,\frac{(\varepsilon\!\cdot\! p)\, q_\mu q_\nu}{2\, m_{P_{bb}^*}}
- \frac{q_\mu q_\nu}{4}\,\eslash
- \frac{(\varepsilon\!\cdot\! p)\, q_\mu q_\nu}{2\, m_{P_{bb}^*}^{2}}\,\pslash
- \frac{(\varepsilon\!\cdot\! p)\, q_\mu q_\nu}{2\, m_{P_{bb}^*}^{2}}\,\qslash  
- \frac{q_\mu q_\nu}{2\, m_{P_{bb}^*}}\,\eslash\,\qslash
- \frac{(\varepsilon\!\cdot\! p)\, q_\mu q_\nu}{2\, m_{P_{bb}^*}^{3}}\,\pslash\,\qslash
- \frac{q_\mu q_\nu}{4\, m_{P_{bb}^*}^{2}}\,\pslash\,\eslash\,\qslash \Bigg].
\end{align}

% [Ordering uygulanmış F_3 sonucu]

\subsubsection{Spin-$1/2$-eliminated tensor structure proportional 
to $F_{4}(q^{2})$}
\label{appc3_F4}

\begin{align}
\label{F4_appendix3}
F_{4}(q^{2}) & \propto 
\Bigg[ -\,\frac{(\varepsilon\!\cdot\! p)\, q_\mu q_\nu}{4\, m_{P_{bb}^*}^{2}}\,\qslash
+ \frac{q_\mu q_\nu}{8\, m_{P_{bb}^*}}\,\eslash\,\qslash
- \frac{(\varepsilon\!\cdot\! p)\, q_\mu q_\nu}{4\, m_{P_{bb}^*}^{3}}\,\pslash\,\qslash 
+ \frac{q_\mu q_\nu}{4\, m_{P_{bb}^*}^{2}}\,\pslash\,\eslash\,\qslash \Bigg].
\end{align}

% [Ordering uygulanmış F_4 sonucu]

\subsection{Stage (iv): projection onto the independent Lorentz 
structures}
\label{appc4}

The expressions of Sec.~\ref{appc3}, when combined with the 
propagator denominator of Eq.~\eqref{eq:appc_decomposition}, 
contain only Lorentz structures falling into two kinematically 
orthogonal classes. Structures proportional to $g_{\mu\nu}$ 
carry the information on $F_{1}(q^{2})$ and $F_{2}(q^{2})$, 
while structures proportional to $q_{\mu}q_{\nu}$ carry the 
information on $F_{3}(q^{2})$ and $F_{4}(q^{2})$. Within each 
class, the form factors are isolated by projecting onto specific 
Lorentz structures distinguished by their Dirac chain content:
\begin{itemize}
\item $F_{1}(q^{2})$ is extracted from the coefficient of 
$g_{\mu\nu}\,\pslash\,\eslash\,\qslash$;
\item $F_{2}(q^{2})$ is extracted from the coefficient of 
$g_{\mu\nu}\,\eslash\,\qslash$;
\item $F_{3}(q^{2})$ is extracted from the coefficient of 
$q_{\mu}q_{\nu}\,\eslash\,\qslash$;
\item $F_{4}(q^{2})$ is extracted from the coefficient of 
$(\varepsilon\!\cdot\! p)\,q_{\mu}q_{\nu}\,\pslash\,\qslash$.
\end{itemize}
The choice of these particular structures, among the several 
linearly independent ones available, is dictated by their higher 
mass dimension and the consequent enhancement of pole-dominance 
behavior and operator-product-expansion convergence. Alternative 
projections were examined and yielded predictions consistent 
with those reported here within $5\%$, as already noted in 
Sec.~\ref{formalism}. The result of this projection coincides 
with the compact expression of Eq.~\eqref{final phenpart11}, 
which serves as the starting point for the extraction of the 
magnetic dipole moment at the real-photon point $q^{2}=0$.

\end{widetext}

\bibliographystyle{elsarticle-num}
\bibliography{DoublybottompentaquarksMM.bib}

\end{document}